\documentclass[12pt]{report}

\usepackage{eqsection,latexsym,epsf}
\usepackage{amsfonts}
\usepackage{amssymb}
\usepackage{epsfig}

\begin{document}

\bibliographystyle{JHEP}

\def\grz{\overline{g}_{\zeta}(r)}
\def\gr2z{\overline{g}_{\zeta}(r/2)}
\def\gxz{\overline{g}_{\zeta}(x)}
\def\sg{\mbox{\boldmath $\sigma$}}
\def\og{\mbox{\boldmath $\omega$}}
\def\pg{\mbox{\boldmath $\pi$}}
\def\tg{\mbox{\boldmath $\tau$}}
\def\jg{\mbox{\boldmath $j$}}
\def\Jg{\mbox{\boldmath $J$}}
\def\phg{\mbox{\boldmath $\phi$}}
\def\psg{\mbox{\boldmath $\psi$}}
\def\dip{\frac{d^2 p}{(2\pi)^2}}
\def\diq{\frac{d^2 q}{(2\pi)^2}}
\def\dirac{(2\pi)^2\delta^{(2)}}
\def\i{\int_{BZ}}
\def\f{\hat{f}}
\def\opl{\left[}
\def\opr{ \right] _{\overline{MS}}}
\def\mbar{\overline\mu }
\def\MMS{{\overline{\rm MS}}}
\def\MS{$\overline{\rm MS}$ }
\newcommand{\reff}[1]{(\ref{#1})}
\newcommand{\beq}{\begin{eqnarray}}
\newcommand{\eeq}{\end{eqnarray}}

% simboli standard 

\newcommand{\be}{\begin{equation}}
\newcommand{\ee}{\end{equation}}
\newcommand{\<}{\langle}
\renewcommand{\>}{\rangle}

%\ltapprox and \gtapprox produce > and < signs with twiddle underneath
\def\spose#1{\hbox to 0pt{#1\hss}}
\def\ltapprox{\mathrel{\spose{\lower 3pt\hbox{$\mathchar"218$}}
 \raise 2.0pt\hbox{$\mathchar"13C$}}}
\def\gtapprox{\mathrel{\spose{\lower 3pt\hbox{$\mathchar"218$}}
 \raise 2.0pt\hbox{$\mathchar"13E$}}}

\def\pb{\overline{p}}
\def\qb{\overline{q}}
\def\Real{{\rm Re}}
\def\Im{{\rm Im}}

\begin{picture}(250,300)
\put(-15,-340){\epsfig{figure=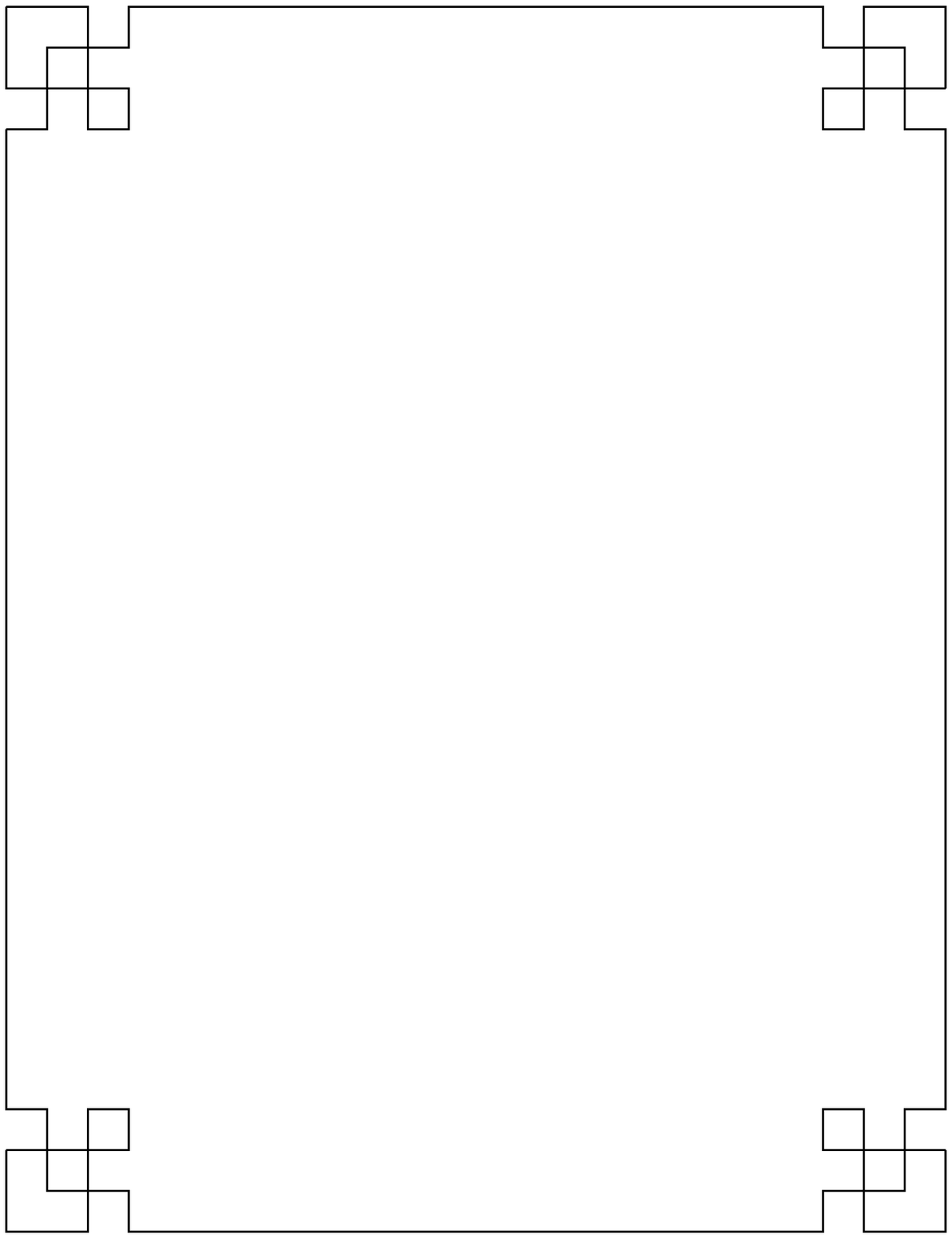,width=16cm,height=22.5cm}}
\end{picture}

\vspace{-9cm}

\begin{center}
\vskip 1.5truecm

{\Large  Andrea Montanari} \\[0.5cm]

\end{center}

\vskip 0.5truecm

\begin{center} 
\LARGE 
Non-Perturbative Renormalization 

in Lattice Field Theory

\end{center}

\vskip 1truecm
\begin{center}
{\large \rm  TESI DI PERFEZIONAMENTO } \\[3cm]
\end{center}

\vspace{6cm}

\begin{center}
{\large SCUOLA NORMALE SUPERIORE }

{\large    Pisa } \\[0.5cm]

\end{center}

\thispagestyle{empty}

\clearpage

\vskip 5cm
.
\thispagestyle{empty}
\clearpage

\abstract{Non-perturbative renormalization of lattice composite
operators plays a crucial role in many applications of lattice
field theory. We sketch the general problems involved in this task
and the methods which are currently used to cope with them.
We present a detailed investigation of a new approach based 
on the operator product expansion.
We test the new method on the two-dimensional $O(3)$ $\sigma$-model
and discuss its advantages and limitations.

Ph.D. thesis at Scuola Normale Superiore, Pisa, Italy.}

\tableofcontents

\chapter{A Short Review on Non-Perturbative Renormalization}
\label{Introduction}
Why do we need renormalization in lattice field theory?
We can broadly distinguish two types of needs: a ``fundamental'' 
one, and a ``phenomenological'' one \cite{Testa:1997ne}.

By ``fundamental'' we mean the tuning of bare lattice parameters  which is 
necessary for recovering the  correct continuum theory. In the case of 
lattice QCD with Wilson fermions a ``minimal'' set of parameters to be tuned 
is given by the gauge coupling and the quark masses. Since the theory is 
asymptotically free, the bare gauge coupling must be sent to zero. The bare 
masses can be obtained by fixing the masses of an appropriate number of
mesons in units of some reference scale, e.g., the string tension.
This type of renormalization is necessary no matter which model we are 
studying. Generally speaking, physical predictions can be 
extracted from the lattice regularized theory without further 
renormalization.
In fact physical quantities are given as matrix elements of the $S$
matrix. Such matrix elements do not depend upon the normalization of 
the interpolating fields.

This ``first principles'' point of view must be modified in many cases
of phenomenological interest. An important example is the study
of QCD corrections to weak interactions. 
Weak interactions cannot be straightforwardly discretized and
simulated on the lattice. 
This happens for two types of reasons:
\begin{itemize}
\item Practical ones: the masses of the weak bosons are much larger than
the currently achievable lattice cutoffs.
\item Theoretical ones: preserving the chiral properties of fermions on the 
lattice is a difficult (and intensively studied) problem.
\end{itemize}
A widespread solution to the above problems consists in adopting the 
{\it effective hamiltonian} approach. Heavy degrees of freedom are
integrated over treating weak interactions in perturbation theory.
Non-perturbative (low energy) QCD contributions are encoded
in the hadronic matrix elements of some basis of composite operators.
Such operators must be properly renormalized. This is what we referred to
as the ``phenomenological'' need for renormalization.

The renormalization of composite operators is a complex and important
problem. 
In this Chapter we shall review the renormalization techniques which 
are used currently, having in mind the type of applications sketched above.

The exposition is organized as follows. In Sec. \ref{GeneralSection}
we add some feature concerning the use of renormalized matrix elements
in phenomenological applications.
In Sec. \ref{SchemesSection} we outline the two available approaches
to non-perturbative renormalization, broadly distinguishing between infinite 
and finite volume schemes. In Sec. \ref{RunningSection} we focus on a property
of renormalized operator (the dependence upon the renormalization scale)
which plays a special role in finite volume schemes. In 
Sec. \ref{RenormalizationViaOPE} we describe the new approach which will be 
tentatively investigated in this Thesis. 
%
%***************************************************************************
%
\section{The General Setting}
\label{GeneralSection}
The general context which we have in mind can be described schematically 
as follows:
\begin{eqnarray}
\mbox{``interesting quantity''}(M)\sim \sum_{\cal O}C_{\cal O}(\mu,M)\,
M^{-\kappa({\cal O})}\,\<h_1|{\cal O}(\mu)|h_2\>\, .
\label{GeneralSetting}
\end{eqnarray}
The ``interesting quantity'' on the left hand side of 
Eq. (\ref{GeneralSetting}) depends upon the application we are considering.
If we are studying weak interactions physics 
\cite{Buchalla:1996vs,Buras:1998ra}, it can be the amplitude for 
a non-leptonic decay as well as a meson mixing amplitude, etc.   
For deep inelastic scattering applications \cite{Petronzio:1999},
it can be the moment of a structure function.
In both cases the aim is to include QCD effects in electroweak
processes.
Widely separated energy scales play a role in such processes.
Among them a ``large energy'' scale $M$ can be usually identified.
We keep track of the $M$ dependence in Eq. (\ref{GeneralSetting}).
In weak interactions physics this energy scale is, typically, the mass of the 
weak vector bosons. In deep inelastic scattering the relevant scale is 
determined by the exchanged four-momentum.
The ${\cal O}$ on the right-hand side are local, gauge-invariant
(under the colour $SU(3)$ gauge group) 
operators\footnote{In general two other classes of operators may
appear \cite{Joglekar:1976nu,Joglekar:1977eb,Joglekar:1977pe}: 
{\it (a)} operators which are BRS-variations;
{\it (b)} operators which vanish by the equation of motions.
However, as long as $|h_1\>$ and $|h_2\>$ are on-shell, physical 
states, the matrix elements of these two classes of operators vanish.}, 
which are renormalized at the scale $\mu$. They are 
evaluated between the  hadronic states $|h_1\>$ and $|h_2\>$.
The series on the right-hand side is asymptotic in the parameter $M^{-1}$.
The exponent $\kappa({\cal O})$ is fixed by naive power counting.
Usually the scale $M$ is much larger than the hadronic scales
involved in the matrix elements on the right-hand side of 
Eq. (\ref{GeneralSetting}).
This allows to neglect all the terms of the series but a few
ones.

The technical tool for obtaining the expansion given
in Eq. (\ref{GeneralSetting}) is the Operator Product Expansion (OPE). 
The OPE has been postulated for the first time by Wilson \cite{Wilson:1969zs}
thirty years ago, has been later proved in perturbation theory by Zimmermann 
\cite{Zimmerman:70}, and is widely thought to hold beyond perturbation theory.
If the theory is asymptotically free the Wilson coefficients 
$C_{\cal O}(\mu, M)$ can be computed in renormalization-group (RG) improved 
perturbation theory.  The result will be reliable as long as $\mu$ is in the
perturbative regime. 

The general idea behind Eq. (\ref{GeneralSetting}) is to divide the energy
scales involved in the process in two regimes. The perturbative regime
from $\mu$ to $M$ is well described by RG improved
perturbation theory. This contribution is kept into account by
the Wilson coefficients $C_{\cal O}(\mu,M)$.
The non-perturbative regime from $0$ to $\mu$
has to be treated with some other method. This contribution is cast into
the matrix elements $\<h_1|{\cal O}(\mu)|h_2\>$, whose computation
is, in many cases, a still unsolved problem. 

For this approach to work the factorization scale $\mu$ must be 
chosen large enough (i.e. in the perturbative regime).
Can we give a quantitative indication for the onset of the
perturbative regime? 
The answer to this question depends upon which observable is
being studied and which truncation in perturbation theory is used. 
As an example, let us consider the running coupling $g^2(\mu)$ and the
running quark masses $m(\mu)$ in quenched QCD. These observables
are very interesting since their running has been computed
non-perturbatively in \cite{Luscher:1994gh,Capitani:1998mq}. 
The perturbative expansions of the beta function and of the 
mass anomalous dimensions are known up to four-loop order. 
We can then compare the non-perturbative and the perturbative running
of these quantities.
Above $1\, GeV$, two-loop perturbative expansions describe the 
non-perturbative results with a systematic error of a few percent. 
We must be very careful here because the systematic error 
is  not well defined unless we know the exact result. 
However, the large energy scale
$M$ is, in all practical applications, far in the perturbative regime.
We can consider it to be an infinite energy, for our purposes.
%
%**************************************************************************
%
\section{Different Renormalization Schemes}
\label{SchemesSection}
\label{DifferentSchemes}
In this section we shall outline two general approaches to the construction 
of renormalized operators. We shall refer to the simple case of
purely multiplicative renormalization (without mixing):
\begin{eqnarray}
{\cal O}_R(\mu) = Z_{\cal O} (\Lambda/\mu,\Lambda a){\cal O}_{LAT}\, ,
\label{MultiplicativeRenormalization}
\end{eqnarray}
where ${\cal O}_{LAT}$ is a bare lattice operator, and ${\cal O}_R(\mu)$
is its renormalized counterpart.
In Eq. (\ref{MultiplicativeRenormalization}) we have emphasized the 
dependence of the renormalization constant upon the various scales of the 
problem: the lattice spacing $a$, the renormalization scale $\mu$ and
the physical scale $\Lambda$ (the so called ``lambda parameter'') 
which breaks the scale invariance of the continuum theory. 
The renormalized operator ${\cal O}_R(\mu)$ is required to
have a finite continuum limit ($\Lambda a\to 0$ at $\Lambda/\mu$ fixed).

In Eq. (\ref{MultiplicativeRenormalization}) we implicitly assumed 
an important simplifying feature: the renormalization 
scheme is mass independent. Such a scheme can always be defined
for QCD.
This is done by computing the renormalization constants at zero quark masses 
\cite{Weinberg:1973ss}. Nevertheless it is often difficult to implement such a
scheme in numerical simulations. 
Simulating QCD with very light quarks implies several complications:
finite volume effects, exceptional configurations, slowing down
of the computation of the quark propagator, etc.
The way this problem is dealt with depends upon the adopted 
renormalization method\footnote{This discussion must be modified
if we want to construct $O(a)$ improved operators
\cite{Luscher:1996sc}.
In this case the $Z_{\cal O}$ includes terms which are linear in
$m_q a$, $m_q$ being the bare quark mass.}.

Before discussing the various renormalization schemes, let us recall 
that, if the operator ${\cal O}_R$  in Eq. 
(\ref{MultiplicativeRenormalization}) is a Noether current,  
the corresponding renormalization constant $Z_{\cal O}$ cannot depend
upon $\mu$.
The  reason is that the values
taken by the  related conserved charges are fixed algebraically.
Examples of such operators are the vector and axial flavour 
currents in QCD.
There exists a well studied method for renormalizing this type of operators 
using the Ward identities. We shall not give more details on this type of 
operators and focus instead on the ``difficult'' case of scale-dependent
operators.
%
%***************************************************************************
%
\subsection{Infinite-Volume Schemes}
The use of infinite-volume non-perturbative renormalization schemes has been 
suggested for the first time in Ref. \cite{Martinelli:1995ty} and 
intensively studied since then\footnote{See, for instance, 
Refs. \cite{Vladikas:1995qt,Rossi:1997yu,Testa:1997ne,Martinelli:1998hz}.}

The proposed procedure mimics what is often done in perturbation theory.
One determines the renormalization constants by requiring that some
well-chosen vertex function takes its tree-level
value when the scale of external momenta is equal to the renormalization scale.
As an example, let us consider the non-singlet pseudo-scalar density:
\begin{eqnarray}
{\cal P}^a_x\equiv \overline{\psi}_x\gamma_5 T^a\psi_x\, ,
\end{eqnarray}
where $T^a$ is a generator of the flavour group $SU(N_f)$.
This is a case of great physical relevance, since it gives
access, through the PCAC relation, to the quark mass renormalization
\cite{Gimenez:1998ue,Becirevic:1998yg,Gockeler:1998ye}.
We define the following Green function with one ${\cal P}^a_x$ insertion at 
zero momentum:
\begin{eqnarray}
G^{\cal P}_{\alpha\beta}(p)\equiv \sum_{x,y;a} e^{ip(x-y)}
\<\overline{\psi}_{\alpha,x}T^a\psi_{\beta,y}P^a_z\>\, .
\label{OperatorInsertion}
\end{eqnarray}
The indices $\alpha$ and $\beta$ in the preceding expression are Dirac 
indices and must not be summed over. Let us denote as $\Gamma^{\cal P}(p)$ 
the corresponding vertex function. This is obtained from
Eq. (\ref{OperatorInsertion}) by amputating the external 
quark legs:
\begin{eqnarray}
G^{\cal P}(p) \equiv S(p)\Gamma^{\cal P}(p) S(p)\, .
\end{eqnarray}
The renormalization condition reads:
\begin{eqnarray}
\frac{1}{4}{\rm Tr}_{\rm Dirac}
\left[\Gamma^{\cal P}(p)\Gamma_0^{\cal P}(p)^{-1}\right]
_{p^2 = \mu^2}= Z_{\psi}(\mu)Z^{-1}_{\cal P}(\mu)\, , 
\label{OffShellCondition}
\end{eqnarray}
where the trace $\rm{Tr}_{\rm{Dirac}}$ has to be taken with respect to the 
Dirac indices and $\Gamma_0(p)$ is the tree-level value for the vertex 
function.   
The constant $Z_{\psi}$ on the right-hand side of Eq. (\ref{OffShellCondition})
is needed in order to renormalize the external quark legs. It can be computed 
by imposing a condition analogous to Eq. (\ref{OffShellCondition}) on the 
two-point quark function.

Notice that the correlation function (\ref{OperatorInsertion}) is not
gauge invariant. In order to avoid a trivial outcome of the above 
computation, a gauge must be fixed. Usually the
Landau gauge is chosen. Moreover,
the condition (\ref{OffShellCondition}) is intended to be imposed at zero 
quark masses. In practice this is done by extrapolating to the chiral limit the
numerical result for $\Gamma^{\cal P}(p;m_q)$, obtained for a non-zero
value $m_q$ of the quark masses. Apart from this extrapolation, 
Eq. (\ref{OffShellCondition}) yields the \underline{exact} renormalization 
constant at the scale $\mu$ up to lattice artifacts which are of order 
$O(\mu a,\Lambda a)$ (or $O(\mu^2a^2,\Lambda^2a^2)$ 
if the theory is improved non-perturbatively).

In order to keep lattice artifacts and finite-size effect under control,
we must consider energy scales in the range:
\begin{eqnarray}
L\gg \frac{1}{\mu a}\gg 1\, ,
\label{ScalingWindow1}
\end{eqnarray}
and bare couplings such that 
\begin{eqnarray}
L\gg \frac{1}{\Lambda a}\gg 1\, ,
\label{ScalingWindow2}
\end{eqnarray}
where $L$ is the linear lattice size in lattice units.
Equations (\ref{ScalingWindow1}) and (\ref{ScalingWindow2}) are  the crucial 
limitations of infinite-volume schemes. 
We shall reconsider them in the next Sections.

Equation (\ref{OffShellCondition}), together with analogous conditions 
for other composite operators and for the quark field, defines 
a particular renormalization scheme. This is often referred to
as the Regularization Independent (RI) scheme.

Once the hadronic matrix elements have been computed 
and renormalized in the RI scheme,
 we would like to use them in Eq. (\ref{GeneralSetting}).
Therefore, we must ``translate'' them in the
same scheme, usually minimal subtraction (MS), used for the 
Wilson coefficients $C_{\cal O}(\mu,M)$. 
This passage can be accomplished by computing a finite
renormalization constant. In our example:
\begin{eqnarray}
\left[{\cal P}^a\right]_{MS}(\mu) = Z^{\cal P}_{RI,MS}(\mu)
\left[{\cal P}^a\right]_{RI}(\mu)\, .
\label{FiniteRenormalization}
\end{eqnarray}
The constant $Z^{\cal P}_{RI,MS}(\mu)$ can be reliably computed in 
perturbation theory as long as 
\begin{eqnarray}
\mu\gg \Lambda\, . 
\label{PerturbativeCondition}
\end{eqnarray}
Notice that this condition follows from the simple 
fact that we compute {\it perturbatively}\footnote{See Refs. 
\cite{Capitani:1998fe,Capitani:1999fm} for 
an alternative proposal.} the Wilson coefficients
$C_{\cal O}(\mu,M)$ in Eq. (\ref{GeneralSetting}).

As discussed in the previous Section,
Eq. (\ref{PerturbativeCondition}) assures that 
the separation between low-energy and high-energy contributions 
in Eq. (\ref{GeneralSetting}) is sensible, and that the result is $\mu$
independent.
The compatibility of the condition in Eq. (\ref{PerturbativeCondition}) 
with the previous ones given by Eq. 
(\ref{ScalingWindow1})-(\ref{ScalingWindow2})
is a serious (and still not completely solved) problem.
%
%**************************************************************************
%
\subsection{Finite-Volume Schemes}
Finite-volume schemes are the practical application of an important 
observation due to Symanzik \cite{Symanzik:1981wd}: renormalizability is not 
spoiled when a field theory is put on a finite space-time manifold.
We can define a finite volume renormalization scheme by using the linear 
size $aL$ of this space-time manifold as the renormalization scale.
Non-perturbative computations are made possible by discretizing
this space time on a lattice of spacing $a$.
The first advantage of this approach, with respect to the 
infinite-volume one, is that the two limitations 
in Eqs. (\ref{ScalingWindow1})-(\ref{ScalingWindow2}) 
reduce to the much weaker:
\begin{eqnarray}
L\gg 1\, ,\quad
\Lambda a \ll 1\, .
\label{FiniteVolumeConditions}
\end{eqnarray}
These constraints correspond to working in the finite-size scaling (FSS) 
regime.
The second advantage is that a proper choice of the boundary conditions 
produces a gap of order $1/L$ in the spectrum of the Dirac-Wilson 
operator $(D+m)$. This gap survives in the chiral limit $m\to m_c$.
One can safely work at $m = m_c(g_0)$ avoiding chiral extrapolations.
Finally, a careful examination shows that nontrivial results
can be obtained without fixing a particular gauge.

How are finite-volume schemes implemented in practice?
A popular geometry is a $L^3\times T$ lattice with twisted boundary 
conditions in the space direction and Dirichelet boundary conditions in 
the time direction. In order to have a unique scale in the problem 
the size of the lattice in the time direction is proportional to the size
in the space direction: to be definite let us say $T=2L$. 
It is convenient to think of the boundary conditions in the time direction
as of ``boundary fields''
(let us denote them $\phi_{\rm bf}(0)$ or $\phi_{\rm bf}(T)$
depending upon which of the two boundaries we are considering) 
which must themselves be properly renormalized.
To be definite let us consider again the example of the 
pseudoscalar density, already treated in the previous Section 
\cite{Capitani:1998mq}.
The renormalization condition is
\begin{eqnarray}
\left.\frac{\<\phi_{\rm bf}(0)\,{\cal P}^a_{z}\>}
{\<\<\phi_{\rm bf}(0)\,\<\phi_{\rm bf}(T)\>^{1/2}}\right|_{z_0 = T/2} 
= Z_{\cal P}^{-1}(\mu a=L^{-1})
\left. \frac{\<\phi_{\rm bf}(0)\,{\cal P}^a_{z}\>}
{\<\<\phi_{\rm bf}(0)\,\<\phi_{\rm bf}(T)\>^{1/2}}
\right|_{z_0 = T/2,{\rm TREE}}\, .
\label{FiniteSizeCondition}
\end{eqnarray}
The ratio on the left-hand side of 
Eq. (\ref{FiniteSizeCondition}) is constructed so that the 
uninteresting boundary-field renormalizations cancel. 
We shall not describe the precise
boundary conditions which are usually adopted, although their careful
choice is quite a relevant point.

The setting outlined above has been dubbed the ``Schr\"odinger functional''
and has been much studied in the last years\footnote{
An incomplete list of references is 
\cite{Luscher:1992an,Luscher:1993zx,Luscher:1994gh,Jansen:1996ck,
Capitani:1998mq,Luscher:1998pe}. See also 
\cite{Rossi:1980jf,Rossi:1980pg,Rossi:1984ag} for an earlier
proposal.}.  It allows to compute non-perturbatively 
the renormalization constants $Z(\Lambda/\mu,\Lambda a)$ in the interesting 
regime $\Lambda/\mu\ll 1$ and $\Lambda a \ll 1$ with the minimum 
of computational work. 

The above handwaving description hides an important 
difficulty, and leaves out the crucial step which is required for solving it.
Let us in fact consider the following trivial identity:
\begin{eqnarray}
\Lambda a = \frac{1}{L}\frac{\Lambda}{\mu}\, ,
\label{DifferentScales}
\end{eqnarray}
and recall that, in order to avoid lattice artifacts, we required 
$L\gg 1$. In practical cases $L = 5\div 12$. 
At the end we would like to use our renormalization constant
$Z_{\cal O}(\Lambda/\mu,\Lambda a)$ for computing physical quantities
from lattice simulations.
As stressed in the previous Sections renormalization scale 
$\mu$ must then be safely in the perturbative regime:
let us say $\mu\gtrsim 10\Lambda$. 
For obtaining physical results we must compute bare  hadronic matrix 
elements near the infinite-volume
limit. This forces us to consider not too fine lattices:
with current computing capabilities we are restricted to 
$\Lambda a\gtrsim  1/10$. The last two conditions imply,
using the identity (\ref{DifferentScales}), $L\lesssim 1$, which 
contradicts the first requirement in Eq. (\ref{FiniteVolumeConditions}). 
There exists a clever solution to this problem: we 
shall explain this solution in the next section.
%
%***************************************************************************
%
\section{The Scale Dependence of Renormalized Operators}
\label{RunningSection}
In the previous Section we stressed the fundamental problems which arise 
with two classes of  non-perturbative renormalization schemes. 
The source of these problems is the requirement of a large separation between 
the different scales: $a^{-1}\gg\mu\gg\Lambda$. In
infinite-volume schemes one tries to realize all this scales on the same 
lattice, keeping all the length scales much smaller than the size of 
the lattice, i.e. $L\gg 1/(\Lambda a)$. 
This is obviously difficult, since a very large lattice is 
required.
In finite-volume schemes one identifies the lattice size with the 
renormalization scale. This produce however a separation between the scales 
$a^{-1}$ and $\Lambda$ which cannot be reproduced when hadronic, 
infinite-volume, matrix elements are computed.

The first step to the solution of this dilemma consists in considering
the ratio:
\begin{eqnarray}
\lim_{a\to 0}\frac{Z_{\cal O}(s\Lambda/\mu,\Lambda a)}
{Z_{\cal O}(\Lambda/\mu,\Lambda a)} \equiv\sigma_{\cal O} (\Lambda/\mu;s)\, .
\label{StepScalingDefinition}
\end{eqnarray}
The existence of the above limit is a scaling hypothesis, and is assured
by the existence of the continuum limit.
The function $\sigma_{\cal O}(\cdot)$ is 
universal\footnote{For a nice numerical verification of this universality 
hypothesis, see Ref. \cite{Guagnelli:1999gu}.}, 
i.e. it does not depend upon the 
precise definition of the lattice operator and of the lattice action, but only
upon the renormalization prescription. Finite lattice spacing corrections
to the limit defined by Eq. (\ref{StepScalingDefinition}) are of order $a$ 
for a general theory with 
fermions. They can be reduced to $O(a^2)$ if the lattice action 
and the operator ${\cal O}$ are non-perturbatively improved.
The function $\sigma_{\cal O}(\cdot)$ is closely related to the 
anomalous dimensions of the operator ${\cal O}$:
\begin{eqnarray}
\sigma_{\cal O} (\Lambda/\mu;s) = \exp\left\{
\int_{g(\mu)}^{g(s\mu)}\frac{\gamma_{\cal O}(x)}{\beta(x)}dx
\right\}\, .
\end{eqnarray}

Let us suppose that we know $\sigma_{\cal O}(\cdot)$.
If we compute the renormalization constant 
$Z_{\cal O}(\Lambda/\mu,\Lambda a)$ for some value of the lattice 
spacing $\Lambda a$ and for some scale $\Lambda/\mu$, then we can use 
the function  $\sigma_{\cal O}(\cdot)$ for computing it at the same lattice 
spacing for any scale of the type $s^{k}\Lambda/\mu$ (the
most common choice is $s=2$). In other words the function
$\sigma_{\cal O} (\Lambda/\mu;s)$ describes the running of the 
renormalized operator ${\cal O}$. Its non-perturbative determination allows 
to bypass the problems encountered in the previous Section, both for finite
and for infinite volume schemes.

Until now we did not specify any renormalization scheme. The second step
consists in noticing that, in a finite-volume scheme, it 
is quite simple to compute
the ``step-scaling function'' $\sigma_{\cal O}(\Lambda/\mu;s)$
non-perturbatively. 
This computation corresponds to the computation of a finite-size
scaling function. 
This is a remarkable feature of finite-size schemes. 
Their peculiarity comes from the fact that they allow to consider 
very small lattice spacings without using huge lattices.
%
%***************************************************************************
%
\section{Renormalized Operators from the OPE}
\label{RenormalizationViaOPE}
In this Section we briefly discuss a recently proposed method 
\cite{Dawson:1997ic,Rossi:1998kc,Testa:1997ne,Martinelli:1998hz} 
for constructing renormalized composite operators in asymptotically 
free theories.
This procedure shares many features of the infinite-volume schemes
described in Section \ref{DifferentSchemes} and in fact it is 
defined in infinite volume. It has the advantage of being more direct
and allowing a simplified treatment of operator mixing.
Moreover, it avoids the evaluation of products of local operators at 
coincident points. Such products must be taken into account with 
infinite-volume schemes as the expression on the right-hand side of 
Eq. (\ref{OperatorInsertion}) shows. 
Avoiding coincident points should reduce lattice artifacts and allow a simpler
implementation of operator improvement. 
Finally, the method we will describe yields renormalized operators in a 
zero mass continuum scheme without the necessity of taking the chiral limit.

We proceed now to describe the general context to which this method apply:
\begin{itemize}
\item Let us consider, for instance, the following simple example of 
Operator Product Expansion in the continuum:
\begin{eqnarray}
{\cal A}(x){\cal B}(-x) \sim C_{\cal O}(x){\cal O}(0)+\dots\, ,
\label{SimpleExample}
\end{eqnarray}
where the dots $\dots$ indicate terms of higher order in $x^2$, 
corresponding to operators of higher canonical dimension.
\item Let us suppose that we know how to construct the renormalized operators
${\cal A}$ and ${\cal B}$ non-perturbatively. This is the case
if ${\cal A}$ and ${\cal B}$  are conserved Noether currents. If the lattice
does not break the corresponding symmetry, then an 
exactly conserved discretized current can be constructed. 
Such a conserved lattice current does not 
renormalize. This happens, in lattice QCD, with the Noether currents of the
vector flavour group.
\item Let us suppose that we are interested in computing some hadronic
matrix element of ${\cal O}$. We denote this matrix element 
$\<h_1|{\cal O}|h_2\>$.
\end{itemize}
The proposed procedure works as follows:
\begin{enumerate}
\item The Wilson coefficient $C_{\cal O}(x)$ in Eq. (\ref{SimpleExample}) 
is calculated using RG 
improved perturbation theory. Any renormalization scheme can be used in 
this step. Let us call $C^{(l)}_{\cal O}(x)$ the resulting ($l$-loop) 
approximation for the Wilson coefficient.
\item The matrix element $\<h_1|{\cal A}(x){\cal B}(-x)|h_2\>$ is computed in 
a numerical simulation for a properly chosen range of $x$,
let us say $\rho \le |x|\le R$. This step gives
a function $G_{\cal AB}(x)$.
\item Finally $G_{\cal AB}(x)$ is fitted using the form 
$C^{(l)}_{\cal O}(x)\cdot\hat{\cal O}$ and keeping $\hat{\cal O}$
as the parameter of the fit.
\end{enumerate}
The outcome of this third step, i.e. the parameter of the fit 
$\hat{\cal O}$, is identified with the matrix element we are looking for,
$\<h_1|{\cal O}|h_2\>$, renormalized in the same scheme and at the same scale
at which we computed the Wilson coefficient $C^{(l)}_{\cal O}(x)$.
This identification will be correct in the limit $R,\rho\to 0$
keeping always $a\ll \rho,R$. Of course in practice
$\rho$ and $R$ must be kept finite, because of the finiteness of the lattice
spacing. We can estimate the
systematic errors to be of the order of the
neglected terms in the perturbative expansion of the Wilson coefficient:
$O(g(\mu)^{l+1})$, $g(\mu)$ being the running coupling at the scale $\mu = R^{-1}$.

Let us conclude by recalling the most important physical motivation
for the use of the above approach \cite{Dawson:1997ic}. The study
of weak decays through the effective-hamiltonian formulation 
is often complicated by operator mixings. The pattern of operator
mixing is dictated by power counting and is greatly restricted by 
the symmetries of the theory. In lattice numerical computations,
the Wilson discretization of the fermion action is usually adopted.
The explicit breaking of chiral symmetries in the Wilson formulation
makes operator mixing much more difficult to be treated.
In the OPE approach outlined above, the {\it continuum} OPE is employed.
As a consequence the right-hand side of Eq. (\ref{SimpleExample}) is
restricted by the symmetries  of the continuum theory.
Irrelevant terms of the action, and, among them, the Wilson term,
manifest themselves as lattice artifacts and
can be disregarded in the continuum limit.
Moreover chiral symmetry is completely restored at short distances:
both the spontaneous symmetry breaking and the fermion masses produce
power corrections to the leading behavior.

\chapter[Perturbative Renormalization of the $O(N)$ $\sigma$-Model]{Perturbative Renormalization and Composite Operators in the
$O(N)$ Non-Linear $\sigma$-Model}

We aim at computing matrix elements of renormalized operators in 
the $O(N)$ non-linear $\sigma$ model. 
In the next Chapters we shall adopt the ``OPE method'', sketched in 
Sec. \ref{RenormalizationViaOPE},
for coping with this task. This computation will
constitute a non-trivial test of this newly proposed approach.

Before proceeding, we shall study 
the renormalization of the model (and in particular the renormalization
of composite operators) in perturbation theory.
This is important for two reasons. First reason: it is commonly supposed 
in non-perturbative renormalization studies, that the structure of 
perturbative renormalization (which is dictated by the symmetries and
by power counting) holds beyond perturbation theory. Second reason:
in applying the OPE method we shall need some perturbative inputs.
A part of these inputs (the composite operators anomalous dimensions)
will be computed in this Chapter.
Moreover it is interesting to compare the outcomes of the
OPE method to the perturbatively renormalized operators.
In order to allow this comparison, we shall compute the lattice 
renormalization constants in perturbation theory.

In Secs. \ref{ModelSection} and \ref{CompositeOperatorsSection} 
we present the  model, its renormalization, and the renormalization 
of composite operators in perturbation theory.  
In Sec. \ref{ConstantsSection} we compute the continuum  
renormalization constants for some interesting operators and in
Sec. \ref{AnomalousSection} we report the corresponding anomalous dimensions.
In Sec. \ref{LatticeOperatorsSection} we repeat some of these computations on
the lattice. The basic definitions for lattice anomalous dimensions
are recalled in Sec. \ref{LatticeAnomalousSection}. 
In Sec. \ref{RenormalizationGroupSection} we recall
how the renormalization group (RG) can be used for improving the perturbative
expressions of the Wilson coefficients. A basic ingredient for evaluating
RG improved Wilson coefficients is the 
running coupling. In Sec. \ref{RunningCoupling} we compare various procedures
for evaluating the running coupling. 
%
%**************************************************************************
%
\section{Renormalization of the Model}
\label{ModelSection}

The $O(N)$ non-linear $\sigma$-model can be defined through the
lattice discretization. The standard lattice action reads:
\begin{eqnarray}
S^{\rm latt}[\sg] = \frac{1}{2g_L}\sum_{x\in\mathbb{Z}^2,\mu}
(\partial_{\mu} \sg)^2_x\, ,
\label{LatticeAction}
\end{eqnarray}
where $(\partial_{\mu}f)_x\equiv f_{x+\mu}-f_x$ is the forward lattice 
derivative, and the spin variables $\sg_x \equiv (\sigma^1_x,\dots,\sigma^N_x)$
are constrained to lie on the unit sphere: $\sg_x^2=1$.
The partition function is obtained by specifying the measure:
\begin{eqnarray}
Z(g_L) = \int\!\prod_xd^N\sg_x\delta(\sg^2_x-1)\exp\{-S^{\rm latt}[\sg]\}\, .
\end{eqnarray}

The continuum counterpart of the above model is obtained, as usual in 
perturbation theory, by writing down the naive continuum limit of the
action (\ref{LatticeAction}), and adopting a continuum regularization
(we shall use dimensional regularization).
The naive continuum action reads $S[\sg] =
1/(2g_B)\int\!dx\,(\partial\sg_B)^2$.
In order to construct the perturbative expansion, the $N$-vector 
field $\sg_B$ is parametrized as follows:
\begin{eqnarray}
\sg_B(x)\equiv (\pg_B(x),\sigma_B(x))\, ,\quad
\sigma_B(x)\equiv\sqrt{1-\pg_B^2(x)}\, ,
\end{eqnarray}
and the fields $\pg_B(x)=(\pi^1_B(x),\dots,\pi^{N-1}_B(x))$  are 
taken as the elementary (independent) degrees of freedom of the
theory.
The perturbative expansion is obtained by expanding the path integral
for small fields $\pg_B(x)$.
This perturbative expansion is plagued by infra-red divergences. 
The problem can be understood by noticing
that the tree-level propagator of the $\pg_B$ fields is $g_B/p^2$
and has no Fourier transform in two dimensions.

A possible approach for treating this problem is to introduce an external 
magnetic field. The continuum action obtained in this approach
can be written as follows in terms of bare fields:
\begin{eqnarray}
S[\sg]&=&\frac{1}{g_B}\int d^dx \left[\frac{1}{2}
(\partial \sg_B(x))^2-h_B \sigma_B\right]\, ,
\label{ContinuumAction}
\end{eqnarray}
with $d=2+\epsilon$.
The magnetic field $h_B$ acts as an infrared regulator but breaks the
$O(N)$ symmetry.
According to the Mermin-Wagner \cite{Mermin:1966fe,Coleman:1973ci}
theorem, $O(N)$ symmetry must be recovered in the $h_B\to 0$ limit.
The restoration of $O(N)$ symmetry manifests itself in perturbation theory  
in a rather peculiar way 
\cite{Jevicki:1977zn,Elitzur:1983ww,David:1981rr}. 
While the perturbative expansion of
$O(N)$ invariant quantities is infrared finite in the $h_B\to 0$ limit,
infrared singularities do not cancel in the perturbative expansion
of non-invariant quantities. The latter can however be re-expressed in terms 
of the former using $O(N)$ symmetry.

The renormalizability of the perturbative expansion described above
has been investigated in Refs. \cite{Brezin:1976sq,Brezin:1976qa} 
and proven in Ref. \cite{Brezin:1976ap}.
In order to make the perturbative expansion ultraviolet finite, the following 
renormalized quantities must be defined:
\begin{eqnarray}
g_B&\equiv&\mu^{2-d}N_d^{-1}Z_g g\, ,\\
\pg_B(x)&\equiv & Z^{1/2}\pg(x)\, ,\\
\sigma_B(x)&\equiv& Z^{1/2}\sigma(x) = \sqrt{1-Z\pg^2(x)}\, ,\\
h_B & = & \frac{Z_g}{Z^{1/2}}h\, .
\end{eqnarray}
The factor $N_d = (4\pi)^{-\epsilon/2}/\Gamma(1+\epsilon/2)$
is introduced to implement naturally the \MS prescription.
The two renormalization constants $Z$ and $Z_g$ defined above are 
known to four-loop order in perturbation theory 
\cite{Hikami:1981hi,Bernreuther:1986js,Wegner:1989ss}.
The beta-function and the anomalous dimension of the field $\sg(x)$
are defined in terms of $Z$ and $Z_g$ as follows:
\begin{eqnarray}
\beta^{\MMS}(g) \equiv\frac{\epsilon g}
{1+g\frac{\partial}{\partial g}\log Z_g}\, ,\quad
\gamma^{\MMS}(g) \equiv \beta^{\MMS}(g)
\frac{\partial}{\partial g}\log Z\, ,
\label{BetaGammaDefinition}
\end{eqnarray}
where we single out the dependence of these functions 
upon the renormalization scheme. Within a different scheme we shall
obtain different RG functions $\beta^{\rm scheme}(g)$
and $\gamma^{\rm scheme}(g)$.
For future use we fix the notation of their perturbative expansion as 
follows:
\begin{eqnarray}
\beta^{\rm scheme}(g) = \epsilon g-\sum_{k=0}^{\infty}
\beta^{\rm scheme}_k g^{k+2}\, ,
\quad 
\gamma^{\rm scheme}(g) = \sum_{k=0}^{\infty}\gamma^{\rm scheme}_k
g^{k+1}\, .
\label{BetaExpansion}
\end{eqnarray}
The first coefficients of these expansions are listed below
\begin{eqnarray}
\beta_0 = \frac{N-2}{2\pi}\, ,\quad \beta_1 = \frac{N-2}{(2\pi)^2}\, ,\\
\gamma_0 = \frac{N-1}{2\pi}\, ,
\end{eqnarray}
where we dropped the superscript ``scheme'' since, as is well known,
the first coefficients $\beta_0$, $\beta_1$, and $\gamma_0$
are scheme-independent.

In the following, we shall often refer to the schemes listed below:
\begin{itemize}
\item The minimal subtraction \MS renormalization scheme, already
used in this Section, see also Secs. \ref{ConstantsSection} and 
\ref{AnomalousSection}. The corresponding beta-function and
anomalous dimensions are $\beta^{\MMS}(g)$ and $\gamma^{\MMS}(g)$.
\item The bare lattice theory, see Secs. \ref{LatticeOperatorsSection}
and \ref{LatticeAnomalousSection}.
In this scheme the RG functions are denoted $\beta^L(g_L)$ and
$\gamma^L(g_L)$.
\item The improved-coupling scheme, which differs from the lattice
theory uniquely in the definition of the coupling constant, see 
Sec. \ref{RunningCoupling} and Eq. (\ref{DressedCouplingEnergy}).
We denote the corresponding functions as $\beta^E(g_E)$ and
$\gamma^E(g_E)$.
\item The finite-size scheme, see Sec. \ref{RunningCoupling} and Eq.
\ref{FiniteSizeCoupling}, whose RG functions are $\beta^R(g_R)$
and $\gamma^R(g_R)$.
\end{itemize}
%
%**********************************************************************
%
\section{The Structure of Renormalized Operators}
\label{CompositeOperatorsSection}
In this Section we describe the structure of renormalization (and mixing)
for composite operators as it emerges in perturbation theory.
The basic task is to understand how $O(N)$ symmetry 
restricts the possible mixings among composite operators.
This problem has been solved in Refs. \cite{Brezin:1976ap,Brezin:1976an}.
We recall here the results of these papers for greater convenience of 
the reader.

The general form of a renormalized composite operator is
\be
[ A ]_R (x)\equiv \sum_B Z_{AB} B(x)\, ,
\label{ContinuumRen}
\ee
where the $B$'s are unrenormalized composite operators, that is products of
$\pg(x)$'s, $\sigma(x)$'s (renormalized fields) and of their derivatives. 
Which operators $B$ must appear on the r.h.s. of Eq. (\ref{ContinuumRen})
for a given  $A$ on the l.h.s.?
The naive answer would be: all the operators which 
transform like $A$ under $O(N)$ and have canonical dimension 
$\rm{dim}[B]\le \rm{dim}[A]$. 
This answer is wrong because of the magnetic field 
in Eq. (\ref{ContinuumAction})
which breaks explicitly the $O(N)$ symmetry\footnote{
A similar problem is encountered in non-abelian gauge theories.
In order to renormalize gauge-invariant operators,
both gauge non-invariant, and BRS non-invariant operators
must be subtracted \cite{Joglekar:1976nu}.}.

In order to give the correct answer,
let us start by considering the non-linear realization of the $O(N)$ 
symmetry on the independent degrees of freedom:
\begin{eqnarray}
\delta_{\omega}\pi^a \equiv
 \sum_{b=1}^{N-1} \omega^{ab}\pi^b+
\omega^{aN}\sqrt{1/Z-\pg^2}\quad;\quad \omega^{ab}+\omega^{ba}=0\, .
\label{TrasformationRule}
\end{eqnarray}
In the following we shall use the convention 
$\delta_{\omega} \equiv \sum_{a,b = 1}^N \omega^{ab}\delta_{ab}$.
The transformation rule defined above is a rotation on a 
sphere of renormalized radius $1/Z$. 
In fact a rotation of radius one (i.e. Eq. (\ref{TrasformationRule})
with the substitution $Z\to 1$) does not leave invariant the
action (\ref{ContinuumAction}) even in the limit $h_B\to 0$. 
We could say that renormalization changes 
the transformation properties of the fields.

Let us now consider a composite operator ${\cal Q}$ and write down
its variation under the rotation (\ref{TrasformationRule}).
If we write ${\cal Q}$ as a function of $h$, $\pg$ and 
its derivatives ${\cal Q}(h;\pg,\partial\pg,\dots)$, we can define the 
variation of ${\cal Q}$  induced by Eq. (\ref{TrasformationRule}) as follows:
\begin{eqnarray}
{\cal Q}
(h;\pg+\delta_{\omega}\pg,\partial\pg+\partial\delta_{\omega}\pg,\dots)
& \simeq & 
{\cal Q}(h;\pg,\dots)+\delta_{\omega}{\cal Q}(h;\pg,\dots)\, .
\end{eqnarray}
The variation $\delta_{\omega}{\cal Q}(h;\pg,\dots)$ can be written more 
explicitly as follows:
\begin{eqnarray}
\delta_{\omega}{\cal Q}(h;\pg,\dots) = \int\!dy\,\left. 
\frac{\delta{\cal Q}}{\delta\pg(y)}\right|_h\cdot \delta_{\omega}\pg(y)\, .
\label{CompositeOperatorsRule}
\end{eqnarray}

We consider now an irreducible multiplet $\{ {\cal O}^A(x),\,A = 1,
\dots,{\cal N}\}$ of composite operators:
\begin{eqnarray}
\delta_{\omega}{\cal O}^A(x) =
\sum_{a,b=1\atop a<b}^N \omega^{ab}\sum_{B=1}^{{\cal N}} M^{ab}_{AB}
{\cal O}^B(x)\, ,
\label{Multiplet}
\end{eqnarray}
where the matrices $M^{ab}$ define a linear irreducible representation of 
(the Lie algebra of) $O(N)$. 
It easy to classify all the multiplets of dimension zero \cite{Brezin:1976an}.
They are the irreducible $O(N)$ tensor of rank $n$:
\begin{eqnarray}
{\cal O}_{(0,n)}^{a_1\dots a_n} = \sigma^{a_1}\dots\sigma^{a_n}-
{\rm traces}\, ,
\end{eqnarray}
where the ``traces'' term assures that ${\cal O}_{(0,n)}^{a_1\dots a_n}$
is traceless and completely symmetric in the indices $a_1,\dots,a_n$.
Another simple example is given by
\begin{eqnarray}
{\cal O}_{(2,n)}^{ab} =\partial\sigma^a\partial\sigma^b-
\frac{\delta^{ab}}{N}(\partial\sg)^2\, ,
\end{eqnarray}
which has dimension $2$, and is a rank-$2$ $O(N)$ tensor.

Let us call $[{\cal O}^A(x)]$ the renormalized counterparts of the 
multiplet (\ref{Multiplet}). The naive expectation would be that 
the renormalized operators transform as follows
\begin{eqnarray}
\delta_{\omega}[{\cal O}^A](x) =
\sum_{a,b=1\atop a<b}^N 
\omega^{ab}\sum_{B=1}^{{\cal N}} M^{ab}_{AB}[{\cal O}^B](x)\, .
\label{NaiveRenormalization}
\end{eqnarray}
Equation (\ref{NaiveRenormalization}) holds for operators of
dimension zero, but it is wrong in the general case. 
In order to give the correct answer in the general case, we introduce 
the following composite operator:
\begin{eqnarray}
\alpha(x)\equiv
\frac{1}{\sigma_B(x)}\left[\partial^2\sigma_B(x)+h_B\right]
=\frac{1}{\sigma(x)}\left[\partial^2\sigma(x)+\frac{Z_g}{Z}h\right]\, ,
\label{AlphaDefinition}
\end{eqnarray}
which can be rewritten as follows:
\begin{eqnarray}
\alpha(x) = h_B\sigma_B(x)-(\partial\sg_B)^2(x)+
g_B\pg_B(x)\cdot\frac{\delta S[\sg]}{\delta \pg_B(x)}\, .
\label{AlphaBareOnShell}
\end{eqnarray}
The operator $\alpha(x)$ is not invariant under 
$O(N)$ transformations, but it is invariant under the unbroken subgroup 
$O(N-1)$.

A generic composite operator ${\cal Q}(h;\pg,\partial\pg,\dots)$ can be 
considered as a function of
$\alpha$, $\pg$ and of the derivatives of $\pg$ according to the 
following rule:
\begin{eqnarray}
{\widetilde {\cal Q}}(\alpha;\pg,\partial\pg,\dots)\equiv
{\cal Q}(ZZ_g^{-1}(\alpha\sigma-\partial^2\sigma);\pg,\partial\pg,\dots)\, ,
\end{eqnarray}
which is a simple change of variables.
This allows us to define a new, and somewhat artificial, transformation rule 
``at fixed $\alpha$'':
\begin{eqnarray}
{\widetilde {\cal Q}}(\alpha;\pg+\delta_{\omega}\pg,\partial\pg+
\partial\delta_{\omega}\pg,\dots)
& \simeq & 
{\cal Q}(h;\pg,\dots)+{\widetilde \delta_{\omega}}{\cal Q}(h;\pg,\dots)
\, .
\end{eqnarray}
Analogously to Eq. (\ref{CompositeOperatorsRule}), we can write
explicitly the action of ${\widetilde \delta_{\omega}}$, as follows:
\begin{eqnarray}
{\widetilde \delta_{\omega}}{\cal Q}(h;\pg,\dots) = \int\!dy\,\left. 
\frac{\delta\widetilde{\cal Q}}{\delta\pg(y)}\right|_{\alpha}
\cdot \delta_{\omega}\pg(y)\, .
\label{CompositeOperatorsModifiedRule}
\end{eqnarray}
Notice that the derivative with respect to $\pg(y)$ on the r.h.s.
of Eq. (\ref{CompositeOperatorsModifiedRule}) is taken at $\alpha$ fixed.
We can now formulate the correct statement regarding the renormalization 
structure of a generic composite operator.
This is obtained by rewriting Eq. (\ref{NaiveRenormalization})
with the substitution $\delta_{\omega}\rightarrow{\widetilde \delta_{\omega}}$:
\begin{eqnarray}
{\widetilde \delta_{\omega}}[{\cal O}^A](x) =
\sum_{a,b=1\atop a<b}^N 
\omega^{ab}\sum_{B=1}^{{\cal N}} M^{ab}_{AB}[{\cal O}^B](x)\, .
\label{Thesis}
\end{eqnarray}
Notice that, since $\alpha(x)$ is invariant under the unbroken $O(N-1)$ 
subgroup, ${\widetilde \delta_{ab}}=\delta_{ab}$ if $a,b = 1,\dots,N-1$.
This is what we expect, since the global unbroken symmetries are 
preserved under renormalization. 

An explicit form for the renormalized
operators $[{\cal O}^A](x)$ can be obtained by noticing that 
the canonical dimension of $\alpha(x)$ is 2.
To be definite let us consider a renormalization scheme
such that the renormalized operators depend only logarithmically
upon the renormalization scale. Minimal subtraction is an example of such a 
scheme.
If we call 
$d_{\cal O} \equiv {\rm dim}({\cal O})$ the canonical dimension of
$[{\cal O}^A](x)$ we get 
\begin{eqnarray}
[{\cal O}^A](x) = \sum_{k=0}^{ d_{\cal O} }
\sum_{\cal Q} Z_{\cal O,Q}^{(k)}\, P_k[\alpha(x)] {\cal Q}_{(k)}^A(x).
\label{GeneralStructure}
\end{eqnarray}
The $P_k[\alpha]$ are local functionals of $\alpha$ of canonical
dimension $k$. 
The $\{{\cal Q}_{(k)}^A,A=1,\dots,{\cal N}\}$ are multiplets of composite
operators of canonical dimension $d_{{\cal Q}_{(k)}}=d_{\cal O}-k$,
transforming like $\{{\cal O}^A,A=1,\dots,{\cal N}\}$ under $O(N)$.
Moreover they do not depend upon $h$.

We shall now give a proof of the previous statement along the lines of 
Refs. \cite{Brezin:1976ap,Zinn-Justin:1992jq}. 
In the following it will be useful to consider $h$ as an external space 
dependent field $h(x)$. Equation (\ref{AlphaDefinition}) will
be modified accordingly with the prescription $h\rightarrow h(x)$.

We shall prove Eq. (\ref{GeneralStructure}) in perturbation theory
by induction over the perturbative order.
Our first step will be formulating the thesis at 
$n$-loop order. Then we shall prove it for any $n$ by showing
that, assuming it as an inductive hypothesis for a given order $n$,
it holds also for the successive order $n+1$.

Given a multiplet of composite operators 
$\{ {\cal O}^A,A=1,\dots,{\cal N}\}$, it is possible to
construct the $n$-loop perturbatively renormalized operators
$\{ [{\cal O}^A]_n\}$ in such a way that:
\begin{enumerate}
\item \label{TreeLevelMatch} 
$[{\cal O}^A]_n\rightarrow {\cal O}^A$ at tree level;
\item \label{Finiteness}
the insertions of $[{\cal O}^A]_n$ are ultraviolet finite up to terms of
order $g^{n+1}$;
\item \label{Unbroken}
$\delta_{ab}[{\cal O}^A]_n(x) = 
\sum_{B=1}^{{\cal N}} M^{ab}_{AB}[{\cal O}^B]_n(x)$ if $a,b = 1,\dots, N-1$;
\item \label{WIcondition} 
$\widetilde{\delta}_{aN}[{\cal O}^A]_n(x) = 
\sum_{B=1}^{{\cal N}} M^{aN}_{AB}[{\cal O}^B]_n(x)$ if $a = 1,\dots,
N-1$.
\end{enumerate}
Before proving the points \ref{TreeLevelMatch}-\ref{WIcondition}, let us make
a few observations. The statement \ref{Unbroken} is trivial
because, as we noticed above, global symmetries are preserved under 
renormalization. 
The point \ref{WIcondition} replaces the naive expectation 
$\delta_{aN}[{\cal O}^A]_n(x) = 
\sum_{B=1}^{{\cal N}} M^{aN}_{AB}[{\cal O}^B]_n(x)$, see 
Eq. (\ref{NaiveRenormalization}),
which would be the simplest extension of point \ref{Unbroken}. 

Let us now elaborate on point \ref{WIcondition}.
We want to rewrite it in a form which can be easily obtained with the
generating functional technique.
More precisely, we formulate it as follows: 
\begin{eqnarray}
S[\pg, h]*_a [{\cal O}^A]_n(x) + \frac{1}{g}\sum_{B=1}^{{\cal N}} 
M^{aN}_{AB}[{\cal O}^B]_n(x) = 0\, ,
\label{WIcondition-Conv}
\end{eqnarray}
 where we define
\begin{eqnarray}
S[\pg, h] & \equiv 
&\frac{Z}{Z_g g}\int dx\left[\frac{1}{2}(\partial\pg)^2+
\frac{1}{2}\left(\partial\sigma\right)^2-\frac{Z_g}{Z}h(x)\sigma\right]\, ,
\label{ActionVariableMass}\\
S[\pg,h]*_a{\cal O}(x) & \equiv &
\int dz\left\{
\frac{\delta S[\pg,h]}{\delta\pi^a(z)}
\frac{\delta{\cal O}(x)}{\delta h(z)}+
\frac{\delta S[\pg,h]}{\delta h(z)}
\frac{\delta{\cal O}(x)}{\delta \pi^a(z)}
\right\}\, .
\label{ActionConvolution}
\end{eqnarray}
The above definition of $S[\pg,h]$ differs from the one given in 
Eq. (\ref{ContinuumAction}) only in the fact that the magnetic field 
$h(x)$ is taken to be position dependent. 
Using Eqs. (\ref{ActionVariableMass}) and (\ref{ActionConvolution}), 
it is easy to show that
\begin{eqnarray}
S[\pg,h]*_a{\cal O}(x) = \frac{1}{g}\int dz\left\{
\frac{Z}{Z_g }[-\partial^2\pi^a(z)+\alpha\pi^a(z)]
\left.\frac{\delta{\cal O}(x)}{\delta h(z)}\right|_{\pi}
-\sigma(z)\left.\frac{\delta{\cal O}(x)}{\delta \pi^a(z)}\right|_{h}
\right\}\, ,\nonumber\\
\end{eqnarray}
whence, using the chain rule:
\begin{eqnarray}
\left.\frac{\delta{\cal O}(x)}{\delta h(z)}\right|_{\pi} & = & 
\int dy \left.\frac{\delta\alpha(y)}{\delta h(z)}\right|_{\pi}
\left.\frac{\delta{\cal O}(x)}{\delta \alpha(y)}\right|_{\pi}=
\frac{Z_g}{Z}\frac{1}{\sigma(z)}
\left.\frac{\delta{\cal O}(x)}{\delta \alpha(z)}\right|_{\pi}\, ,
\label{Chain1}\\
\left.\frac{\delta{\cal O}(x)}{\delta \pi^a(z)}\right|_{h} & = &
\left.\frac{\delta{\cal O}(x)}{\delta \pi^a(z)}\right|_{\alpha}+
\int dy \left.\frac{\delta\alpha(y)}{\delta \pi^a(z)}\right|_{h}
\left.\frac{\delta{\cal O}(x)}{\delta \alpha(y)}\right|_{\pi} = 
\label{Chain2}\\
& = & \left.\frac{\delta{\cal O}(x)}{\delta \pi^a(z)}\right|_{\alpha}
+\frac{\pi^a(z)}{\sigma^2(z)}\alpha(z)
\left.\frac{\delta{\cal O}(x)}{\delta \alpha(z)}\right|_{\pi}-
\frac{\pi^a(z)}{\sigma(z)}\partial^2_z\left[
\frac{1}{\sigma(z)}\left.\frac{\delta{\cal O}(x)}
{\delta \alpha(z)}\right|_{\pi}\right]\, ,\nonumber
\end{eqnarray}
we finally obtain, using Eqs. (\ref{CompositeOperatorsModifiedRule})
and (\ref{TrasformationRule}):
\begin{eqnarray}
S[\pg,h]*_a{\cal O}(x) = -\frac{1}{g}\int \!dz\,\sigma(z)
\left.\frac{\delta{\cal O}(x)}{\delta \pi^a(z)}\right|_{\alpha} = 
-\frac{1}{g}{\widetilde \delta_{aN}}{\cal O}(x)\, .
\end{eqnarray}
Equation (\ref{WIcondition-Conv}) is then equivalent to 
the point \ref{WIcondition} of our thesis.

The proof of \ref{TreeLevelMatch}-\ref{WIcondition} proceeds by induction on 
$n$. The thesis is obviously true for $n=0$ by taking 
$[{\cal O}^A]_0 = {\cal O}^A$.
Let us suppose the thesis to be true for a generic integer $n$.
We want to construct new renormalized operators
$[{\cal O}^A]_{n+1}(x)$ which satisfy \ref{TreeLevelMatch}-\ref{WIcondition}
with $n\rightarrow n+1$.
Equations (\ref{Chain1}) and (\ref{Chain2}) allow us to derive a relation
between ${\widetilde \delta_{aN}}$ and ${\delta_{aN}}$ which will turn out 
to be useful in the following:
\begin{eqnarray}
\delta_{aN}{\cal O}(x) = {\widetilde \delta_{aN}}{\cal O}(x)+
\frac{Z}{Z_g}\int\!dy\,\left[-\partial^2\pi^a(y)+\alpha\pi^a(y)\right]
\left.\frac{\delta{\cal O}(x)}{\delta h(y)}\right|_{\pi}\, .
\label{OperatorTrasformation}
\end{eqnarray}
We are interested in the behavior of the renormalized operators
$[{\cal O}^A]_n(x)$. Because of the induction hypothesis \ref{WIcondition},
we get
\begin{eqnarray}
\delta_{aN}[{\cal O}^A]_n(x) = \sum_{B=1}^{{\cal N}} 
M^{aN}_{AB}[{\cal O}^B]_n(x)+
\frac{Z}{Z_g}\int\!dy\,\left[-\partial^2\pi^a(y)+\alpha\pi^a(y)\right]
\left.\frac{\delta[{\cal O}^A]_n(x)}{\delta h(y)}\right|_{\pi}\, .
\label{OperatorTrasformationSimplified}
\end{eqnarray}
This identity characterize the behavior of the renormalized operators
under an $O(N)$ rotation. Notice that the second term on the r.h.s.
is a quantum correction which receives contributions only from the $k\neq 0$
terms on the r.h.s. of Eq. (\ref{GeneralStructure}).

Our next step will be to prove a Ward identity related to the explicitly 
broken symmetries. We shall adopt the generating functional technique.
The final outcome is given by Eq. (\ref{WIOperators}). 
In order to prove it, we define the effective action $\Gamma_{[n]}$
as follows:
\begin{eqnarray}
&&\exp\left\{\frac{1}{g}W_{[n]}[h,J,K]\right\} \equiv
\label{DefinitionGamma1}\\
&&\equiv \int \!d[\pg]\exp\left\{
-S[\pg,h]+
\frac{1}{g}\int\! dx\left[\Jg(x)\cdot\pg(x)+
\sum_{A=1}^{\cal N}K_A(x)[{\cal O}^A]_n(x)\right]
\right\}\nonumber\\
&&\Gamma_{[n]}[\pg,h,K]\equiv \left.\int\! dx\,\Jg(x)\cdot\pg(x) -
W_{[n]}[h,\Jg,K]\right|_{\pg =\frac{\delta W}{\delta \Jg}}\, ,
\end{eqnarray}
where we added the sources $K_A(x)$ coupled to the composite operators
$[{\cal O}^A]_n(x)$.
The $O(N)$ invariant integration measure is formally defined
as:
\begin{eqnarray}
\mbox{``}d[\pg]\equiv\prod_{x\in\mathbb{R}^d}
\frac{d\pg(x)}{\sqrt{1-\pg^2(x)}}\mbox{''}\, .
\end{eqnarray}
The usual generating functional is recovered when the fields 
$K_A(x)$ vanish. The 
vertex functions without composite operators insertions 
are obtained by taking the derivative with respect to $\pi$ at $K^A=0$:
\begin{eqnarray}
\Gamma^{(k)}(x_1,a_1;\dots;x_k,a_k) = 
\left.\frac{\delta^k\Gamma_{[n]}[\pg,h,0]}
{\delta\pi^{a_1}(x_1)\dots\delta\pi^{a_k}(x_k)}\right|_{\pi=0}\, .
\end{eqnarray}

We are interested in renormalizing
vertex functions with only one inserted composite operator. This 
is enough for making finite
all the correlation functions with an arbitrary number of
composite operators  at separate positions.
The vertex functions with one composite operator insertion 
are determined by the term of $\Gamma_{[n]}[\pg,h,K]$ linear in $K_A(x)$:
\begin{eqnarray}
\Gamma_{[n]}[\pg,h,K] = \Gamma[\pg,h]-\sum_{A=1}^{\cal N}\int\!dx\,
K_A(x)\Gamma_{[n]A}[x;\pg,h]+O(K^2)\, .
\end{eqnarray}
Notice that the zeroth-order term of the above expansion 
$\Gamma[\pg,h] = \Gamma_{[n]}[\pg,h,0]$ is finite to any order in 
perturbation theory. In fact the coupling constant, the elementary fields 
$\pg(x)$, and the magnetic field have been properly renormalized in Eq. 
(\ref{ActionVariableMass}).
Because of the inductive hypothesis \ref{Finiteness}, the first order 
terms $\Gamma_{[n]A}[x;\pg,h]$ are finite up to divergences of order 
$g^{n+1}$.

Let us check the above definitions at tree level. The loop expansion reads:
\begin{eqnarray}
\Gamma_{[n]}[\pg,h,K] = \sum_{l=0}^{\infty}\Gamma^{(l)}_{[n]}[\pg,h,K]\,g^l
\, ,
\end{eqnarray}
and, using the inductive hypothesis \ref{TreeLevelMatch}, we obtain
\begin{eqnarray}
\Gamma^{(0)}_{[n]}[\pg,h,K] &= &
\int\!dx\left[\frac{1}{2}(\partial\pg)^2+\frac{1}{2}(\partial\sqrt{1-\pg^2})^2
-h(x)\sqrt{1-\pg^2}\right]-\\
&&-\sum_{A=1}^{\cal N}\int\!dx\,K_A(x){\cal O}^A(x)\nonumber\, ,
\end{eqnarray}
which yields the correct insertions of the operators $[{\cal O}^A]_n(x)$
at tree level.

Now we have set up the effective functional formalism and we can proceed
to prove the relevant Ward identity. 
Let us consider the following symmetry transformation:
\begin{eqnarray}
{\widehat \delta_{\omega}}\pi^a(x) = 
\left\{
\sqrt{1/Z-\pg^2(x)}+\sum_{A=1}^{\cal N} \int \!dy\, K_A(y)
\frac{\delta [{\cal O}^A]_n(y)}{\delta h(x)}
\right\}\,\omega^{aN}\, .
\label{SymmetryTransformation}
\end{eqnarray}
This is a modification of Eq. (\ref{TrasformationRule}) and can be thought 
as a ``modified rotation'' on a sphere of radius: 
\begin{eqnarray}
\frac{1}{Z'}=\frac{1}{Z}+2\sigma(x)\sum_{A=1}^{\cal N} 
\int\!dy\, K_A(y)
\frac{\delta [{\cal O}^A]_n(y)}{\delta h(x)}+O(K^2)\, .
\end{eqnarray}
The modification is required by the introduction of the source terms
$K_A(x)[{\cal O}^A]_n(x)$ in Eq. (\ref{DefinitionGamma1}), 
just as the introduction of counterterms in the action 
requires a modification of the radius $1\rightarrow 1/Z$.  
The symmetry transformation  (\ref{SymmetryTransformation})
follows the general prescription of  Ref. \cite{Brezin:1976ap}:
if a term $S'[\pg,h]$ is added to the action (\ref{ActionVariableMass}),
the modified transformation rule will be
\begin{eqnarray}
\delta'_{\omega}\pi^a(x) = -g\,\frac{\delta(S+S')}{\delta h(x)}\,\omega^a
\, .
\end{eqnarray}
In the following we shall work at order $O(K)$. Using 
Eq. (\ref{OperatorTrasformationSimplified}) it is easy to obtain
the identity:
\begin{eqnarray}
\int\! dx\left\{
\frac{\delta \Gamma_{[n]}}{\delta h(x)}
\frac{\delta \Gamma_{[n]}}{\delta\pi^a(x)}
+h(x)\pi^a(x)
+\sum_{A,B=1}^{\cal N} K_A(x) M^{aN}_{AB}
\frac{\delta\Gamma_{[n]}}{\delta K_B(x)}
\right\}=O(K^2)\, .
\end{eqnarray}
The term linear in $K_A(x)$ of the above identity reads
\begin{eqnarray}
\Gamma[\pg,h]*_a\Gamma_{[n]A}[x;\pg,h]=
-\sum_{B=1}^{\cal N} M_{AB}^{aN}\,\Gamma_{[n]A}[x;\pg,h]\, .
\label{WIOperators}
\end{eqnarray}
Equation (\ref{WIOperators}) is the Ward identity related to the explicitly 
broken $O(N)$ symmetry. It gives a constraint on the insertions
of composite operators and, in particular, on their divergences.

We remark that $\Gamma_{[n]A}[x;\pg,h]$  diverges 
at order $g^{n+1}$. Since Eq. (\ref{WIOperators}) is linear in $\Gamma_{[n]A}$
and valid for any value of the cutoff we can choose a decomposition
\begin{eqnarray}
\Gamma_{[n]A}[x;\pg,h] = \Gamma'_{[n]A}[x;\pg,h]+
\Gamma^{\rm div}_{[n]A}[x;\pg,h]\, ,
\label{Decomposition}
\end{eqnarray}
such that: (i) $\Gamma'_{[n]A}$ is finite; 
(ii) $\Gamma^{\rm div}_{[n]A}[x;\pg,h]
\equiv \sum_{l=n+1}^{\infty} g^l\,\Gamma^{{\rm div}(l)}_{[n]A}[x;\pg,h]= 
O(g^{n+1})$; (iii) both $\Gamma'_{[n]A}$ and $\Gamma^{\rm div}_{[n]A}$ satisfy
Eq. (\ref{WIOperators}). 
Moreover the decomposition (\ref{Decomposition}) can be chosen 
such that  $\Gamma^{{\rm div}(n+1)}_{[n]A}[x;\pg,h]$ is local, i.e. a
function of $\pg(x),\partial\pg(x),\dots$.
This is a general theorem on the renormalization of composite operators
and it is independent of the symmetry properties of the theory 
\cite{Zimmerman:70}.

Given such a decomposition we get, from Eq. (\ref{WIOperators}):
\begin{eqnarray}
S_0[\pg,h]*_a\Gamma^{{\rm div}(n+1)}_{[n]A}[x;\pg,h]=
-\frac{1}{g}\sum_{B=1}^{\cal N} M_{AB}^{aN}\,\Gamma^{{\rm div}(n+1)}_{[n]A}
[x;\pg,h]\, ,
\label{AlmostThesis}
\end{eqnarray}
where $S_0[\pg,h] = \Gamma^{(0)}[\pg,h]/g$ is the first term of the
perturbative expansion of the action (\ref{ActionVariableMass}):
$S[\pg,h] \equiv\sum_{l=0}^{\infty}g^{l-1}S_l[\pg,h]$. 
Equation (\ref{AlmostThesis}) is ``almost'' what we need.
We would like that $S[\pg,h]$ appeared on the l.h.s. of 
Eq. (\ref{AlmostThesis}), instead of $S_0[\pg,h]$.
However this problem can be overcome by adding $O(g^{n+2})$ terms
to $\Gamma^{{\rm div}(n+1)}_{[n]A}[x;\pg,h]$.
More precisely, it is possible to find a local functional
\begin{eqnarray}
R^A[x;\pg,h] = g^{n+1}\Gamma^{{\rm div}(n+1)}_{[n]A}[x;\pg,h]+
\sum_{l=n+2}^{\infty}g^l\,R^A_l[x;\pg,h]\, .
\end{eqnarray}
which satisfies Eq. (\ref{AlmostThesis}) with the substitution
$S_0[\pg,h]\rightarrow S[\pg,h]$. Such a functional can be found 
by solving recursively
\begin{eqnarray}
S_0[\pg,h]*_a R^A_k[x;\pg,h]+\sum_{B=1}^{\cal N} M^{aN}_{AB}
R^B_k[x;\pg,h] =
-\sum_{l=1}^{k-n-1}S_l[\pg,h]*_a R^A_{k-l}[x;\pg,h]\nonumber\\
\end{eqnarray}
for $k = n+2, n+3, \dots$.

If is now easy to verify that the renormalized operators can be defined
at $n+1$ loops as follows
\begin{eqnarray}
[{\cal O}^A]_{n+1}(x)\equiv[{\cal O}^A]_n(x)-R^A[x;\pg,h]\, .
\label{Construction}
\end{eqnarray}
The definition (\ref{Construction})
satisfies the inductive thesis, i.e. \ref{TreeLevelMatch}-\ref{WIcondition}
with the substitution $n\rightarrow n+1$.
%
%***********************************************************************
%
\section{Renormalization Constants}
\label{ConstantsSection}

In this Section we apply the general results of Sec. 
\ref{CompositeOperatorsSection} to a few interesting cases.
We describe the mixing structure for various composite operators and
give the corresponding renormalization constants.
In order to distinguish the various renormalization constants we adopt the 
following convention. Given a basis $\{{\cal O}_1,\dots,{\cal O}_{\cal
N}\}$ of 
composite operators, they can be characterized through
their canonical dimension $d$, and their transformation properties
under $O(N)$ rotations. In particular they will be 
$O(N)$ tensors of rank $s$.
We shall denote the corresponding renormalization constants 
as $Z^{(d,s)}_{AB}$. Renormalized operators are obtained as follows
from bare ones:
$\opl {\cal O}_A\opr \equiv \sum_j Z^{(d,s)}_{AB}{\cal O}_B$. 
%
%*****************************************************
%
\subsection{$O(N)$ Invariant Operators of Dimension 2}
\label{D2IS0}

All the $O(N)$ invariant operators of dimension 2 can be expressed as 
linear combinations\footnote{Notice that $(\partial \sg)^2$ is  obtained
from $\partial_{\mu}\sg\cdot\partial_{\nu}\sg$ by taking the trace
over the space-time indices. They are not linearly
independent. However, as is well known, minimal subtraction does not
commute with taking the trace over the space-time indices. 
We must then consider the two operators as distinct elements of the basis.}
of $\partial_{\mu}\sg\cdot\partial_{\nu}\sg$
and $(\partial \sg)^2$. Under renormalization they mix with the operator 
$\alpha$ defined in Eq. (\ref{AlphaDefinition}):
\begin{eqnarray}
\opl \partial_\mu   \sg\cdot   \partial_{\rho}\sg \opr & = &
Z^{(2,0)}_{11}\partial_\mu   \sg\cdot   \partial_{\rho}\sg+
Z^{(2,0)}_{12}(\partial\sg)^2\delta_{\mu   \rho}
+Z^{(2,0)}_{13}\alpha\delta_{\mu   \rho}\, ,\\
\opl (\partial\sg)^2\opr & = & Z^{(2,0)}_{22}(\partial\sg)^2+
Z^{(2,0)}_{23}\alpha\, , \label{S2}\\
\opl \alpha\opr & =  &Z^{(2,0)}_{32}(\partial\sg)^2+
Z^{(2,0)}_{33}\alpha\, . \label{AlphaRenormalization}
\end{eqnarray}
The computation of the renormalization constants $Z^{(2,0)}_{AB}$
is pretty simple. 
For this particular set of operators they can be expressed in terms 
of the field and coupling constant renormalizations $Z$ and $Z_g$.

We start by noticing that a particular linear combination of
$\partial_{\mu}\sg\cdot\partial_{\nu}\sg$ and $(\partial\sg)^2$ 
yields the 
energy-momentum tensor\footnote{The reader will notice that Eq.
(\ref{EnergyMomentum}) gives the correct energy-momentum tensor
only in the limit $h\to 0$, see Eq. (\ref{ContinuumAction}). 
For nonzero magnetic field a term 
$h\sigma/g$ must be added. However, since this term is finite, our discussion 
does not need any modification.}
\begin{eqnarray}
T_{\mu \nu}\equiv\frac{1}{g}\opl \partial_\mu   \sg\cdot   \partial_{\nu}\sg-
\delta_{\mu  \nu}\frac{1}{2}(\partial\sg)^2\opr 
= Z_{T,T}\frac{1}{g} 
\left(\partial_\mu   \sg\cdot   \partial_{\nu}\sg 
-\frac{1}{2}\delta_{\mu   \nu}(\partial\sg)^2\right)\, .
\label{EnergyMomentum}
\end{eqnarray}
From energy-momentum conservation it follows that $T_{\mu   \nu}$ is finite
if we replace the renormalized fields and the renormalized coupling constant 
in Eq. (\ref{EnergyMomentum}) with the bare ones.
It follows that:
\begin{eqnarray}
Z_{TT} = Z^{(2,0)}_{11} = \frac{Z}{Z_g} = 1+\frac{1}{2\pi\epsilon}g+O(g^2)\, .
\label{EMRen}
\end{eqnarray}
Then we remark that differentiating the action 
(\ref{ContinuumAction}) with respect to renormalized parameters yields 
finite operators \cite{Collins:1984xc}. 
Upon differentiation with respect to $g$ (keeping 
$d$, $h$ and $\pg$ constants), we  obtain a linear 
combination of $(\partial \sg)^2$ and $\alpha$ as on the right-hand side
of (\ref{S2}), with the following coefficients:
\begin{eqnarray}
Z^{(2,0)}_{22} & = & \frac{Z}{Z_g}-g\frac{\partial}{\partial g}
\left(\frac{Z}{Z_g}\right)= 1+O(g^2)\, ,\\
Z^{(2,0)}_{23} & = & -\frac{1}{Z_g}g\frac{\partial}{\partial g}\log Z=
-\frac{N-1}{2\pi\epsilon}g+O(g^2)\, .
\label{D2Ren2}
\end{eqnarray}
We can unambiguously identify the above coefficients 
with the correct \MS renormalization constants because they
are series of poles in $\epsilon$.

The next useful observation is that the equations of motion, obtained
by varying the action (\ref{ContinuumAction}) with respect to $\pg$,
do not need renormalization. It follows that the combination
\begin{eqnarray}
\frac{Z}{Z_g}(\partial\sg)^2+\frac{1}{Z_g}\alpha = h\sigma+g\pg
\cdot\frac{\delta S}{\delta \pg}
\end{eqnarray}
is finite, cf. Eq. (\ref{AlphaBareOnShell}). Combining this result with 
the previous ones, we get
\begin{eqnarray}
Z^{(2,0)}_{32} & = & g\frac{\partial}{\partial g}\left(
\frac{Z}{Z_g}\right)=\frac{1}{2\pi\epsilon}g+O(g^2)\, ,\\
Z^{(2,0)}_{33} & = & \frac{1}{Z_g}\left(1+ g\frac{\partial}{\partial g}
\log Z\right)=1+\frac{1}{2\pi\epsilon}g+O(g^2)\, .
\label{AlphaRen}
\end{eqnarray}
The remaining renormalization constants can be computed by using 
Eq. (\ref{EnergyMomentum})--(\ref{D2Ren2}):
\begin{eqnarray}
Z^{(2,0)}_{12} & = & -\frac{1}{2}g\frac{\partial}{\partial g}
\left(\frac{Z}{Z_g}\right) = -\frac{1}{4\pi\epsilon}g+O(g^2)\, ,\\
Z^{(2,0)}_{13} & = & -\frac{1}{2Z_g}g\frac{\partial}{\partial g}
\log Z = -\frac{N-1}{4\pi\epsilon}g+O(g^2)\, .
\end{eqnarray}

We finally notice that the identities derived above are 
true only if a ``consistent'' renormalization scheme is adopted.
In particular there must be consistency between the prescriptions for
renormalizing the action and the composite operators.
Examples of such schemes are minimal subtraction or zero momentum (BPHZ)
subtraction.

The above results allow us to rewrite Eq. 
(\ref{AlphaBareOnShell}) in terms of renormalized operators:
\begin{eqnarray}
\opl \alpha\opr (x) = h\sigma(x)-\opl (\partial\sg)^2\opr (x)+
g\pg\cdot\frac{\delta S}{\delta\pg(x)}\, .
\label{AlphaOnShell}
\end{eqnarray}
In the following we shall be interested in on-shell matrix elements 
of composite operators at zero magnetic field ($h\to 0$).
Equation (\ref{AlphaOnShell}) allows us to eliminate
the contribution of ``spurious'' operator $\alpha$
in such matrix elements.

The renormalization constants computed in this Section can be used for
accomplishing a simple exercise: the computation of the trace 
of the energy-momentum tensor. Using Eq. (\ref{EnergyMomentum}), we have
\begin{equation}
T \equiv \delta_{\mu\nu} T_{\mu\nu} = - {\epsilon\over2} Z_{TT} 
   {1\over g} (\partial\sg)^2\, .
\end{equation}
Then, using Eqs. (\ref{S2}) and (\ref{AlphaRenormalization}), we can rewrite 
\begin{equation}
(\partial\sg)^2 =\, 
 {Z_g\over Z\epsilon} \left\{ \gamma(g) \opl \alpha\opr + 
   \left({\beta(g)\over g} + \gamma(g)\right)
\opl(\partial\sg)^2\opr\right\}\, ,
\end{equation}
where we have expressed $\partial Z/\partial g$ and $\partial Z_g/\partial g$ 
in terms of 
$\beta(g)$ and $\gamma(g)$ using Eqs. (\ref{BetaGammaDefinition}),
and dropped the superscripts \MS on $\beta(g)$ and $\gamma(g)$. 
We finally obtain
\begin{equation}
T = - {1\over 2g^2}[\beta(g) + g \gamma(g)] \opl(\partial\sg)^2\opr - 
      {\gamma(g)\over 2 g} \opl \alpha\opr\, .
\label{EnergyMomentumTrace}
\end{equation}
%
%
%*********************************************************
%
\subsection{Antisymmetric Rank 2 Operators}

We shall now consider operators which are tensors of $O(N)$
of rank two (i.e. they have two $O(N)$ indices). 
Let us begin from operators which are antisymmetric 
under the indices exchange. Obviously, there exists no such operator of 
dimension zero. There exists a unique antisymmetric operator of 
dimension $1$:
\begin{eqnarray}
j^{(a,b)}_\mu   \equiv\frac{1}{g}\opl\sigma^{a}\partial_\mu   \sigma^{b}-
\sigma^{b}\partial_\mu   \sigma^{a}\opr & = &
Z^{(1,1)}\frac{1}{g}\left(\sigma^{a}\partial_\mu   \sigma^{b}-
\sigma^{b}\partial_\mu   \sigma^{a}\right)\, ,
\end{eqnarray}
which is the Noether current associated to the $O(N)$ symmetry
Conservation of $j^{(a,b)}_\mu   $ implies
\begin{eqnarray}
Z^{(1,1)} = \frac{Z}{Z_g} = 1+\frac{1}{2\pi \epsilon}g+O(g^2)\, .
\end{eqnarray}

We can classify the antisymmetric, rank $2$  operators of dimension $2$
according to their Lorentz\footnote{
Here and in the following we denote as Lorentz transformations
the ordinary $O(2)$ rotations of the two-dimensional 
euclidean space-time. These must be distinguished from the internal
$O(N)$ rotations of the fields $\sg$.} symmetry.
Let us list them in order of increasing spin (respectively 0, 1, 2):
\begin{eqnarray}
A^{(0)} & \equiv &
\sigma^a\partial^2\sigma^b-\sigma^b\partial^2\sigma^a\, ,
\label{AntiSymmetric0}\\
A^{(1)}_{\mu \nu} & \equiv & \partial_{\mu}\sigma^a\partial_{\nu}\sigma^b-
\partial_{\mu}\sigma^b\partial_{\nu}\sigma^a\, ,
\label{AntiSymmetric1}\\
A^{(2)}_{\mu \nu} & \equiv & \sigma^a\partial_\mu   \partial_{\nu}\sigma^b-
\sigma^b\partial_\mu   \partial_{\nu}\sigma^a -\frac{1}{2}\delta_{\mu\nu}
(\sigma^a\partial^2\sigma^b-\sigma^b\partial^2\sigma^a)\, .
\label{AntiSymmetric2}
\end{eqnarray}
The three operators defined above can be written as linear functions 
of $\partial_{\mu}j^{(a,b)}_{\nu}$. As a consequence they renormalize
as the current itself and do not mix.
%
%******************************************************************
%
\subsection{Symmetric Rank 2 Operators}
\label{SymmetricZetaSection}

Finally we shall consider rank-two $O(N)$ tensors
which are symmetric and traceless in the two $O(N)$ indices. 
The unique dimension $0$ operator with this symmetry is:
\begin{eqnarray}
\opl \sigma^a\sigma^b-\frac{1}{Z}\frac{\delta^{ab}}{N} \opr & = & 
Z^{(0,2)}\left( \sigma^a\sigma^b-\frac{1}{Z}\frac{\delta^{ab}}{N} \right)\, .
\label{DefinitionDimension0Symmetric}
\end{eqnarray}
The corresponding renormalization constant has been calculated in 
Ref. \cite{Bernreuther:1986js,Wegner:1989ss} up to four-loop
order.
We give below the first two terms of the perturbative result:
\begin{eqnarray}
Z^{(0,2)} & = &
1-\frac{1}{2\pi\epsilon}g-\frac{N-3}{8\pi^2\epsilon^2}g^2
+O(g^3)\, .
\label{ZetaDimension0Symmetric}
\end{eqnarray}

There are $7$ linearly independent operators of dimension $2$.
A simple basis for these operators is the following:
\begin{eqnarray}
S_{1\mu   \rho}&\equiv& \frac{1}{2}\left(\partial_\mu 
\sigma^a\partial_{\rho}\sigma^b+
\partial_\mu   \sigma^b\partial_{\rho}\sigma^a\right)-\frac{\delta^{ab}}{N}
\partial_\mu   \sg\cdot   \partial_{\rho}\sg\, ,
\label{Symmetric1}\\
S_{2\mu   \rho}&\equiv & 
\frac{1}{2}\left(\sigma^a\partial_\mu   \partial_{\rho}\sigma^b+
\sigma^b\partial_\mu   \partial_{\rho}\sigma^a\right)-\frac{\delta^{ab}}{N}
\sg\cdot   \partial_\mu   \partial_{\rho}\sg\, ,\\
S_3 & \equiv & \partial\sigma^a\cdot   \partial\sigma^b-\frac{\delta^{ab}}{N}
(\partial\sg)^2\, ,\\
S_4 & \equiv & \frac{1}{2}
\left(\sigma^a\partial^2\sigma^b+\sigma^b\partial^2\sigma^a\right)+
\frac{\delta^{ab}}{N}(\partial\sg)^2\, ,\\
S_{5\mu   \rho}&\equiv&
\left(\sigma^a\sigma^b-\frac{1}{Z}\frac{\delta^{ab}}{N}\right)
\partial_\mu   \sg\cdot   \partial_{\rho}\sg\, ,\\
S_6&\equiv&
\left(\sigma^a\sigma^b-\frac{1}{Z}\frac{\delta^{ab}}{N}\right)
(\partial\sg)^2\, ,\label{Symmetric6}\\
S_7&\equiv&
\left(\sigma^a\sigma^b-\frac{1}{Z}\frac{\delta^{ab}}{N}\right)\alpha
\, . 
\label{Symmetric7}
\end{eqnarray}
We computed the corresponding renormalization matrix at two-loop
order in perturbation theory.
In the \MS scheme we get:
\begin{eqnarray}
Z^{(2,2)} & = & 1-\frac{1}{4\pi\epsilon}g A +\frac{1}{16\pi^2\epsilon^2}g^2 B
+ \frac{1}{32\pi^2\epsilon}g^2 C +O(g^3)\, ,
\label{RenConst2.2}
\end{eqnarray}
with
{\footnotesize
\begin{eqnarray}
A & = & \left[
\begin{array}{ccccccc}
 0  &  0  &  1  &  0  & -2  &  0  &  -1\\
 2  &  2  & -1  &  0  &  2  &  0  &   1\\
 0  &  0  &  2  &  0  &  0  & -2  &  -2\\
 0  &  0  &  0  &  2  &  0  &  2  &   2\\
 0  &  0  &  0  &  0  &  0  &  1  &  (N-1)\\
 0  &  0  &  0  &  0  &  0  &  2  & 2(N-1)\\
 0  &  0  &  0  &  4  &  0  &  2  &   0
\end{array}
\right]\, ,
\label{RenConst2.2.A}\\
B & = & \left[
\begin{array}{ccccccc}
    0   &    0    & -(N-3) & -2 &  2(2N-3) & -3 & -2  \\ 
-2(N-3) & -2(N-3) & (N-3)  &  2 & -2(2N-3) & 3  &  2 \\
0 & 0 & -2(N-3) & -4      & 0 &  4(N-3) & -4 \\
0 & 0 &    0    & -2(N-5) & 0 & -4(N-3) &  4 \\ 
0 & 0 & 0 & 2(N-1) & 0 & 2 &  2(N-1) \\
0 & 0 & 0 & 4(N-1) & 0 & 4 &  4(N-1) \\
0 & 0 & 0 & -4(N-3) & 0 &  -4(N-3) & 2(N+1)
\end{array}
\right]\nonumber\, ,\\
\label{RenConst2.2.B}\\
C & = & \left[
\begin{array}{ccccccc}
-(N-1) & -2(N-2) &  \frac{(3N-11)}{2} & -N & -(6N-11) & -\frac{(2N+5)}{2}& 0 \\
 (N-1) &  2(N-2) & -\frac{(3N-11)}{2} &  N &  (6N-11) &  \frac{(2N+5)}{2}& 0 \\
0 & 0 & -8 & -4(N-2) & 0 & -8(N-1) & 0 \\
0 & 0 &  8 & 4(N-2) & 0 &  8(N-1) & 0 \\
-8 & -4(N-2) & 0   & 2 & -8(2N-3) & 2(2N-3) & 0 \\
 0 &  0     & -8  & -8(N-2) & 0 & -4(3N-4) & 0 \\
0 & 0 & 8 & 8(N-2) & 0 & 8(N-1) & -4(N-2)
\end{array}
\right]\nonumber\, .\\
\label{RenConst2.2.C}
\end{eqnarray}
}

While the operators $S_{1\mu\nu}$,\dots,$S_6$ are 
ordinary rank two $O(N)$ tensors, 
$S_{7}$ is the product of the non $O(N)$ invariant operator 
$\alpha$ times a rank two $O(N)$ tensor.
This mixing pattern agrees with  the general results of 
Sec. \ref{CompositeOperatorsSection}.

Analogously to what we did  in Sec. \ref{D2IS0},  
we can eliminate the non $O(N)$ covariant operator $S_7$
in the limit $h\to 0$ and on-shell.
Let us prove this statement in the general case. 
We consider Eq. (\ref{AlphaBareOnShell})
and multiply it by a generic finite local operator ${\cal Q}(x)$:
\begin{eqnarray}
{\cal Q}(x)\alpha(x) = h_B{\cal Q}(x)\sigma_B(x)-{\cal Q}(x)
(\partial\sg_B)^2(x)+g_B{\cal Q}(x)
\pg_B(x)\cdot\frac{\delta S[\sg]}{\delta \pg_B(x)}\, .
\label{AlphaBareTimesOperator}
\end{eqnarray}
We would like to renormalize the previous Equation.
We notice that, on general grounds \cite{Collins:1984xc}, 
the last term on the r.h.s. of Eq. 
(\ref{AlphaBareTimesOperator}) is finite. Applying minimal
subtraction to both sides of Eq. (\ref{AlphaBareTimesOperator}) amounts 
to subtracting equal quantities from the left-hand and right-hand sides.
In fact Eq. (\ref{AlphaBareTimesOperator}) holds for any value 
of $\epsilon =d-2$, and thus holds between the poles in $\epsilon$.
It follows that:
\begin{eqnarray}
\opl {\cal Q}(x)\alpha(x)\opr  = h \opl {\cal Q}(x)\sigma(x)\opr -
\opl {\cal Q}(x)(\partial\sg_B)^2\opr (x)+g{\cal Q}(x)
\pg(x)\cdot\frac{\delta S[\sg]}{\delta \pg(x)}\, .\nonumber\\
\label{AlphaTimesSomething}
\end{eqnarray}
Finally we remark that the last term on the r.h.s. of this Equation vanishes 
on-shell. The contact term due to ${\cal Q}(x)$ does not contribute in
dimensional regularization because of the 
well known identity $\delta^d(0)=\int\! d^dp\, 1 = 0$.
Using Eq. (\ref{AlphaTimesSomething}) we can eliminate 
$S_7$ in favour of $S_6$
(on-shell, in the limit $h\to 0$).

We have chosen the basis defined in 
Eqs. (\ref{Symmetric1}),\dots,(\ref{Symmetric7})
for sake of simplicity.
However the structure of the mixing matrix (\ref{RenConst2.2})
becomes more transparent if we use the following new 
basis\footnote{This definition holds for bare operators.
If minimal subtraction is adopted, the renormalized basis reads:
$\opl S_{1\mu \nu}\opr$, $\opl S_3\opr$; $\opl S_{5\mu \nu}\opr$,
$\opl R_1\opr\equiv \opl S_4\opr +\opl S_6\opr$, 
$\opl R_2 \opr \equiv \opl S_6\opr +\opl S_7\opr$, 
$\opl D_{1\mu \nu}\opr \equiv \opl S_{1\mu\nu}\opr +\opl S_{2\mu\nu}\opr$, 
$\opl D_2\opr \equiv \opl S_3\opr +\opl S_4\opr$.}: 
$S_{1\mu \nu}$, $S_3$; $Z S_{5\mu \nu}$,
$R_1\equiv S_4+ZS_6$, $R_2 \equiv ZS_6+S_7$, 
$D_{1\mu \nu} \equiv S_{1\mu\nu}+S_{2\mu\nu}$, $D_2 \equiv S_3+S_4$. 
Notice that $D_{1\mu\nu}$ and $D_2$ are space derivatives and therefore their
insertion at zero momentum vanishes. Moreover $R_1$ and $R_2$ are
proportional to $h$ up to contact terms:
\begin{eqnarray}
R_1 & =  & Z_g h\left[\sigma^a\sigma^b\sigma-
\frac{1}{2Z}(\sigma^a\delta^{bN}+\delta^{aN}\sigma^b)\right]
+\mbox{ contact terms}\, ,
\label{SymmetricEqMot1}\\
R_2 & = & Z_g h\sigma\left(\sigma^a\sigma^b-
\frac{1}{Z}\frac{\delta^{ab}}{N}\right)+\mbox{ contact terms}\, .
\label{SymmetricEqMot2}
\end{eqnarray}
From the above observations we can deduce the following structure of the
mixing matrix:
\begin{eqnarray}
\left(\begin{array}{c}
\opl S\opr\\
\opl D\opr\\
\opl R\opr
\end{array}
\right) =
\left[\begin{array}{ccc}
Z_{SS} & Z_{SD} & Z_{SR}\\
0      & Z_{DD} & 0     \\
0      & 0      & Z_{RR}
\end{array}
\right]
\left(\begin{array}{c}
 S\\
 D\\
 R
\end{array}
\right)\, ,
\label{BlockStructure}
\end{eqnarray}
where we used a block notation.

The structure (\ref{BlockStructure}) 
can be easily verified on the two-loop expressions given in 
Eqs. (\ref{RenConst2.2.A})-(\ref{RenConst2.2.C}). In the
new basis the renormalization matrix of Eq. (\ref{RenConst2.2}) becomes
\begin{eqnarray}
Z'_{(2,2)} & = & 1-\frac{1}{4\pi\epsilon}g A' 
+\frac{1}{16\pi^2\epsilon^2}g^2 B'
+ \frac{1}{32\pi^2\epsilon}g^2 C' +O(g^3)\, ,
\end{eqnarray}
where
{\footnotesize
\begin{eqnarray}
A' & = &\left[\begin{array}{ccccccc}
0 & 2 & -2 & 0 & -1 & 1 & -1 \\ 
0 & 2 & 0 & 0 & 0 & 0 & -2 \\ 
0 & -(N-2) & 2(N-1) & 0 & (N-2) & -(N-2) & (N-1) \\ 
0 & 0 & 0 & 2 & 0 & 0 & 0\\ 
0 & 0 & 0 & 0 & 2 & 0 & 0 \\ 
0 & 0 & 0 & 0 & 0 & 2 & 2N \\ 
0 & 0 & 0 & 0 & 0 & 4 & 2(N-1)  
\end{array}
\right]\, ,\\ 
B' & = &\left[\begin{array}{ccccccc}
0 & -(N-4) & -2 & 0 & -1 & -1 & -2 \\
0 & -2(N-3)& 0 & 0 & 0 & -4 & -4 \\
0 & -2(N-2)& 2(N-1)  & 0 & 2(N-2)  & 2 & 2(N-1) \\
0 & 0 & 0 & -2(N-3) & 0 & 0 & 0 \\
0 & 0 & 0 & 0 & -2(N-3) & 0 & 0 \\
0 & 0 & 0 & 0 & 0 & 2(N+3)  & 4N\\
0 & 0 & 0 & 0 & 0 & 8 & 2(3N-1) 
\end{array}
\right]\, ,\\
C' &=& \left[\begin{array}{ccccccc}
N-3 & \frac{3N-16}{2}& -(6N-11)& -2(N-2) & \frac{5}{2} & -\frac{2N+5}{2} & 0 \\
0 & -4(N+2) & 0 & 0 & 4N & -8(N-1) & 0 \\
4(N-4) & 4(N-2) & -8(2N-3)& -4(N-2) & -4(N-2)& 2(2N-3) & 0 \\
0 & 0 & 0 & 0 & 0 & 0 & 0 \\
0 & 0 & 0 & 0 & 0 & 0 & 0 \\ 
0 & 0 & 0 & 0 & 0 & -4(N-2)& 0 \\
0 & 0 & 0 & 0 & 0 & 0 & -4(N-2)
\end{array}
\right]\nonumber\, .\\
\end{eqnarray}
}
%
%************************************************************************
%
\section{Anomalous Dimensions}
\label{AnomalousSection}
In this Section we list the anomalous dimensions for some of the 
composite operators treated in the previous one. 

First of all we shall recall the basic definitions.
This will be useful in order to fix our notations. 
Let us consider a set of operators which is ``complete'' under mixing:
\begin{eqnarray}
\opl {\cal O}_i\opr (x) = \sum_{i=1}^{\cal N} Z^{\cal O}_{ij} 
{\cal O}_j(x)\, .
\label{MixingGen}
\end{eqnarray}
We adopt the convention of Ref. \cite{Collins:1984xc}: the operators on
the r.h.s. of Eq. (\ref{MixingGen}) are local functions
of the \underline{renormalized} fields $\pg$, $\sigma$, and of the 
\underline{renormalized} parameters $h$ and $g$.
We can rewrite Eq. (\ref{MixingGen}) in a matrix notation as follows:
\begin{eqnarray}
\opl {\cal O}\opr (x)  =  Z^{\cal O} {\cal O}(x) \, ,
\end{eqnarray} 
where ${\cal O}$ $\equiv$  $[ {\cal O}_1,\dots,{\cal O}_{\cal N} ]^T$,
and $Z^{\cal O}$ $\equiv$  $[Z^{\cal O}_{ij}]_{1\le i,j\le {\cal N}}$.

The anomalous dimension matrix of ${\cal O}$ is defined 
implicitly by the following relation:
\begin{eqnarray}
\left.\mu   \frac{d}{d\mu   }\right|_{g_B,h_B,\epsilon}\opl {\cal O}\opr (x) = 
\gamma^{{\cal O}}(g)\opl {\cal O}\opr (x)\, .
\label{OperatorRunning} 
\end{eqnarray}
Equation (\ref{OperatorRunning}) is a shorthand for the 
RG equation:
\begin{eqnarray}
\left[\mu \frac{\partial}{\partial \mu   }+\beta(g)
\frac{\partial}{\partial g}+\gamma_h(g)h\frac{\partial}{\partial h}
-\frac{n}{2}\gamma(g)\right]\Gamma_{{\cal O}}^{(n)}
=\gamma^{{\cal O}}(g)\;\Gamma_{{\cal O}}^{(n)}\, ,
\label{OperatorRunning2} 
\end{eqnarray}
where $\Gamma_{{\cal O}}^{(n)}=\Gamma_{{\cal O}}^{(n)}(p_i;g,h,\mu   )$ 
is the vertex function with one $\opl{\cal O} \opr$ insertion
and $n$ $\pg$-``legs''.  The running of the magnetic field is
given by the anomalous dimensions defined below:
\begin{eqnarray}
\gamma_h(g) \equiv \frac{1}{2}\gamma(g)+\frac{1}{g}\beta(g)\, .
\end{eqnarray}

In Eq. (\ref{MixingGen}) and in the previous Section 
we used the convention of defining renormalized operators as
products of renormalized fields times an operator renormalization 
constant. This convention makes it clear the origin of
composite-operator renormalization. This is necessary 
because of short-distance singularities which arise when
we consider two or more fields at the same space-time point.
However, our convention is unpractical for giving explicit formulae for 
the anomalous dimensions defined in Eq. (\ref{OperatorRunning}).
The renormalization of the fields ($\pg$ and $\sigma$),
which is implicit on the r.h.s.
of Eq. (\ref{MixingGen}), must be taken into account.
Let us consider the case of a set of operators ${\cal O}$ which are 
the product of $n({\cal O})$ fields ($\pg$, $\sigma$ or their derivatives 
$\partial_{\mu}\pg$, $\partial^2\sigma$, etc.).
Schematically we could write  ${\cal O}  =
\mbox{`` }\partial^{d_{\cal O}}\sg^{n({\cal O})}\mbox{ ''} $.
In this case the following formula can be obtained from
Eqs. (\ref{MixingGen}) and (\ref{OperatorRunning}):
\begin{eqnarray}
\gamma^{{\cal O}}(g) = \left(\left.\mu 
\frac{\partial}{\partial\mu   } \right|_{g_B,\epsilon}
Z^{{\cal O}}\right)(Z^{{\cal O}})^{-1}-
\frac{n({\cal O})}{2}\gamma(g) = 
\left(\beta(g)
\frac{\partial}{\partial g  }
Z^{{\cal O}}\right)(Z^{{\cal O}})^{-1}-
\frac{n({\cal O})}{2}\gamma(g)\, .\nonumber\\
\label{ExplicitAnomalous}
\end{eqnarray}
A simple example of Eq. (\ref{ExplicitAnomalous}) is the computation 
of the anomalous dimension of the elementary field $\sg$ itself.
In this case ${\cal O} = \sg$, $n({\cal O}) = 1$ and $Z^{\cal O} = 1$.
From Eq. (\ref{ExplicitAnomalous}) it follows that
\begin{eqnarray}
\gamma_{\sg}(g) = -\frac{1}{2} \gamma(g)\, ,
\end{eqnarray}
a well-known result which agrees with the definition Eq. 
(\ref{OperatorRunning}), see also Eq. (\ref{OperatorRunning2}).

Equation (\ref{OperatorRunning}) can be solved by introducing the 
renormalization-group invariant (RGI) operators $[{\cal O}]_{RGI}$ as follows:
\begin{eqnarray}
[{\cal O}]_{RGI} \equiv g^{-\gamma_0^{{\cal O}}/\beta_0}
{\rm Gexp}\left\{\int_0^g\left(
z^{\gamma_0^{{\cal O}}/\beta_0} 
\frac{\gamma^{{\cal O}}(z)}{\beta(z)}z^{-\gamma_0^{{\cal O}}/\beta_0}+
\frac{\gamma_0^{{\cal O}}}{\beta_0 z}
\right)dz\right\}\opl{\cal O} \opr\, ,
\label{RGIOpDef}
\end{eqnarray}
where we expanded $\gamma^{\cal O}(g) = \sum_{k=0}^{\infty}
\gamma^{\cal O}_k g^{k+1}$, 
and ${\rm Gexp}$ denotes the $g$-ordered exponential:
\begin{eqnarray}
{\rm Gexp}\left\{\int_0^g \!\!\!\!dz\,\,f(z)\right\} = 
1+\int_0^g\!\!\!\! dz\,\, f(z)
+\int_0^g\!\!\!\! dz_1\!\int_0^{z_1}\!\!\!\!
dz_2\,\,f(z_1)\,f(z_2)+\dots\,  .
\end{eqnarray}
If ${\cal O}$ renormalizes multiplicatively, then $\gamma^{\cal O}(g)$ is a 
scalar function and Eq. (\ref{RGIOpDef}) simplifies to:
\begin{eqnarray}
[{\cal O}]_{RGI} = g^{-\gamma^{\cal O}_0/\beta_0}
\exp\left\{\int_0^g\left(\frac{\gamma^{\cal O}(z)}{\beta(z)}+
\frac{\gamma_0^{\cal O}}{\beta_0 z}
\right)dz\right\}\opl{\cal O} \opr\, .
\end{eqnarray}

The new operator $[{\cal O}]_{RGI}$ is renormalization-group invariant,
i.e. its insertions satisfy:
\begin{eqnarray}
\left[\mu   \frac{\partial}{\partial \mu   }+\beta(g)
\frac{\partial}{\partial g}+\gamma_h(g)h\frac{\partial}{\partial h}-
\frac{n}{2}\gamma(g)\right]
\Gamma_{{\cal O}_{RGI}}^{(n)} = 0\, .
\label{RGIRunning}
\end{eqnarray}
Moreover, unlike $\opl{\cal O}\opr$, $[{\cal O}]_{RGI}$ is scheme-independent. 

Notice that, in general, we are not interested in the whole set of
operators which mix under renormalization, see Eq. (\ref{MixingGen}).
A part of them is proportional to the non $O(N)$-covariant operator
$\alpha(x)$, defined in Eq. (\ref{AlphaDefinition}).
In general we shall consider
on-shell correlation functions in the $h\to 0$ limit. In this case Eq. 
(\ref{AlphaTimesSomething})
can be used to eliminate the non-covariant operators. 

Does any RG equation hold for the ``reduced'' basis
which is obtained after the elimination of $\alpha(x)$?
The answer is positive. First we choose an operator basis such that
the operator $\alpha(x)$ always appears in the combination 
$[\alpha+(\partial\sg_B)^2]$. Because of the general structure
of composite operators outlined in Sec. \ref{CompositeOperatorsSection},
we can distinguish two classes of operators:
$(i)$ $O(N)$-covariant operators, let us denote them collectively as
${\cal A}$; $(ii)$ products of $O(N)$-covariant operators
times some power of $[\alpha+(\partial\sg_B)^2]$, 
let us denote them as ${\cal B}$. 
Because of Eq. (\ref{AlphaBareTimesOperator})  the operators 
${\cal B}$ vanish on-shell, for $h\to 0$. Therefore the renormalization
matrix has the following triangular form:
\begin{eqnarray}
\left(\begin{array}{c}
\opl {\cal A} \opr \\
\opl {\cal B} \opr
\end{array}
\right)=
\left[
\begin{array}{cc}
Z_{\cal AA} & Z_{\cal AB}\\
0 & Z_{\cal BB}
\end{array}
\right]
\left(\begin{array}{c}
{\cal A}\\
{\cal B}
\end{array}
\right)\, .
\label{TriangForm}
\end{eqnarray}
It follows that the operators ${\cal A}$
satisfy the following RG equation (in shorthand 
notation):
\begin{eqnarray}
\left.\mu   \frac{d}{d\mu   }\right|_{g_B,h_B,\epsilon}\!\!\!
\opl {\cal A}\opr^{{\rm on\, shell},h\to 0} = 
\gamma_{{\cal AA}}(g)\opl {\cal A}\opr^{{\rm on\, shell},h\to 0}\, .
\label{RGEquationOnShell}
\end{eqnarray}
The anomalous dimension matrix $\gamma_{\cal AA}(g)$
to be used in Eq. (\ref{RGEquationOnShell})
is the same as if there were no mixing with the operators ${\cal B}$.

We come now to the compilation of our list of anomalous dimensions, 
classifying them, as in the previous Section, according to their 
$O(N)$ symmetry and canonical dimension.
%
%*****************************************************
%
\subsection{$O(N)$ Invariant Operators of Dimension 2}
\label{Anomalous20Section}
In this particular case, we were able to express the renormalization matrix
$Z^{(2,0)}$ in terms of the field and 
the coupling constant renormalizations, $Z$ and $Z_g$. This allows us to write
the corresponding anomalous dimension matrix in terms of the beta function 
and the field anomalous dimensions, $\beta(g)$ and $\gamma(g)$.
The result is given below:
\begin{eqnarray}
\gamma^{(2,0)}(g) = \frac{1}{g}\left[
\begin{array}{ccc}
\beta(g) & \frac{1}{2}[\beta(g)-g\beta'(g)-g^2\gamma'(g)] & 
-\frac{1}{2}g^2\gamma'(g)\\
0 & 2\beta(g) - g\beta'(g)-g^2\gamma'(g) & -g^2\gamma'(g) \\
0 & -\beta(g)+g\beta'(g)+g^2\gamma'(g) & \beta(g)+g^2\gamma'(g)
\end{array}
\right]\, .
\end{eqnarray}
In particular from Eqs. (\ref{EnergyMomentum}) and (\ref{EMRen}), we 
get the anomalous dimensions of the energy-momentum
tensor\footnote{This 
result cannot be obtained by applying directly 
Eq.  (\ref{ExplicitAnomalous}), since this is valid only for operators
which are products of renormalized fields (and their derivatives)
without explicit $g$ dependence. Equation (\ref{ExplicitAnomalous})
can be applied, for instance, to $\widehat{T}_{\mu\nu} \equiv gT_{\mu\nu}$.} 
$\gamma_T(g) = 0$. This is a well known fact: 
conserved currents are RGI, see Eq. 
(\ref{RGIRunning}).
%
%*********************************************************
%
\subsection{Antisymmetric Rank 2 Operators}

We can apply to the Noether current $j^{(a,b)}_{\mu}$ the same observations
concerning the energy-momentum tensor made in the previous Subsection.
We get $\gamma_{j}(g)=0$. 

The three operators $A^{(n)}$, $n=0,1,2$ defined in Eqs. 
(\ref{AntiSymmetric0})-(\ref{AntiSymmetric2}) have $\gamma_A(g)=\beta(g)/g$,
and it is easy to see that this implies
$[A^{(n)}]_{RGI} = 1/g\, \opl A^{(n)}\opr$. This is a trivial consequence 
of the fact that the $A^{(n)}$ are linear combinations of 
$\partial_{\mu}j^{(a,b)}_{\nu}$.
%
%******************************************************************
%
\subsection{Symmetric Rank 2 Operators}
\label{SymmetricAnomalousContinuum}

The anomalous dimensions of the dimension-0 operator 
(\ref{DefinitionDimension0Symmetric}) have been computed up to
four-loop  order in Refs. \cite{Bernreuther:1986js,Wegner:1989ss}. Using Eqs. 
(\ref{ZetaDimension0Symmetric}) and (\ref{ExplicitAnomalous}) with
$n({\cal O})=2$, we can write the first two terms of the perturbative
expansion:
\begin{eqnarray}
\gamma^{(0,2)}(g) = \frac{N}{2\pi}g+O(g^3) \, .
\end{eqnarray}

Let us now consider the anomalous dimensions matrix of the dimension-2,
symmetric traceless rank-2 $O(N)$ tensors.
It is convenient to choose 
a basis which simplifies the solution of the RG equations.
Such a basis can be constructed from the $(S,R,D)$ operators, defined 
in the text above Eqs. (\ref{SymmetricEqMot1}) and (\ref{SymmetricEqMot2}). 
The new basis will contain operators of definite spin (0 or 2)
and is defined as follows:
\begin{eqnarray}
Q^{(1)R}_{\mu\nu} & \equiv & \opl S^{(1)}_{\mu\nu}\opr - 
\frac{1}{d}\delta_{\mu\nu}\delta^{\rho\sigma}
\opl S^{(1)}_{\rho\sigma}\opr\, ,
\label{Q1Def}\\
Q^{(2)R} & \equiv & \opl S^{(3)} \opr\, ,\\
Q^{(3)R}_{\mu\nu} & \equiv & \opl S^{(5)}_{\mu\nu} \opr -
\frac{1}{d}\delta_{\mu\nu}\delta^{\rho\sigma}
\opl  S^{(5)}_{\rho\sigma}\opr\, ,\\
Q^{(4)R}_{\mu\nu} & \equiv & \opl D^{(1)}_{\mu\nu} \opr -
\frac{1}{d}\delta_{\mu\nu}\delta^{\rho\sigma}
\opl  D^{(1)}_{\rho\sigma}\opr\, ,\\
Q^{(5)R} &\equiv& \opl D^{(2)} \opr\, ,\\
Q^{(6)R} &\equiv& \opl R^{(1)} \opr\, ,\\
Q^{(7)R} &\equiv& \opl R^{(2)} \opr\, .\label{Q7Def}
\end{eqnarray}
Among the above operators, $Q^{(2)R}$, $Q^{(5)R}$, $Q^{(6)R}$ and $Q^{(7)R}$
are Lorentz scalars (i.e. they have spin 0), while $Q^{(1)R}_{\mu\nu}$,
$Q^{(3)R}_{\mu\nu}$ and $Q^{(3)R}_{\mu\nu}$ are rank-2, symmetric, traceless
Lorentz tensors (i.e. they have spin 2). These two classes do not
mix under renormalization. Notice that, in the above definitions, we did not 
use the notation $\opl\, \cdot\, \opr$, since, for instance
\begin{eqnarray}
Q^{(1)R}_{\mu\nu}\neq\opl  S^{(1)}_{\mu\nu} - 
\frac{1}{d}\delta_{\mu\nu}\delta^{\rho\sigma} S^{(1)}_{\rho\sigma}\opr\, .
\end{eqnarray}

Because of the observations made above and in Sec. \ref{SymmetricZetaSection},
the structure of the anomalous dimension matrix is
\begin{eqnarray}
\gamma^{(2,2)}(g) = \left[\begin{array}{ccccccc}
\gamma^{(2,2)}_{11} & 0 & \gamma^{(2,2)}_{13} & \gamma^{(2,2)}_{14} & 
0 & 0 & 0 \\
0 & \gamma^{(2,2)}_{22} & 0 & 0 & \gamma^{(2,2)}_{25} & \gamma^{(2,2)}_{26} &
\gamma^{(2,2)}_{27} \\
\gamma^{(2,2)}_{31} & 0 & \gamma^{(2,2)}_{33} & \gamma^{(2,2)}_{34} & 
0 & 0 & 0 \\
0 & 0 & 0 & \gamma^{(2,2)}_{44} & 0 & 0 & 0 \\
0 & 0 & 0 & 0 & \gamma^{(2,2)}_{55} & 0 & 0 \\
0 & 0 & 0 & 0 & 0 & \gamma^{(2,2)}_{66} & \gamma^{(2,2)}_{67} \\
0 & 0 & 0 & 0 & 0 & \gamma^{(2,2)}_{76} & \gamma^{(2,2)}_{77}
\end{array}
\right]\, . 
\label{AnomalousForm22}
\end{eqnarray}
Moreover, since both $Q^{(4)R}_{\mu\nu}$ and $Q^{(5)R}$ are total
space-time derivatives of the dimension zero operator 
(\ref{DefinitionDimension0Symmetric}), we get 
\begin{eqnarray}
\gamma^{(2,2)}_{44}(g) = \gamma^{(2,2)}_{55}(g) = \gamma^{(0,2)}(g)\, .
\label{DerivativesOf02}
\end{eqnarray}

Both the statements (\ref{AnomalousForm22}), and
(\ref{DerivativesOf02}), are easily verified on the two-loop result:
\begin{eqnarray}
\gamma_{(2,2)}(g) & = & -\frac{1}{4\pi}\mathfrak{A} g+
\frac{1}{16\pi}\mathfrak{B} g^2+O(g^3)\, ,
\label{Gamma22Expansion}
\end{eqnarray}
where 
{\footnotesize
\begin{eqnarray}
\mathfrak{A} & = & \left[\begin{array}{ccccccc}
2(N-1) & 0 & -2 & 0 & 0 & 0 & 0 \\
0 & 2N & 0 & 0 & 0 & 0 & -2 \\
0 & 0 & 4(N-1) & 0 & 0 & 0 & 0 \\
0 & 0 & 0 & 2N & 0 & 0 & 0 \\
0 & 0 & 0 & 0 & 2N & 0 & 0 \\
0 & 0 & 0 & 0 & 0 & 2N & 2N \\
0 & 0 & 0 & 0 & 0 & 4 & 4(N-1)
\end{array}
\right]\, ,\\
\mathfrak{B} & = & \left[\begin{array}{ccccccc}
N-3 & 0 & -(6N-11) & -2(N-2) & 0 & 0 & 0 \\
0 & -4(N+2)  & 0 & 0 & 4N & -8(N-1) & 0 \\
4(N-4) & 0 &-8(2N-3) & -4(N-2) & 0 & 0 & 0 \\
0 & 0 & 0 & 0 & 0 & 0 & 0 \\
0 & 0 & 0 & 0 & 0 & 0 & 0 \\
0 & 0 & 0 & 0 & 0 & -4(N-2) & 0 \\
0 & 0 & 0 & 0 & 0 & 0 & -4(N-2)
\end{array}
\right]\, .\nonumber\\
\label{Gamma22g2}
\end{eqnarray}
}
%
%************************************************************************
%
\section{Lattice Model and Lattice Composite Operators}
\label{LatticeOperatorsSection}

We shall now consider the  non--perturbative lattice definition 
of the theory which is formally described by
Eq. (\ref{ContinuumAction}).
Moreover we shall study some examples of lattice composite operators
in order to understand how
renormalization must be adapted to the lattice regularization.
In perturbation theory, there are three main modifications 
concerning this subject:
\begin{itemize}
\item The lattice breaks the symmetry of the action 
(\ref{ContinuumAction}) under space--time transformations. 
As a consequence the operators $A$ and $B$ on the left and right-hand
sides of Eq. (\ref{ContinuumRen}), may belong to different 
representations of the Lorentz group.
\item In general there exist different lattice counterparts of a
given continuum operator. All of them differ by ``irrelevant'' terms.
After renormalization, these lattice discretization are supposed to 
converge to the same 
continuum limit with corrections of order $O(a^2\log^ka)$.
Better convergence rates can be obtaining by applying the Symanzik
improvement program. If, for instance, action and operator improvement
is non--perturbatively implemented up to $O(a^2)$, then 
we expect an improved convergence rate $O(a^4\log^k a)$. 
\item As for any cutoff regularization (dimensional regularization
is somehow an exception in this respect), correlation functions
of composite operators may have power-like divergences
of the type $a^{-2p}\log^k a$. This requires power subtractions.
Considering again Eq. (\ref{ContinuumRen}), it may happen that 
${\rm dim}[B]< {\rm dim}[A]$, and that $Z_{AB}\sim a^{-2p}\log^k a$
as $a\to 0$.
\end{itemize}
Examples of the above remarks can be given in perturbation theory.
Nevertheless they are widely believed to hold beyond perturbation theory.
The last one, in particular, implies severe difficulties in the
non--perturbative renormalization of some composite operators.

The simplest discretization of the action (\ref{ContinuumAction})
is given by Eq. (\ref{LatticeAction}).
For a discussion of other equivalent choices we refer to
\cite{Caracciolo:1998ir}.

As in the continuum, lattice perturbation theory can be made infrared 
finite by adding an external magnetic field:
\begin{eqnarray}
S^{\rm latt}[\sg] = \frac{1}{2g_L}\sum_{
x\in {\mathbb Z}^2\!\!,\,\mu}
(\partial_{\mu}\sg)_x^2-\frac{h_L}{g_L}\sum_{x\in {\mathbb
Z}^2}\sigma^N_x\, .
\label{LatticeActionPlusMagneticField}
\end{eqnarray}
Moreover, one must keep track of the measure
contribution.
This can be done by adding a new term to the action
(\ref{LatticeActionPlusMagneticField}), yielding:
\begin{eqnarray}
S^{\rm latt}_{\rm TOT}[\sg] = S^{\rm latt}[\sg]+\sum_{x\in
\mathbb{Z}^2} \log \sigma_x^N\, .
\label{LatticeActionTotal}
\end{eqnarray}
The procedure outlined above is by no means unique. There exist
alternative possibilities for regularizing the infrared singularities
which plague perturbation theory. A theoretically appealing 
approach consists in putting the model in a finite box 
\cite{Hasenfratz:1984jk}.
The independence of the perturbative series upon the infrared regularization
has been questioned in \cite{Patrascioiu:1995pf}.

The correlation functions of $\pg_x$'s and $\sigma_x$'s fields,
have a finite continuum limit 
if the bare parameters and fields are properly renormalized:
\begin{eqnarray}
g_L & = & Z_{g,L} g\, ,\label{LatticeRenModel1}\\
\pg_x & = & (Z_L)^{1/2}\pg(x)\, ,\\
\sigma_x & = & (Z_L)^{1/2}\sigma(x)\, ,\\
h_L & = & \frac{Z_{g,L}}{Z_L^{1/2}}h\, .\label{LatticeRenModel4}
\end{eqnarray}
In general the constants $Z_{g,L}$ and $Z_L$ are fixed by imposing
appropriate renormalization conditions. As long as perturbation theory
is concerned, we may choose $Z_{g,L}$ and $Z_L$ in such a way that 
renormalized lattice correlation functions match continuum \MS ones.
We shall therefore consider the renormalized fields and parameters
on the r.h.s. of Eqs. (\ref{LatticeRenModel1})--(\ref{LatticeRenModel4}), 
as \MS quantities.

The perturbative expansion of lattice renormalization constants has the 
general form:
\begin{eqnarray}
Z(g,\mu   a) = 
\sum_{l = 0}^{\infty}\sum_{n = 0}^l Z_n^{(l)} g^l \log^n \!\mu  a\, .
\end{eqnarray}
The renormalization constants $Z^L$ and $Z^L_g$ 
have been computed up to three-loop order in
perturbation theory \cite{Caracciolo:1995bc,Shin:1998bh,Alles:1999fh}. 
For greater convenience of the reader we report here the 
one-loop result:
\begin{eqnarray}
Z_{g,L}  =  1+\frac{N-2}{4\pi}g_L\log (\frac{\mbar^2 a^2}{32})-
\frac{1}{4}g_L+O(g_L^2)\, ,\quad 
Z_L = 1+\frac{N-1}{4\pi}g_L\log (\frac{\mbar^2 a^2}{32})+O(g_L^2)
\, .\vspace{-2.0cm}\nonumber\\
\end{eqnarray}
The lattice beta-function $\beta^L(g_L)$ and anomalous dimensions 
$\gamma^L(g_L)$ are obtained from $Z_{g,L}$ and $Z_L$ as follows:
\begin{eqnarray}
\beta^L(g_L) = \frac{-a\frac{\partial}{\partial a}\log Z_{g,L}}
{1-g_L\frac{\partial}{\partial g_L}\log Z_{g,L}}g_L\, ,\quad
\gamma^L(g_L) = \left(a\frac{\partial}{\partial a}-
\beta^L(g_L)\frac{\partial}{\partial g_L}\right)\log Z_{g,L}\, .
\end{eqnarray}

Let us now consider composite-operator renormalization.
The role of the $O(N)$ symmetry in restricting operator mixing is the same
as in the continuum. The proof in Sec. \ref{CompositeOperatorsSection}
does not rely on any
regularization scheme as long as $O(N)$ symmetry is broken uniquely
by the external magnetic field $h$. 
On the r.h.s of Eq. (\ref{ContinuumRen}) we must consider products
of operators transforming like $A$, times powers of $\alpha_L$.
The form of $\alpha_L$, i.e. the lattice counterpart of $\alpha(x)$,
see Eq. (\ref{AlphaDefinition}), is dictated by the $O(N)$ transformation
properties of the lattice action. For the
discretization (\ref{LatticeActionPlusMagneticField}), we get
\begin{eqnarray}
\alpha^L_x &\equiv & \frac{1}{\sigma_x}\left[h_L+\partial^2\sigma_x-
g_L\frac{1}{\sigma_x}\right]\, ,
\label{AlphaLatticeDefinition}
\end{eqnarray}
where $\partial^2 \equiv \sum_{\mu} \partial^-_{\mu}\partial_{\mu}$
is the lattice laplacian. The last term in
Eq. (\ref{AlphaLatticeDefinition}) is not present in the continuum and
arises because of the measure term, see Eq. (\ref{LatticeActionTotal}).
As in the continuum renormalized theory, we can get rid of the
operators proportional to $\alpha_L$ in the $h\to 0$ limit, as long 
as on-shell matrix elements are considered. In  particular, 
the following identity follows from
Eqs. (\ref{AlphaLatticeDefinition}) and (\ref{LatticeActionTotal}):
\begin{eqnarray}
\alpha_x^L = h \sigma_x  +\sg_x\cdot \partial^2 \sg_x + 
   g_L \pg_x\cdot {\delta S^{\rm latt}_{\rm TOT}\over \delta \pg_x}-
   g_L\, .
\label{AlphaLatticeOnShell}
\end{eqnarray}

We conclude this overview by a simple remark. 
In Sec. (\ref{CompositeOperatorsSection}) we discussed the
construction of renormalized composite operators.
This construction renormalizes the correlation functions 
of $\pg_x$ and $\sigma_x$ 
fields with a single composite operator insertion. This is enough
for renormalizing correlation functions with multiple operator
insertions as long as the insertions are made at distinct 
{\it physical} positions.
When lattice correlation functions are studied, for instance
a two-point function $\<{\cal O}_x{\cal Q}_y\>$, we must
keep different composite operators at distances $|x-y|=O(\xi)$.
This implies that $|x-y|\to\infty$ in the continuum limit (recall 
that we set the lattice spacing $a=1$).
%
%*****************************************************
%
\subsection{$O(N)$ Invariant Operators of Dimension 2}
\label{D2IS0LATT}

As we explained above, there is some freedom in choosing lattice
composite operators, if we are not interested in improving their
approach to the continuum limit. 
We make the choice of discretizing the operators
of Sec. \ref{D2IS0} by substituting the space-time derivative
$\partial_{\mu}$
with the symmetric lattice derivative $\overline{\partial}_{\mu}$,
where $(\overline{\partial}_{\mu}f)_x = (f_{x+\mu}-f_{x-\mu})/2$.
Moreover, we shall neglect contributions coming from the identity operator.
Indeed such terms cancels in connected correlation functions.
The renormalization structure is given below:
\begin{eqnarray}
\opl \partial_\mu   \sg\cdot   \partial_{\nu}\sg\opr (x)
&=& Z^{L(2,0)}_{11}(\partial_\mu   \sg\cdot   \partial_{\nu}\sg)_x
+Z^{L(2,0)}_{12}\delta_{\mu   \nu}(\overline{\partial}_\mu   \sg)_x^2+
\nonumber\\
&&+Z^{L(2,0)}_{13}\delta_{\mu   \nu}
(\overline{\partial}\sg)_x^2
+Z^{L(2,0)}_{14}\delta_{\mu   \nu}\alpha^L_x\label{Ren20.1}\, ,\\
\delta_{\mu   \nu}\opl(\partial_\mu   \sg)^2\opr (x)& = & 
Z^{L(2,0)}_{22}\delta_{\mu   \nu}(\overline{\partial}_\mu   \sg)_x^2
+Z^{L(2,0)}_{23}\delta_{\mu   \nu}
(\overline{\partial}\sg)_x^2+\nonumber\\
&&+Z^{L(2,0)}_{24}\delta_{\mu   \nu}\alpha^L_x\, ,
\label{Ren20.2}\\
\opl(\partial \sg)^2\opr (x) & = & Z^{L(2,0)}_{33}
(\overline{\partial}\sg)_x^2+Z^{L(2,0)}_{34}\alpha^L_x\, ,\\
\opl \alpha \opr (x) & = & Z^{L(2,0)}_{43}
(\overline{\partial}\sg)_x^2+Z^{L(2,0)}_{44}\alpha^L_x\, .\label{Ren20.4}
\end{eqnarray}
Notice that the operator 
$\delta_{\mu   \nu}(\overline{\partial}_\mu   \sg)_x^2$
does not renormalize as $(\partial_\mu   \sg\cdot   \partial_{\nu}\sg)_x$
because of the lack of Lorentz invariance.

We now list the one-loop perturbative expressions for the constants entering in
Eqs. (\ref{Ren20.1})--(\ref{Ren20.4}). Some of these constants have
been already computed in \cite{Buonanno:1995us}.
We give them here in order to correct a few misprints.
\begin{eqnarray}
Z^{L(2,0)}_{11} & = & 1-\frac{N-2}{4\pi}g_L\log (\frac{\mbar^2 a^2}{32})+
\frac{1}{\pi}g_L+O(g_L^2)\, ,\label{Ren20.11.one-loop}\\
Z^{L(2,0)}_{12} & = &
\left(\frac{1}{2}-\frac{3}{2\pi}\right)g_L+O(g_L^2)\, ,\\
Z^{L(2,0)}_{13} = Z^{L(2,0)}_{23} & = & 
-\frac{1}{8\pi}g_L\log (\frac{\mbar^2
a^2}{32})-\frac{1}{2\pi}g_L+O(g_L^2)\, ,\\
Z^{L(2,0)}_{14} = Z^{L(2,0)}_{14} & = & 
-\frac{N-1}{8\pi}g_L\log (\frac{\mbar^2
a^2}{32})-\frac{N-1}{8}g_L+O(g_L^2)\, ,\\
Z^{L(2,0)}_{33} & = & 1-\frac{N-1}{4\pi}g_L\log (\frac{\mbar^2 a^2}{32})+
\left(\frac{1}{2}-\frac{5}{4\pi}\right)g_L+O(g_L^2)\, ,\\
Z^{L(2,0)}_{34} & = & -\frac{N-1}{4\pi}g_L\log (\frac{\mbar^2 a^2}{32})+
(N-1)\left(\frac{1}{4\pi}-\frac{1}{4}\right)g_L+O(g_L^2)\, ,\\
Z^{L(2,0)}_{43} & = & \frac{1}{4\pi}g_L\log (\frac{\mbar^2 a^2}{32})+
\frac{1}{4}g_L+O(g_L^2)\, ,\\
Z^{L(2,0)}_{44} & = & 1+\frac{1}{4\pi}g_L\log (\frac{\mbar^2 a^2}{32})+
\frac{1}{4}g_L+O(g_L^2)\, ,
\end{eqnarray}
and $Z^{L(2,0)}_{22}= Z^{L(2,0)}_{11}+Z^{L(2,0)}_{12}$. Notice that we
were not able to express the above constants 
in terms of the field and coupling renormalization constants 
$Z$ and $Z_g$, as we did in the continuum, see Sec. \ref{D2IS0}.
This happens because of two reasons.
In Eqs. (\ref{Ren20.1})--(\ref{Ren20.4}) we used the symmetric
lattice derivative $\overline{\partial}_{\mu}$. As a consequence 
$(\overline{\partial}\sg)^2_x$ is not directly related to the lattice
action (\ref{LatticeAction}).
The second reason is that, since translation invariance explicitly broken,
there exists no exactly conserved energy-momentum tensor on the
lattice.
 
A naive discretization of the energy-momentum tensor can be written in
terms of the operators appearing in Eqs. (\ref{Ren20.1})--(\ref{Ren20.4}):
\begin{eqnarray}
T^{L}_{\mu   \nu,x}&\equiv&\frac{1}{g_L} 
\left(\overline{\partial}_\mu   \sg_x\cdot   \overline{\partial}_{\nu}\sg_x 
-\delta_{\mu   \nu}(\overline{\partial}\sg)_x^2\right)\, .
\label{NaiveEMT}
\end{eqnarray}
We can write down an energy-momentum tensor which is conserved 
up to lattice artifacts.
This can be done by mixing the naively discretized version 
(\ref{NaiveEMT}) with the other operators appearing in Eqs.  
(\ref{Ren20.1})--(\ref{Ren20.4}). We get
\begin{eqnarray}
T_{\mu   \nu}(x)&=& Z^{L(2,0)}_{TT}T^{L}_{\mu   \nu,x}+
Z^{L(2,0)}_{T2}
\frac{1}{g_L}\delta_{\mu   \nu}(\overline{\partial}_\mu   \sg)_x^2+
\label{EnergyMomentumRenormalization}\\
&&+Z^{L(2,0)}_{T3}\frac{1}{g_L}\delta_{\mu   \nu}
(\overline{\partial}\sg)_x^2
+Z^{L(2,0)}_{T4}\frac{1}{g_L}\delta_{\mu   \nu}\alpha^L_x\nonumber\, ,
\end{eqnarray}
where $T_{\mu \nu}(x)$ is the continuum renormalized energy-momentum
tensor, and the relevant renormalization constant are
\begin{eqnarray}
Z^{L(2,0)}_{TT} & = &
1+\left(\frac{1}{\pi}-\frac{1}{4}\right)g_L+O(g_L^2)\, ,
\label{EMZLatticePT}\\
Z^{L(2,0)}_{T2} & = & \left(\frac{1}{2}-\frac{3}{2\pi}\right)g_L+O(g_L^2)\, ,\\
Z^{L(2,0)}_{T3} & = & \left(\frac{5}{8\pi}-\frac{1}{4}\right)g_L+O(g_L^2)\, ,\\
Z^{L(2,0)}_{T4} & = & -\frac{N-1}{8\pi}g_L+O(g_L^2)\, .
\end{eqnarray}
Since the continuum energy-momentum tensor on the l.h.s. of Eq.
(\ref{EnergyMomentumRenormalization}) is conserved, the corresponding
renormalization constants satisfy:
\begin{eqnarray}
\left.\frac{\partial}{\partial a}\right|_{g_L}Z^{L(2,0)}_{Tj}=0\, .
\end{eqnarray}

Finally, notice that we can eliminate $\alpha_x^L$ from 
Eqs. (\ref{Ren20.1})--(\ref{Ren20.4}) using
Eq. (\ref{AlphaLatticeOnShell}).
However, $-\sg_x\cdot \partial^2 \sg_x$ is different from 
$(\overline{\partial} \sg)^2_x$, which appears in 
Eqs. (\ref{Ren20.1})--(\ref{Ren20.4}). 
The two operators are related by a finite renormalization:
\be
-\sg_x\cdot \partial^2 \sg_x = \zeta_1 (\overline{\partial} \sg)^2_x + 
                     \zeta_2 \alpha_x^L.
\ee
At one loop
\be
\zeta_1 = 1 + {g_L\over 2} \left(1 - {5\over 2\pi}\right)+O(g_L^2), 
\qquad\qquad
\zeta_2 = {N-1\over4} \left({1\over \pi} - 1\right) g_L+O(g_L^2)\, .
\ee
Then, considering only connected correlation functions, 
on-shell and for $h\to 0$, we have
\be
\alpha_x^L =\, - {\zeta_1\over 1 + \zeta_2} (\overline{\partial}
\sg)^2_x
\, .
\ee
Using this relation, we can write the energy-momentum tensor 
(\ref{EnergyMomentumRenormalization}) 
in terms of $O(N)$ invariant operators (as always on-shell,
and in the $h\to 0$ limit) as follows:
\begin{eqnarray}
T_{\mu \nu}(x) =
  Z_{TT}^{L,(2,0)} T^L_{\mu \nu,x} + 
  Z_{T2}^{L,(2,0)} {1\over g_L} \delta_{\mu \nu} 
      (\overline{\partial}_\mu \sg)^2_x + 
  \widetilde{Z}_{T3}^{L,(2,0)} {1\over g_L} \delta_{\mu \nu}
      (\overline{\partial} \sg)^2_x \, ,
\end{eqnarray}
where
\be
\widetilde{Z}_{T3}^{L,(2,0)} = Z^{L(2,0)}_{T3} - 
    {\zeta_1\over 1 + \zeta_2} Z^{L(2,0)}_{T4}.
\ee
%
%*********************************************************
%
\subsection{Antisymmetric Rank-2 Operators}

Since $O(N)$ symmetry is not broken by the lattice discretization, 
renormalization is quite simple for these operators.
One can explicitly construct the lattice Noether currents.
With the lattice action (\ref{LatticeAction}) we get the simple expression
\begin{eqnarray}
j^{L,ab}_{\mu,x   }\equiv\frac{1}{g_L}\left(\sigma^{a}_x\partial_\mu 
\sigma^{b}_x-\sigma^{b}_x\partial_\mu   \sigma^{a}_x\right)\, .
\label{LatticeCurrent}
\end{eqnarray}
The lattice current (\ref{LatticeCurrent}) 
satisfies (exactly) the following Ward identity:
\begin{eqnarray}
\partial^{-}_\mu   \langle j^{L,ab}_{\mu,x   } {\cal O} \rangle &=&
\langle \delta^{ab}_x {\cal O}\rangle\, ,
\label{WILattice}
\end{eqnarray}
where ${\cal O}$ is a generic composite operator (or a product of
composite operators), and $\delta^{ab}_x {\cal O}$ is the variation
of this operator induced by a rotation of the spin $\sg_x$:
\begin{eqnarray}
\delta^{ab}_x \sigma^c_y &= &\delta_{x,y}(\delta^{a,c}\sigma^b - 
\delta^{b,c}\sigma^a)\, .
\end{eqnarray}
Equation (\ref{WILattice})  guarantees that the lattice current 
(\ref{LatticeCurrent}) has the correct normalization, i.e.
$j^{ab}_\mu   (x) = j^{L,ab}_{\mu,x}$ (up to lattice artifacts).

We can construct lattice discretizations of 
the rank-2 antisymmetric operators 
(\ref{AntiSymmetric0})--(\ref{AntiSymmetric2}) in terms 
of  derivatives of the lattice current 
$j_\mu  = j^{L,ab}_{\mu,x}$.
\begin{eqnarray}
A^{L(0)} & \equiv &  g_L \sum_\mu   \partial^-_\mu   j_\mu\, ,\\
A^{L(1)}_{\mu \nu} & \equiv & \frac{1}{2}g_L\left(
\partial^-_\mu   j_{\nu}-\partial^-_\nu   j_{\mu}\right)\, , \\
A^{L(2)}_{\mu \nu} & \equiv & \frac{1}{2}g_L\left(
\partial^-_\mu   j_{\nu}+\partial^-_\nu j_{\mu}\right)
-\frac{1}{2}g_L\delta_{\mu\nu}\sum_{\rho} \partial^-_{\rho}   j_{\rho}\, .
\end{eqnarray}
Since the lattice current $j^{L,ab}_{\mu,x   }$ does not need 
any renormalization, we must renormalize only the coupling constant which
explicitly appears in front of the above definitions:
\begin{eqnarray}
\opl A^{(n)}\opr = Z_{g,L}^{-1}\, A^{L(n)}\, .
\end{eqnarray}
%
%
%******************************************************************
%
\subsection{Symmetric Rank 2 Operators}

We limit ourselves to the symmetric rank 2 operator of dimension 0.
We consider the natural discretization appearing below:
\begin{eqnarray}
\opl \sigma^a\sigma^b-\frac{1}{Z}\frac{\delta^{ab}}{N} \opr(x) & = & 
Z^{L(0,2)}\left( \sigma^a_x\sigma^b_x-\frac{\delta^{ab}}{N} \right)\, .
\end{eqnarray}
The relevant renormalization constant has been computed in Ref. 
\cite{Caracciolo:1995bc} up to two loops in perturbation theory:
\begin{eqnarray}
Z^{L(0,2)}& = &1-\frac{N}{4\pi}g_L\log (\frac{\mbar^2 a^2}{32})+
+g_L^2\left\{\frac{N(N-1)}{16\pi^2}
\log^2 (\frac{\mbar^2 a^2}{32})-\frac{N}{16\pi}
\log (\frac{\mbar^2 a^2}{32})\right\}\, .\nonumber\\
\end{eqnarray}
%
%****************************************************************
%
\section{Lattice Anomalous Dimensions}
\label{LatticeAnomalousSection}

In this Section we recall the basic definitions of 
the anomalous dimensions for {\it bare} lattice composite operators.

Let us consider, once again, the general mixing pattern
(\ref{MixingGen}):
\begin{eqnarray}
\opl {\cal O}_i\opr (x)=\sum_{j=1}^{\cal N}Z^{L,{\cal O}}_{ij}{\cal
O}^L_{j,x}\, ,
\end{eqnarray}
or its matrix formulation $\opl {\cal O}\opr = Z^{L,{\cal O}}{\cal
O}^L$. In  the previous formulae, we specified the regularization
scheme to be the lattice one. 

As for renormalized operators, the anomalous dimensions of the 
lattice operators ${\cal O}^L$ can be implicitly defined through a 
RG equation:
\begin{eqnarray}
-a\left.\frac{\partial}{\partial a}\right|_{g,\mu}{\cal O}^L =
\gamma^{L}_{\cal O}(g_L) {\cal O}^L\, .
\label{ShortRGELatt}
\end{eqnarray}
The derivative with respect to the cutoff $a$ has to be taken keeping 
the renormalized parameters (for instance the coupling $g$, and the
scale $\mu$) fixed.

Equation (\ref{ShortRGELatt}) is a shorthand for 
\begin{eqnarray}
\left[-a\frac{\partial}{\partial a}+\beta^L(g_L)\frac{\partial}{\partial g_L}
+\gamma_h^L(g)h
\frac{\partial}{\partial h}-\frac{n}{2}\gamma^L(g_L)\right]
\Gamma^{(n)}_{L,{\cal O}} =\gamma^{L}_{\cal O}(g_L)
\Gamma^{(n)}_{L,{\cal O}}\, ,
\end{eqnarray} 
where $\Gamma^{(n)}_{L,{\cal O}}$ is the vertex with one ${\cal O}^L$
insertion, and $\gamma^L_h(g_L) = \gamma^L(g)/2+\beta^L(g)/g$.

An explicit formula for the anomalous dimensions
$\gamma^L_{\cal O}(g_L)$ 
is easily obtained from the above definitions:
\begin{eqnarray}
\gamma^L_{\cal O}(g_L)\equiv \left(Z^{{\cal O},L}\right)^{-1}
\left[\left.a\frac{\partial}{\partial a}\right|_{g,\mu   }
Z^{{\cal O},L}\right]=\left(Z^{{\cal O},L}\right)^{-1}
\left[\left( a\frac{\partial}{\partial a}-\beta^L(g_L)
\frac{\partial}{\partial g_L}\right)
Z^{{\cal O},L}\right]\, .
\end{eqnarray} 
The knowledge of the continuum anomalous dimensions can be of great help
when computing their lattice cousins.
Indeed the following relation holds
\begin{eqnarray}
\gamma_{\cal O}^L(g_L) = (Z^{L,{\cal O}})^{-1}
\gamma^{\MMS}_{{\cal O}}(g_L/Z_{g,L})Z^{L,{\cal O}}-
(Z^{{\cal O},L})^{-1}\beta^L(g_L)
\frac{\partial}{\partial g_L}Z^{{\cal O},L}\, .
\label{LatticeContinuumAnomalous}
\end{eqnarray}
This equation yields $\gamma^L_{\cal O}(g_L)$ at $l$-loop order,
once $Z^{L,{\cal O}}$ is known at $(l-1)$ loops. In particular, if 
we write the perturbative expansion of $\gamma^L_{\cal O}(g_L)$ as 
\begin{eqnarray}
\gamma^L_{\cal O}(g_L) = \sum_{k=0}^{\infty}\gamma^{L,{\cal O}}_k
g_L^k\, ,
\end{eqnarray}
Eq. (\ref{LatticeContinuumAnomalous}) implies that 
$\gamma^{L,{\cal O}}_0= \gamma^{\cal O}_0$.
%
%**************************************************************
%
\section{Renormalization-Group Equations}
\label{RenormalizationGroupSection}
Let us now discuss how RG can be used
for ``resumming'' the perturbative expansions of the Wilson coefficients.
We consider here the OPE for renormalized operators. The extension
to the case of bare lattice operators is straightforward.

The most general short-distance expansion has the form:
\begin{eqnarray}
\opl {\cal A}\opr (x)\opl {\cal B}\opr (-x)
\sim\sum_{{\cal O}}W_{\cal O}(x;g,\mu)\opl {\cal O}\opr (0)\, .
\label{OPEGen}
\end{eqnarray}
The operators ${\cal O}$ appearing on the r.h.s. of Eq. (\ref{OPEGen}) are 
the same which would mix with the product ${\cal A}\cdot   {\cal B}$ under 
renormalization. 
Let us assume, for sake of simplicity, that the operators
${\cal A}$ and ${\cal B}$  renormalize multiplicatively.
In general $W_{\cal O}(x;g,\mu)$ will have some non-trivial Lorentz
structure.
However we can always factorize out one (or more) 
homogeneous function of $x$, carrying both the canonical dimensions and
the tensor structure of $W_{\cal O}(x;g,\mu)$.
We can therefore restrict ourselves to the case of  Wilson coefficients
with zero canonical dimension and depending upon $x$ only through 
its modulus $r\equiv|x|$.

With a slight abuse of notation we denote these ``reduced'' Wilson
coefficients as $W_{\cal O}(r;g,\mu)$.
They satisfy the following RG equation
\begin{eqnarray}
\left[\mu   \frac{\partial}{\partial \mu   }+\beta(g)
\frac{\partial}{\partial g}+\gamma^{{\cal O}}_W(g)\right]
W_{\cal O}(r;g,\mu   ) = 0\, ,
\label{RGWilson}
\end{eqnarray}
where
\begin{eqnarray}
\gamma^{{\cal O}}_W(g) \equiv \left(\gamma^{{\cal O}}(g)\right)^T-
\left(\gamma^{{\cal A}}(g)+\gamma^{{\cal B}}(g)\right) \, .
\end{eqnarray}
In the case of operator mixing, see Eq. (\ref{MixingGen}),
we must consider $\gamma^{\cal O}_W(g)$
as an ${\cal N}\times{\cal N}$ matrix, and $W_{{\cal O}}(r;g,\mu)$ as a
column vector of length ${\cal N}$.

The perturbative expansion for $W$ has the usual structure:
\begin{eqnarray}
W(r;g,\mu   ) =\sum_{l=0}^{\infty}\sum_{n=0}^l W^{(l)}_n\, g^l 
\log^n\!\mu r\, .
\label{PerturbativeExpansion}
\end{eqnarray}
where we dropped, for sake of simplicity, the superscript ${\cal O}$.
Equation (\ref{RGWilson}) implies a recursive relation 
between the coefficients $\{ W^{(l)}_n\}$:
\begin{eqnarray}
(n+1)W^{(l+1)}_{n+1}=\sum_{k=n}^l
(k\beta_{l-k}-\gamma^W_{l-k})W^{(k)}_n\, .
\label{RecursiveRelation}
\end{eqnarray}
We could use this relation for resumming the perturbative expansion.
Once $W^{(0)}_0$ (which is given by a tree-level calculation) is
known, Eq. (\ref{RecursiveRelation}) allows us to sum up all the terms
$g^n\log^n\!\mu r$ ({\it leading-log approximation}). 
The calculation of $W^{(1)}_0$ yields the sum of the terms
$g^n\log^{n-1}\!\!\mu r$ ({\it next-to-leading log}), and so on.

However, it is more convenient (both practically and conceptually)
to solve Eq. (\ref{RGWilson}) and use the perturbative calculation 
as a ``boundary condition''.
The solution has the well known form:
\begin{eqnarray}
W(r;g,\mu) = U(g)W_{RGI}(\Lambda r)\, .
\label{RGSolution}
\end{eqnarray} 
In the case of operator mixing $U(g)$ has to be interpreted as an
${\cal N}\times{\cal N}$ matrix, and $W_{RGI}(\Lambda r)$ as a column vector of
length ${\cal N}$. 

$U(g)$ satisfies the ordinary differential equation
\begin{eqnarray}
\beta(g)\frac{\partial}{\partial g}U(g) = -\gamma_W(g) U(g)\, ,
\label{URGEquation}
\end{eqnarray}
Let us write the perturbative expansion of $\gamma_W(g)$ as
$\gamma_W(g)=\sum_{k=0}^{\infty}\gamma^W_k g^{k+1}$.
The solution of Eq. (\ref{URGEquation}) can be formally written as follows:
\begin{eqnarray}
U(g) = g^{\gamma_0^W/\beta_0}
{\rm Gexp}\left\{-\int_0^g\left(
z^{\gamma_0^W/\beta_0} \frac{\gamma^W(z)}{\beta(z)}z^{-\gamma_0^W/\beta_0}+
\frac{\gamma_0^W}{\beta_0 z}
\right)dz\right\}\, .
\label{UFormalSolution}
\end{eqnarray}

Here $\Lambda$ is the intrinsic scale of the model (the so-called 
{\it lambda-parameter}):
\begin{eqnarray}
\Lambda\equiv \mu   \; e^{-\frac{1}{\beta_0 g}}(\beta_0 g)^{-\beta_1/\beta_0^2}
\exp\left\{-\int_0^g \left(\frac{1}{\beta(z)}+
\frac{1}{\beta_0z^2}-\frac{\beta_1}{\beta_0^2z}\right)dz\right\}\, .
\label{LambdaDefinition}
\end{eqnarray}
It is useful to define the dimensionless function $\lambda(g)$ through
the identity $\Lambda = \mu\lambda(g)$. The function $\lambda(g)$
clearly depends upon the renormalization scheme through the
beta-function. When necessary we shall indicate the particular scheme through a
subscript. Within the four schemes listed in Sec. \ref{ModelSection}, 
we shall write, respectively, $\lambda_{\MMS}(g)$, $\lambda_L(g_L)$,
$\lambda_E(g_E)$, $\lambda_R(g_R)$.
The explicit definition in a generic scheme is:
\begin{eqnarray}
\lambda_{\rm scheme}(g) =
\; e^{-\frac{1}{\beta_0 g}}(\beta_0 g)^{-\beta_1/\beta_0^2}
\exp\left\{-\int_0^g \left(\frac{1}{\beta^{\rm scheme}(z)}+
\frac{1}{\beta_0z^2}-\frac{\beta_1}{\beta_0^2z}\right)dz\right\}\, .
\label{lambdadig}
\end{eqnarray}
In this Section we shall drop the subscript.
The lambda-parameter depend upon the scheme too. 
We have four lambda parameters corresponding to the four schemes
listed in Sec. \ref{ModelSection}: $\Lambda_{\MMS}$, $\Lambda_L$,
$\Lambda_E$, and $\Lambda_R$. In order
to match two different schemes, the corresponding lambda-parameters
must be in a fixed ($g$-independent) ratio. 
This ratio is easily obtained through a one-loop calculation.

Notice that the prefactor $U(g)$ can be readsorbed with a redefinition of 
the operators. In particular, we can get rid of it by replacing 
the renormalized operators $\opl{\cal A}\opr$, $\opl{\cal B}\opr$,
and $\opl{\cal O}\opr$ in Eq. (\ref{OPEGen}) with their
RGI counterparts, see Eq. (\ref{RGIOpDef}).

Using the perturbative expansion (\ref{PerturbativeExpansion}) and 
the solution (\ref{RGSolution}), (\ref{UFormalSolution}) 
of the RG equation, we derive an expansion for 
$W_{RGI}(\Lambda r)$:
\begin{eqnarray}
W_{RGI}(\Lambda r) = \overline{g}(\Lambda r)^{\gamma^W_0/\beta_0}
\sum_{k=0}^{\infty} W_{RGI}^{(k)}\; \overline{g}(\Lambda r)^k\, .
\label{RGIExpansion}
\end{eqnarray}
The expansion is written in terms of $\overline{g}(\Lambda r)$
(the coupling at the energy scale $1/r$) which is implicitly
defined as follows:
\begin{eqnarray}
\Lambda r = \lambda(\overline{g}(\Lambda r))\, .
\label{CouplingAtScaleX}
\end{eqnarray}
The definition of the coupling $\overline{g}(\Lambda r)$,
and, consequently, of the expansion (\ref{RGIExpansion}),
is by no means unique. If we knew the whole expansion
(\ref{RGIExpansion}), the resulting $W_{RGI}(\Lambda r)$ would not
depend upon the particular choice. 
In practice we shall compute the expansion (\ref{RGIExpansion})
in perturbation theory, truncating it to some finite order.
We shall use the dependence of the truncated Wilson 
coefficient $W_{RGI}(\Lambda r)$
upon the definition of the coupling $\overline{g}(\Lambda r)$,
in order to assess the reliability of perturbation theory.
We refer to Sec. \ref{FieldsNumericalResults} for further discussion
on this point. Hereafter we shall use both the notations $W_{RGI}(\Lambda r)$
and $W_{RGI}(\overline{g}(\Lambda R))$ .

As in any asymptotically free theory $\overline{g}(\Lambda r)\sim 1/|\log
\Lambda r|$ as $r\to 0$.
Is we define $z=-\log\Lambda r$, we can write down an expansion of
$\overline{g}(\Lambda r)$ in inverse powers of $z$:
\begin{eqnarray}
\overline{g}(\Lambda r) =\frac{1}{\beta_0 z}-
\frac{\beta_1\log z}{\beta_0^3 z^2}
-\frac{\beta_1^2\log^2 z-
\beta_1^2\log z-\beta_1^2+\beta_2\beta_0}{\beta_0^5 z^3}+
O(\log^3 z/z^4)\, .
\label{GxExpansion}
\end{eqnarray}
We know the beta-function at four-loop order, both on the
lattice \cite{Caracciolo:1995bc,Shin:1998bh,Alles:1999fh},
in continuum \MS scheme 
\cite{Hikami:1981hi,Bernreuther:1986js,Wegner:1989ss},
and in the Finite Volume (FV) \cite{Shin:1996gi} scheme. 
This allows to add one more
term (of order $z^{-4}$) to the expansion (\ref{GxExpansion}).
Equation (\ref{GxExpansion}) implies that the expansion
(\ref{RGIExpansion}) is asymptotically good as $r\to 0$.
More precisely $l$-loop perturbation theory gives an estimation of
Wilson coefficients with a systematic error of order $|\log\Lambda
r|^{-l-1}$.

The coefficients $W^{(k)}_{RGI}$ are obtained by plugging Eqs.
(\ref{UFormalSolution}), (\ref{RGIExpansion}) and
(\ref{LambdaDefinition}) in Eq. (\ref{RGSolution}), 
expanding it in powers of $g$ 
and matching this expansion with Eq. (\ref{PerturbativeExpansion}).
The expressions for $W^{(k)}_{RGI}$ are simple if ${\cal O}$
renormalizes multiplicatively. In the case of general operator mixing,
they are quite involved.

For the general case, see Eq. (\ref{MixingGen}), we give the 
expressions of the first two coefficients of the expansion 
(\ref{RGIExpansion}):
\begin{eqnarray}
W^{(0)}_{RGI} & = & W^{(0)}_0\, ,\\
W^{(1)}_{RGI} & = & W^{(1)}_0+K_1 W^{(0)}_0\, .
\end{eqnarray}
In the next Chapters we shall not need higher-order coefficients.
$K_1$ is a ${\cal N}\times{\cal N}$ matrix, determined
by the following linear equation:
\begin{eqnarray}
\beta_0^2 K_1+K_1\beta_0\gamma^W_0-\beta_0\gamma^W_0K_1
= \beta_1\gamma^W_0-\beta_0\gamma^W_1\, .
\end{eqnarray}
This is a rather implicit formula for $K_1$.
In order to obtain a more explicit expression,
let us consider a change of basis which diagonalizes $\gamma^W_0$. 
If we define  $\gamma^W_0 \equiv V\gamma^D V^{-1}$ with 
$\gamma^D = {\rm diag}(\gamma^D_1,\gamma^D_2,\dots)$, then we get:
\begin{eqnarray}
(V^{-1}K_1 V)_{ij}=\frac{\beta_1}{\beta_0^2}\gamma^D_i\delta_{ij}-
\frac{(V^{-1}\gamma^W_1V)_{ij}}{\beta_0-\gamma^D_i+\gamma^D_j}\, .
\end{eqnarray}
Notice that the r.h.s. is not well defined if there exist two
eigenvalues $\gamma^D_i$ and $\gamma^D_j$ of $\gamma^W_0$
which satisfy $\gamma^D_i-\gamma^D_j =\beta_0$. 
Such an unlucky case is called a {\it resonance}\footnote{
More generally we would have a resonance if $\gamma^D_i-\gamma^D_j
=n\beta_0$, with $n$ a positive integer.}
in the theory of ordinary differential equations \cite{Arnold,Wasow}.
We will encounter a resonance in Sec. \ref{ResonanceSection}.
It turns out that, in such a case, non-analytic terms of the type 
$g^n\log^k g$ must be added to the expansion (\ref{RGIExpansion}).

Things simplify if the operator ${\cal O}$, see Eq. (\ref{OPEGen}),
renormalizes multiplicatively.
In this case $K_1$ becomes a number:
\begin{eqnarray}
K_1 = \frac{\beta_1\gamma_0-\beta_0\gamma_1}{\beta_0^2}\, ,
\end{eqnarray}
and it is easy to write down the three-loop coefficient in the
expansion (\ref{RGIExpansion})
\begin{eqnarray}
\hspace{-1.5cm}
W^{(2)}_{RGI} &=& W^{(2)}_0+K_1 W^{(1)}_0+K_2W^{(0)}_0\, ,\\
\hspace{-1.5cm}
K_2 &=& \frac{ -\beta_0 \beta_1^2 \gamma_0 + \beta_0^2 \beta_2 \gamma_0 + 
\beta_1^2 \gamma_0^2 + \beta_0^2 \beta_1 \gamma_1 - 
 2 \beta_0 \beta_1 \gamma_0 \gamma_1 + \beta_0^2 \gamma_1^2 - 
\beta_0^3 \gamma_2}{2\beta_0^4}\, .
\end{eqnarray}
%
%********************************************************************
%
\section{On the Evaluation of the Running Coupling Constant}
\label{RunningCoupling}
The determination of the running coupling constant is a key ingredient in 
the application of RG-improved perturbation theory to any
asymptotically free theory. 
If we use the lattice OPE, the coupling is $g_L$,
one of the input parameters of our numerical calculations. However,
perturbation theory in $g_L$ is poorly behaved, so that one expects 
a poor agreement with the numerical data. It is known that it is much
more convenient to use perturbative expansions in the \MS scheme. 
Perturbative coefficients are smaller, so that truncations in the number 
of loops give smaller systematic errors. For these reasons, it is important
to relate the \MS coupling to the bare coupling $g_L$. 
Given $g_L$, we fix the scale $\mu a$ and then compute the 
coupling $g_{\overline{MS}}$. In principle, 
it is a function of $g_L$ and $\mu a$,
but, because of the RG equations, it can be written as a function of 
the single variable $\mu /m$, where $m$ is the mass gap.

Here, we shall outline several different procedures---all of them
are exact in the continuum limit---and we shall compare 
their efficiency. We consider the following methods:
\begin{enumerate}
\item \label{NaivePerturbative}
The {\em naive perturbative} method. 

In this approach, one computes 
the continuum coupling as a function of the lattice coupling
by matching the continuum and the lattice perturbative expansion of
some physical quantity, e.g., of the two-point correlation function.
At $l$-loops one obtains a truncated series of the form
$g_{\overline{MS}} (\mu a,g_L ) = g_L + \sum_{k=2}^{l+1} c_k(\mu a) g_L^k$.
We shall call $g_{\overline{MS}}^{{\rm np},l}(\mu a,g_L )$ 
the value obtained in this way. The relevant perturbative expansions 
are known to three loops \cite{Caracciolo:1995bc,Alles:1999fh}.
Note that the $l$-loop approximation does not satisfy the exact RG equations
and thus this approximation is not a function of $\mu /m$ only.

\item \label{RGImprovedPerturbative}
The {\em RG improved perturbative} method. 

The idea of this method---and also of those that will be presented below---is
to compute $g_{\overline{MS}}(\mu )$ starting from some quantity that can 
be computed
numerically at the given value of the bare lattice coupling constant.

In this case, we consider the RG prediction for the mass gap
$m$ in the \MS scheme:
\begin{eqnarray}
m & = & \widehat{C}_N\Lambda_{\MMS}(\mu
,g_{\overline{MS}}(\mu /m ))=\widehat{C}_N\, \mu\,\lambda_{\MMS}(g_{\MMS}(\mu /m )
,
\label{MassGap}
\end{eqnarray}
see Eqs. (\ref{LambdaDefinition}) and (\ref{lambdadig}). 
The constant $\widehat{C}_N$ is not known for a general theory. For the 
two-dimensional $\sigma$-model it has been computed 
\cite{Hasenfratz:1990ab,Hasenfratz:1990zz} using the 
thermodynamic Bethe ansatz. The result, in the \MS scheme,
reads:
\begin{eqnarray}
\widehat{C}_N= \left(\frac{8}{e}\right)^{\frac{1}{N-2}}
\frac{1}{\Gamma(1+(N-2)^{-1})}\; .
\label{BetheAnsatzPrediction}
\end{eqnarray}
The method works as follows: For a given value of the lattice coupling 
$g_L$, compute numerically (for instance, by means of a Monte Carlo simulation)
the mass gap $ma$. 
Then, fix $\mu a$ and solve numerically
Eq. \reff{MassGap}, obtaining $g_{\overline{MS}}$, which is a function 
of $\mu /m$ only. Note that,
since the $\beta$-function is known only to a finite
order in perturbation theory, we have to substitute the function 
$\lambda_{\MMS}(g)$ with its truncated perturbative expansion.
There is some arbitrariness in this truncation. We shall make the
simplest choice
\begin{eqnarray}
\lambda_{\rm scheme}^{(l)}(g)\equiv    \; e^{-\frac{1}{\beta_0 g}}
(\beta_0 g)^{-\beta_1/\beta_0^2}\left[1+\sum_{k=1}^{l-2}
\lambda^{\rm scheme}_k g^k\right],
\label{LambdaTruncatedDefinition}
\end{eqnarray}
where the coefficients $\lambda_k$ are obtained by expanding perturbatively
Eq. (\ref{lambdadig}). Equation
(\ref{LambdaTruncatedDefinition}) gives the $l$-loop
approximation of the $\Lambda$-parameter. The solution of the
corresponding Eq. (\ref{MassGap}) will be denoted as
$g^{{\rm rgp},l}_{\overline{MS}}(\mu /m )$.
The perturbative expansion of the \MS $\beta$-function is known
to four loops \cite{Bernreuther:1986js}.
\item \label{FiniteSizeNonPerturbative}
The {\em finite-size non-perturbative} method. 

This method,
due to L{\"u}scher \cite{Luscher:1982aa}, was initially 
tested in the two-dimensional $O(3)$ $\sigma$-model
\cite{Luscher:1991wu}.
Recently, it has been successfully
employed in the computation of the $\Lambda$-parameter in quenched QCD
\cite{Luscher:1994gh}. The idea is to consider the  
theory in a finite box and to define a ``finite-size scheme''
in which the renormalization scale is the size of the box. 
For the $\sigma$-model, Ref. \cite{Luscher:1991wu} introduces a 
coupling $g_R(a/L,g_L)$ defined as follows:\footnote{This definition 
is by no means unique. For instance, one could also use 
$g_R(a/L,g_L) = [m(L)L]^2/(N-1)$, where $m(L)$ is the inverse 
of the second-moment correlation length on a square lattice of size $L/a$. 
The corresponding universal finite-size scaling function---i.e. 
the function that gives the correspondence between 
$g_R(a/L,g_L)$ and $mL$---has been determined numerically
in \cite{Caracciolo:1995ud}.}
\be
g_R(a/L,g_L) =\, {2 m(L) L\over N-1},
\label{FiniteSizeCoupling}
\ee
where $m(L)$ is the mass gap in a strip of width $L$. Standard finite-size 
scaling theory indicates that $g_R$ is a universal function 
of $mL$, where $m$ is the {\em infinite-volume} mass gap: 
$g_R(a/L,g_L) = g_R(m L)$. Such a 
function can be computed non-perturbatively by means of 
Monte Carlo simulations with a good control of the systematic 
errors. If we set\footnote{The constant $c$ is arbitrary.
In Ref. \cite{Luscher:1991wu} $c=1$ was used together with the 
minimal subtraction scheme. Here we will use the \MS scheme, 
and, in order to be consistent with previous results, we set
$c = \Lambda_{\overline{MS}}/\Lambda_{MS} = \sqrt{4 \pi e^{-\gamma}}$.}
$\mu =c/L$, $g_R$ defines a running coupling 
constant that is a function of $\mu /m$. The function $g_R$ 
can also be computed in perturbation theory in a different perturbative scheme.
This provides the 
connection between $g_R$ and any other perturbative scheme. 

We will now present two different methods of computing 
$g_{\overline{MS}}(\mu /m )$. First, we will 
compute the $l$-loop approximation to the 
${\overline{MS}}$ coupling $g_{\overline{MS}}(\mu /m )$ 
by using its perturbative expansion in terms 
of  $g_R$ at the {\em same} scale $\mu /m$:
$g^{{\rm fs1},l}_{\overline{MS}}(\mu /m ) = 
 g_R(\mu /m )+\sum_{k=2}^{l+1} d_k g_R^{k}(\mu /m )$. 
The perturbative expansion of 
$g_R$ is known to three-loop order \cite{Shin:1996gi}.

A different method (see, e.g., \cite{Capitani:1998mq,Luscher:1998pe}) 
works as follows. First we compute $\Lambda_R$, using its 
expression truncated at $l$-loops (see Eq. \reff{LambdaTruncatedDefinition})
and $g_R(\mu /m)$. Then, we derive $\Lambda_{\overline{MS}}$ using 
\be
\Lambda_{\overline{MS}} =\, \sqrt{4\pi e^{-\gamma}} \Lambda_R,
\ee
and finally we solve Eq. \reff{LambdaTruncatedDefinition}, obtaining 
$g^{{\rm fs2},l}_{\overline{MS}}(\mu /m )$. 

As we will discuss below, the two methods are essentially equivalent, 
and therefore in our numerical work we have always used 
$g^{{\rm fs1},l}_{\overline{MS}}(\mu /m )$ because of its simplicity.

The finite-size scaling method does not provide---at least in the 
implementation of Ref. \cite{Luscher:1991wu}---the coupling 
$g_R(\mu /m )$ for any $\mu /m $, but only on a properly chosen 
mesh of values, say $\{\mu _i/m\}_{i=1,\dots}$. 
Therefore, the methods described above provide 
$g_{\overline{MS}}(\mu /m)$ only for selected values of $\mu /m$. 
We want now to explain how to determine the coupling for generic 
values of the scale. In principle, one could use perturbation theory,
generalizing the definition $g^{{\rm fs1},l}_{\overline{MS}}(\mu /m )$. 
Indeed, we could simply define 
$g^{{\rm fs1},l}_{\overline{MS}}(\mu /m ) = g_R(\mu _i/m) + 
 \sum_{k=2}^l d_k(\mu /\mu _i) g_R(\mu _i/m)^k$. 
However, this definition does not work well, because of the presence of 
logarithms of $\mu /\mu _i$. A RG-improved version can be obtain using the 
RG equations.
Since the mass gap is a
RG-invariant quantity, at order $l$, we may require
\begin{eqnarray}
\Lambda_{\overline{MS}}^{(l)}(\mu _i,g_{\overline{MS}}(\mu _i/m)) =
\Lambda_{\overline{MS}}^{(l)}(\mu ,g_{\overline{MS}}(\mu /m )) .
\label{HybridDefinition}
\end{eqnarray} 
Using $g_{\overline{MS}}(\mu _i/m)$, one can then obtain 
$g_{\overline{MS}}(\mu /m)$ for any given $\mu /m$. 
We shall call $g_{\overline{MS}}^{\rm hybr}(\mu /m)$
the running coupling obtained by this procedure.
\item \label{DressedCoupling}
The {\em improved-coupling} method. 

Method 
\ref{NaivePerturbative} does not work well because lattice
perturbation theory is not ``well behaved'': Perturbative 
coefficients are large, giving rise to large truncation errors.
Parisi \cite{Parisi:1980aa,Martinelli:1981tb} noticed that much smaller
coefficients are obtained if one expands in terms of 
``improved" (or ``boosted") couplings 
defined using ``short-distance" observables. In the $\sigma$-model one can 
define a new coupling in terms of the energy density
\begin{eqnarray}
g_E \equiv \frac{4}{N-1}(1-\<\sg_x\cdot \sg_{x+\mu }\>),
\label{DressedCouplingEnergy}
\end{eqnarray}
which is then related to $g_{\overline{MS}}$ perturbatively. At order $l$,
we can write
$g_{\overline{MS}}^{{\rm dc},l}(a\mu ,g_L )=
 g_E+\sum_{k=2}^{l+1} c^E_k(\mu a) g_E^k$. 
In practice the method works as follows: Given $g_L$,
one computes numerically $g_E$; then, given $\mu a$,
one uses the previous perturbative 
expansion to determine the \MS coupling constant. This method is expected to 
be better than the naive one. Indeed, one expects 
$|c^E_k(\mu a)|\ll |c_k(\mu a)|$, so that truncation errors should be 
less important. The perturbative coefficients $c_k^E$ can be computed 
up to $l=3$ using the results of 
\cite{Caracciolo:1995bc,Alles:1999fh,Alles:1997rr}. 
Notice that the $l$-loop approximation is not a function of $\mu /m$ only
at variance with methods (B) and (C).
\end{enumerate}
Notice that the list above is by no means exhaustive. For instance, an 
alternative non-perturbative coupling may be defined 
using off-shell correlation functions:
\begin{eqnarray}
g_{R}(\mu = l^{-1}) = \xi ^d
\frac{\<(\sg_{x+l/a}-\sg_x)^2\>}{\sum_x \sg_0\cdot \sg_x}\; .
\label{RunningOffShell}
\end{eqnarray}
Something similar has been proposed in Refs. 
\cite{Becirevic:1999uc,Becirevic:1999sc,Becirevic:1999hj}, with the purpose of
computing the QCD $\Lambda$-parameter. This approach opens the Pandora
box of possible definitions of the running coupling 
in substitution of Eq. (\ref{RunningOffShell}). A scheme that
has been intensively studied in the context of QCD employs the
three-gluon vertex 
(see Refs. \cite{Becirevic:1999rv,Becirevic:1999sc,Henty:1995gv,
Parrinello:1994fu,Parrinello:1994wd,Parrinello:1997wm,Alles:1997fa}).
\begin{table}
\hspace*{-0.2truecm}
\begin{tabular}{|c|c|c|c|c|c|c|}
\hline
$g_R(\mu /m)$ & $\frac{m}\mu $ & $\frac{1}{\mbar a}$ &
$g^{{\rm fs1},3}_{\overline{MS}}$ & 
$g^{{\rm rgp},4}_{\overline{MS}}$ &
$g^{{\rm np},3}_{\overline{MS}}$ & 
$g^{{\rm dc},3}_{\overline{MS}}$\\
\hline
\hline
0.5372&0.00071(11) & 0.0097(15) &0.5870[$-$3]&  
 0.5892[$-$1] & 0.574[$-$5]  & 0.5902[$-$28]\\
0.5747&0.00143(11) & 0.0195(15) &0.6321[$-$4]&
 0.6351[$-$1] & 0.617[$-$6]  & 0.6354[$-$31]\\
0.6060&0.00237(15) & 0.0323(20) &0.6703[$-$5]&
 0.6736[$-$2] & 0.652[$-$6]  & 0.6737[$-$16]\\
0.6553&0.00478(15) & 0.0652(20) &0.7312[$-$7]&  
 0.7362[$-$2] & 0.708[$-$8]  & 0.7364[+5]\\
0.6970&0.00794(19) & 0.1083(25) &0.7835[$-$9]&  
 0.7900[$-$3] & 0.755[$-$12] & 0.7903[+6]\\
0.7383&0.01231(15) & 0.1678(20) &0.8361[$-$11]&  
 0.8437[$-$4] & 0.800[$-$16] & 0.8436[$-$13]\\
0.7646&0.01589(15) & 0.2166(20) &0.8701[$-$13]&  
 0.8788[$-$5] & 0.828[$-$20] & 0.8777[$-$35]\\
0.8166&0.02481(22) & 0.3382(30) &0.9382[$-$16]&  
 0.9486[$-$7] & 0.881[$-$28] & 0.9434[$-$99]\\
0.9176&0.04958(38) &0.6759(51)  &1.0742[$-$26]& 
 1.0852[$-$13] &  0.975[$-$45] & 1.0623[$-$275]\\
1.0595&0.1033(6) &1.4082(77) &1.2743[$-$47]& 
 1.2895[$-$27] & 1.090[$-$73] & 1.2135[$-$593]\\
1.2680&0.2092(5) &2.8519(67) &1.5886[$-$96]& 
 1.5963[$-$68] & 1.217[$-$108]& 1.3863[$-$1056]\\
\hline
\end{tabular}
\caption{The \MS running coupling constant. We use here 
several different methods as explained in the text,
and $(ma)^{-1} = 13.632(6)$ at $1/g_L=1.54$.
The errors on the second and third columns are statistical.}
\label{RunningComparison1}
\end{table} 

Let us compare the different methods. In
Tab. \ref{RunningComparison1} we compare the procedures 
\ref{NaivePerturbative}, \ref{RGImprovedPerturbative},
\ref{FiniteSizeNonPerturbative}, and \ref{DressedCoupling}.
In the first column we report a collection of values of $g_R$.
A subset of the values given in the table have been considered for the first 
time in Ref. \cite{Luscher:1991wu}. Later, the mesh was enlarged by
Hasenbusch  \cite{Hasenbusch-unpublished}.
For these values of $g_R$, Hasenbusch 
computed the corresponding value of $m/\mu $ which is reported in the second 
column. Note a peculiarity of the finite-size approach: 
usually, one fixes $\mu /m$
and then determines the running coupling constant. Here, the 
running coupling constant is fixed at the beginning and the value of the 
scale is determined numerically. In the third column we report
the scale in lattice units for $g_L = 1/1.54$, the value of the lattice 
coupling at which we have done most of our simulations. The results
are obtained by using $(ma)^{-1} = 13.632(6)$. The error
reported there corresponds to the error on $m/\mu $, the error on $(ma)$
being negligible. In column 4 we report the estimate of 
$g_{\overline{MS}}^{\rm fs1,3} (\mu /m)$ obtained by using $g_R$ and three-loop 
perturbation theory \cite{Shin:1996gi}. In brackets we report the difference 
$g_{\overline{MS}}^{\rm fs1,2} (\mu /m) - 
 g_{\overline{MS}}^{\rm fs1,3} (\mu /m)$. 
In the next column we report the four-loop coupling 
$g_{\overline{MS}}^{\rm rgp,4}(\mu /m)$ 
obtained by using the value of $m/\mu $ given in the second column.
Again, in brackets we report 
$g_{\overline{MS}}^{\rm rgp,3}(\mu /m) - g_{\overline{MS}}^{\rm rgp,4}(\mu /m)$. 
In the last two columns 
we report the results obtained by using three-loop lattice perturbation theory
\cite{Caracciolo:1995bc,Alles:1999fh,Alles:1997rr}. In the fifth
column we use $g_L=1/1.54$ as the expansion parameter.
In the sixth column the
improved coupling defined by Eq. (\ref{DressedCouplingEnergy}) is used.
The connection with the bare coupling is obtained by using the perturbative 
expressions given in Ref. \cite{Alles:1997rr}.
The relevant expectation value has been evaluated in a Monte Carlo
simulation at $g_L=1/1.54$ on a lattice $128\times 256$ with
statistics $N_{\rm stat} =10000$, yielding 
$g_E = 0.768133(49)$. 

\begin{table}
\begin{center}
\begin{tabular}{|c|c|c|c|c|c|}
\hline
$\frac{1}{\mbar a}$ & $g^{{\rm rgp},4}_{\overline{MS}}(\mu /m)$ &
$g^{{\rm hybr},A1}_{\overline{MS}}(\mu /m)$ & 
$g^{{\rm hybr},B1}_{\overline{MS}}(\mu /m)$ &
$g^{{\rm hybr},A2}_{\overline{MS}}(\mu /m)$ & 
$g^{{\rm hybr},B2}_{\overline{MS}}(\mu /m)$\\
\hline
\hline
1 & 1.18422 & 1.171(2)  & 1.1804(6) & 1.171(2) & 1.1793(6) \\
2 & 1.42282 & 1.403(3)  & 1.417(1)  & 1.402(3) & 1.415(1) \\
3 & 1.62532 & 1.598(4)  & 1.617(1)  & 1.597(4) & 1.615(1) \\
4 & 1.81923 & 1.783(5)  & 1.809(2)  & 1.782(6) & 1.806(2) \\
5 & 2.01713 & 1.970(7)  & 2.003(2)  & 1.969(7) & 1.999(2) \\
6 & 2.22905 & 2.167(9)  & 2.211(3)  & 2.166(9) & 2.206(3) \\
7 & 2.46722 & 2.385(12) & 2.443(4)  & 2.384(12) & 2.436(4) \\
8 & 2.75208 & 2.637(16) & 2.717(6)  & 2.636(17) & 2.708(6) \\
\hline
\end{tabular}
\end{center}
\caption{The \MS running coupling constant.
In the second column we report the RG coupling 
obtained using $(ma)^{-1} = 13.632(6)$ at $1/g_L=1.54$.
The couplings reported in the last four columns 
are obtained using the interpolation scheme \reff{HybridDefinition}:
the columns differ in the choice of
the ``boundary condition'' $g_{\overline{MS}}(\mu _i/m)$, see text. 
The reported error is due to the error on $\mu _i/m$ appearing in the l.h.s. of
Eq. (\ref{HybridDefinition}), see Tab. \ref{RunningComparison1}, 
second column.}
\label{RunningComparison2}
\end{table}

In Table \ref{RunningComparison1} we have used the first definition for the 
finite-size coupling, $g_{\overline{MS}}^{\rm fs1,3} (\mu /m)$, but 
completely equivalent results are obtained adopting the second procedure.
For instance, for $m/\mu = 0.00071$ (resp. 0.2092) we obtain
$g_{\overline{MS}}^{\rm fs2,3} (\mu /m) = 0.5870[4]$ (resp. 1.5864[258]). 
Clearly, the two procedures are equivalent for $l=3$.

In Tab. \ref{RunningComparison2} we compare, on a broad range of scales, 
the outcome of RG-improved perturbation theory and the 
interpolation procedure \reff{HybridDefinition}. In both cases four-loop 
perturbation theory is used.
The couplings differ in the ``boundary
condition'' for the RG interpolation, that
is in the value used in the left hand side of Eq. (\ref{HybridDefinition}).
The couplings A1 and A2 have been obtained using $m/\mu _i = 0.04958$.
The coupling A1 was determined using 
$g_{\overline{MS}}^{\rm fs1,3} (\mu _i/m)$, while A2 was computed starting 
from $g_{\overline{MS}}^{\rm fs2,3} (\mu _i/m)$. Analogously 
the couplings B1 and B2 have been obtained using 
$g_{\overline{MS}}(\mu _i/m)$ for $m/\mu _i = 0.2092$.
In all cases we fixed $(ma)^{-1} = 13.632(6)$. 

What do we learn from this comparison? First of all, lattice (naive)
perturbation theory (sixth column of Tab. \ref{RunningComparison1})
is a very bad tool. Even at energies as high as $50$ times the mass gap
$g_{\overline{MS}}(\mu /m)$
is affected by a $\sim 5\%$ systematic error. 
However, it is reassuring that the expansion tells us its own unreliability.
Indeed, the observed discrepancy is of the order (at most twice as large) 
of the difference between the two-loop and the three-loop result.
The perturbative expansion in terms of the improved coupling
is much better. The results are quite precise up to $\mu \approx 10m$. 
For smaller values of $\mu $ the discrepancy increases, but it is nice
that it is again of the order of the difference between 
the two- and the three-loop result. Perturbative RG
supplemented with the prediction (\ref{BetheAnsatzPrediction})  gives 
results which are in agreement with the non-perturbative ones obtained 
using the finite-size scaling method within a few
percent for all the energy scales given in Tab. \ref{RunningComparison1}.
The accuracy remains good (if the comparison is made with the ``interpolation'' 
procedure \reff{HybridDefinition}) 
also for scales of the order of the mass gap. 

Up to now we have discussed the \MS scheme and how to obtain the value 
of the \MS coupling. However, as we already mentioned above,
reasonably good results can also be obtained if we use 
the coupling $g_E$. In this scheme we introduce the $\Lambda_E$ parameter
as follows
\be
\Lambda_{E} (a,g_E) = {1\over a}\ \lambda_E(g_E) \, ,
\label{LambdaEdef}
\ee
where $\lambda_E(g_E)$ is defined by Eq. (\ref{lambdadig}) in terms
of the corresponding beta-function $\beta_E(g_E)$. 
The beta-function $\beta_E(g_E)$ is related to the lattice one through
a simple change of variables: if
$g_L = f(g_E)$, then $\beta_E(g_E) = \beta^{L}(f(g_E))/f'(g_E)$. 
The mass gap is invariant and thus $m = C_{N,E} \Lambda_{E} (a,g_E)$,
where 
\be
C_{N,E} = \left( {8\over e} \right)^{1/(N-2)}
{1\over \Gamma\left(1+{1\over N-2}\right)} 2^{5/2}
  \exp\left[{\pi \over 4(N-2)}\right].
\label{costanteHasenfratz_schemaE}
\ee

Finally, to be exhaustive, we give the formulae for the
lambda-parameter and for the mass gap in the bare lattice theory.
Analogously to the previous case, we have 
$\Lambda_L (a,g_L) = (1/a)\, \lambda_L(g_L)$, and 
\begin{eqnarray}
m = C_N\Lambda_L(a,g_L)\, ,
\label{MassGapLattice}
\end{eqnarray}
where
\begin{eqnarray}
C_N = \left( {8\over e} \right)^{1/(N-2)}
{1\over \Gamma\left(1+{1\over N-2}\right)} 2^{5/2} 
  \exp\left[{\pi\over 2(N-2)}\right].
\label{HasenfratzconstLattice}
\end{eqnarray}
\chapter{Operator Product Expansion for Conserved Currents}
In this Chapter we present our perturbative and numerical results
concerning the OPE of $O(N)$ Noether currents in the non-linear
$\sigma$-model.
Since $O(N)$ currents are exactly conserved on the lattice, they do
not need to be renormalized. This makes it simpler to verify
the validity of the OPE on the lattice.

We consider one-particle matrix elements of 
the current product. Moreover, we keep only the leading term of the 
OPE. In brief, we shall study the following example of OPE:
\begin{eqnarray}
\langle \pb|j(x)j(0)|\qb\rangle\sim W(x)\langle\pb|{\cal
O}|\qb\rangle\, ,
\label{OPECurrentsGen}
\end{eqnarray}
where $\langle\pb |$ and $|\qb\rangle$ are one-particle states with
spatial momentum $\pb$ and $\qb$ (respectively).
We shall compute the left-hand side of Eq. (\ref{OPECurrentsGen})
for $r=|x|\ltapprox \xi$.
In the above equation we adopted a loose notation, omitting 
both $O(N)$ and Lorentz indices. The particular choices of these
indices will be specified in Sec. \ref{NumericalCurrentsSection}.
Finally, we often consider the angular average of 
Eq. (\ref{OPECurrentsGen}), i.e. the average over $x$ at fixed 
$r$.
Moreover, we shall compute the {\it renormalized} 
matrix elements $\langle\pb|{\cal O}|\qb\rangle$ without relying on
the OPE approach. This makes it possible
to compare the two sides of Eq. (\ref{OPECurrentsGen}), yielding a
stringent test of the validity of the OPE.

The procedure outlined above is quite different from what would be
done in more physical (QCD) applications. In this case the matrix element
on the r.h.s. of Eq. (\ref{OPECurrentsGen}) would be unknown.
In this Chapter we focus mostly on the validity of the OPE, and on
the reliability of the perturbative calculation of the Wilson
coefficients.
We would like to get an idea of the window of $r$ for which the OPE
works.
Moreover, we will investigate different procedures 
resumming perturbation theory using the RG. 
We will try to assess the goodness of the various procedures.

A preliminary account of this work has been presented
at the Lattice conference in 1998 \cite{Caracciolo:1998gf}.

The organization of this Chapter is quite simple.
In Sec. \ref{PerturbativeCurrentsSection} we write down the structure
of the OPE for two different products of Noether currents, and we list
the one-loop results for the Wilson coefficients.
In Sec. \ref{NumericalCurrentsSection} we gives the details of our
Monte Carlo simulations and compare the results withe the OPE 
prediction. Finally, we summarize the outcomes of our investigation
in Sec. \ref{SummaryCurrentsSection}
%
%********************************************************************
%
\section{Perturbative Calculation of the Wilson Coefficients}
\label{PerturbativeCurrentsSection}

A general product of two $O(N)$ currents reads 
$j^{ab}_{\mu}(x)j_{\nu}^{cd}(y)$. This is a reducible rank-4
$O(N)$-tensor. 
We shall decompose it into irreducible parts and consider uniquely the
two simplest sectors, namely the $O(N)$-scalar, and the antisymmetric 
rank 2 $O(N)$-tensor. According to the general considerations of 
Sec. \ref{CompositeOperatorsSection}, the operators appearing in
the OPE will be either $O(N)$-tensors in the same representation 
or products of such tensors times some power of $\alpha(x)$ (see
Eq. (\ref{AlphaDefinition}) for the definition of $\alpha(x)$).

We shall present the results both in the continuum \MS renormalization scheme
and for the lattice bare theory. 
We recall that the OPE holds on the lattice (and in particular in
lattice perturbation theory) as long as we keep distinct lattice
operators at non-zero physical separations in the continuum limit.
This means taking $\xi\to \infty$ and $|x-y|\to\infty$ at
$|x-y|/\xi$ fixed. Next one can consider the short-distance regime
$|x-y|/\xi\ll 1$.  
The OPE will be valid up to scaling corrections of relative order 
$1/\xi^2$ (such corrections cannot be seen in perturbation theory),
$1/|x-y|^2$, etc.
The only difference between lattice and continuum OPE is related
to space-time symmetries. 
In fact, while the Lorentz invariance strongly restricts the OPE in the
continuum, it is lost on the lattice.
%
%********************************************************************
%
\subsection{Continuum}
\label{PerturbativeContinuumCurrentsSection}
\subsubsection{Scalar Sector}

We begin by considering the OPE for the product of two currents
in the scalar sector. There exists an unique manner of combining two
currents to make a scalar.
The general form of the OPE, neglecting $O(x\log^p x)$ terms, is:
\begin{eqnarray}
\frac{1}{2}\jg_\mu   (x)\cdot   \jg_{\rho}(-x)&\equiv&
\frac{1}{2}\sum_{a,b}j^{ab}_\mu   (x)
j^{ab}_{\rho}(-x)= 
\nonumber \\
& = & \left[\frac{\delta_{\mu   \rho}x_{\nu}x_{\sigma}}{x^2}W_1(x)+
\frac{x_\mu   x_{\rho}x_{\nu}x_{\sigma}}{(x^2)^2}W_2(x)+
\frac{x_\mu   x_{\nu}\delta_{\rho\sigma}+
x_{\rho}x_{\sigma}\delta_{\mu   \nu}}{x^2}W_3(x)+\right.\nonumber\\
&&\left.+\frac{\delta_{\mu   \nu}\delta_{\rho\sigma}+\delta_{\mu   \sigma}
\delta_{\rho\nu}}{2}
W_4(x)\right]\frac{1}{g} \opl T_{\nu\sigma}\opr (0)+\nonumber\\
&&+\left[\frac{x_\mu   x_{\rho}}{x^2}W_5(x)+\delta_{\mu   \rho}W_6(x)\right]
\frac{1}{g^2}\opl(\partial\sg)^2\opr(0)+\nonumber\\
&&+\left[\frac{x_\mu   x_{\rho}}{x^2}W_7(x)+\delta_{\mu   \rho}W_8(x)\right]
\frac{1}{g^2}\opl\alpha\opr(0)+\nonumber\\
&&+\frac{1}{x^2}W_{0,\mu   \rho}(x)\,\frac{1}{g}\mbox{\boldmath $1$}\, ,
\label{scalcorrcont}
\end{eqnarray}
where  $W_{0,\mu   \rho}(x)$ and $W_1(x),\dots, W_8(x)$
are functions of $x$, of the \MS 
coupling $g$, and of the renormalization scale $\mu $. Explicit one-loop
expressions are reported below, see Eqs. 
(\ref{ScalCorrOneLoop.0})--(\ref{ScalCorrOneLoop.8}).

We are interested in the $O(N)$-symmetric limit $h\to 0$.
Moreover we shall consider on-shell matrix elements of the 
operator product on the left-hand side of Eq. (\ref{scalcorrcont}).
In this case, as we explained in Sec. \ref{D2IS0}, we can express the 
non $O(N)$-invariant operator $\opl\alpha\opr$, 
appearing in the right-hand side of 
Eq. (\ref{scalcorrcont}), in terms of $O(N)$ invariant operators. 
After eliminating  $\opl\alpha\opr$ through Eq. (\ref{AlphaOnShell}),
we recover an $O(N)$-invariant expansion:
\begin{eqnarray}
\frac{1}{2}\jg_\mu   (x)\cdot   \jg_{\rho}(-x)
& = & \left[\frac{\delta_{\mu   \rho}x_{\nu}x_{\sigma}}{x^2}W_1(x)+
\frac{x_\mu   x_{\rho}x_{\nu}x_{\sigma}}{(x^2)^2}W_2(x)+
\frac{x_\mu   x_{\nu}\delta_{\rho\sigma}+
x_{\rho}x_{\sigma}\delta_{\mu   \nu}}{x^2}W_3(x)+\right.\nonumber\\
&&\left.+\frac{\delta_{\mu   \nu}\delta_{\rho\sigma}+\delta_{\mu   \sigma}
\delta_{\rho\nu}}{2}
W_4(x)\right]\frac{1}{g}\opl T_{\nu\sigma}\opr(0)+\nonumber\\
&&+\left[\frac{x_\mu   x_{\rho}}{x^2}W'_5(x)+\delta_{\mu   \rho}W'_6(x)\right]
\frac{1}{g^2}\opl(\partial\sg)^2\opr(0)+\nonumber\\
&&+\frac{1}{x^2}W_{0,\mu   \rho}(x)\, \frac{1}{g}\mbox{\boldmath $1$}
\, ,
\label{ONInvariantScalCorr}
\end{eqnarray}
with
\be
W'_5(x)\equiv W_5(x)-W_7(x)\, , \qquad\qquad
W'_6(x)\equiv W_6(x)-W_8(x)\, .
\ee
The Wilson coefficients satisfy the following RG equations:
\begin{eqnarray}
\left[\mu   \frac{\partial}{\partial\mu   }+\beta(g)\frac{\partial}{\partial g}
-\frac{\beta(g)}{g}\right]W_{0,\mu   \rho}(x;g,\mu   )&=&0, \\
\left[\mu   \frac{\partial}{\partial\mu   }+\beta(g)\frac{\partial}{\partial g}
-\frac{\beta(g)}{g}\right]W_i(x;g,\mu   )&=&0,\;\;\;\; i = 1,\dots,4,\\
\left[\mu   \frac{\partial}{\partial\mu   }+\beta(g)\frac{\partial}{\partial g}
-\frac{\beta(g)}{g}-g\frac{\partial}{\partial g}
\left(\frac{\beta(g)}{g}\right)\right]W'_i(x;g,\mu   )&=&0,\;\;\;\; i = 5,6,
\end{eqnarray}
where $\beta(g)$ is the \MS $\beta$-function. 
These equations can be derived from the general formulae of Sec. 
\ref{RenormalizationGroupSection},
using the anomalous-dimension matrix given in Sec. 
\ref{Anomalous20Section}.
Notice that, as we explained in Sec. \ref{AnomalousSection}, 
we can write RG equations for the ``reduced'' $O(N)$-invariant
expansion (\ref{ONInvariantScalCorr}), without taking care
of on-shell vanishing terms.

The explicit one-loop expression for the Wilson coefficients
$W_{0,\mu\nu}(x)$ and $W_1(x),\dots,W_8(x)$ appearing 
in Eq. (\ref{scalcorrcont}) are given below:
\begin{eqnarray}
W_{0,\mu   \nu}(x)&=& \delta_{\mu   \nu}\frac{N-1}{8\pi}\left[
1-\frac{N-2}{2\pi}g(\gamma+\log(\mu x))\right]-\nonumber\\
&&-x_\mu   x_{\nu}\frac{N-1}{4\pi x^2}\left[1-\frac{N-2}{2\pi}g\left(\gamma+
\log(\mu x)+\frac{1}{2}\right)\right] + O(g^2)\, ,
\label{ScalCorrOneLoop.0}\\
W_1(x) & = & -\frac{N-2}{4\pi}g + O(g^2)\, ,\\
W_2(x) & = & \frac{N-2}{2\pi}g + O(g^2)\, ,\\
W_3(x) & = & \frac{N-2}{2\pi}g + O(g^2)\, ,\\
W_4(x) & = & 1-\frac{N-2}{2\pi}g(\gamma+\log(\mu x)) + O(g^2)\, ,\\
W_5(x) & = & \frac{3N-5}{4\pi}g + O(g^2)\, ,\\
W_6(x) & = & \frac{1}{2}\left[ 1-\frac{N-2}{4\pi}g-\frac{N-3}{2\pi}g
\left(\gamma+\log(\mu x)\right)\right]  + O(g^2)\, ,\\
W_7(x) & = & \frac{N-1}{4\pi}g + O(g^2)\, ,\\
W_8(x) & = & \frac{N-1}{4\pi}g\left(\gamma+\log(\mu x)\right) +
O(g^2)\, .
\label{ScalCorrOneLoop.8}
\end{eqnarray}

\subsubsection{Antisymmetric Sector}

We consider now the OPE of the antisymmetric product of currents.
As in the previous case, there exists a unique manner
of constructing a rank-2 antisymmetric $O(N)$-tensor 
from the product of two Noether currents. 
Neglecting terms of order $O(x \log^p x)$, we have\footnote{
Note that one could also add a contribution proportional to 
$(x_\mu   x_{\alpha}\delta_{\nu\beta}-
  x_\nu   x_{\alpha}\delta_{\mu \beta})/x^2 
\left[\partial_\alpha j_\beta^{ab}(0) -
                 \partial_\beta j_\alpha^{ab}(0)\right]$. 
However, in two dimensions, 
$(x_\mu   x_{\alpha}\delta_{\nu\beta}-
  x_\nu   x_{\alpha}\delta_{\mu \beta} - (\alpha \leftrightarrow\beta) ) =
  x^2 (\delta_{\mu \alpha} \delta_{\nu\beta} - 
       \delta_{\nu\alpha} \delta_{\mu \beta})$, and thus 
this term is equivalent to that proportional to $U_1(x)$.} 
\begin{eqnarray}
&& \hskip -1truecm
\sum_c\left[j^{ac}_\mu   (x)j^{bc}_{\nu}(-x)-
j^{bc}_\mu   (x)j^{ac}_{\nu}(-x)\right]  =  
\nonumber \\
&& \left[{x_\mu x_\nu x_\alpha\over (x^2)^2} U_{00}(x) + 
      {\delta_{\mu \nu} x_\alpha\over x^2} U_{01}(x) + 
      {\delta_{\mu \alpha}x_\nu + \delta_{\nu\alpha} x_\mu \over x^2} U_{02}(x)
\right]
\frac{1}{g}j^{ab}_{\alpha}(0)
\nonumber \\
&&+ (\delta_{\mu   \alpha}\delta_{\nu\beta}-
\delta_{\mu   \beta}\delta_{\nu\alpha})U_1(x)
\frac{1}{4 g}\left[\partial_\alpha j_\beta^{ab}(0) - 
                 \partial_\beta j_\alpha^{ab}(0)\right]
\nonumber\\
&&+\frac{x_\mu   x_{\alpha}\delta_{\nu\beta}-
x_{\nu}x_{\alpha}\delta_{\mu   \beta}}{x^2}U_2(x)\frac{1}{2 g}
\left[\partial_\alpha j_\beta^{ab}(0) +
                 \partial_\beta j_\alpha^{ab}(0)\right].
\label{anticorrcont}
\end{eqnarray}
The coefficients $U_i(x)$ and $U_{0i}(x)$ satisfy the RG equations:
\begin{eqnarray}
\left[\mu   \frac{\partial}{\partial\mu   }+\beta(g)\frac{\partial}{\partial g}
-\frac{\beta(g)}{g}\right]U(x;g,\mu   )&=&0.
\label{RG-Wilson-antisymmetric}
\end{eqnarray}
The meaning of this equation is very simple: $(1/g)\, U(x;g,\mu)$ is 
RG invariant, i.e. 
$(1/g)\, U(x;g,\mu) = U_{RGI}(\Lambda_{\MMS} x)$. This could be easily
understood from Eq. (\ref{anticorrcont}), since neither
$j^{ab}_{\mu}$, nor its space-time derivatives must be renormalized.

The coefficients appearing in Eq. (\ref{anticorrcont}) are given by:
\begin{eqnarray}
U_{00}(x) & = & \frac{N-2}{2\pi}g + O(g^2),  \\
U_{01}(x) & = & - \frac{N-2}{4\pi}g + O(g^2), \\
U_{02}(x) & = &  \frac{N-2}{4\pi}g + O(g^2), \\
U_1(x) & = &  1-\frac{N-2}{2\pi}g(\gamma+ \log(\mu x)) + \frac{N-6}{4\pi}g 
     + O(g^2),\\
U_2(x) & = & -\frac{N-2}{4\pi}g + O(g^2).
\end{eqnarray}
%
%
%**************************************************************************
%
\subsection{Lattice}
\label{PerturbativeLatticeCurrentsSection}

\subsubsection{Scalar Sector}
If we write Eq. (\ref{scalcorrcont}) in terms of lattice operators we get:
\begin{eqnarray}
\frac{1}{2}\jg^L_{\mu   ,x}\cdot   \jg^L_{\rho,-x}&=&
\left[\frac{\delta_{\mu   \rho}x_{\nu}x_{\sigma}}{x^2}W^L_1(x)+
\frac{x_\mu   x_{\rho}x_{\nu}x_{\sigma}}{(x^2)^2}W^L_2(x)+
\frac{x_\mu   x_{\nu}\delta_{\rho\sigma}+
x_{\rho}x_{\sigma}\delta_{\mu   \nu}}{x^2}W^L_3(x)+\right.\nonumber\\
&&\left.+\frac{\delta_{\mu   \nu}\delta_{\rho\sigma}+\delta_{\mu   \sigma}
\delta_{\rho\nu}}{2}
W^L_4(x)\right]\frac{1}{g_L}T^L_{\nu\sigma,0}+\nonumber\\
&&\left[\frac{\delta_{\mu   \rho}x_{\nu}x_{\sigma}}{x^2}{\widehat W^L_1}(x)+
\frac{x_\mu   x_{\rho}x_{\nu}x_{\sigma}}{(x^2)^2}{\widehat W^L_2}(x)+
\frac{x_\mu   x_{\nu}\delta_{\rho\sigma}+
x_{\rho}x_{\sigma}\delta_{\mu   \nu}}{x^2}{\widehat W^L_3}(x)+\right.\nonumber\\
&&\left.+\frac{\delta_{\mu   \nu}\delta_{\rho\sigma}+\delta_{\mu   \sigma}
\delta_{\rho\nu}}{2}
{\widehat W^L_4}(x)\right]\frac{1}{g_L^2}\delta_{\nu\sigma}
(\overline{\partial_{\nu}}\sg)^2_0+\nonumber\\
&&+\left[\frac{x_\mu   x_{\rho}}{x^2}W^L_5(x)+\delta_{\mu   \rho}W^L_6(x)\right]
\frac{1}{g_L^2}(\partial\sg)_0^2+\nonumber\\
&&+\left[\frac{x_\mu   x_{\rho}}{x^2}W^L_7(x)+\delta_{\mu   \rho}W^L_8(x)\right]
\frac{1}{g_L^2}\alpha_0+\nonumber\\
&&+\frac{1}{x^2}W^L_{0,\mu   \rho}(x)\frac{1}{g_L}\mbox{\boldmath $1$} 
\, ,\label{scalcorrlatt}
\end{eqnarray}
where $T^L_{\nu\sigma}$ is the naive lattice energy momentum
tensor, see Eq. (\ref{NaiveEMT}).
Notice the appearance of the non-Lorentz covariant operator 
$\delta_{\nu\sigma}(\overline{\partial_{\nu}}\sg)^2$. 
The Wilson coefficient of this operator is of order $a^0\log^p a$ 
(i.e. non vanishing) in the continuum limit. 
One could suspect that Lorentz invariance is
lost even in the continuum limit. Of course this is not the case since
the terms proportional to
$\delta_{\nu\sigma}(\overline{\partial_{\nu}}\sg)^2$ are readsorbed
in the renormalization of the energy-momentum tensor $T_{\nu\sigma}$,
see Sec. \ref{D2IS0LATT}.

The one-loop expressions for the Wilson coefficients are easily
obtained by writing the \MS renormalized operators appearing in Eq. 
(\ref{scalcorrcont}) in terms of bare lattice operators. The formulae of
Sec. \ref{D2IS0LATT} for the renormalization constants can be used. 
Alternatively one can use directly lattice perturbation theory
and proceed as in the continuum. The result is:
\begin{eqnarray}
\hspace{-2.0cm}
W^L_{0,\mu   \rho}(x)&=& \delta_{\mu   \rho}\frac{N-1}{8\pi }\left[
1-\frac{N-2}{4\pi}g_L(2\gamma+\log(32 x^2))-\frac{1}{4}g_L\right]-\nonumber\\
&&-x_\mu   x_{\rho}\frac{N-1}{4\pi x^2}\left[1-\frac{N-2}{4\pi}g\left(2\gamma+
\log(32 x^2)+1\right)-\frac{1}{4}g_L\right] + O(g^2_L), 
\\
\hspace{-2.0cm}W^L_1(x) & = & -\frac{N-2}{4\pi}g_L + O(g^2_L),\\
\hspace{-2.0cm}W^L_2(x) & = & \frac{N-2}{2\pi}g_L + O(g^2_L),\\
\hspace{-2.0cm}W^L_3(x) & = & \frac{N-2}{2\pi}g_L + O(g^2_L),\\
\hspace{-2.0cm}W^L_4(x) & = & 1-\frac{N-2}{4\pi}g_L(2\gamma+\log(32 x^2))+\left(\frac{1}{\pi}
-\frac{1}{2}\right)g_L + O(g^2_L),\\
\hspace{-2.0cm}\widehat{W^L_1}(x) & = & O(g^2_L),\\
\hspace{-2.0cm}\widehat{W^L_2}(x) & = & O(g^2_L),\\
\hspace{-2.0cm}\widehat{W^L_3}(x) & = & O(g^2_L),\\
\hspace{-2.0cm}\widehat{W^L_4}(x) & = & \left(\frac{1}{2}-\frac{3}{2\pi}\right) g_L + 
              O(g^2_L), \\
\hspace{-2.0cm}W^L_5(x) & = & \frac{3N-5}{4\pi}g_L + O(g^2_L),\\
\hspace{-2.0cm}W^L_6(x) & = & \frac{1}{2}\left\{1-\frac{N-2}{4\pi}g_L-\frac{N-3}{4\pi}g_L
\left(2\gamma+\log(32 x^2)\right)-\frac{1}{4}g_L\right\} + O(g^2_L),\\
\hspace{-2.0cm}W^L_7(x) & = & \frac{N-1}{4\pi}g_L + O(g^2_L),\\
\hspace{-2.0cm}W^L_8(x) & = & \frac{N-1}{8\pi}g_L\left(2\gamma+\log(32 x^2)-\pi\right)
+ O(g^2_L)\, .
\end{eqnarray}
We shall need the RG equations uniquely for the terms proportional to 
the energy-momentum, cf. Eq. (\ref{scalcorrlatt}):
\begin{eqnarray}
\left[-a\frac{\partial}{\partial a}+\beta^L(g_L)\frac{\partial}{\partial
g_L}+\gamma^{L,T}_W(g_L)\right] W^L_i(x;g_L,a) 
=0 \, ,\quad i=1,\dots,4\, .\nonumber\\
\end{eqnarray}
Notice that these equations decouple from the ones for the other 
Wilson coefficients because of the block triangular form of the 
renormalization matrix $Z^{L(2,0)}$, see Eqs. 
(\ref{Ren20.1})--(\ref{Ren20.1}).
Moreover, $\gamma^{L,T}_W(g_L)$ is determined by the following simple formula: 
\begin{eqnarray}
\gamma^{L,T}_W(g_L)= -\beta^L(g_L)\frac{\partial}{\partial g_L}
\log Z^{L(2,0)}_{11}-\frac{\beta^L(g_L)}{g_L}\, .
\end{eqnarray}
Using the one-loop result for $Z^{L(2,0)}_{11}$, see Eq. 
(\ref{Ren20.11.one-loop}), we get 
\begin{eqnarray}
\gamma^{L,T}_W(g_L)= \frac{N-2}{2\pi}g_L+
\frac{N-2}{2\pi}\left(\frac{3}{2\pi}-\frac{1}{4}\right)g_L^2+O(g_L^3)\, .
\end{eqnarray}

\subsubsection{Antisymmetric Sector}
The OPE in the antisymmetric sector has an even simpler structure:
\begin{eqnarray}
&& \hskip -1truecm
\sum_c\left[j^{L,ac}_{\mu   ,x}j^{L,bc}_{\nu,-x}-
j^{L,bc}_{\mu   ,x}j^{L,ac}_{\nu,-x}\right]  =  
\nonumber \\ 
&&
\left[{x_\mu x_\nu x_\alpha\over (x^2)^2} U_{00}^L(x) +
      {\delta_{\mu \nu} x_\alpha\over x^2} U_{01}^L(x) +
      {\delta_{\mu \alpha}x_\nu + \delta_{\nu\alpha} x_\mu \over x^2} 
      U_{02}^L(x)
\right]
\frac{1}{g_L}j^{L,ab}_{\alpha,0}
\nonumber \\[1mm]
&&+ \left(\delta_{\mu   \alpha}\delta_{\nu\beta}-
\delta_{\mu   \beta}\delta_{\nu\alpha}\right) U_1^L(x)
\frac{1}{4 g_L}\left[(\partial_\alpha^- j_\beta^{L,ab})_0 -
                   (\partial_\beta^- j_\alpha^{L,ab})_0\right]
\nonumber\\[1mm]
&&+\frac{x_\mu   x_{\alpha}\delta_{\nu\beta}-
x_{\nu}x_{\alpha}\delta_{\mu   \beta}}{x^2}U_2^L(x)\frac{1}{2 g_L}
\left[(\partial_\alpha^- j_\beta^{L,ab})_0 +
      (\partial_\beta^- j_\alpha^{L,ab})_0\right].
\label{anticorrlatt}
\end{eqnarray}
It is easy to write th RG equations which hold for the 
Wilson coefficients 
$U^L_{0i}(x)$ and $U^L_i(x)$. It is easier to guess the solution of
these equations without writing them. Since $j^{L,ab}_{\mu,x}$ does
not renormalize, we have $(1/g_L) U^L(x;g_L) = U^L_{RGI}(\Lambda_L x)$.

The one-loop results for the Wilson coefficients are 
\begin{eqnarray}
U^L_{00}(x) & = & \frac{N-2}{2\pi}g_L + O(g_L^2)\, , \\
U^L_{01}(x) & = & - \frac{N-2}{4\pi}g_L + O(g_L^2)\, , \\
U^L_{02}(x) & = &  \frac{N-2}{4\pi}g_L + O(g_L^2)\, , \\
U^L_1(x) & = &  1-\frac{N-2}{4\pi}g_L(2\gamma+
\log(32 x^2))-\frac{1}{4}g_L + \frac{N-6}{4\pi}g_L + O(g_L^2)\, ,\\
U^L_2(x) & = & -\frac{N-2}{4\pi}g_L + O(g_L^2)\, .
\end{eqnarray}
%
%*************************************************************************
%
\section{Constraints on the OPE Coefficients} \label{secB.2}

In this Section we want to derive the constraints on the coefficients
of the OPE due to the current conservation. First, we need 
the Ward identity related to the $O(N)$ invariance in the presence of
a magnetic term $h$. A simple calculation gives:
\be
\< \partial_\mu j_\mu ^{ab}(x) {\cal O}\> = \,
  {h\over g} \< (\delta^{Na} \sigma^b(x) - \delta^{Nb} \sigma^a(x)) {\cal O}\>
  + \left \< {\delta {\cal O}\over \delta \sigma^a(x)} \sigma^b(x) -
       {\delta {\cal O}\over \delta \sigma^b(x)} \sigma^a(x) \right\>.
\label{Wardid-conmassa}
\ee
Then, we need the OPE of $j_\mu ^{ab}(x)\sigma^c(0)$. 
The leading term for $x\to 0$ has the form 
\be
j_\mu ^{ab}(x)\sigma^c(0) = {x_\mu \over x^2} f(\mu x; g) 
   (\delta^{ac} \sigma^b(0) - \delta^{bc}\sigma^a (0) ).
\ee
Using Eq.
\reff{Wardid-conmassa} and noticing that $\sigma^a(x)\sigma^b(0)\sim O(1)$
for $x\to0$, we have $\partial f(\mu x; g)/\partial x^2 = 0$. Thus, 
$f(\mu x; g)$ is a function 
of $g$ only. But the Wilson coefficient satisfies the RG equation
\be
\left( \mu {\partial\over\partial\mu } + \beta(g){\partial\over\partial g}\right)
   f = 0.
\ee
Thus, if it is independent of $x$, and therefore of $\mu $, it is also 
independent of $g$. A simple calculation at tree level gives then
\be
j_\mu ^{ab}(x)\sigma^c(0) = {1\over 2\pi} {x_\mu \over x^2}
   (\delta^{ac} \sigma^b(0) - \delta^{bc}\sigma^a (0) ).
\label{OPE-j-sigma}
\ee
The same result has been obtained in \cite{Luescher:1978} using the 
canonical formalism. 

We now consider the OPE of the scalar product of currents. 
Using the Ward identity \reff{Wardid-conmassa} and Eq. \reff{OPE-j-sigma}
we have for $x\to 0$
\be
{1\over2} \sum_{ab} \partial_\mu j^{ab}_\mu (x) j^{ab}_\nu(0) = 
   {h\over g} \sigma^b(x) j_\nu^{Nb} (0) =
   {h\over g} {N-1\over 2\pi} {x_\nu\over x^2} \sigma^N(0),
\ee
where we have discarded contact terms. Then, using Eq. \reff{AlphaOnShell} 
and discarding again contact terms, we obtain for $x\to 0$
\be
{1\over2} \sum_{ab} \partial_\mu j^{ab}_\mu (x) j^{ab}_\nu(0) =
 {N-1\over 2\pi g} {x_\nu\over x^2} \left\{
 \opl\alpha\opr(0) + \opl (\partial\sg)^2\opr (0) \right\}.
\ee
This equation implies the following relations on the Wilson coefficients:
\begin{eqnarray}
&& x^2 \partial_\mu W_{0,\mu \rho} (x) = 2x_\mu W_{0,\mu \rho} (x) ,
\\
&& 2 x^2 {\partial\over \partial x^2}\left[
   W_1(x) + W_2(x) + W_3(x) \right] = 2 W_1(x) - W_2(x) + 2 W_3(x),
\\
&& x^2 {\partial\over \partial x^2}\left[
   W_3(x) + W_4(x) \right] = - W_1(x) - W_3(x),
\\
&& 2 x^2 {\partial\over \partial x^2}\left[
   W_5(x) + W_6(x) \right] = 
   {N-1\over 2\pi} g + {1\over 2g} \left[\beta(g) + g\gamma(g)\right] W_3(g) 
  - W_5(x),
\\
&& 2 x^2 {\partial\over \partial x^2}\left[
   W_7(x) + W_8(x) \right] = 
   {N-1\over 2\pi} g + {1\over 2} \gamma(g) W_3(g) 
  - W_7(x).
\end{eqnarray}
In the derivation we used Eq. \reff{EnergyMomentumTrace}
for the trace of the energy-momentum tensor. 

Now let us consider the antisymmetric case. Using the Ward identity
\reff{Wardid-conmassa} and Eq. \reff{OPE-j-sigma}, we obtain for $x\to 0$
\be
\sum_c \partial_\mu j_\mu ^{ac}(x) j_\nu^{bc}(0) - 
\partial_\mu j_\mu ^{bc}(x) j_\nu^{ac}(0) =\,
 {N-2\over 2\pi} {x_\nu\over x^2} {h\over g} 
 \left(\delta^{Na}\sigma^b(0) - \delta^{Nb}\sigma^a(0)\right) = 
 {N-2\over 2\pi} {x_\nu\over x^2} \partial_\mu j_\mu ^{ab}(0).
\ee
Again, contact terms have been discarded in the derivation.
Using this relation, we obtain the following constraints 
on the Wilson coefficients\footnote{
Equations \reff{anti-constr-eq1}--\reff{anti-constr-eq5} have been
derived in Ref. \cite{Luescher:1978}. Eq. \reff{anti-constr-eq6} 
is due to the matching of the terms proportional to 
$\partial_\mu j_\mu ^{ab}(0)$. It is also a simple consequence 
of Eqs. \reff{anti-constr-eq2} and \reff{anti-constr-eq4} and of the 
RG equations.}
\begin{eqnarray}
&& x^2 {\partial\over \partial x^2}
  \left[ U_{00}(x) + U_{01}(x) + U_{02}(x) \right] = 
  U_{01}(x) + U_{02}(x),
\label{anti-constr-eq1} \\
&& 2 x^2 {\partial\over \partial x^2} U_{02}(x) = 
  - U_{01}(x)  - U_{02}(x),
\label{anti-constr-eq2} \\
&& x^2 {\partial\over \partial x^2}
  \left[ U_1(x) + U_{02}(x)\right] = - U_{02}(x),
\label{anti-constr-eq3} \\
&& x^2 {\partial\over \partial x^2}
  \left[ U_2(x) + U_{02}(x)\right] = - U_{01}(x)  - U_{02}(x),
\label{anti-constr-eq4} \\
&& 2 x^2 {\partial\over \partial x^2}
  \left[ - U_2(x) + U_{00}(x) + U_{01}(x) + U_{02}(x) \right] =
\nonumber \\
   && \qquad\qquad - 2 U_2(x) - U_{00}(x) + 2 U_{01}(x) + 2 U_{02}(x),
\label{anti-constr-eq5} \\
&& U_{02}(x) - U_2(x) = {N-2\over2\pi} g.
\label{anti-constr-eq6} 
\end{eqnarray}
Using Eqs. \reff{anti-constr-eq1} and \reff{anti-constr-eq2} 
we obtain immediately
\be
{\partial\over \partial x^2}
   \left[ U_{00}(x) + U_{01}(x) + 3 U_{02}(x)\right] =\, 0,
\ee
which implies that this combination is $x$ and $\mu $ independent. 
By making use of the RG equations \reff{RG-Wilson-antisymmetric}
one proves that this combination 
is determined uniquely by its one-loop value. Then, using the 
results of the previous Section, we obtain:
\be
U_{00}(x) + U_{01}(x) + 3 U_{02}(x) = {N-2\over \pi} g.
\label{vincolo1}
\ee
Thus, using \reff{vincolo1} and \reff{anti-constr-eq2}, 
$U_{00}(x)$ and $U_{01}(x)$ are uniquely determined by $U_{02}(x)$. Moreover,
using Eqs. \reff{vincolo1} and \reff{anti-constr-eq6} one immediately verifies 
that Eqs. \reff{anti-constr-eq4} and \reff{anti-constr-eq5} are 
equivalent to Eq. \reff{anti-constr-eq2}. 

Finally, consider \reff{anti-constr-eq3}. We will now show that this equation
provides the two-loop estimate of $U_{02}(x)$. Indeed, since 
\be
\left(\mu {\partial\over \partial\mu } + 
      \beta(g) {\partial\over \partial g} - {\beta(g)\over g}\right)
   [U_1(x) + U_{02}(x)] = 0
\ee
and $U_1(x) + U_{02}(x) = 1 + O(g)$, cf. previous Section, we have
\be
U_1(x) + U_{02}(x) = 1 - (\beta_0 \log \mu x + a_0) g -
                      (\beta_1 \log \mu x + a_1) g^2 + O(g^3),
\ee
where $\beta(g) = - g^2 \sum_{k=0}\beta_k g^k$, and $a_0$, $a_1$ 
are constants that are not fixed by the RG equation. Plugging 
this expression into \reff{anti-constr-eq3}, we obtain
\be
U_{02}(x) = {\beta_0\over2} g + {\beta_1\over2} g^2 + O(g^3) = \, 
   {N-2\over 4\pi} g + {N-2\over 8 \pi^2} g^2 + O(g^3).
\ee
Of course, the result at order $g$ agrees with the expression reported
in Sec. \ref{PerturbativeContinuumCurrentsSection}. Correspondingly we obtain
\begin{eqnarray}
U_{00}(x) &=&   {N-2\over 2\pi} g - {N-2\over 4 \pi^2} g^2 + O(g^3), \\
U_{01}(x) &=& - {N-2\over 4\pi} g - {N-2\over 8 \pi^2} g^2 + O(g^3), \\
U_{2}(x)  &=& - {N-2\over 4\pi} g + {N-2\over 8 \pi^2} g^2 + O(g^3).
\end{eqnarray}
Let us finally mention that in Ref. \cite{Luescher:1978} it was argued 
that the functions $U_{0i}(x)$ are one-loop exact, in the sense that 
there are no corrections of order $g^k$, $k\ge 2$. As we have shown above
and it has also been recognized by the 
author,\footnote{In Ref. \cite{Luescher:1978Erratum}
it was also shown that, even though the expressions for the Wilson
coefficients were incorrect, one could still modify the argument
so that the main result (existence of a conserved charge) of 
Ref. \cite{Luescher:1978} remains true.}
this is inconsistent with the RG equations.

%
%**************************************************************************
%
\section{Numerical Results} 
\label{NumericalCurrentsSection}

In this Section we present our numerical computations. 
We evaluated short-distance products of the type (\ref{OPECurrentsGen})
through Monte Carlo simulations. We considered several different
specifications of the Lorentz and $O(N)$ indices, omitted in Eq. 
(\ref{OPECurrentsGen}), as well as of the external one-particle states 
$|\qb\>$ and $\<\pb|$. 

The typical procedure we adopt is the following, see Eq. (\ref{OPECurrentsGen}):
\begin{enumerate}
\item We compute a matrix element
$\<\pb|j(x)j(0)|\qb\>$ by measuring a suitable lattice 
correlation function in Monte Carlo simulations, and taking the
on-shell limit for the one-particle states.
\item We compute the renormalized matrix element
$\<\pb|{\cal O}|\qb\>$ appearing in the short distance expansion 
either exactly (this is possible in most of the cases), or numerically 
by means of some different numerical non-perturbative technique.
\item We divide $\<\pb|j(x)j(0)|\qb\>$ by the OPE prediction
$W_{\cal O}(x)\<\pb|{\cal O}|\qb\>$. If more than one operator 
appears on the r.h.s. of Eq. (\ref{OPECurrentsGen}), we of course 
sum over them. 
\end{enumerate}
The goal is to see if there is a window of values of $|x|$ in which the 
OPE works, i.e. the result of step 3 is 1 independently of $|x|$.
For the cases we will consider here, the OPE gives an accurate
description (at the level of 5-10\%) of correlation functions for distances 
$2\lesssim |x|\lesssim\xi=m^{-1}$ (remember that, when not
explicitly stated, we take $a=1$).
This result is quite encouraging for future applications of this method.

\subsection{The Observables} \label{sec4.1}

We have simulated the 
$O(3)$ $\sigma$-model with action (\ref{LatticeAction})
using a Swendsen-Wang cluster algorithm 
with Wolff embedding 
\cite{Wolff:1989uh,Wolff:1989kw,Wolff:1990dm,Caracciolo:1993nh}. 
We did not try to optimize the updating procedure:
Most of the CPU time was employed in evaluating the relevant observables
(four-point functions) 
on the spin configurations of the ensemble. 
In order to estimate the scaling corrections,
we simulated three different lattices, of size $T\times L$, using 
in all cases periodic boundary conditions:
\renewcommand{\theenumi}{(\Alph{enumi})}
\begin{enumerate}
\item \label{Lattice64x128}
Lattice of size $128\times 64$ with $g_L = 1/1.40$. 
\item \label{Lattice128x256}
Lattice of size $256\times 128$ with $g_L = 1/1.54$. 
\item \label{Lattice256x512}
Lattice of size $512\times 256$ with  $g_L = 1/1.66$.
\end{enumerate}
The algorithm is extremely efficient---the dynamic critical exponent 
$z$ is approximately 0---and the autocorrelation time is very small. 

%
%**************************************************************************
%
\begin{figure}
\hspace{0.0cm}
\centerline{
\epsfig{figure=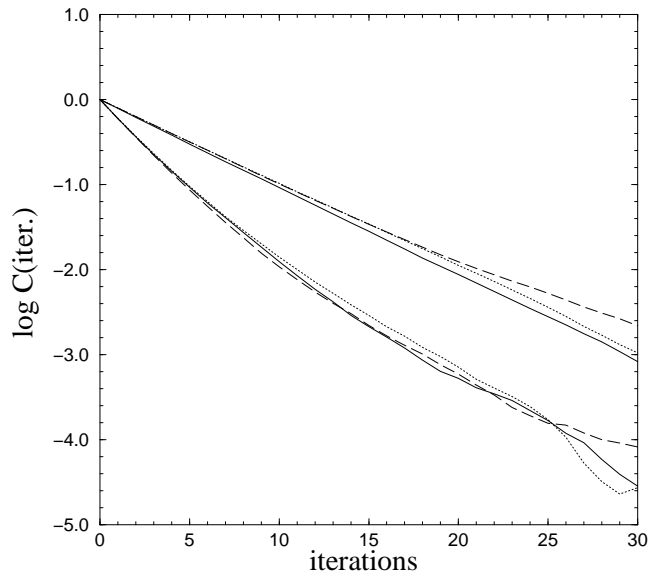,angle=-90,
width=0.6\linewidth}
}
\caption{Logarithm of the autocorrelation function $\log A(i)$
for the three lattices employed. Continuous lines refer to lattice
\ref{Lattice64x128}, dotted lines to lattice \ref{Lattice128x256}
and dashed lines to lattice \ref{Lattice256x512}. The upper curves
refer to the short-distance observable ${\cal O}_d$, $d=1$.
The lower curves refer to the long-distance observable
${\cal O}_d$ for $d\approx \xi^{\rm exp}$:
$d=8$ for lattice \ref{Lattice64x128}, 
$d=15$ for lattice \ref{Lattice128x256} and
$d=30$ for lattice \ref{Lattice256x512}.}
\label{Autocorrelation}
\end{figure}
%
%**************************************************************************
%
We performed a preliminary study in order to determine how many iterations 
are needed to obtain independent configurations. For this purpose we measured 
the normalized autocorrelation function
\be
A(j) = {{\cal N}\over {\cal N} -j} {\sum_{i=1}^{{\cal N}-j} 
     ({\cal O}_i - \overline{{\cal O}})
     ({\cal O}_{i+j} - \overline{{\cal O}}) \over 
           \sum_{i=1}^{{\cal N}} ({\cal O}_i - \overline{{\cal O}})^2 }
\ee
for different observables $\cal O$. Here 
${\cal N}$ is the number of Monte Carlo iterations, ${\cal O}_i$ is the 
value of $\cal O$ at the $i$-th iteration, and 
$\overline{{\cal O}}$ the sample mean of $\cal O$:
$\overline{\cal O} = (1/{\cal N})\, \sum_{i=1}^{\cal N} {\cal O}_i$.
In Fig. \ref{Autocorrelation} we report $A(\tau)$ for the observable
\begin{eqnarray}
{\cal O}_{d} &\equiv& {1\over 2 L T} \sum_{x} 
( \sg_x\cdot\sg_{x+v}+ \sg_x\cdot\sg_{x+w})\, ,
\end{eqnarray}
where $v = (d,0)$ and $w = (0,d)$. 
We have considered $d=1$ ---a short-distance 
observable---and $d\approx \xi^{\rm exp}=m^{-1}$
(more precisely $d=8,\, 15,\, 30$, on lattices
\ref{Lattice64x128},\ref{Lattice128x256} and \ref{Lattice256x512},
respectively). 
As expected, local observables have a slower dynamics---this is due to the 
fact that the dynamics is nonlocal---than long-distance ones. 
In any case, for all observables the autocorrelation function shows a fast 
decay: indeed, for $d=1$ we have $\tau^{\rm exp}\approx 10$, while for 
$d\approx \xi^{\rm exp}$ we have $\tau^{\rm exp}\approx 5$. Moreover,
$A(\tau)$ is independent of the lattice used, confirming the fact that 
$z\approx 0$ (as we shall report below, all lattices have the same 
$L/\xi^{\rm exp}$). 

Since the measurement of the observables is quite CPU-time
consuming, we evaluated the observables listed in the next paragraphs, 
and in particular the short-distance products of Noether currents,
only every 15 iterations.
This should be enough to make the measurements independent. 
Nonetheless, most of the CPU time is employed in
evaluating the observables on each spin configuration of the ensemble.
The updating time is a small fraction of the total computing time.
We computed the product $j(x)j(0)$ of two Noether currents
for distances $x$ smaller than some fixed fraction of
the correlation length: $|x|\lesssim k\xi$.
If the physical size $L/\xi$ of the lattice is kept constant (as we did)
we expect the CPU time to scale as $L^4$.
The CPU time per iteration turns out to be roughly independent
of the particular product considered.
As an example we give the CPU time per measurement for the
simulation in which we compute the antisymmetric product of two currents 
between states with opposite momentum, see Sec. \ref{sec4.5}. 
For the three different lattices, on an SGI Origin2000, we have:
$\tau_{128\times 64}\approx 5.4\, \rm{sec}$,
$\tau_{256\times 128}\approx 71\, \rm{sec}$,
$\tau_{512\times 256}\approx 1100\, \rm{sec}$.

We measure several different observables. First, we measure the two-point 
function $C(\pb;t)$ (here and in the following the ``temporal" direction
is the first one, of extent $T$)
\begin{eqnarray}
C(\pb;t)\equiv \frac{1}{LT}\sum_{t_1=1}^T\sum_{x_1,x_2=1}^L
   e^{i\pb(x_1-x_2)} \<\sg_{t_1,x_1}\cdot\sg_{t+t_1,x_2}\>.
\label{TwoPointCorrelationFunction}
\end{eqnarray}
We computed the correlation function $C(\pb;t)$ on the lattices
\ref{Lattice128x256} and \ref{Lattice256x512} for momenta
$\pb=2\pi n/L,\, n=0\dots 3$ and times separations
$0\le t\le 100$; on lattice \ref{Lattice64x128} we considered the same set of
momenta and time separations $0\le t\le 40$. 
The number of independent configurations we generated is: 
$N_{\rm conf}\simeq 6\cdot 10^6$
for lattice \ref{Lattice64x128}; $N_{\rm conf}=590000$ for
lattice \ref{Lattice128x256}; $N_{\rm conf}= 180000$ for
lattice \ref{Lattice256x512} and $\pb=0$; finally
$N_{\rm conf}= 139000$ for lattice \ref{Lattice256x512} and $\pb\neq 0$. 

A  check of our simulation is provided by the results of Ref. 
\cite{Luscher:1990ck}, who computed, among other things, the mass gap for 
lattices \ref{Lattice64x128} and \ref{Lattice128x256}.
For the exponential correlation length $\xi^{\rm exp} = m^{-1}$ 
we obtain 
\be
\xi^{\rm exp} = 6.878(2), \, 13.638(10), \, 27.054(25) ,
\label{XiValues}
\ee
for lattices  \ref{Lattice64x128}, \ref{Lattice128x256}, \ref{Lattice256x512}
respectively. They are in good agreement with the 
results of Ref. \cite{Luscher:1990ck}: they obtain 
$\xi^{\rm exp} = 6.883(3)$ and $\xi^{\rm exp}=13.632(6)$ 
for the first two lattices.
The three lattices we simulate 
have approximately the same physical size,
$L/\xi\sim 9$, which is large enough to make finite-size effects 
much smaller than our statistical errors.
Finite-size corrections are indeed supposed to be exponentially 
small in the physical size (i.e. of order $\exp(-L/\xi)$). This is consistent
with the analysis of \cite{Caracciolo:1995ed}.
In Ref. \cite{Luscher:1990ck} the authors verified the smallness
of finite-size effects on
lattices \ref{Lattice64x128} and \ref{Lattice128x256}.
\begin{table}
\begin{center}
\begin{tabular}{|c|c|c|c|}
\hline 
$p$ 
& ${\cal O} = (\overline{\partial}_0\sg)^2$
& ${\cal O} = (\overline{\partial}_0\sg\cdot\overline{\partial}_1\sg)$
& ${\cal O} = (\overline{\partial}_1\sg)^2$ \\
\hline
$0$ &$34.619(25)$&$0.00053(48)$ & $34.663(25)$\\
$2\pi/L$ & $34.707(18)$ &$0.03065(26)$ & $34.776(18)$\\
$4\pi/L$ & $34.735(18)$ &$0.06080(39)$ & $34.857(18)$\\
$6\pi/L$ & $34.741(26)$ &$0.09026(64)$ & $34.923(26)$\\
\hline
\end{tabular}
\end{center}
\caption{Estimates of $\sum_a\widehat{C}^{aa}_{\cal O}(p,p;10)$ for different 
operators measured on lattice \ref{Lattice128x256}. 
For $(\overline{\partial}_0\sg\cdot\overline{\partial}_1\sg)$ we report 
the imaginary part, the real part being zero. 
The matrix element of the other two operators is real.}
\label{DiagonalMatrixElement}
\end{table}
\begin{table}
\begin{center}
\begin{tabular}{|c|c|c|c|}
\hline 
$p$  
& ${\cal O} = (\overline{\partial}_0\sg)^2$
& ${\cal O} = (\overline{\partial}_0\sg\cdot\overline{\partial}_1\sg)$
& ${\cal O} = (\overline{\partial}_1\sg)^2$ \\
\hline
$2\pi/L$ & $0.25434(60)$ &
$ 0.01452(47) $&$ 0.20241(68) $\\
$4\pi/L$ & $0.25153(85)$ &
$ 0.02824(61) $&$ 0.18394(94) $\\
$6\pi/L$ & $0.25597(128)$ & 
$ 0.04024(96) $&$ 0.16643(140) $\\
\hline
\end{tabular}
\end{center}
\caption{Estimates of $\sum_a\widehat{C}^{aa}_{\cal O}(p,0;20)$ 
for different operators measured on 
lattice \ref{Lattice128x256}. We report here
the real part for $(\overline{\partial}_0\sg)^2$ and 
$(\overline{\partial}_1\sg)^2$, and the
imaginary part for $(\overline{\partial}_0\sg\cdot\overline{\partial}_1\sg)$.}
\label{OutOfDiagonalMatrixElement}
\end{table}

In order to verify the OPE, we need the values of the matrix elements
which appear in the r.h.s. of Eq. (\ref{OPECurrentsGen}).
Matrix elements of lattice operators can be computed from properly
defined three-point correlation functions. If
${\cal O}_{t,x}$ is a lattice operator,
we define the correlation function 
\begin{eqnarray}
C^{ab}_{\cal O} (\pb,\qb;2t) \equiv 
{1\over L T} \sum_{t_0=1}^T \sum_{x_0=1}^L \sum_{x_1,x_2=1}^L
e^{i\pb x_1-i\qb x_2}\<\sigma^a_{t_0-t,x_0+x_1}{\cal O}_{t_0,x_0}
\sigma^b_{t_0+t,x_0+x_2}\>,
\end{eqnarray}
and the corresponding normalized correlation
\be
\widehat{C}^{ab}_{\cal O} (\pb,\qb;2t) \equiv 
   {C^{ab}_{\cal O} (\pb,\qb;2t)\over 
    \sqrt{C(\pb;2t) C(\qb;2t)} }.
\label{NormalizedCorrelation}
\ee
The function $\widehat{C}^{ab}_{\cal O} (\pb,\qb;2t)$ has a finite limit
for $t\to\infty$. 
This limit gives access to the one-particle matrix elements of ${\cal
O}$, see Sec. \ref{sec4.3}. For this reason, we shall look for a
plateau in the large-$t$ behavior of 
$\widehat{C}^{ab}_{\cal O} (\pb,\qb;2t)$.

In this Chapter we will only need the matrix elements of the 
naive lattice energy-momentum tensor (\ref{NaiveEMT}). For this reason,
we have computed $C^{aa}_{\cal O} (\pb,\pb;2t)$ with 
${\cal O} = \overline{\partial}_{\mu}\sg\cdot\overline{\partial}_{\rho}\sg$.
Such a correlation function has been computed on lattice \ref{Lattice128x256},
using $N_{\rm conf} = 320000$ configurations, for
$t=5,\dots,20$ and $\pb = 2n\pi/L$, $n=0,\dots,3$. 
For these observables $\widehat{C}^{ab}_{\cal O} (\pb,\pb;2t)$ 
is independent of $t$, within the statistical errors, already at $t=5$.
The results  obtained for $t=5$ are reported in Table 
\ref{DiagonalMatrixElement}. For 
${\cal O} = \overline{\partial}_0\sg\cdot\overline{\partial}_1\sg$, 
statistical errors are dominated by the error on the evaluation of 
$C^{ab}_{\cal O} (\pb,\pb;2t)$. On the other hand,
for ${\cal O} = (\overline{\partial}_\mu\sg)^2$, 
the statistical error of the numerator
in Eq. (\ref{NormalizedCorrelation}) is roughly equal to that of 
the denominator. The reason is clear: since 
in the continuum limit $(\overline{\partial}_\mu\sg)^2$ is proportional
to the identity operator, we are computing
essentially (up to $O(a^2)$ corrections) the same quantity in the numerator
and in the denominator, with approximately the same statistics. 
The reported errors
on the ratios are obtained using the independent error formula. 
For $(\overline{\partial}_\mu\sg)^2$ smaller error bars could have been 
obtained by taking into account the statistical correlations between
numerator and denominator.

We also measured $C^{aa}_{\cal O}(0,\pb;2t)$ for the same operators 
on lattice \ref{Lattice128x256}, using $N_{\rm conf} = 62000$ independent
configurations, for $\pb = 2n\pi/L$ with $n=1,\dots,3$ and 
$t=5,\dots,10$. The normalized three-point function shows a plateau for
$t\gtrsim 10$ when $\mu = \nu $ and for $t\gtrsim 6$ when $\mu\neq\nu$.
The results obtained for $t=10$ 
are reported in Tab. \ref{OutOfDiagonalMatrixElement}. 
In this case the statistical errors are dominated by
the uncertainty on $C^{aa}_{\cal O}(0,\pb;2t)$.

In order to obtain one-particle matrix elements of 
products of Noether currents, we proceed in the
same manner as above. The only difference is that we must
consider four-point (instead of three-point) functions.
In particular, let us define:
\begin{eqnarray}
\hspace{-1.0cm}
G^{(s)}(t,x;\pb,\qb;2t_s)&\equiv& \frac{1}{2}
\sum_{x_1,x_2}\<(\jg_{0,(0,0)}^L\cdot
\jg_{1,(t,x)}^L)(\sg_{-t_s,x_1}\cdot \sg_{t_s,x_2})\>e^{i\pb x_1-i\qb
x_2}\! ,
\\
\hspace{-1.0cm}
G^{(a)}_{\mu\nu}(t,x;\pb,\qb;2t_s) &\equiv&
\sum_{x_1,x_2}\sum_{abc}\<(j^{L,ac}_{\mu,(0,0)}j^{L,bc}_{\nu,(t,x)}-
j^{L,bc}_{\mu,(0,0)}j^{L,ac}_{\nu,(t,x)})
\sigma^a_{-t_s,x_1} \sigma^b_{t_s,x_2}\>\,e^{i\pb x_1-i\qb x_2}\! ,
\end{eqnarray}
where $j_{\mu,x}^{L,ab}$ is the lattice Noether current defined in
Eq. \reff{LatticeCurrent}. Of course, we averaged over lattice translations.

Here is a list of the quantities we measured in our Monte Carlo simulations:
\begin{itemize}
\item[a)] $G^{(s)}(t,x;\pb,\pb;2t_s)$ using 
 $N_{\rm conf} \simeq 1.3\cdot 10^6$
 configurations on lattice \ref{Lattice64x128} and using 
 $N_{\rm conf} = 58350$ independent configurations on lattice
 \ref{Lattice128x256}.
\item[b)] ${\rm Im}\, G^{(s)}(t,x;\pb,0;2t_s)$ using $N_{\rm conf} = 112000$
independent configurations on lattice \ref{Lattice128x256}.
\item[c)] 
${\rm Re}\, G^{(a)}_{11}(t,x;\pb ,-\pb ;2t_s)$
using $N_{\rm conf}\simeq 2.4\cdot 10^6$
configurations on lattice \ref{Lattice64x128},
$N_{\rm conf}= 69500$ independent configurations on
lattice \ref{Lattice128x256}, and $N_{\rm conf}= 31550$ configurations
on lattice \ref{Lattice256x512}.
\item[d)]
${\rm Im}\, G^{(a)}_{01}(t,x;\pb ,0 ;2t_s)$
using $N_{\rm conf}= 41750$ independent configurations on
lattice \ref{Lattice128x256}. 
\end{itemize}
In all cases we consider $\pb=2\pi n/L$, $n=1,2,3$; 
$t_s=7,8,9$ and $|t|\le 5,|x|\le 5$ on lattice \ref{Lattice64x128},
$t_s=10,11,12$ and $|t|\le 8,|x|\le 8$ on lattice \ref{Lattice128x256}, and
$t_s=20,23, 26$ and $|t|\le 12,|x|\le 12$ on lattice \ref{Lattice256x512}.

Using the four-point correlation function determined above, we constructed
the normalized ratios
\begin{eqnarray}
\widehat{G}^{(\cdot)} (t,x;\pb,\qb;2t_s) \equiv 
   {G^{(\cdot)} (t,x;\pb,\qb;2t_s)\over 
    \sqrt{C(\pb;2t_s) C(\qb;2t_s)} },
\label{NormalizedFourPoint}
\end{eqnarray}
which have a finite limit for $t_s\to \infty$.
We verified that $\widehat{G}^{(\cdot)} (t,x;\pb,\qb;2t_s)$ is independent
of $t_s$ in the range considered, and thus we have taken the result
obtained at the lowest considered value of $t_s$ as an estimate
of $\widehat{G}^{(\cdot)} (t,x;\pb,\qb;\infty)$.

In the paper we will usually consider averages over two-dimensional 
rotations, i.e., given a function $f(z_t,z_x)$, we consider 
\begin{eqnarray}
\overline{f}(r) \equiv 
   {\sum_{z\in\mathbb{Z}^2} f(z)\, \Theta_r(z)  \over 
    \sum_{z\in\mathbb{Z}^2} \Theta_r(z) },\quad
  \Theta_r(z) \equiv  \theta\left(|z| - r+\frac{1}{2}\right) 
                      \theta\left(r+\frac{1}{2} -|z|\right) ,
\label{definizione-mediarotazionale}
\end{eqnarray}
with $|z|\equiv\sqrt{z_t^2+z_x^2}$ and $r=n+1/2$, $n$ integer.

%
%*******************************************************************
%
\subsection{One-Particle States} \label{sec4.2}

In the conventional picture
the lowest states of the model are one-particle
states transforming as $O(N)$ vectors. On a lattice of finite spatial 
extent $L$, we normalize the states and the fields as follows:
\begin{eqnarray}
\<\pb,a|\qb,b\> &=& 2\omega(\pb)L\,\delta^{ab}\delta_{\pb,\qb}, 
\label{OneParticleNormalization}\\
\<\pb,a|\hat{\sigma}^b_x|0\> &=& 
 \sqrt{\frac{Z(\pb)}{N}}\,\delta^{ab}e^{ i\pb x}.
\end{eqnarray}
The function $\omega(\pb)$, which is the energy of the state $|\pb,a\>$, and 
the field renormalization $Z(\pb)$ can be determined from the 
large-$|t|$ behavior of the two-point function $C(\pb;t)$:
\begin{eqnarray}
C(\pb;t)\sim \frac{Z(\pb)}{2\omega(\pb)}e^{-\omega(\pb)t}
\qquad
\mbox{for}
\quad
t\gg 1.
\label{AsymptoticCorrelation}
\end{eqnarray}
In the continuum (scaling) limit we have $Z(\pb)=Z$ independent of 
$\pb$ and $\omega(\pb) = \sqrt{\pb^2 + m^2}$. In Table \ref{LatticeSpectrum}
we report our results for the three lattices. 
In order to evaluate $\omega(\pb)$ and $Z(\pb)$,
we determined effective values at time $t$ by solving the equations
\begin{eqnarray}
\frac{C(\pb;t+1)}{C(\pb;t)}&\equiv&
\frac{\cosh[\omega_{\rm eff}(\pb,t)(t+1-T/2)]}
{\cosh[\omega_{\rm eff}(\pb,t)(t-T/2)]}, 
\\
C(\pb;t)&\equiv &\frac{Z_{\rm eff}(\pb,t)}{2\omega_{\rm eff}(\pb,t)}
\left\{ e^{-\omega_{\rm eff}(\pb,t)t}
+e^{-\omega_{\rm eff}(\pb,t)(T-t)}\right\}.
\end{eqnarray}
Then we looked for a plateau in the plot of 
$\omega_{\rm eff}(\pb,t)$ and $Z_{\rm eff}(\pb,t)$ versus $t$. 
Both functions become independent of $t$ for $t\gtrsim \xi$.
The values reported in Table \ref{LatticeSpectrum} correspond to 
one particular value of $t$ of order $\xi^{\rm exp}$: 
$t = 8, 16, 20$ respectively for
lattice \ref{Lattice64x128}, \ref{Lattice128x256},
\ref{Lattice256x512}.

%
%*************************************************************
%
\begin{figure}
\centerline{
\hspace{-1.5cm}
\epsfig{figure=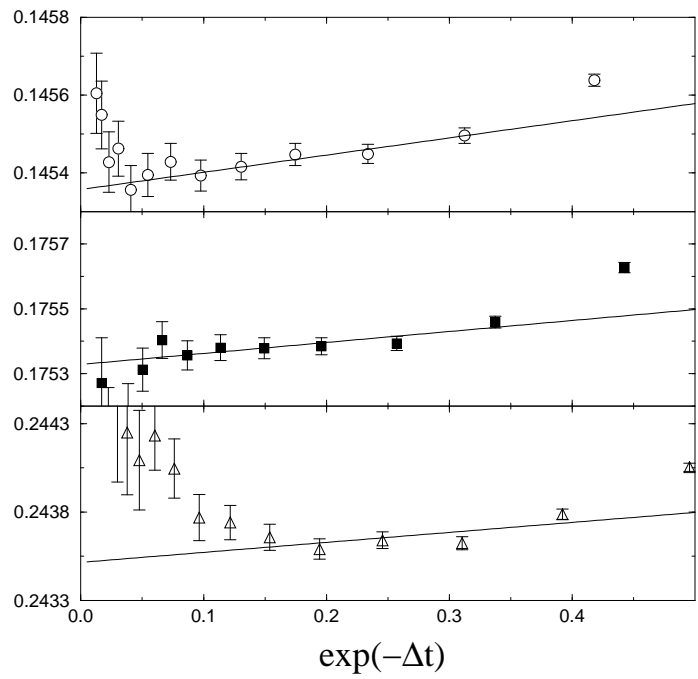,angle=-90,
width=0.6\linewidth}
}
\caption{The asymptotic behavior of $\omega_{\rm eff}(\pb,t)$ as 
$t\to\infty$ on lattice \ref{Lattice64x128}.
Empty circles refer to $\pb = 0$, filled squares to $\pb = 2\pi/L$ 
and triangles to $\pb = 6\pi/L$. The continuous lines are the best fitting
curves of the form (\ref{SpectrumFit}). For sake of clarity we show
the numerical results only for $t\le 15$.}
\label{MassAsymptotia}
\end{figure}
In principle
we should take the limit: $\omega(\pb) = \lim_{t\to\infty}\omega_{\rm
eff}(\pb,t)$, and $Z(\pb)=\lim_{t\to\infty}Z_{\rm eff}(\pb,t)$.
Our procedure consists in using one particular value of $t$
rather than trying an extrapolation. This gives good
results as long as the systematic error (due to the finiteness of $t$)
is of the same order as the statistical one.
The expected large-$t$ behavior of $\omega_{\rm eff}(\pb,t)$ is
\begin{eqnarray}
\omega_{\rm eff}(\pb,t) = \omega(\pb)+\eta(\pb;t)e^{-\Delta(\pb)t}+\dots\, ,
\end{eqnarray}
where we neglected  terms of
order $e^{-\omega(\pb)T}$ and multi-particle states involving more than
three particles. 
The ``standard wisdom'' prediction for the gap $\Delta(\pb)$ is:
\begin{eqnarray}
\Delta(\pb) = \sqrt{(3m)^2+\pb^2}-\sqrt{m^2+\pb^2}\, .
\label{Gap}
\end{eqnarray}
In the thermodynamic ($L\to\infty$) limit,
the coefficient $\eta(\pb;t)$ is a slowly varying (power-like)
function of $t$.

In Fig. \ref{MassAsymptotia} we plot $\omega_{\rm eff}(\pb,t)$, versus
$\exp(-\Delta(\pb)t)$. together with best fitting (least squares)
curves of the form:
\begin{eqnarray}
\omega_{\rm eff}(\pb,t) = \omega^*(\pb)+\eta^*(\pb)e^{-\Delta(\pb)t}\,
. \label{SpectrumFit}
\end{eqnarray}
There is rough agreement between this form and the numerical data. 
The $t$ dependence of $\eta(\pb,t)$ cannot be appreciated due to the 
statistical errors.
From the fit (\ref{SpectrumFit}) we get an idea of the systematic error
on $\omega(\pb)$, namely $\eta^*(\pb)\exp(-\Delta(\pb)t)$.
The estimated systematic error, corresponding to the curves in 
Fig. \ref{MassAsymptotia}, is about $0.3\div 0.5\cdot 10^{-5}$, which is
of the same order as the statistical error.
The above analysis can be easily extended to $Z_{\rm eff}(\pb,t)$.

%
%**************************************************************************
%
\begin{figure}
\centerline{
\epsfig{figure=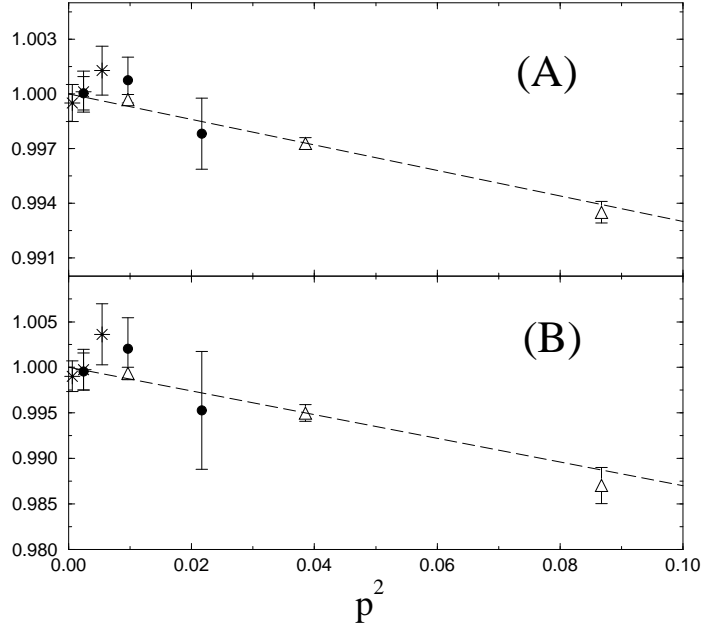,angle=-90,
width=0.6\linewidth}
}
\caption{Corrections to scaling in the one-particle spectrum.
In graph (A) we plot the ratio $\omega(\pb)/\sqrt{\pb^2+m^2}$,
in graph (B) $Z(\pb)/Z(0)$. 
Empty triangles ($\triangle$) refer to lattice \ref{Lattice64x128},
filled circles ($\bullet$) to lattice \ref{Lattice128x256},
and stars ($\ast$) to lattice \ref{Lattice256x512}.
The dashed lines are of the form $1-c\pb^2$.}
\label{Spectrum}
\end{figure}
%
%**************************************************************************
%
Let us now look at scaling corrections in the one-particle spectrum.
In the continuum limit we expect $\omega(\pb)\to\sqrt{m^2+\pb^2}$,
and $Z(\pb)\to Z$. In Fig. \ref{Spectrum} we plot the ratios
$\omega(\pb)/\sqrt{\pb+m^2}$ (graph (A)), and $Z(\pb)/Z(0)$
(graph (B)). Both this quantities all well described
by the same form, see Ref. \cite{Luscher:1990ck}:
\begin{eqnarray}
\frac{\omega(\pb)}{\sqrt{\pb^2+m^2}} & \simeq & 1 - c_{\omega}\pb^2\, ,
\label{SpectrumCorrections1}\\
\frac{Z(\pb)}{Z(0)} & \simeq & 1-c_{Z}\pb^2\, .
\label{SpectrumCorrections2}
\end{eqnarray}
The fitting lines in Fig. \ref{Spectrum} have been obtained using
$c_{\omega} = 0.07$ and $c_{Z} = 0.13$. The fitting
form in Eqs. (\ref{SpectrumCorrections1}) and (\ref{SpectrumCorrections2})
follows from the general behavior of scaling corrections.
Let us consider, for instance, the ratio $Z(\pb)/Z(0)$.
Since the continuum limit of this quantity is equal to one, 
it behaves as follows:
\begin{eqnarray}
\frac{Z(\pb)}{Z(0)} = 1+\frac{1}{L^2}\Delta_{Z}(mL,\pb L;L)+O(L^{-4})\, ,
\end{eqnarray}
with $\Delta_Z(\dots;L)$ weakly (logarithmically) depending upon $L$,
at $mL$ and $\pb L$ fixed.
Since $Z(0)/Z(0)=1$ and $Z(\pb)$ is even in $\pb$, it follows that
$\Delta_Z(mL,\pb L;L) = \Delta^{(1)}_Z(mL;L)(\pb L)^2+O(\pb^4)$, whence
\begin{eqnarray}
\frac{Z(\pb)}{Z(0)} = 1+\Delta^{(1)}_{Z}(mL;L)\pb^2+O(\pb^4)\, ,
\end{eqnarray}
which coincides with Eq. (\ref{SpectrumCorrections2}).
Our three lattices \ref{Lattice64x128}, \ref{Lattice128x256}
and \ref{Lattice256x512} have approximatively the same physical size
$mL$. 
Because of the logarithmic dependence of $\Delta^{(1)}_{Z}(mL;L)$
upon $L$, the fitting coefficients $c_{\omega}$ and $c_{Z}$
should not be independent of the lattice at $mL$ fixed.
However our statistical errors are too large to detect this weak 
dependence. Finally, let us notice that the behavior described 
in Eqs. (\ref{SpectrumCorrections1}) and (\ref{SpectrumCorrections2})
can be recovered in lattice perturbation theory \cite{Caracciolo:1998ir}.
It is easy to obtain $c_{\omega} = 1/12+O(g_L^2)$
(one-loop perturbation theory) and $c_{Z} =1/6+O(g_L)$ 
(tree-level perturbation theory). Both these results are in rough agreement
with our numerical data.
We conclude that for the two largest lattices the spectrum  
scales at the error-bar level. 
Instead, for lattice \ref{Lattice64x128} 
there are tiny (and essentially under control) 
scaling corrections. These corrections are so small 
(at most 1\%) that we can neglect them in the following discussion.

One can also investigate asymptotic 
scaling, i.e. the dependence of $\omega(\pb)$ and $Z$ on $g_L$. 
The dependence of $\omega(\pb)$ can be determined from 
Eq.~(\ref{MassGapLattice}). 
There exists also an exact prediction for $Z$,
including the non-perturbative constant 
\cite{Balog:NPB97,Campostrini:PLB97}. However, as is well known, 
lattice perturbation theory is not predictive at these values of the 
correlation length, and indeed, the perturbative four-loop predictions 
show large discrepancies compared to the numerical data.
The agreement is instead quite good \cite{Caracciolo:1995ah,Campostrini:PLB97}
if one uses the improved coupling $g_E$ defined in 
Eq. \reff{DressedCouplingEnergy}.
\begin{table}
\footnotesize
\begin{center}
\begin{tabular}{|c|c|c|c|c|c|c|}
\hline
 &\multicolumn{2}{|c|}{lattice \ref{Lattice64x128}}&
\multicolumn{2}{|c|}{lattice \ref{Lattice128x256}}&
\multicolumn{2}{|c|}{lattice \ref{Lattice256x512}}\\
\hline
$\pb$ & $\omega(\pb)$ & $Z(\pb)$ & $\omega(\pb)$ & $Z(\pb)$
& $\omega(\pb)$ & $Z(\pb)$\\
\hline
$0$      & $  0.145393(40)$ & $1.6593(8)$
         & $  0.073327(55)$ & $1.3563(18)$       
         & $  0.036963(34)$ & $1.1295(14)$\\
$2\pi/L$ & $  0.175380(40)$ & $1.6582(8)$
         & $  0.088244(67)$ & $1.3557(21)$
         & $  0.044348(35)$ & $1.1284(13)$\\
$4\pi/L$ & $  0.243657(74)$ & $1.6510(13)$
         & $  0.12263(15)$  & $1.3591(42)$
         & $  0.061456(66)$ & $1.1292(21)$\\
$6\pi/L$ & $  0.326327(192)$& $1.6378(32)$
         & $  0.16415(32) $ & $1.3499(86)$ 
         & $  0.082494(109)$& $1.1336(35)$\\
 \hline
\end{tabular}
\end{center}
\caption{The one-particle spectrum and the field normalization
for lattices \ref{Lattice64x128}, \ref{Lattice128x256}, \ref{Lattice256x512}.}
\label{LatticeSpectrum}
\end{table}

%
%**********************************************************************
%
\subsection{Non-Perturbative Renormalization of the Lattice 
Energy-Momentum Tensor} \label{sec4.3}
%
%
%**************************************************************************
%
\begin{figure}
\centerline{
\epsfig{figure=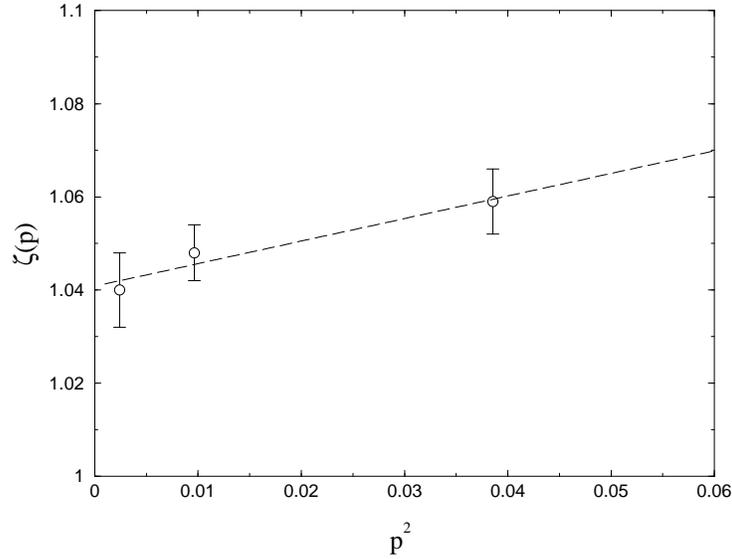,angle=-90,
width=0.6\linewidth}
}
\caption{Corrections to scaling for the renormalization constant 
of the energy-momentum tensor, cf. Eq. (\ref{ZetapDef}).
The circles correspond the numerical results obtained on lattice
\ref{Lattice128x256}.
The dashed line is the best fitting curve of the form 
$\zeta(\pb)=\widehat{\zeta}_0+\widehat{\zeta}_1\pb^2$.}
\label{Zetap}
\end{figure}
%
%**************************************************************************
%
In this Section we want to compute non-perturbatively the renormalization
constant $Z^{L(2,0)}_{TT}$ of the lattice energy-momentum tensor,
see Eq. (\ref{EnergyMomentumRenormalization}). 
In general, given an operator $\cal O$ on the lattice, we define
its matrix element by
\be
\<\pb,a|\hat{\cal O}|\qb,b\> \equiv 
    N \sqrt{4\omega(\pb)\omega(\qb)} 
  \lim_{t\to\infty} \widehat{C}^{ab}_{\cal O}(\pb,\qb;2t).
\ee
For $\overline{\partial}_\mu\sg\cdot \overline{\partial}_\nu\sg$ 
the matrix elements can be obtained from the results given in Tables 
\ref{DiagonalMatrixElement} and \ref{OutOfDiagonalMatrixElement}.
The matrix elements of $(\overline{\partial}_\mu\sg)^2$ are dominated 
by the mixing with the identity operator and thus, in order to define
the renormalized operator for $\mu=\nu$, we should perform a 
non-perturbative subtraction of the large $1/a^2$ term, which is practically
impossible.\footnote{In general we expect 
$\<\pb,a|(\overline{\partial}_0\sg)^2|\pb,b\> = 
2AL\sqrt{\pb^2+m^2}+B - C(\pb^2 + m^2)$ and 
$\<\pb,a|(\overline{\partial}_1\sg)^2|\pb,b\> = 
2AL\sqrt{\pb^2+m^2}+B + C \pb^2$. The quantity
we are interested in is $C$. However, from Table \ref{DiagonalMatrixElement}
we immediately realize that much smaller errors are required 
to really observe the momentum dependence of the matrix elements 
and thus to determine the constant $C$. }
Therefore, we only compute the renormalized operator for $\mu\not=\nu$,
which amounts to determining the constant $Z^{L,(2,0)}_{TT}$. 
This constant is obtained by requiring
\be
\<\pb,a| T_{01,0} |\pb,b\> =\, 
2 i \,\pb \sqrt{\pb^2 + m^2}\, \delta^{ab}.
\label{EnergyMomentumCondition}
\ee
In practice, we first compute an effective (momentum-dependent) 
renormalization constant
\be
\zeta(\pb) \equiv {2\, \pb\, \omega(\pb)\,g_L \over 
     {\rm Im}\, \<\pb,a|(\overline{\partial}_0\sg\cdot
\overline{\partial}_1\sg)  |\pb,a\>},
\label{ZetapDef}
\ee
which, in the continuum limit, becomes independent of $\pb$ and converges to
$Z^{L(2,0)}_{TT}$. Using the data of Table \ref{DiagonalMatrixElement},
on lattice \ref{Lattice128x256}, 
we obtain $\zeta(\pb) = 1.040(8)$, $1.048(6)$, $1.059(7)$ 
for $\pb = 2\pi/L$, $4\pi/L$, $6\pi/L$ respectively.

Corrections to scaling are expected to produce the following
dependence upon the external momentum:
$\zeta(\pb)=\zeta_0(m)+\zeta_1(mL;m)\pb^2+O(\pb^4)$.
In Fig. \ref{Zetap} we report the numerical results for $\zeta(\pb)$,
together with the best fitting curve of the form 
$\zeta(\pb)=\widehat{\zeta}_0+\widehat{\zeta}_1\pb^2$.
Clearly the scaling corrections are small: we can estimate 
$Z^{L(2,0)}_{TT} = 1.05(2)$ on this lattice.
This compares very well with the result 
of one-loop lattice perturbation theory given in Eq. \reff{EMZLatticePT},
which yields $Z^{L(2,0)}_{TT} \simeq 1.044357$, 
and with the result of ``improved'' (sometimes called ``boosted") 
perturbation theory in terms of the coupling $g_E$, cf. Eq. 
\reff{DressedCouplingEnergy}, 
$Z^{L(2,0)}_{TT} = 1.052471(3)$ (the error is due to the error on $g_E$).
%
%***********************************************************************
%

\subsection{OPE for the Scalar Product of Currents} \label{sec4.4}

In this Section we consider the product $j_{0,0}^{ab} j_{1,x}^{ab}$ averaged 
over rotations. Using Eq. \reff{scalcorrcont}, we have in the continuum
scheme 
\be
\frac{1}{2}\,\overline{j_{0,0}^{ab} j_{1,x}^{ab}} = 
  \left[{1\over4} W_2(r/2) + W_3(r/2) + W_4(r/2)\right]
  {1\over g} \opl  T_{01}(0) \opr \, .
\ee
Notice that, since the currents are exactly conserved, 
both on the lattice and in the 
continuum, there is no need to make a distinction between lattice and 
\MS-renormalized operators.
All other contributions vanish after the angular average. Using the results
of Sec. \ref{PerturbativeContinuumCurrentsSection} we have at one loop 
in the \MS scheme
\be
\frac{1}{2}\,\overline{j_{0,0}^{ab} j_{1,x}^{ab}} = \,
\left[1 - {N-2\over 2\pi} g \left(\log\left({{\mu} r\over2}\right) + 
      \gamma - {5\over 4}\right) + O(g^2)\right] 
  {1\over g} \opl  T_{01}(0) \opr,
\ee
where $\mu$ is the renormalization scale and $\gamma$ Euler's constant. 
We will not use this form of the OPE expansion,
but instead the RG-improved Wilson coefficients. 
Thus, cf. Sec. \ref{RenormalizationGroupSection}, we write 
\be
\frac{1}{2}\,\overline{j_{0,0}^{ab} j_{1,x}^{ab}} = 
   W_{RGI}(\overline{g}(\Lambda_\MMS r))  \opl  T_{01}(0) \opr,
\ee
where 
\be 
W_{RGI}(\overline{g}) = \frac{1}{\overline{g}}
\left[1 + {5(N-2)\over 8\pi} \overline{g}\right] ,
\label{Wilson1ms}
\ee
and $\overline{g}(\Lambda_\MMS r)$ is the running coupling constant defined 
by
\be
\lambda_{\MMS}(\overline{g}) = \, \Lambda_{\MMS} 
   {r e^{\gamma}\over 2}.
\label{gdix}
\ee
This expression coincides with the one of Sec. 
\ref{RenormalizationGroupSection}, see Eq. (\ref{CouplingAtScaleX}), 
apart from the factor 
$e^{\gamma}/2$ which has been inserted for future convenience.
Using the perturbative expression 
\reff{GxExpansion}, we can also rewrite \reff{Wilson1ms} as 
\begin{eqnarray}
W_{RGI}(r) =
{N-2\over2\pi} z + \left[{1\over 2\pi}\log z + \frac{5(N-2)}{8\pi}\right]\, ,
\label{Wilson2ms}
\end{eqnarray}
where $z = -\log(\Lambda_{\MMS} r e^{\gamma}/2)$.

We will also use the lattice Wilson coefficients, see Sec. 
\ref{PerturbativeLatticeCurrentsSection}. Using the one-loop results 
of Sec. \ref{PerturbativeLatticeCurrentsSection} 
and the general expressions reported in Sec. \ref{RenormalizationGroupSection},
proceeding as before, we obtain the prediction
\be
\frac{1}{2}\,\overline{j_{0,0}^{L,ab} j_{1,x}^{L,ab}} = 
U(g_L) W^L_{RGI}(\overline{g}_L(\Lambda_L r))   T^L_{01,0},
\ee
where, at one loop, 
\begin{eqnarray}
W^L_{RGI}(\overline{g}_L) & = & \frac{1}{\overline{g}_L}\left[ 1 + 
    \left({5N-2\over 8\pi}-\frac{1}{4}\right) \overline{g}_L\right],
\label{WRGIlatt}\\
U(g_L) & = & 1+\left(\frac{1}{\pi}-\frac{1}{4}\right)g_L,
\label{Ulatt}
\end{eqnarray}
and $\overline{g}_L(\Lambda_L r)$ is the running coupling 
constant defined by
\be
\lambda_L(\overline{g}_L) =   
  \, \Lambda_L\, r e^\gamma \sqrt{8},
\label{eq:gbarraL}
\ee
where $\Lambda_L$ is the lattice 
$\Lambda$-parameter, see Sec. \ref{RenormalizationGroupSection}.

Finally we shall test perturbation theory in the ``improved'' expansion 
parameter $g_E$ defined in Eq. \reff{DressedCouplingEnergy}. 
The OPE becomes:
\be
\frac{1}{2}\,\overline{j_{0,0}^{L,ab} j_{1,x}^{L,ab}} = 
U(g_E) W^E_{RGI}(\overline{g}_E(\Lambda_E r))   T^L_{01,0},
\ee
where
\begin{eqnarray}
W^E_{RGI}(\overline{g}_E) & = & \frac{1}{\overline{g}_E}\left[ 1 + 
    \left({5(N-2)\over 8\pi}-\frac{1}{8}\right) \overline{g}_E\right] \ ,
\label{WRGIE}
\end{eqnarray}
$U(\cdot)$ is the same as in Eq. (\ref{Ulatt}), 
and $\overline{g}_E(\Lambda_E r)$ is the running coupling constant  defined by
\be
\lambda_E(\overline{g}_E) =   
  \, \Lambda_E \, r e^\gamma \sqrt{8},
\label{eq:gbarraE}
\ee
where $\Lambda_E$ is the $\Lambda$-parameter \reff{LambdaEdef}. 

We have tested the validity of the OPE by considering matrix elements
between one-particle states. The matrix elements of the product of 
the currents
can be determined in terms of $G^{(s)}(t,x;\pb,\qb;2t_s)$, since
\begin{eqnarray}
\frac{1}{2}\,\<\pb,c|\jg_{0,(0,0)}\cdot\jg_{1,(t,x)}|\qb,c\> =
N \sqrt{4\omega(\pb)\omega(\qb)}
\lim_{t_s\to\infty}
  \widehat{G}^{(s)}(t,x;\pb,\qb;2t_s).
\end{eqnarray}

\begin{figure}
\begin{tabular}{c}
\hspace{2cm}
\epsfig{figure=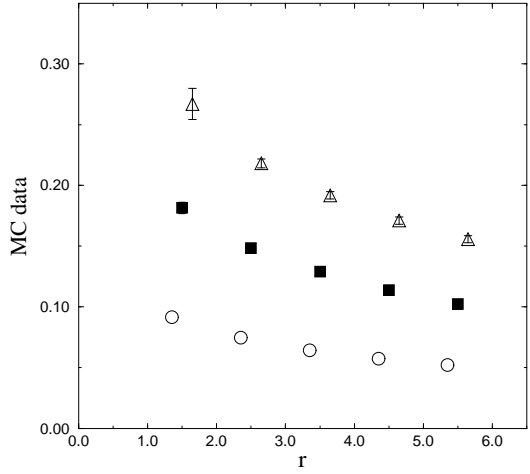,angle=-90,
width=0.5\linewidth} \\
\hspace{2cm}
\epsfig{figure=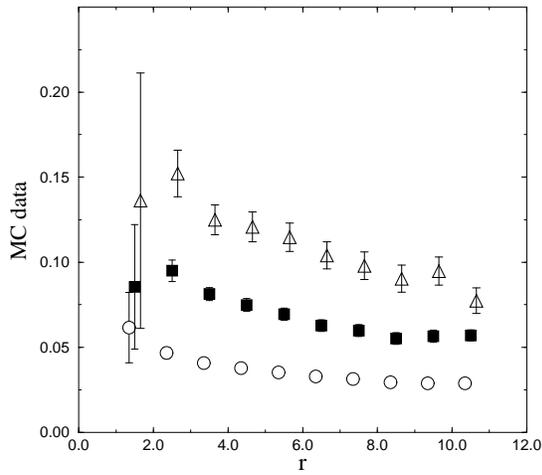,angle=-90,
width=0.5\linewidth}
\end{tabular}
\caption{Estimates of $\widehat{G}^{(s)}(t,x;\pb,\pb;2 t_s)$
averaged over rotations on lattice \ref{Lattice64x128} (upper graph,
here $t_s = 10$) and
\ref{Lattice128x256} (lower graph, $t_s = 6$). Circles,
filled squares, and triangles correspond to $\pb=2\pi/L$, $4\pi/L$,
and $6\pi/L$ respectively. }
\label{ScalarCurrentsMCData}
\end{figure}

\begin{figure}
\begin{tabular}{cc}
\hspace{0.0cm}\vspace{-2.5cm}\\
\hspace{-0.5cm}
\epsfig{figure=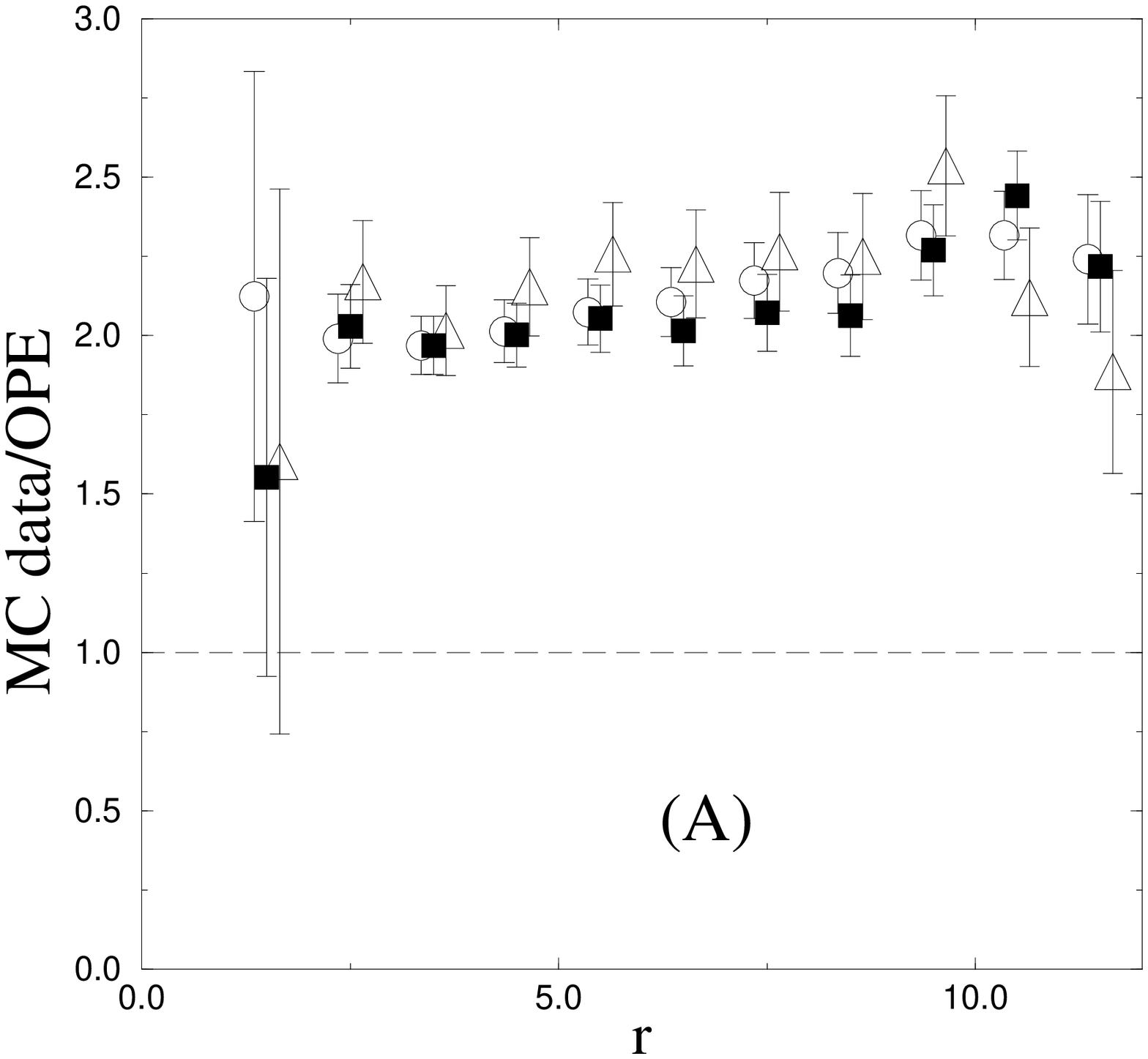,angle=-90,
width=0.5\linewidth}&\hspace{-0.5cm}
\epsfig{figure=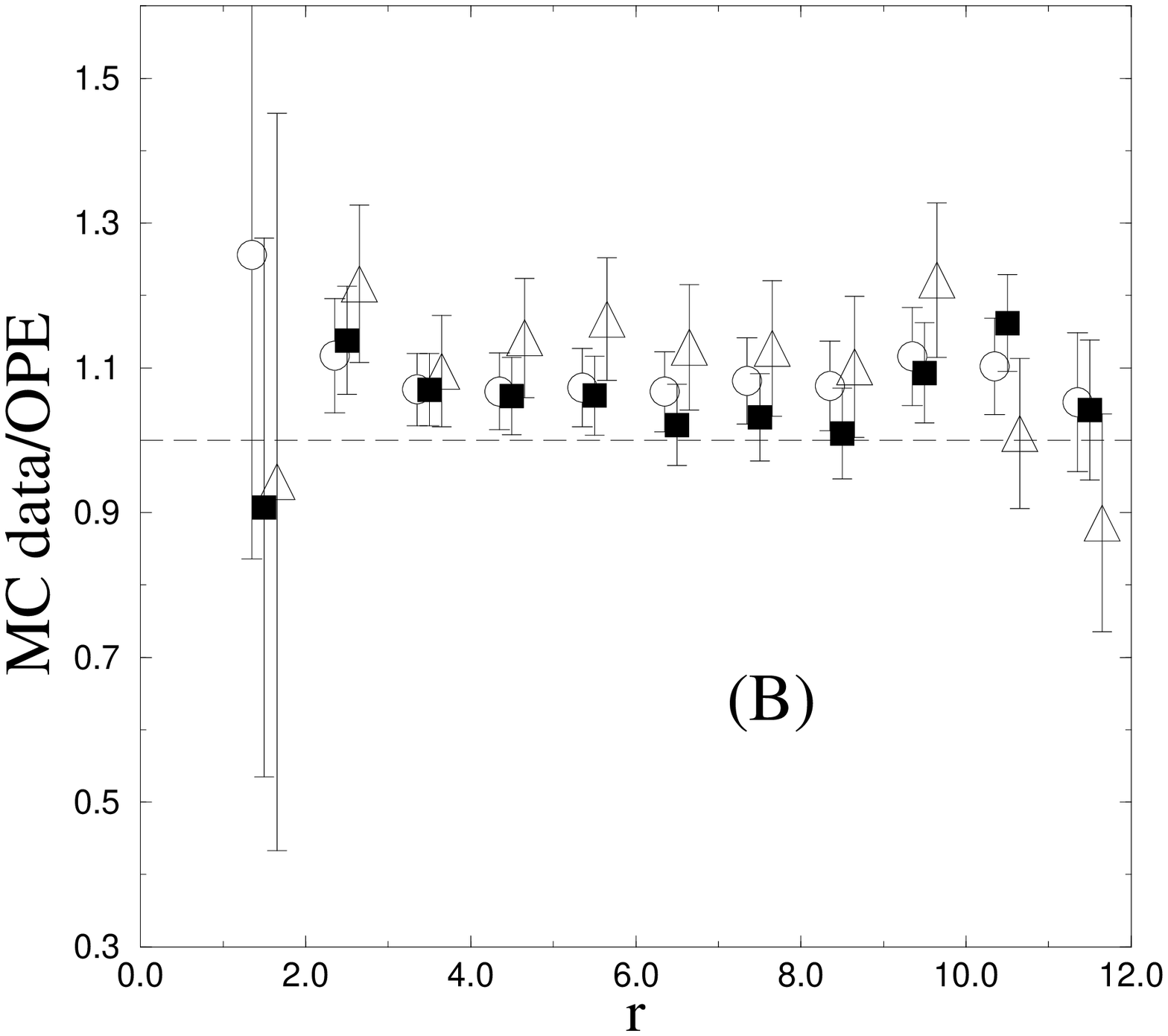,angle=-90,
width=0.5\linewidth}\vspace{-1cm}\\
\hspace{-0.5cm}
\epsfig{figure=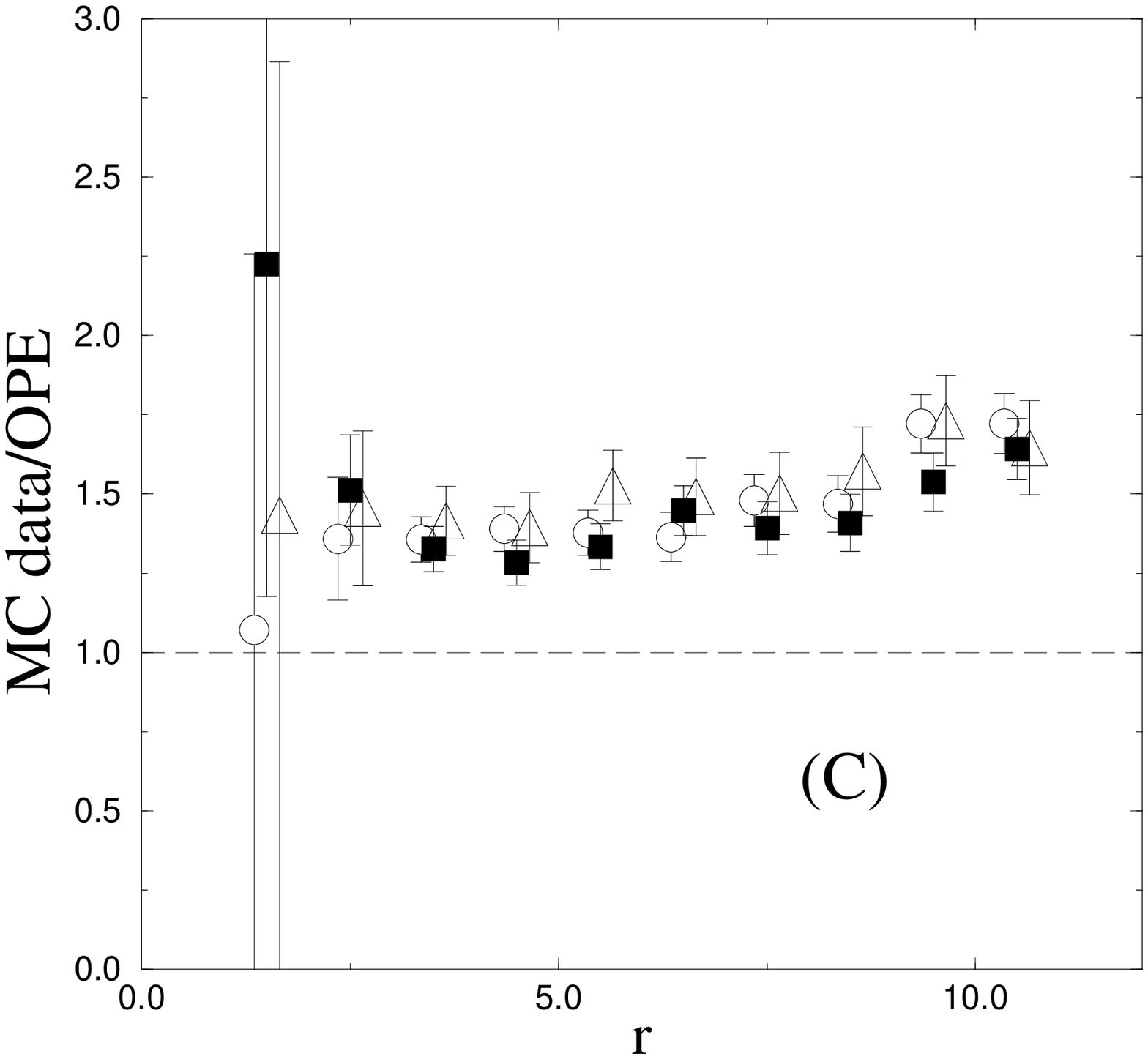,angle=-90,
width=0.5\linewidth}&\hspace{-0.5cm}
\epsfig{figure=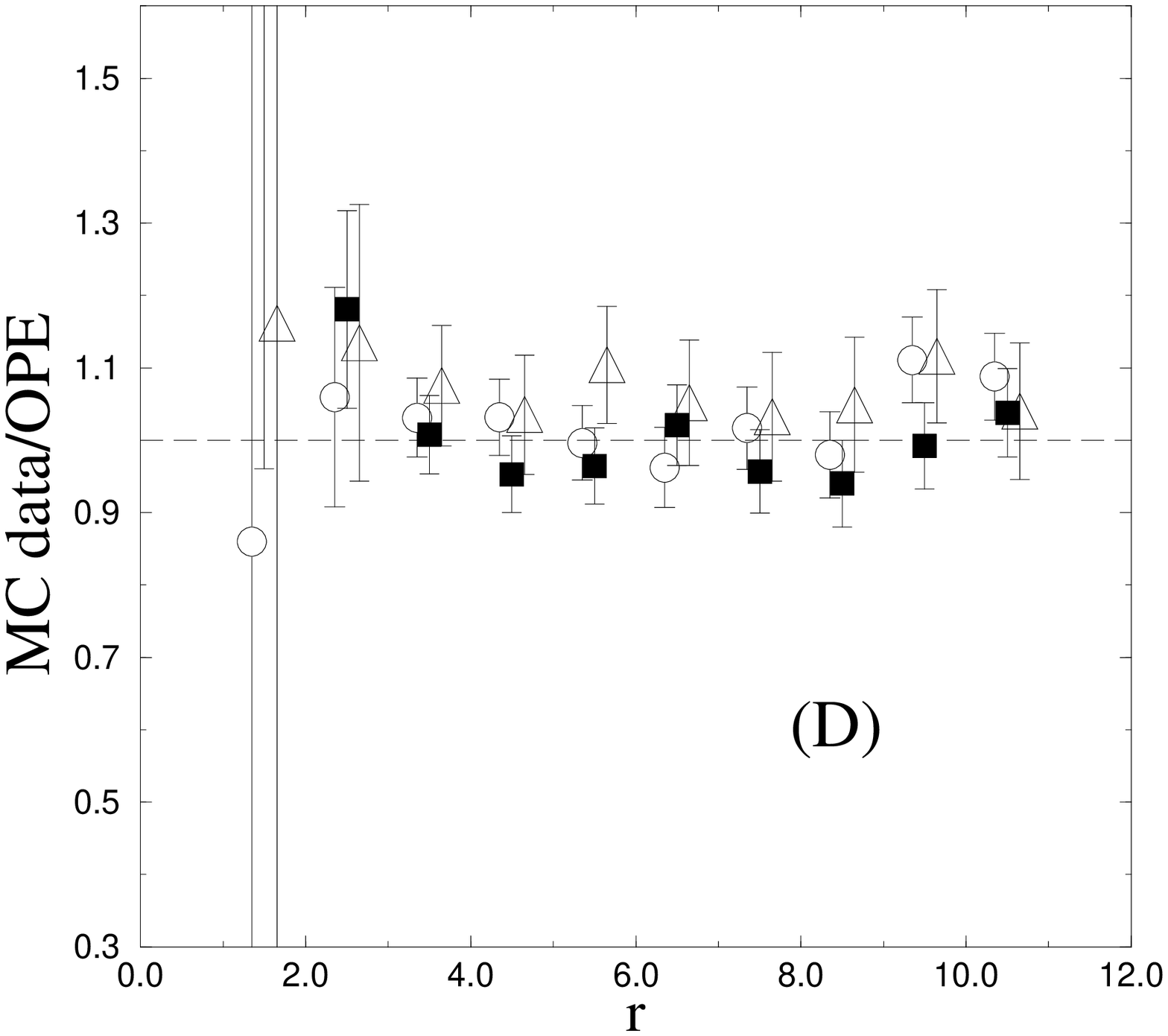,angle=-90,
width=0.5\linewidth}\vspace{-1cm}\\
\hspace{-0.5cm}
\epsfig{figure=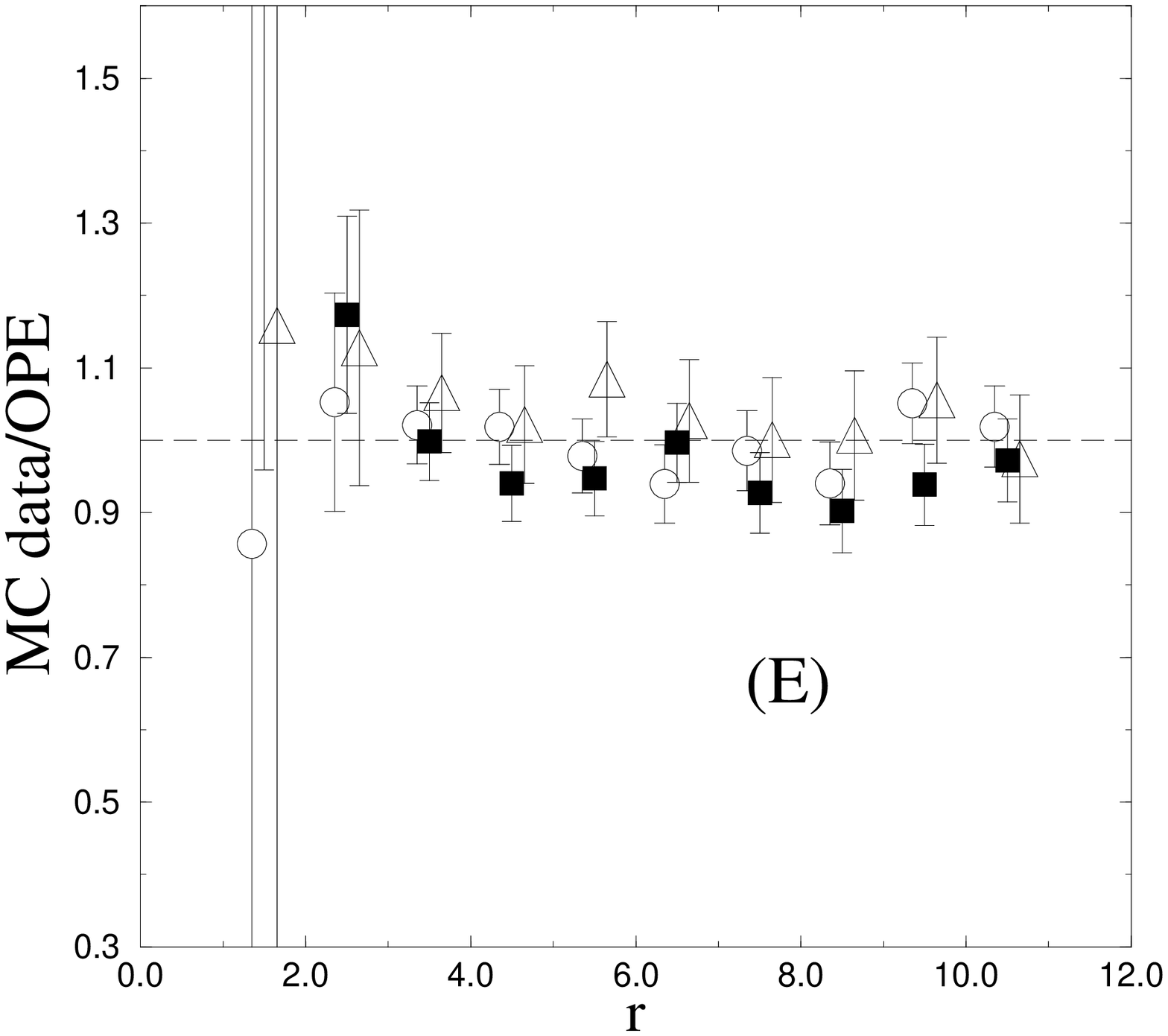,angle=-90,
width=0.5\linewidth}&\hspace{-0.5cm}
\epsfig{figure=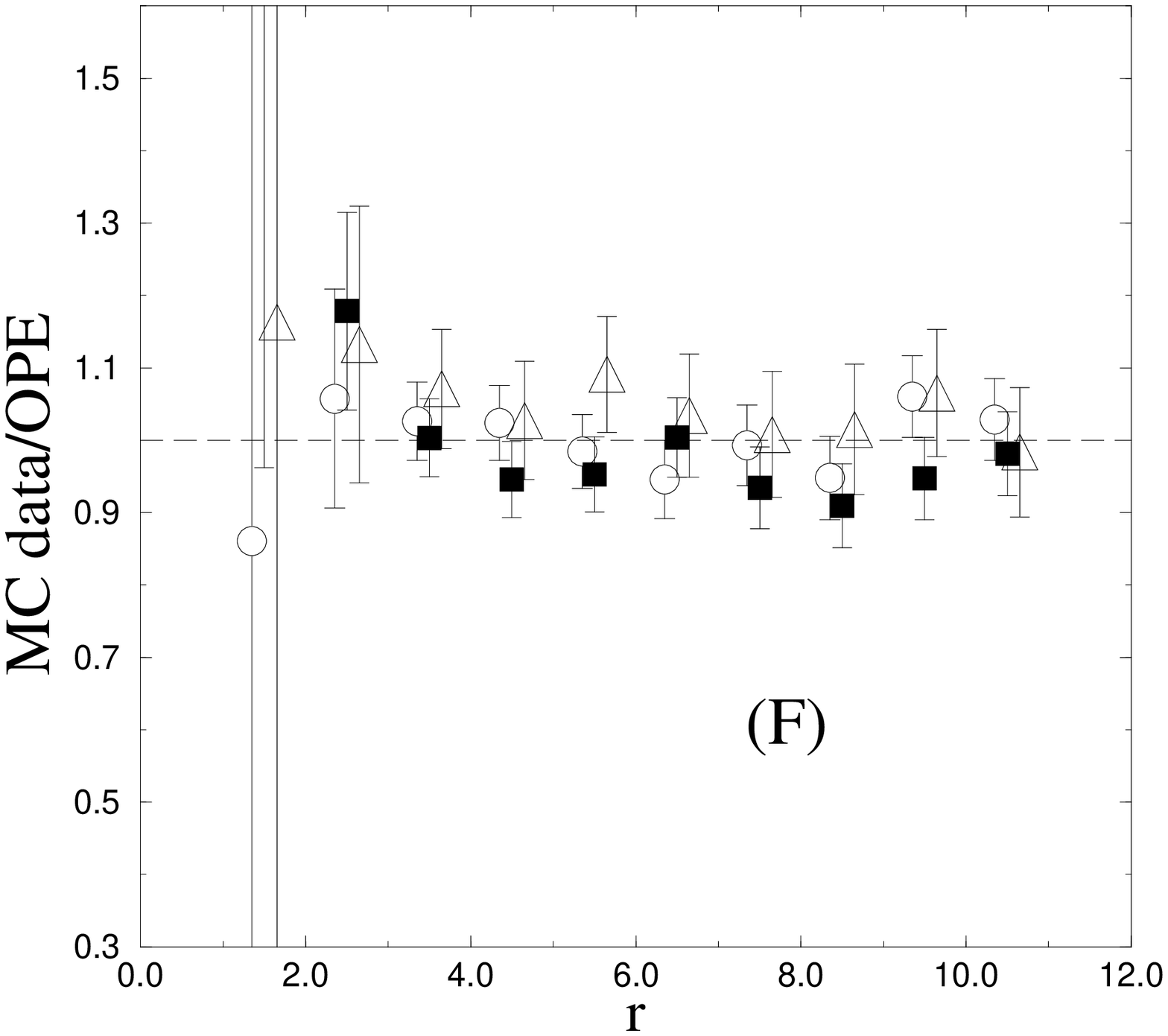,angle=-90,
width=0.5\linewidth}
\end{tabular}
\caption{The scalar product of two Noether currents compared with the OPE 
prediction: graphs of $R(r)$, cf. \protect\reff{defRatioR}, obtained
using \MS RG-improved perturbation theory. Circles,
filled squares, and triangles correspond to $\pb=2\pi/L$, $4\pi/L$,
and $6\pi/L$ respectively. The data are for lattice 
\ref{Lattice128x256}, $\xi^{\rm exp} = 13.636(10)$.}
\label{ScalarCurrentsMS}
\end{figure}

\begin{figure}
\begin{tabular}{cc}
\hspace{-2.5cm}
\epsfig{figure=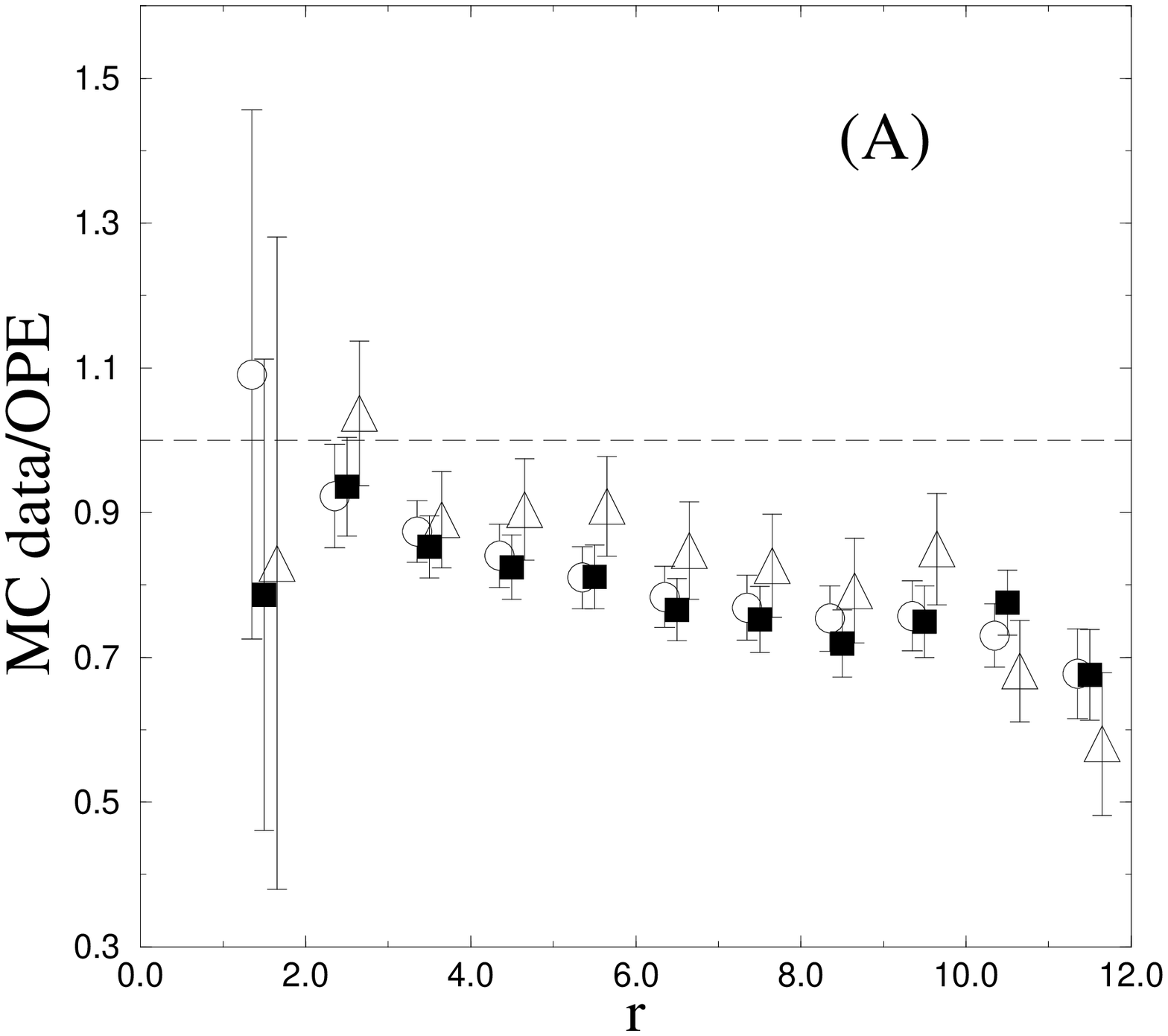,angle=-90,
width=0.6\linewidth}&
\hspace{-0.5cm}
\epsfig{figure=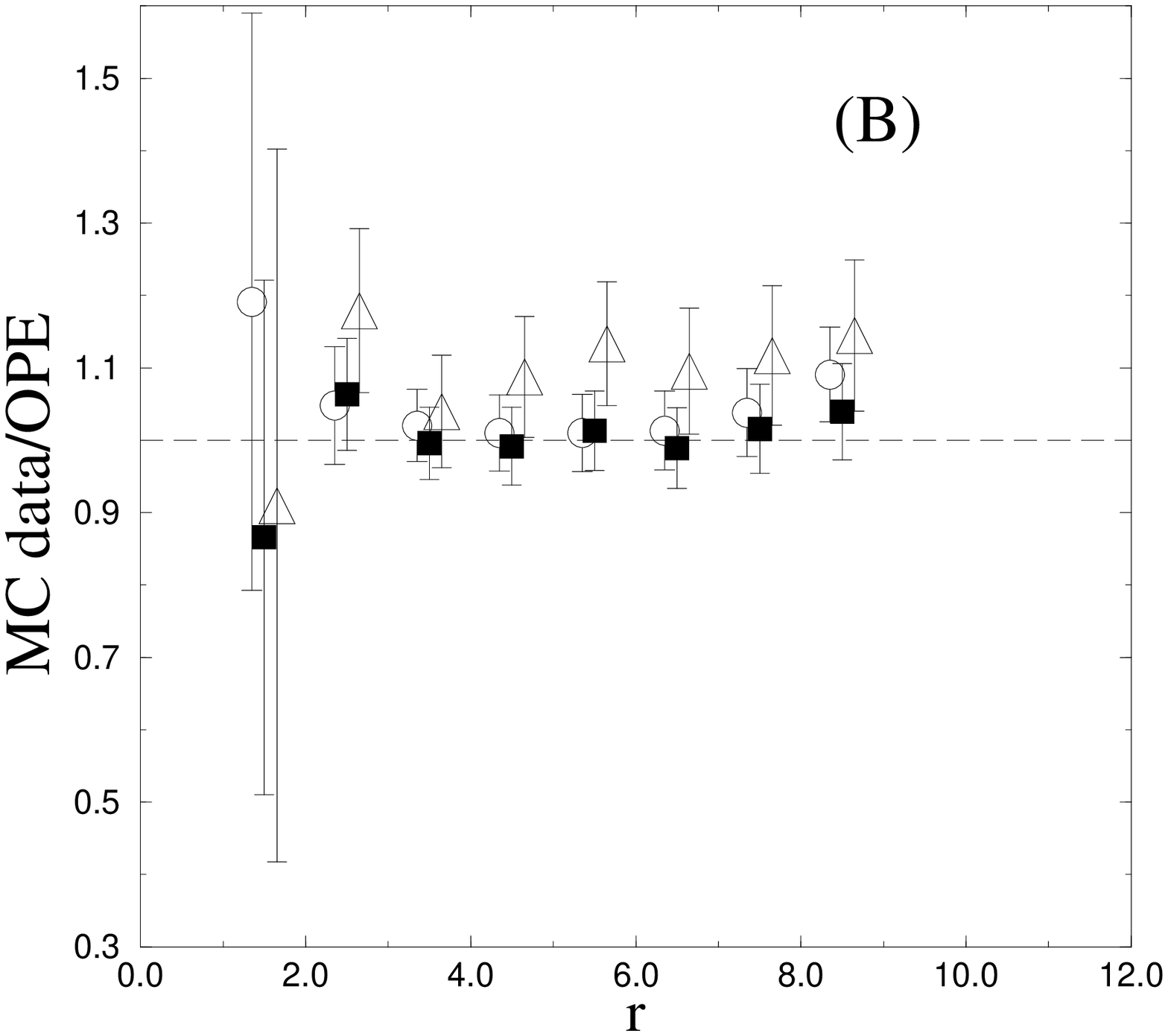,angle=-90,
width=0.6\linewidth}\vspace{-1cm}\\
\hspace{-2.5cm}
\epsfig{figure=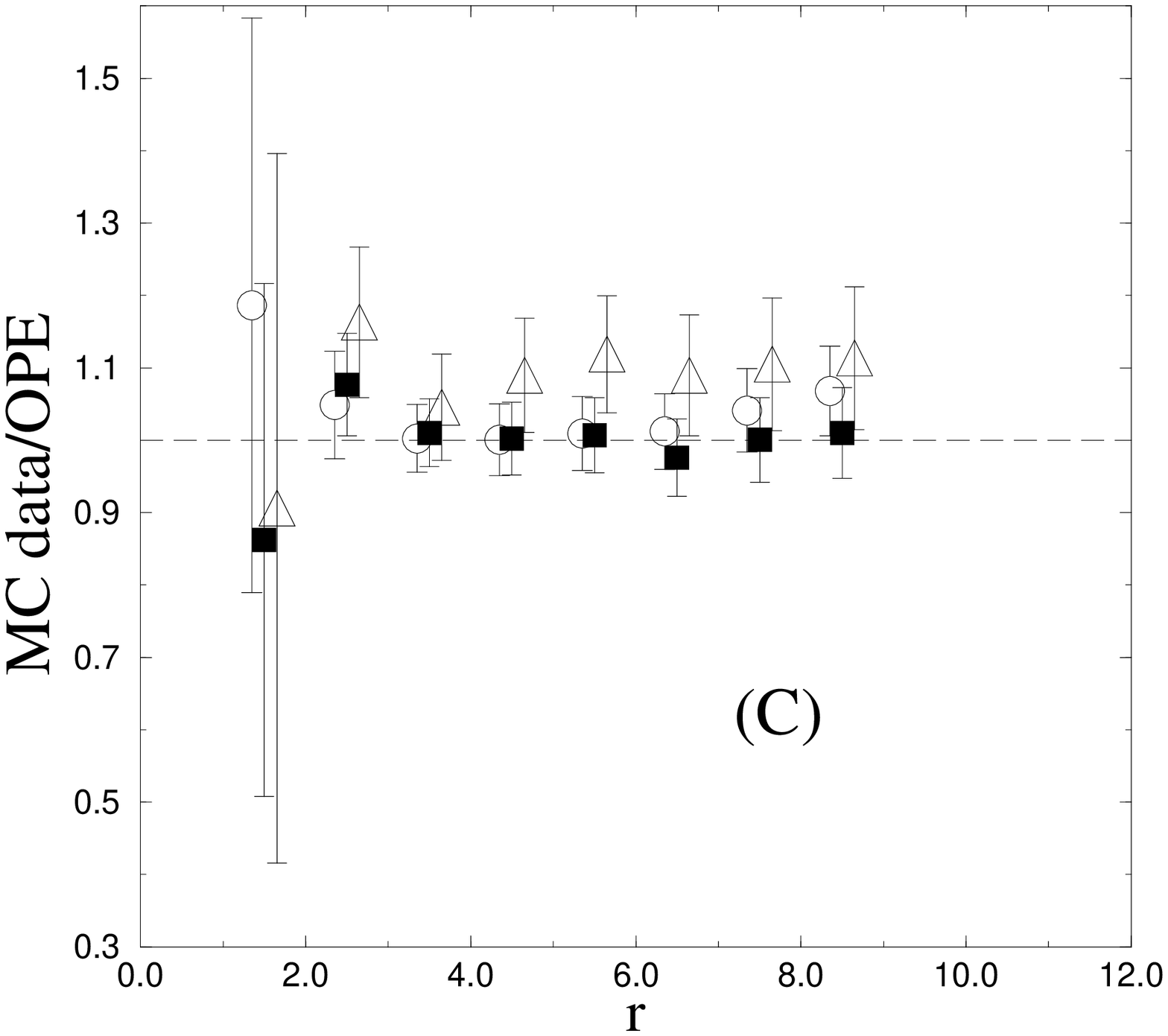,angle=-90,
width=0.6\linewidth}&
\hspace{-0.5cm}
\epsfig{figure=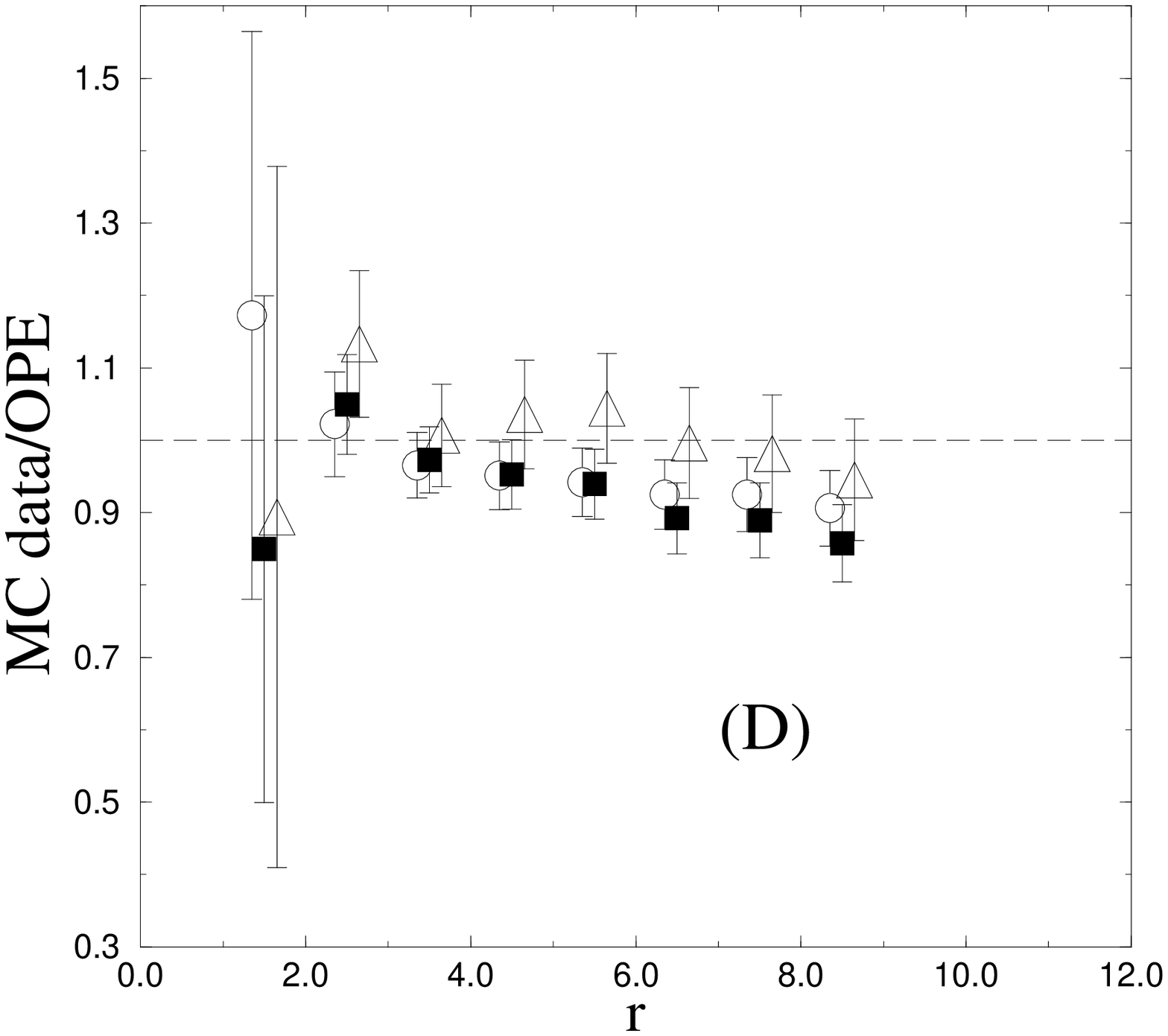,angle=-90,
width=0.6\linewidth}
\end{tabular}
\caption{The scalar product of two Noether currents compared with the OPE 
prediction: graphs of $R^{\rm latt}(r)$, cf. \protect\reff{defRatioRlatt}, 
obtained using RG-improved perturbation theory in the coupling 
$g_L$ and in the improved coupling $g_E$. Circles,
filled squares, and triangles correspond to $\pb=2\pi/L$, $4\pi/L$,
and $6\pi/L$ respectively. The data are for lattice 
\ref{Lattice128x256}, $\xi^{\rm exp} = 13.636(10)$.}
\label{ScalarCurrentLatticePT}
\end{figure}

\begin{figure}
\begin{tabular}{cc}
\hspace{-1.5cm}
\epsfig{figure=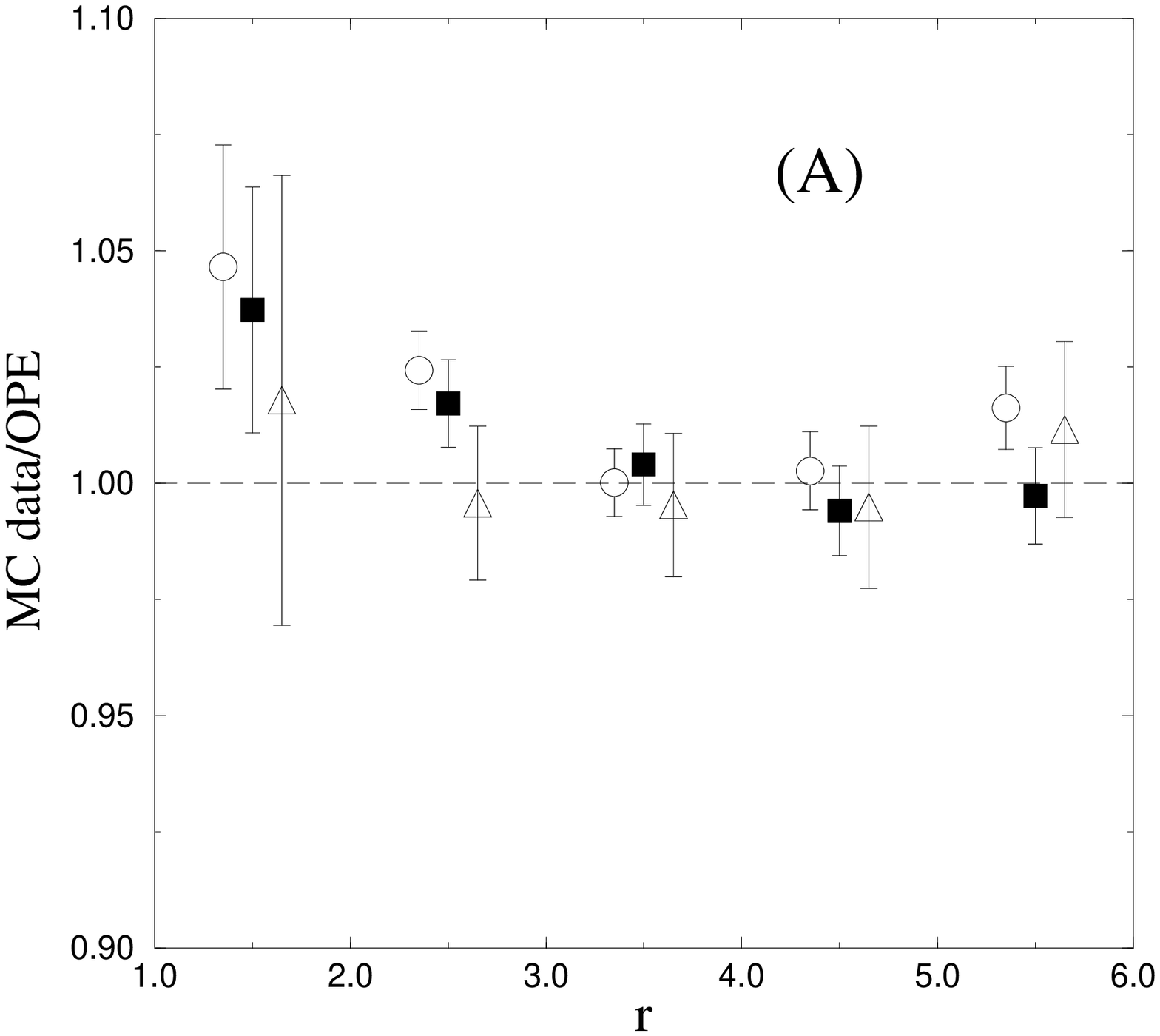,angle=-90,
width=0.55\linewidth}
&
\hspace{-0.5cm}
\epsfig{figure=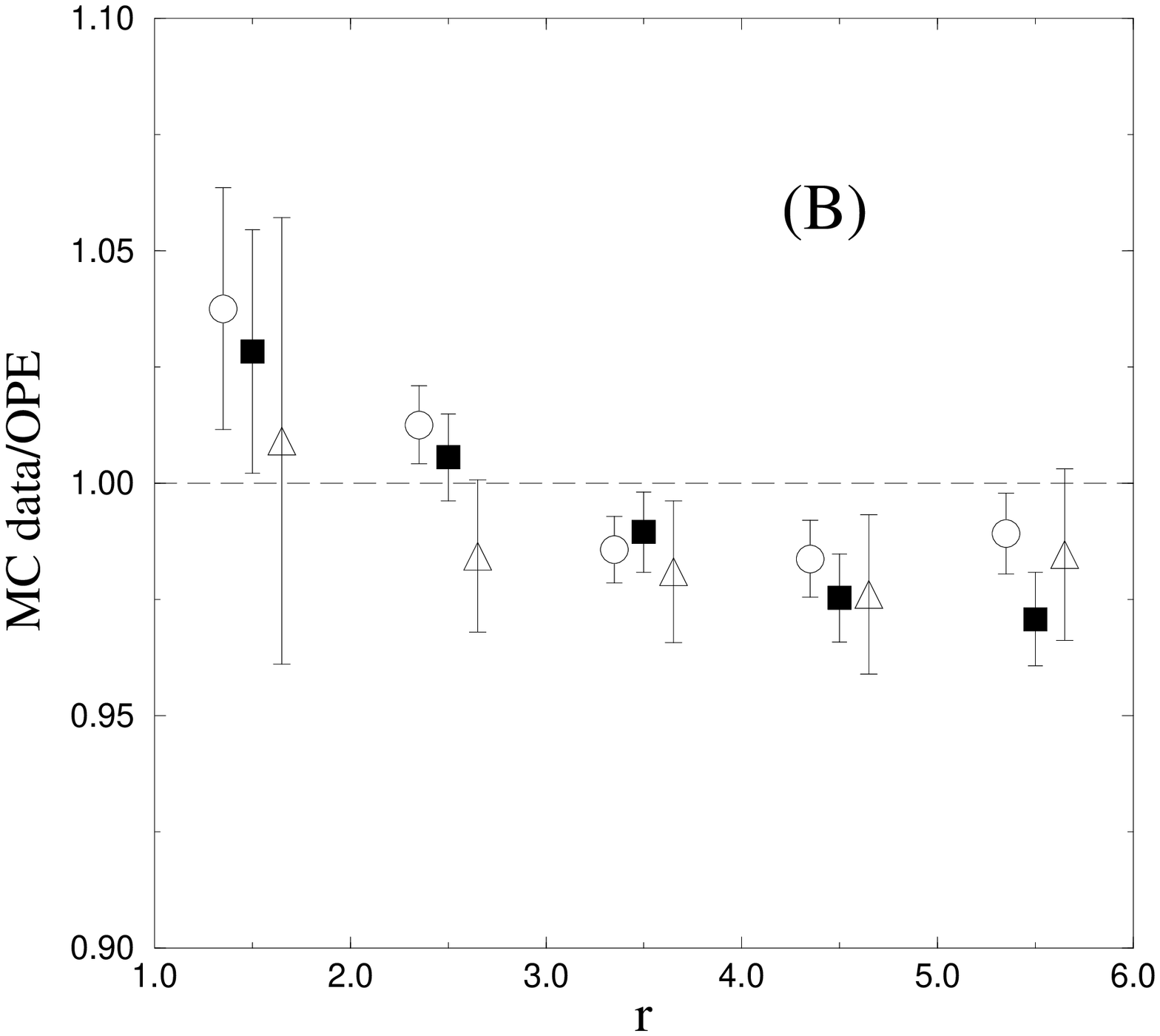,angle=-90,
width=0.55\linewidth} \\
\hspace{-1.5cm}
\epsfig{figure=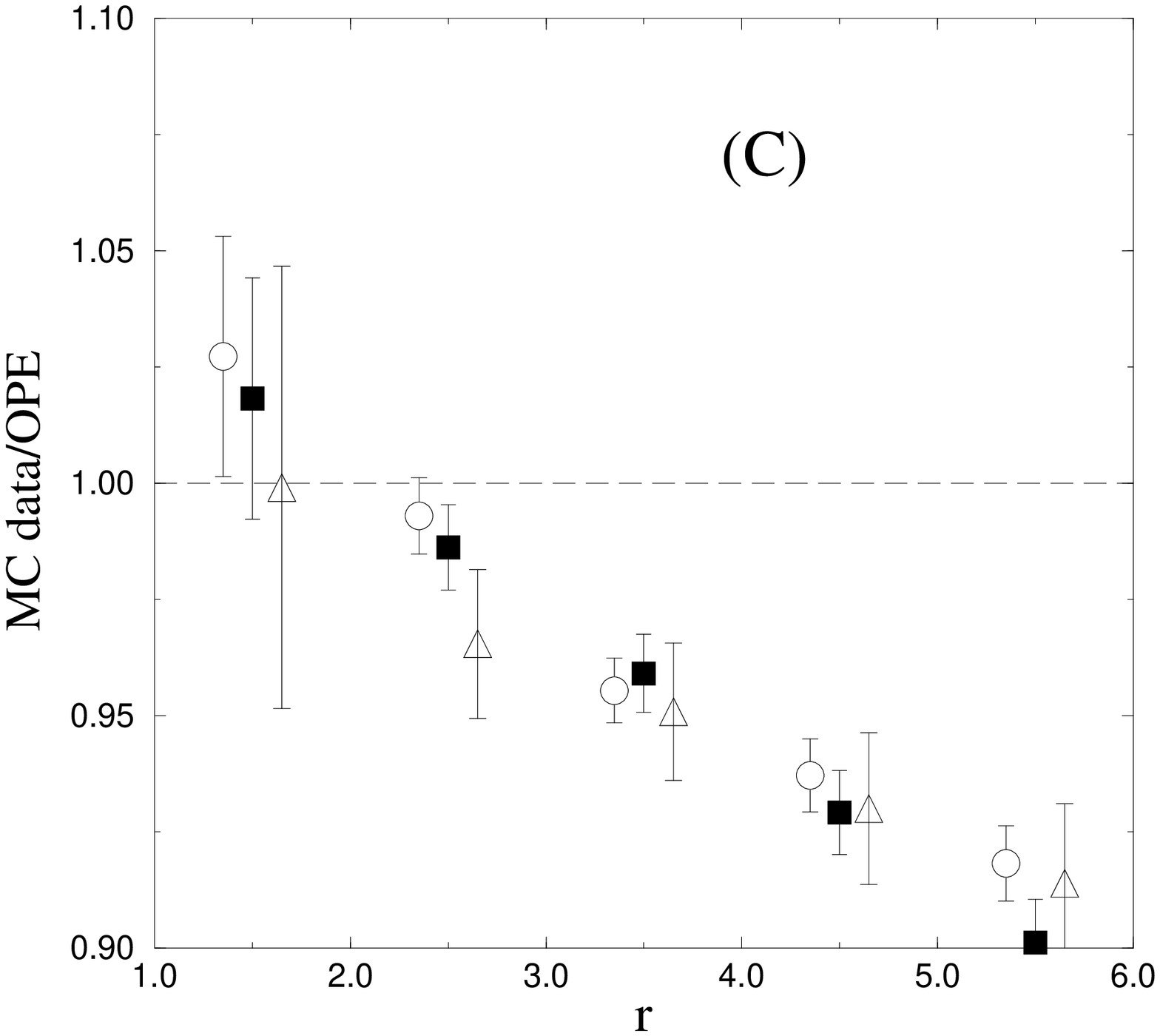,angle=-90,
width=0.55\linewidth} 
&
\hspace{-0.5cm}
\epsfig{figure=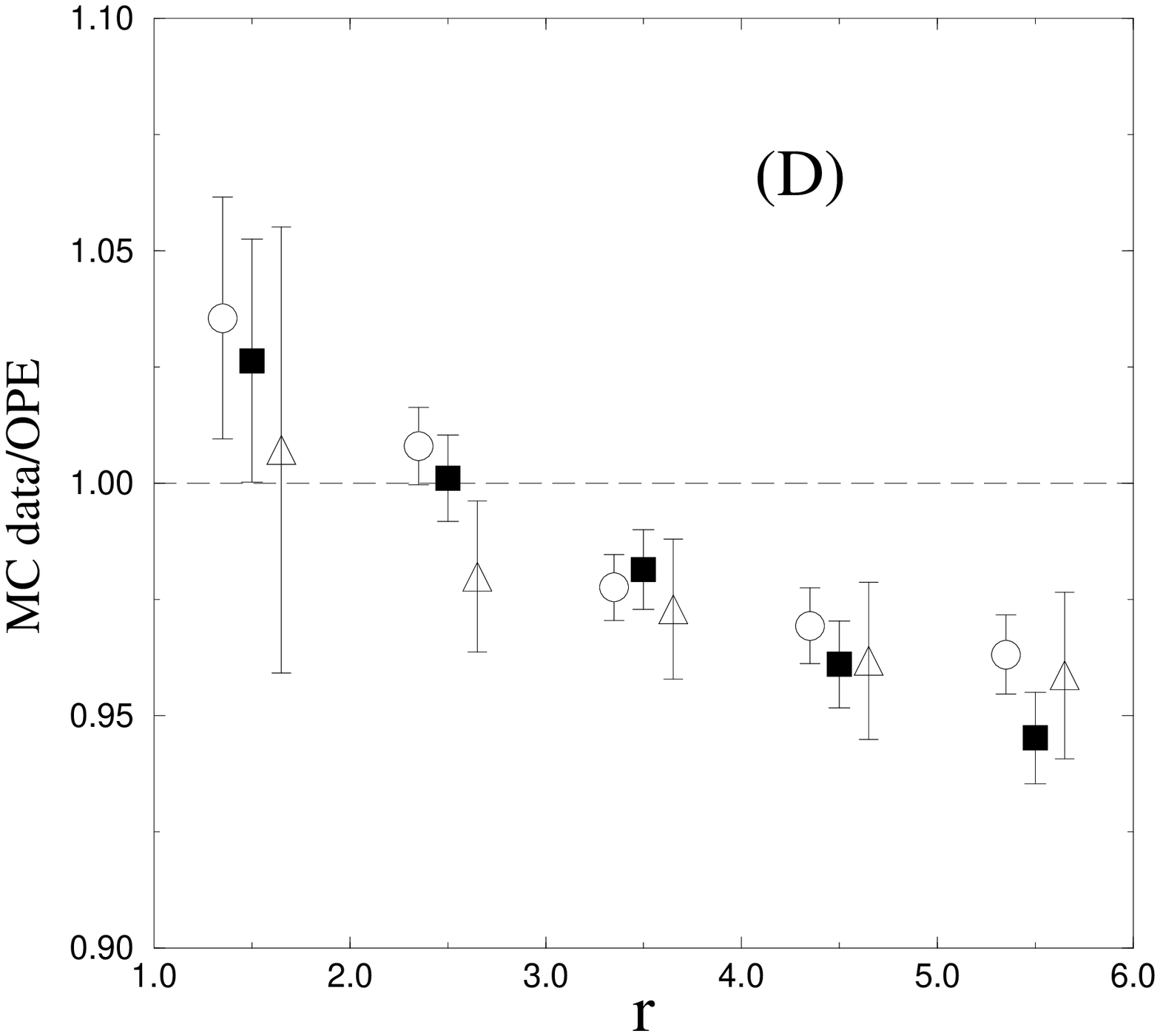,angle=-90,
width=0.55\linewidth}
\end{tabular}
\caption{The scalar product of two Noether currents compared with the OPE 
prediction: graphs of $R(r)$, cf. \protect\reff{defRatioR}, obtained
using \MS RG-improved perturbation theory. Circles,
filled squares, and triangles correspond to $\pb=2\pi/L$, $4\pi/L$,
and $6\pi/L$ respectively. The data are for lattice 
\ref{Lattice64x128}, $\xi^{\rm exp} = 6.878(3)$. Notice the change of vertical 
scale compared to Figs. \ref{ScalarCurrentsMS}, \ref{ScalarCurrentLatticePT}.}
\label{ScalarCurrentLattice64x128}
\end{figure}

In Fig. \ref{ScalarCurrentsMCData} we report\footnote{
On lattice \ref{Lattice128x256} 
$\widehat{G}^{(s)}(t,x;\pb,\pb;2t_s)$ has only been measured in 
${\cal D} = \{(t,x):|t|,|x|\le 8\}$. The points with $r > 8$ appearing in the 
figure correspond to ``partial" angular averages, i.e. they have 
been obtained using Eq. \reff{definizione-mediarotazionale} and 
restricting $z$ to $\cal D$. The 
same comment applies also to the subsequent figures.}
the angular average 
of ${\rm Im}\ \widehat{G}^{(s)}(t,x;\pb,\pb;2t_s)$ for lattices 
\ref{Lattice64x128} and \ref{Lattice128x256}: here $t_s=6$, $10$
for the two lattices respectively.
In Figs. \ref{ScalarCurrentsMS}, \ref{ScalarCurrentLatticePT}, and 
\ref{ScalarCurrentLattice64x128} we compare 
these numerical data with the predictions of perturbation theory. 

In Fig. \ref{ScalarCurrentsMS} we use continuum RG-improved
perturbation theory in the \MS scheme. 
In this case, the matrix element of the energy-momentum tensor is immediately 
computed: $\<\pb,a|T_{01}|\pb,b\> = 2 i \pb \omega(\pb) \delta^{ab}$.
Therefore, we consider the ratio
\be
R(r) = \frac{1}{2}
 {{\rm Im}\<\pb,c|\overline{\jg_{0,(0,0)}\cdot\jg_{1,(t,x)}}|\pb,c\> \over
   2 \pb \omega(\pb) W_{RGI}(\overline{g}(\Lambda_\MMS r))},
\label{defRatioR}
\ee
which should approach 1 in the short-distance limit.
In Fig. \ref{ScalarCurrentsMS} we present several determinations 
of $R(r)$ that differ in the way in which the running coupling 
constant $\overline{g}(\Lambda_\MMS r)$ and the \MS coupling $g$ are determined.

There are several different methods that can be used to compute the 
\MS coupling $g$ and the strictly related $\overline{g}(\Lambda_\MMS x)$.
They have been compared in detail in Sec. \ref{RunningCoupling}.
It turned out that the finite-size scaling method proposed by L\"uscher 
\cite{Luscher:1982aa,Luscher:1991wu} and what we call ``the 
RG-improved perturbative method" are essentially equivalent, see, e.g.,
Table \ref{RunningComparison1}. Therefore,
we can use either of them,\footnote{In QCD the RG-improved perturbative method
cannot be used since no exact prediction for the mass gap exists.
In this case the finite-size scaling method would be the method of choice.
Note that for large values of the scale also the improved-coupling
method \cite{Parisi:1980aa,Martinelli:1981tb}, which can be  
generalized to QCD
\cite{Lepage:1993xa}, works well, see Sec. \ref{RunningCoupling}.} 
obtaining completely equivalent results. In Fig. \ref{ScalarCurrentsMS} 
we have used the finite-size scaling method to be consistent with what
we would do in QCD. Elsewhere, we have used the RG improved perturbative method
because of its simplicity. 

The first step in the computation of $R(r)$ is the determination of 
$\overline{g}(\Lambda_\MMS r)$. This is obtained as follows: 
we fix $\mu/m$, and using the finite-size scaling results 
reported in Table \ref{RunningComparison1}, 
we compute $g_\MMS(\mu)$. Then, using Eq. (\ref{LambdaTruncatedDefinition}) 
we determine $\Lambda_\MMS = \mu\lambda_{\MMS}(g_\MMS)$ at $l$-loops. Finally, 
$\overline{g}(\Lambda_\MMS r)$ is obtained either by solving 
Eq. (\ref{gdix}), again using Eq. (\ref{LambdaTruncatedDefinition})
for $\lambda_\MMS(\overline{g})$ appearing in the right-hand side, 
or by using Eq. \reff{GxExpansion}. As we discussed
in Sec. \ref{RunningCoupling}, the final result should be independent of the 
chosen value of $\mu/m$ and therefore we can evaluate 
the systematic error on $\overline{g}(\Lambda_\MMS x)$ by considering 
different values of $\mu/m$. If we fix $m/\mu = 0.00071$, 
cf. first row of Table \ref{RunningComparison1},
we have $\overline{g}(\Lambda_\MMS x) = 1.498$, $2.206$, $3.363$ 
respectively for $m x = 0.2$, $0.5$, $0.8$, while if we fix $m/\mu = 0.1033$, 
cf. 10th row of Table \ref{RunningComparison1} in App. \ref{RunningCoupling},
we have $\overline{g}(\Lambda_\MMS x) = 1.490$, $2.185$, $3.273$
at the same distances. 
The dependence is tiny and, as expected, it increases for larger values of 
$mx$. In practice, it does not play any role, the main source of 
error being instead the truncation of the OPE coefficients.
Notice that the independence on $\mu$ 
is obvious if we use the RG-improved perturbative method.

Let us now describe the various graphs appearing in 
Fig. \ref{ScalarCurrentsMS}.  In graphs (A) and (B) 
we fix $\mu = m/0.00071$ and then, using the results 
presented in Table \ref{RunningComparison1}, cf. first row, we obtain 
$g_{\MMS}(\mu) = 0.587016$. We then compute 
$\Lambda_{\MMS}(\mu,g_{\MMS})$ 
using the four-loop expression (\ref{LambdaTruncatedDefinition})
with $l=4$. Finally, the Wilson coefficient is given by \reff{Wilson2ms}:
in graph (A) we use the leading term only,
while in graph (B) we include the next-to-leading one. 

In graphs (C) and (D) we compute $\Lambda_{\MMS}$ as in (A) and (B),
choosing a different scale, $\mu = m/0.1033$, cf. $10^{\rm th}$ row
of Table \ref{RunningComparison1}. Then we compute $\overline{g}(x)$ 
by solving numerically Eq.  (\ref{gdix}) using the four-loop expression 
\reff{LambdaTruncatedDefinition}
with $l=4$. Finally, the Wilson coefficient is obtained using
Eq. (\ref{Wilson1ms}). While in graph (C) we keep only the leading 
term $1/\overline{g}(x)$, in graph (D) we use the complete expression.

Finally, graphs (E) and (F) are identical to graph (D) except that 
$\Lambda_{\MMS}$ and $\overline{g}(x)$ are computed using the 
two-loop expression (\ref{LambdaTruncatedDefinition}) with $l=2$.
The two graphs correspond to different choices of $\mu$:
$\mu = m/0.00071$ (graph (E)) and $\mu = m/0.1033$ (graph (F)).

Comparing the different graphs, we immediately see that the choice of 
scale $\mu$ and the order of perturbation theory used for 
$\Lambda_\MMS$ (two loops or four loops) are not relevant with the 
present statistical errors.
Much more important is the role of the Wilson coefficients. If one 
considers only the leading term (graphs (A) and (C)) there are indeed 
large discrepancies and in the present case one would obtain estimates of the 
matrix elements with an error of 50--100\%. Inclusion of the next-to-leading
term---this amounts to considering one-loop Wilson coefficients and two-loop
anomalous dimensions---considerably improves the results, and now the 
discrepancy is of the order of the statistical errors, 
approximately 10\%. For the practical application of the method,
it is important to have criteria for estimating the error on the
results. It is 
evident that the flatness of the ratio of the matrix element by the 
OPE prediction is not, in this case, a good criterion: The points in
graph (A) show a plateau---and for a quite large set of values of $r$---even
if the result is wrong by a factor of two. However, this may just be a 
peculiarity of the case we consider, in which the $r$-dependence of the 
data and of the OPE coefficients is very weak. On the other hand, 
the comparison of the results obtained using 
different methods for determining $\overline{g}(x)$ seems to provide 
reasonable estimates of the error bars. For instance, if only the leading term
of the Wilson coefficients were available, we could have obtained 
a reasonable estimate of the error by comparing graphs (A) and (C).

In Fig. \ref{ScalarCurrentLatticePT} we consider lattice
RG-improved perturbation theory, computing
\be
R^{\rm latt} (r) = \frac{1}{2} 
 {{\rm Im}\, \<\pb,c|\overline{\jg_{0,(0,0)}\cdot\jg_{1,(t,x)}}|\pb,c\> \over
   U(g_L) W_{RGI}^L(\bar{g}_L){\rm Im}\, \<\pb,c|T^L_{10}|\pb,c\>} .
\label{defRatioRlatt}
\ee

In graph (A) we use $g_L$ as an expansion parameter. We compute 
$\Lambda_L$ using the value of the mass gap 
$m^{-1} = 13.632(6)$ and Eq.~\reff{MassGapLattice}, with the 
non-perturbative constant \reff{HasenfratzconstLattice}. Then, we determine 
$\overline{g}_L(\Lambda^{\rm latt} r)$ by solving numerically 
\reff{eq:gbarraL} and using 
for $\lambda_L(\overline{g}_L)$ appearing in the left-hand side
its truncated four-loop expression \reff{LambdaTruncatedDefinition}. 
Finally,  we use 
Eq. \reff{WRGIlatt} for the Wilson coefficient. The results are quite 
poor: there is a downward trend as a function of $r$ and the data are 
far too low. Naive lattice perturbation theory 
is unable to provide a good description of the numerical data.

The results improve significantly if we use the improved coupling $g_E$:
The estimates obtained using this coupling, graphs (B), (C), (D), are 
not very different from those obtained using \MS continuum perturbation
theory.
Graph (B) has been obtained exactly as graph (A), except that in this 
case we used $g_E$ as an expansion parameter. The $\Lambda$-parameter 
is defined in \reff{LambdaEdef}, $m/\Lambda_E$ in 
\reff{costanteHasenfratz_schemaE}, and the relevant Wilson coefficient
is given in Eq. (\ref{WRGIE}). Graphs 
(C) and (D) are analogous to graph (B). The difference is in the 
determination of $\Lambda_E$. We do not compute it
non-perturbatively by using the mass gap
but we determine it directly from the 
perturbative expression \reff{LambdaEdef}. In this case we use 
the perturbative expression truncated at four loops 
(C) and two loops (D).

In Figs. \ref{ScalarCurrentsMS} and \ref{ScalarCurrentLatticePT} we
have checked
the validity of the OPE on lattice \ref{Lattice128x256}. If one has in mind
QCD applications this is quite a large lattice since 
$\xi^{\rm exp} \approx 14$. For this reason, we have tried to understand 
if the nice agreement we have found survives on a smaller lattice,  
by repeating the computation on lattice \ref{Lattice64x128}.
The results are reported in Fig. \ref{ScalarCurrentLattice64x128}. 
Graphs (A) and (B) should be compared with graph (D) of 
Fig. \ref{ScalarCurrentsMS}.
In (A) we compute $\Lambda_{\MMS}$ from Eq. \reff{MassGap}, with the 
non-perturbative constant \reff{BetheAnsatzPrediction} and  $m^{-1} = 6.878(3)$.
Then, we compute $\overline{g}(x)$ solving numerically
Eq.  (\ref{gdix}) using the four-loop expression 
\reff{LambdaTruncatedDefinition}
with $l=4$. The Wilson coefficient is obtained from Eq. \reff{Wilson1ms}.
In (B) we repeat the same calculation as in Fig. \ref{ScalarCurrentsMS}
graph (D) at the scale $\mu = m/0.00071$.
In (C) and (D) we repeat the calculation of graph (B) using 
the two-loop and the three-loop $\beta$-function 
for the  determination of the coupling $\overline{g}$. 
In all graphs (B), (C), (D)  we use the result $m^{-1} = 6.878(3)$.

Graphs (A) and (B) show a nice flat behavior and for 
$2\ltapprox r\ltapprox \xi$, the 
ratio $R(r)$ is approximately 1 with 2--3\% corrections (notice that 
the vertical scale in Figs. \ref{ScalarCurrentsMS} and 
\ref{ScalarCurrentLattice64x128} is different): The OPE works 
nicely even on this small lattice. (A) and (B) differ in the 
method used in the determination of the \MS coupling. 
As we explained in App. \ref{RunningCoupling} and it appears clearly
from the two graphs, the effect is very small. 
Graph (C)---and to a lesser extent graph (D)---shows instead significant 
deviations: Clearly, $\overline{g}(\Lambda_\MMS x)$ must be determined
using four-loop perturbative expansions in order 
to reduce the systematic error to a few percent. Notice that such 
discrepancies are probably present also for lattice 
\ref{Lattice128x256}: however, in this case, the statistical errors 
on $R(r)$ are large---approximately 5-6\% (for $\pb = 2\pi/L$
and  $\pb = 4\pi/L$) and 9\% (for $\pb = 6\pi/L$)---and thus they
do not allow to observe this effect.

\begin{figure}
\hspace{2cm}
\epsfig{figure=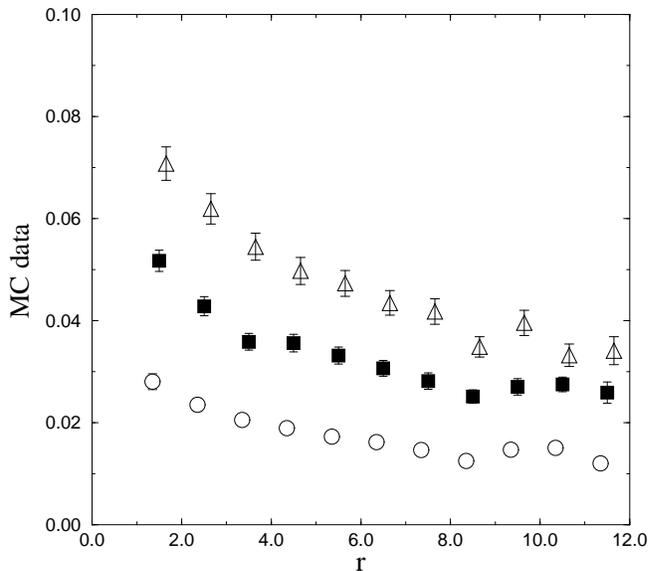,angle=-90,
width=0.6\linewidth}
\caption{Estimates of $\widehat{G}^{(s)}(t,x;\pb,0;20)$
averaged over rotations on lattice \ref{Lattice128x256}. Circles,
filled squares, and triangles correspond to $\pb=2\pi/L$, $4\pi/L$,
and $6\pi/L$ respectively. }
\label{ScalarCurrentsOutOfDiagonalData}
\end{figure}

\begin{figure}
\hspace{2cm}
\epsfig{figure=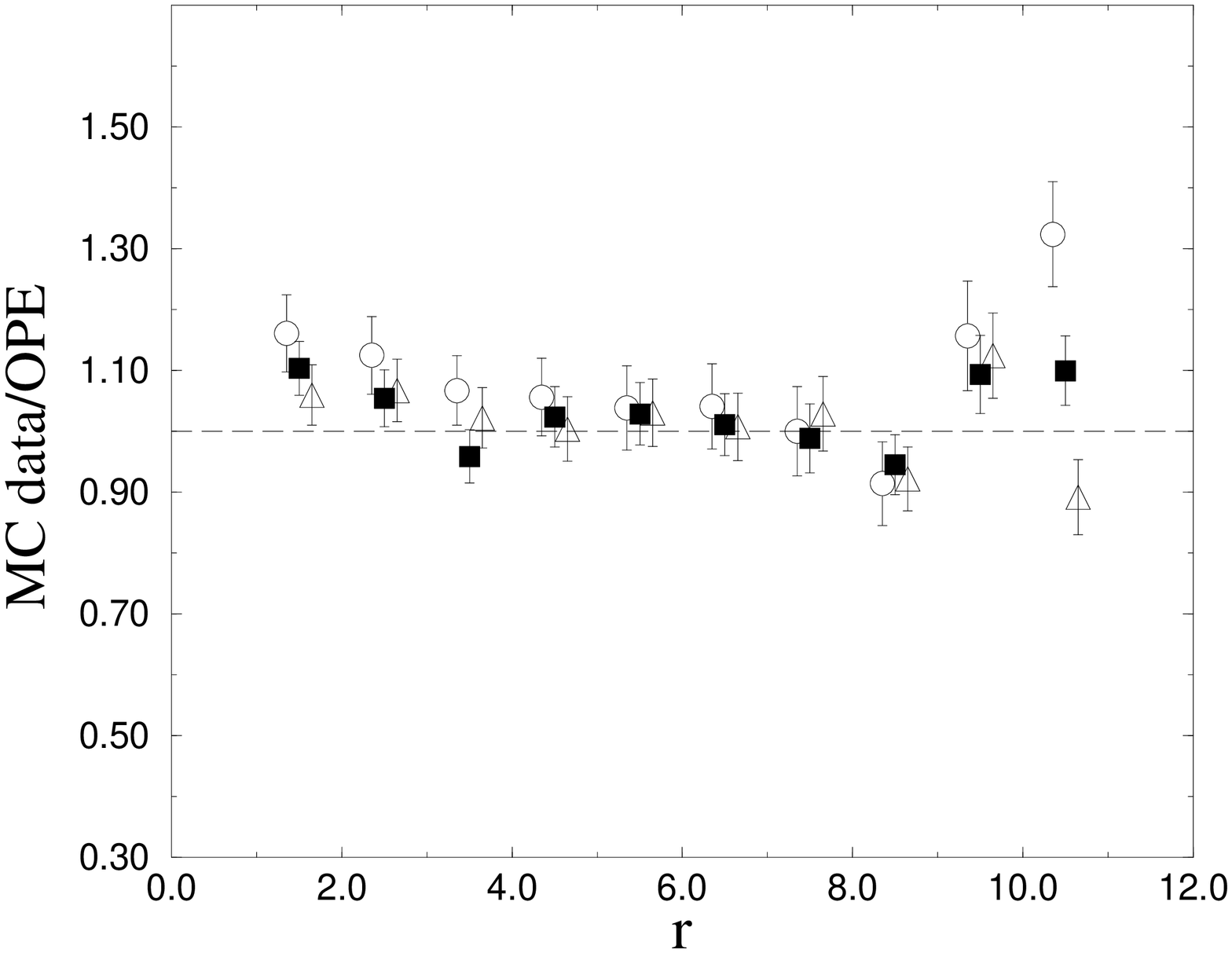,angle=-90,
width=0.6\linewidth}
\caption{The scalar product of two Noether currents compared with the OPE 
prediction: graphs of $S(r)$, cf. \protect\reff{defRatioS}, obtained
using \MS RG-improved perturbation theory. Circles,
filled squares, and triangles correspond to $\pb=2\pi/L$, $4\pi/L$,
and $6\pi/L$ respectively. The data are for lattice 
\ref{Lattice128x256}, $\xi^{\rm exp} = 13.636(10)$.}
\label{ScalarCurrentsMSOutOfDiagonal}
\end{figure}

As a further check we considered matrix elements between states of 
different momentum.
In Fig. \ref{ScalarCurrentsOutOfDiagonalData} we report the angular 
average of ${\rm Im}\ \widehat{G}^{(s)}(t,x;0,\pb;20)$ for lattice 
\ref{Lattice128x256}.
In Fig. \ref{ScalarCurrentsMSOutOfDiagonal} we compare the numerical data 
with the OPE prediction, by considering 
\be
S (r) = \frac{1}{2}
 {{\rm Im}\, \<0,c|\overline{\jg_{0,(0,0)}\cdot\jg_{1,(t,x)}}|\pb,c\> \over
   W^{RGI}(\bar{g}){\rm Im}\<0,c|T_{01}^{\rm latt}|\pb,c\>},
\label{defRatioS}
\ee
where $T_{\mu\nu}^{\rm latt}$ is defined in Eq. 
(\ref{EnergyMomentumRenormalization})
and $Z_{TT}^{L,(2,0)} $ is computed in Sec. \ref{sec4.3}.
In Fig. \ref{ScalarCurrentsMSOutOfDiagonal} we report $S(r)$ for 
lattice \ref{Lattice128x256}. The Wilson coefficients are computed as 
in graph (A) of Fig. \ref{ScalarCurrentLattice64x128}, using 
$m^{-1} = 13.636(10)$. The numerical data are again well described by the 
OPE prediction for a quite large set of values of $r$.
%

%**********************************************************************
%
\subsection{OPE for the Antisymmetric Product of Currents} \label{sec4.5}

In this Section we consider the antisymmetric product of two currents
and compare our numerical results with the perturbative predictions. 
With respect to the previous case, we have here a better knowledge 
of the perturbative Wilson coefficients---some of them are known to two 
loops---and moreover we have an exact expression for the 
one-particle matrix elements of the current $j_{\mu,x}$. Indeed,
we have \cite{Karowski:1978a,Luscher:1990ck}:
\be
\< \pb, c |j^{ab}_{\mu,(t,x)} |\qb, d \> = -
 i (p_\mu + q_\mu)\, G(k)\,
(\delta^{ac}\delta^{bd}-\delta^{ad}\delta^{bc})\, e^{i(p-q)\cdot x} ,
\label{form-factor}
\ee
where $p\cdot x\equiv p_0 t + p_1 x$, 
$p_\mu \equiv (i\sqrt{\pb^2+m^2},\pb)$, 
$k \equiv\frac{1}{2}\sqrt{(p_0-q_0)^2+(p_1-q_1)^2}$, 
and, for $N=3$, 
\be
   G(k) = \frac{\theta}{2\tanh\theta/2}\cdot\frac{\pi^2}{\pi^2+\theta^2},
\ee
where the rapidity variable $\theta$ is defined by $k = m\sinh\theta/2$.

\begin{figure}
\hspace{2cm}
\epsfig{figure=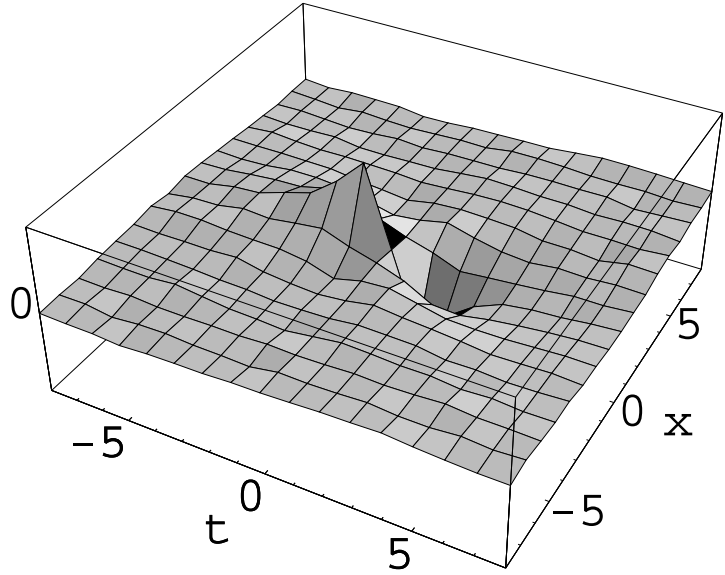,width=0.6\linewidth}
\caption{Estimate of $\widehat{G}^{(a)}_{11}(t,x;\pb,-\pb;20) $
on lattice \ref{Lattice128x256}. 
Here $\pb =2\pi/L$.}
\label{AntisymmetricCurrentsLeadingData}
\end{figure}

\begin{figure}
\hspace{2cm}
\epsfig{figure=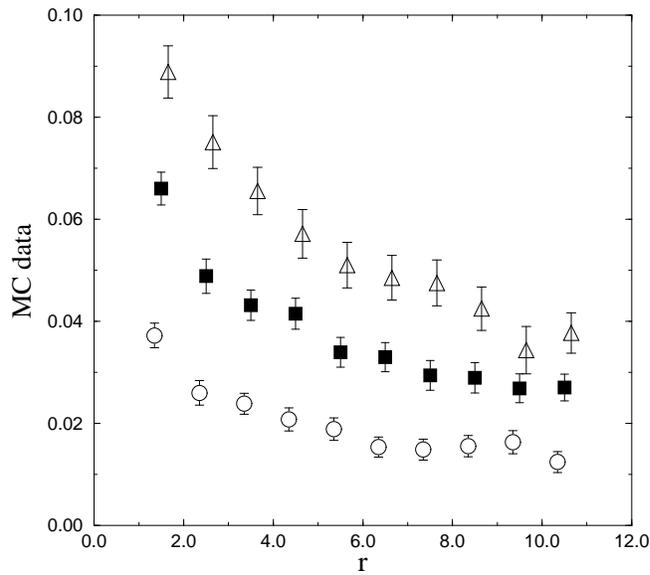,angle=-90,
width=0.6\linewidth}
\caption{Angular average of 
$\widehat{G}^{(a)}_{01}(t,x;\pb,0;20)$
on lattice \ref{Lattice128x256}.
Circles, filled squares, and triangles correspond to $\pb=2\pi/L$, $4\pi/L$,
and $6\pi/L$ respectively.}
\label{AntisymmetricCurrentsNextToLeadingData}
\end{figure}

We first consider the product $(j^{ac}_{\mu,(0,0)}j^{bc}_{\nu,(t,x)}-
j^{bc}_{\mu,(0,0)}j^{ac}_{\nu,(t,x)})$ for $\mu=\nu=1$ and $x=0$. 
The OPE of such a product can be determined from Eq. \reff{anticorrcont}.
Using the results of App. \ref{secB.2}, we have for $t\to 0$,
\be
\sum_c
j^{ac}_{1,(0,0)}j^{bc}_{1,(t,0)}-
j^{bc}_{1,(0,0)}j^{ac}_{1,(t,0)} = 
  {1 \over t} W_{RGI} \left(\overline{g}(\Lambda_\MMS t)\right) 
\left(j_0^{ab}(0) + 
   {t\over2} \partial_0 j_0^{ab}(0)\right),
\label{OPEx=0}
\ee
where, at two loops,
\be
W_{RGI}(\overline{g}) = {N-2\over 2\pi} + 
   {N-2\over 4\pi^2} \overline{g},
\label{Wilson_x=0}
\ee
and $\overline{g}(\Lambda_\MMS t)$ is defined in Eq. \reff{gdix}.
We also consider the angular average of the product of the currents 
for $\mu=0$ and $\nu = 1$. 
Using the results of Sec. \ref{PerturbativeContinuumCurrentsSection}, 
we have for $r\to 0$
\begin{eqnarray}
&& \hskip -1truecm
{\cal I}^{ab}(r) \equiv \overline{j^{ac}_{0,(0,0)}j^{bc}_{1,(t,x)}-
                      j^{bc}_{0,(0,0)}j^{ac}_{1,(t,x)}} =
\\
&& \hskip -1truecm \qquad
 -\,{3(N-2)\over 16\pi}\,
[\partial_0 j_1^{ab} (0) + \partial_1 j_0^{ab} (0)] + 
{1\over 2 \overline{g}}\left(1 + {N-6\over 4\pi} \overline{g}\right) 
\, [\partial_0 j_1^{ab} (0) - \partial_1 j_0^{ab} (0)]\, .\nonumber
\label{sec4.5:SOPE}
\end{eqnarray}
Again, we have tested the validity of the OPE by considering matrix elements 
between one-particle states.
The matrix elements of the product of the currents are obtained from 
\be
\sum_{abc}
\<\pb ,a|j^{(ac)}_{(0,0),\mu}j^{(bc)}_{(t,x),\nu}-
j^{(bc)}_{(0,0),\mu}j^{(ac)}_{(t,x),\nu}|\qb ,b\> = \,
N \sqrt{4\omega(\pb )\omega(\qb )} \lim_{t_s\to\infty }
\widehat{G}^{(a)}_{\mu\nu}(t,x;\pb ,\qb ;2t_s).
\label{AsymptoticAntiSymmetric}
\ee
In Fig. \ref{AntisymmetricCurrentsLeadingData} we show a plot 
of ${\rm Re}\, \widehat{G}^{(a)}_{11}(t,x;\pb ,-\pb ;2t_s)$
obtained on 
lattice \ref{Lattice128x256} for $\pb = 2\pi/L$ and $t_s = 10$, and  
in Fig. \ref{AntisymmetricCurrentsNextToLeadingData} we show 
the angular average of 
${\rm Im}\, \widehat{G}^{(a)}_{01}(t,x;\pb ,0 ;2t_s)$
on the same lattice and again for 
$t_s = 10$. Comparing these graphs with those for the scalar product
of the currents, one sees that the matrix elements show here a larger 
variation with distance, and thus this should provide a 
stronger test of the validity of the OPE.

\begin{figure}
\begin{tabular}{cc}
\hspace{0.0cm}\vspace{-1.5cm}\\
\hspace{-2.5cm}
\epsfig{figure=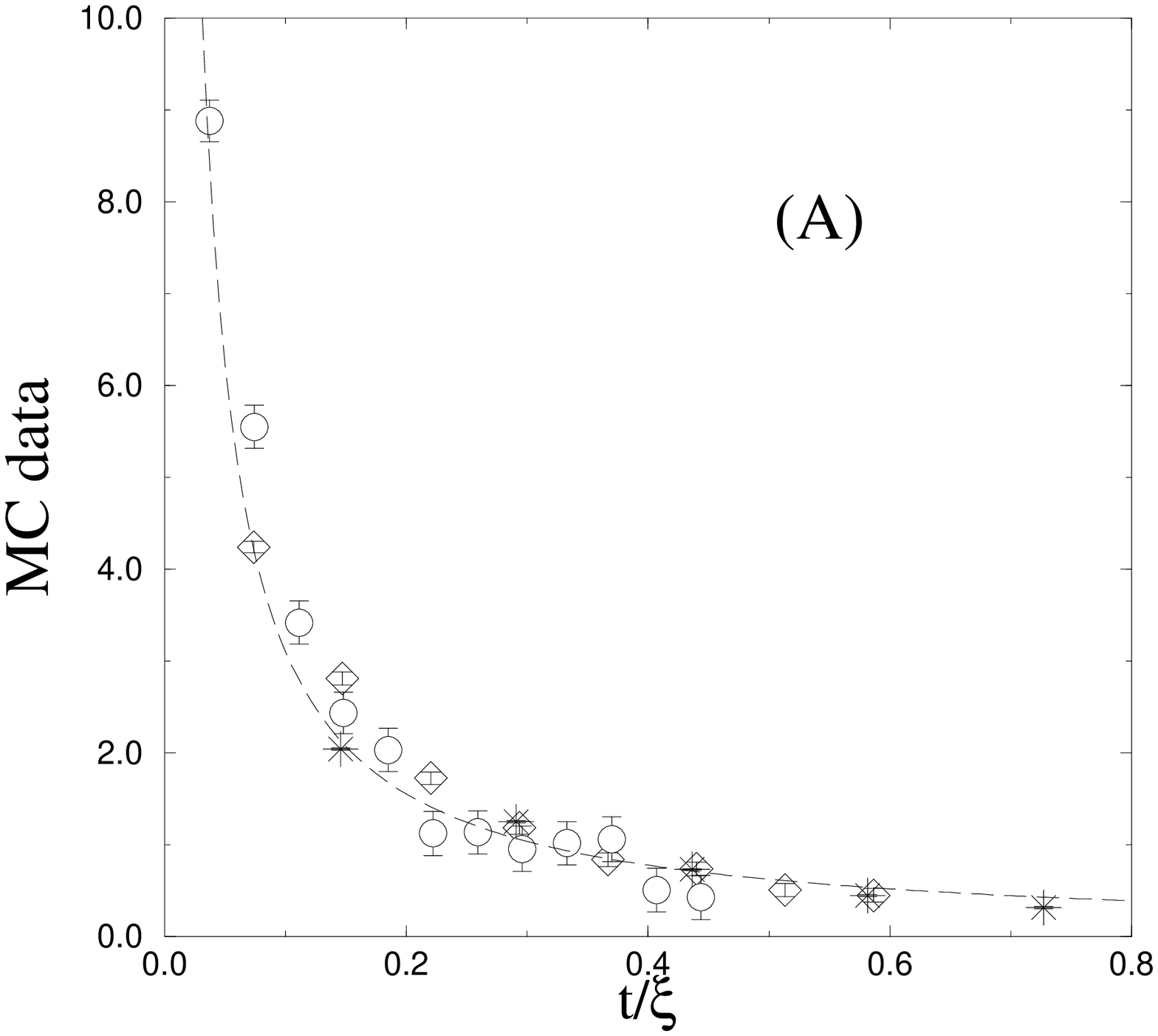,angle=-90,
width=0.6\linewidth}&\hspace{-0.5cm}
\epsfig{figure=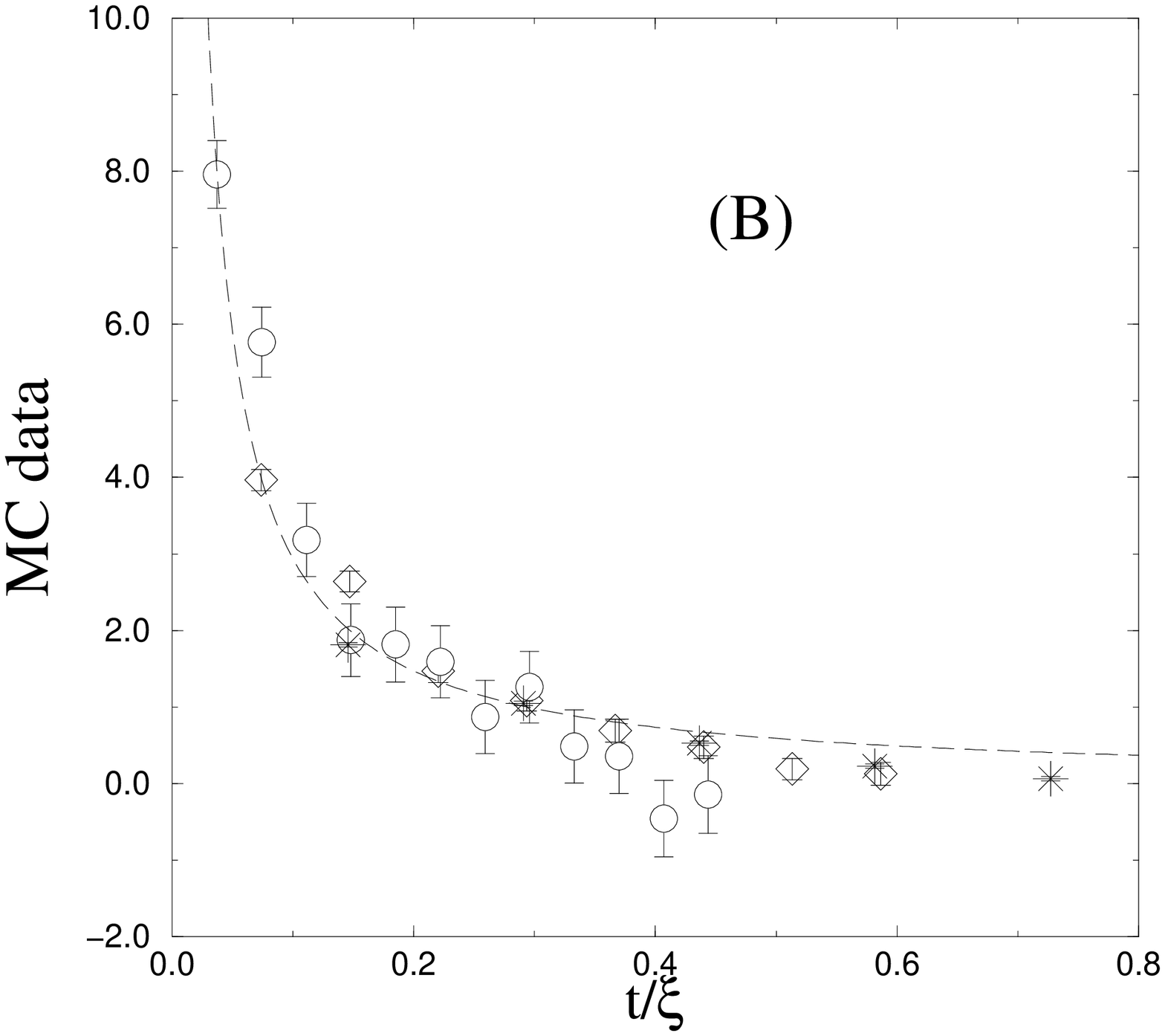,angle=-90,
width=0.6\linewidth}\\
\hspace{0.0cm}\vspace{-1.5cm}\\
\hspace{-3.5cm}
\epsfig{figure=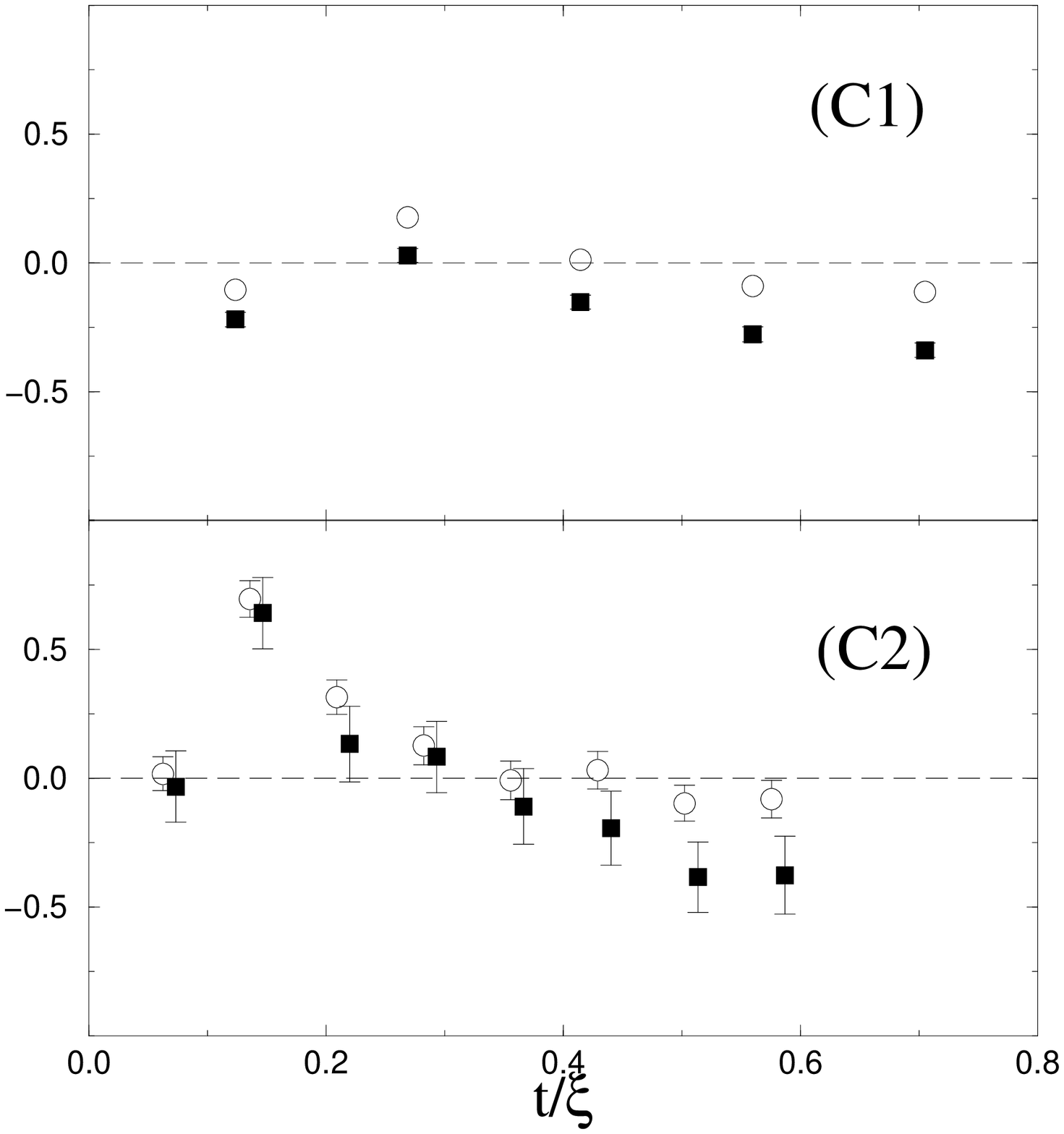,angle=-90,
width=0.6\linewidth}&\hspace{-1.5cm}
\epsfig{figure=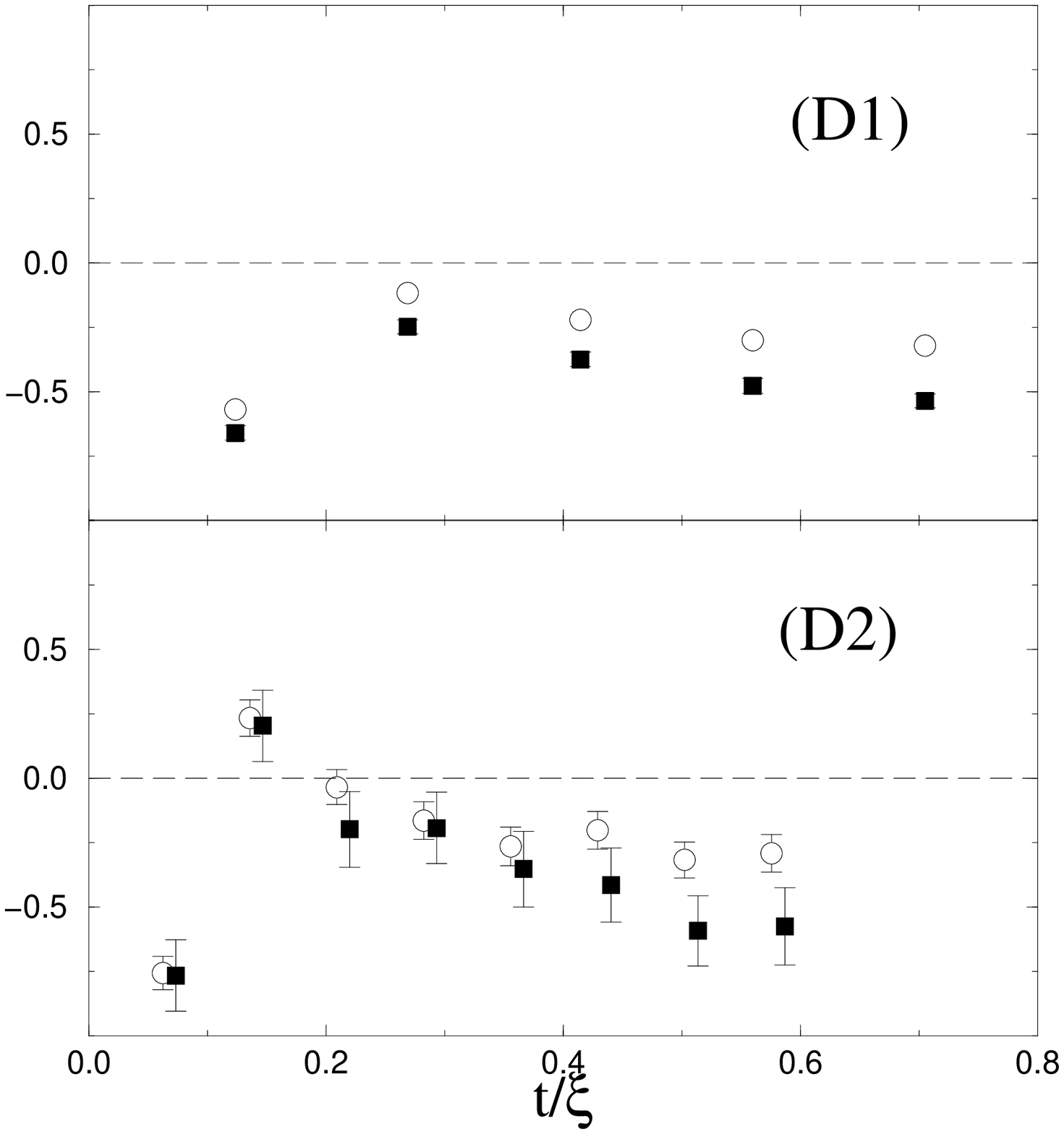,angle=-90,
width=0.6\linewidth}
\end{tabular}
\caption{Antisymmetric
product of two Noether currents for $x=0$ and $t\neq 0$: in graphs (A) and (B)
we report estimates  of $V(t)$, cf. \reff{ScalingCombination}, 
and of the OPE prediction
$V^{\rm OPE}(t)$, cf. Eq. \reff{Vope}, The numerical data correspond to
$\pb = 2\pi/L$ (graph (A)) and $\pb = 4\pi/L$ (graph (B)). 
Stars, diamonds, and circles refer to lattices \ref{Lattice64x128},
\ref{Lattice128x256}, \ref{Lattice256x512} respectively.
In graphs (C) and (D) we show $V(t) - V^{\rm OPE}(t)$.
Empty circles refer to $\pb = 2\pi/L$ while filled squares to 
$\pb = 4\pi/L$.}
\label{AntisymmetricCurrentsLeadingDatavsTheory}
\end{figure}

In Fig. \ref{AntisymmetricCurrentsLeadingDatavsTheory}
we compare the results for 
${\rm Re}\, \widehat{G}^{(a)}_{11}(t,x;\pb ,-\pb ;2t_s)$
with the OPE perturbative predictions. 
Here, as always in this Section, we use the RG-improved perturbative method 
to compute $\overline{g}$, using the four-loop expression for the $\beta$
function. As we explained at length in the previous Section, no significant
difference is observed if one uses the finite-size scaling method, 
or improved lattice perturbation theory. 

In graphs (A) and (B)  we show the combination 
\begin{equation}
V(t) \equiv \frac{1}{2}\, \xi^{\rm exp}\, \left[
\widehat{G}^{(a)}_{11}(t,0;\pb ,-\pb ;2t_s)-
\widehat{G}^{(a)}_{11}(-t,0;\pb ,-\pb ;2t_s)
\right]
\, ,
\label{ScalingCombination}
\end{equation}
for two different values of $\pb$.
In the scaling limit $V(t)$ is a function of $\pb\xi^{\rm exp}$ and
of $t/\xi^{\rm exp}$. As it can be seen from the graphs our results 
show a very nice scaling: The data corresponding to the three different 
lattices clearly fall on a single curve. In the same graphs we also 
report the OPE prediction \reff{OPEx=0}, i.e.
\be
V^{\rm OPE}(t) \equiv 2 N \omega(\pb) {\xi^{\rm exp}\over t} 
   W_{RGI}(\overline{g}(\Lambda_\MMS t))
  \<\pb,a|j^{ab}_0(0)|-\pb,b\>.
\label{Vope}
\ee
Note that $\<\pb,a|\partial_0j^{ab}_0|-\pb,b\> = 0$, so that 
the corrections due to higher-order terms in the OPE expansion are of order $t$.
In graph (A) and (B) we use only the one-loop Wilson coefficient 
for $W_{RGI}(\overline{g})$.
There is a good agreement between the OPE prediction and the numerical 
data: quite surprisingly the agreement extends up to  2 
lattice spacings. 

To better understand the discrepancies, in 
graphs (C) and (D) we report $V(t)-V^{\rm OPE}(t)$.
In graphs (C1) and (C2) we use the one-loop Wilson coefficient, and
in (D1) and (D2) the two-loop Wilson coefficient given in Eq. 
\reff{Wilson_x=0}. The numerical data
refer to lattice \ref{Lattice64x128} for (C1) and (D1) and to
lattice \ref{Lattice128x256} for (C2) and (D2).
There is clearly agreement, although here 
deviations are quantitatively somewhat large, since $V(t)$ is strongly varying. 
Let us consider for instance the data obtained on lattice \ref{Lattice128x256}
for $\pb = 2\pi/L$. If we evaluate the matrix element of the
Noether current using the numerical estimate
of $V(t)$ with $t = 3,4,5,6$ we obtain the result with a systematic error
of $2\%, 14\%, 32\%, 27\%$ respectively.

\begin{figure}
\hspace{2cm}
\epsfig{figure=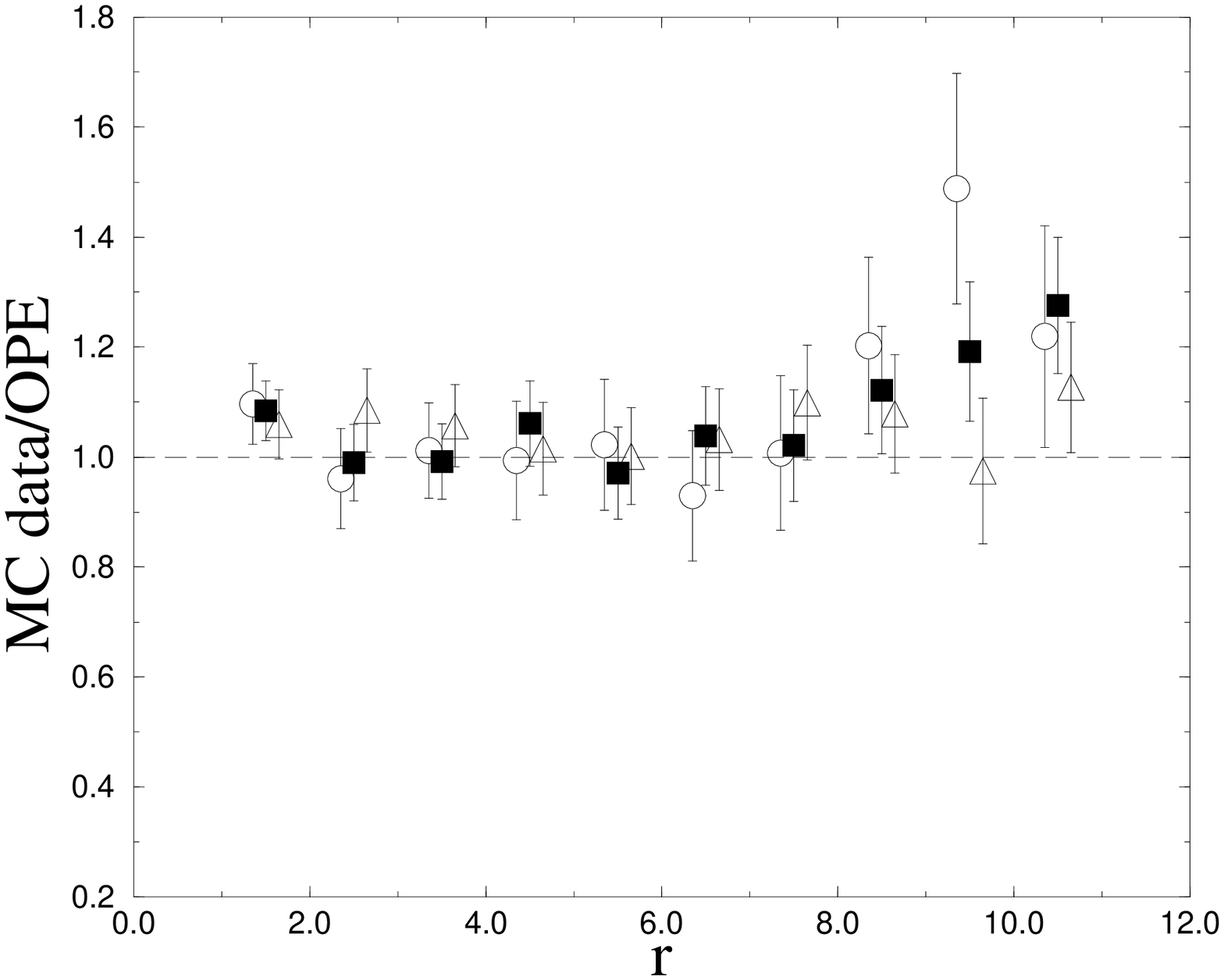,angle=-90,
width=0.6\linewidth}
\caption{The antisymmetric product of two Noether currents compared with 
the OPE prediction: graphs of $Y(r)$, cf. \protect\reff{defRatioY}, obtained
using \MS RG-improved perturbation theory. The circles,
filled square, and triangles correspond to $\pb=2\pi/L$, $4\pi/L$,
and $6\pi/L$ respectively. The data are for lattice 
\ref{Lattice128x256}, $\xi^{\rm exp} = 13.636(10)$.}
\label{AntisymmetricCurrentsNextToLeadingOPE}
\end{figure}

In Fig. \ref{AntisymmetricCurrentsNextToLeadingOPE} we compare the angular 
average of ${\rm Im}\, \widehat{G}^{(a)}_{01}(t,x;\pb ,0 ;2t_s)$
(cf. Fig. \ref{AntisymmetricCurrentsNextToLeadingData})
with the OPE prediction, by defining 
\be
Y(r) = {\<\pb,c|{\cal I}^{ab}(r)|0,c\>\over 
        \<\pb,c|{\cal I}^{OPE,ab}(r)|0,c\>},
\label{defRatioY}
\ee
where ${\cal I}^{OPE,ab}(r)$ is the OPE one-loop prediction
\reff{sec4.5:SOPE}.
Here we use the form-factor prediction \reff{form-factor} for the
matrix elements of the currents, but no significant difference 
would have been observed if the matrix elements of the currents 
had been determined numerically. Use of the form-factor prediction allows 
only a reduction of the statistical errors and thus gives the 
opportunity for a stronger check of the OPE. Again we observe 
a nice agreement and a very large window in which the data are well 
described by perturbative OPE.
The systematic error is below the statistical one (which is approximately 
$10\%$) as soon as $r>2$.
%
%**********************************************************************
%
\section{First Answers}
\label{SummaryCurrentsSection}   

From the examples of short-distance products studied in this 
Chapter we can draw some first conclusions. 

We considered products of the type $j(x)j(0)$. In most of the cases
the leading term of the OPE was of order $r^0$ (with $r=|x|$),
and the first correction, after averaging over rotations, was of 
order $r^2$. In these cases, we can make the following statements, 
which are valid
within the statistical accuracy of our numerical simulations (about
$5\div10\%$, depending upon the particular example):
\begin{itemize}
\item[1)] We extracted one-particle matrix elements of the type
$\<\pb|j(x)j(0)|\qb\>$ from well chosen four-point correlation
functions. These correlation functions show a nice scaling behavior
as soon as $r\gtapprox 2a$. On general grounds we would expect 
scaling corrections of order $1/r^2$ (among the others).
Such scaling corrections cannot be seen in our data.
The only relevant lattice artifacts occur at $r\ltapprox 2a$.
This is easily understood if we remember that the lattice currents
$j^{L,ab}_{\mu,x}$ and $j^{L,cd}_{\nu,0}$
have a spatial extension of one lattice
spacing, see Eq. (\ref{LatticeCurrent}).
The lattice artifacts at $r\ltapprox 2a$ are due to contacts between the two
operators  $j^{L,ab}_{\mu,x}$ and $j^{L,cd}_{\nu,0}$.

Such a good scaling behavior allows to use the OPE on rather coarse
lattices, e.g. on lattice \ref{Lattice64x128} which has 
a correlation length $\xi^{\exp} = 6.878(2)$.
\item[2)] Power corrections (i.e. terms of order $r^2$) are negligible
for $r\ltapprox \xi$. Indeed we did not find evidence for them in our
numerical results.
\item[3)] The running coupling $\overline{g}(\Lambda r)$ can be
accurately determined. This determination makes use of the four-loop
beta function, and of the lambda parameter.
In the $O(N)$ non-linear $\sigma$-model, the lambda parameter 
can be computed either with the finite-size method, or by using the
exact prediction for the mass gap. In QCD there exists no
exact prediction fixing the lambda parameter in terms
of some low-energy quantity. However, the finite-size method
has already given a precise determination of $\Lambda_{QCD}$
for the quenched theory.

In our case we can compute $\overline{g}(\Lambda r)$ 
with a few percent accuracy, over all the range $r\ltapprox\xi$.

\item[4)] The perturbative computation of the Wilson coefficients
seems to be the weakest point of the whole procedure.
The use of the leading-log approximation gives grossly inexact 
(by more than $50\%$) results in most of the cases considered.
The next-to-leading-log approximation (which requires the computation of
one-loop Wilson coefficients and two-loop anomalous dimensions)
yields results with about a $5\%$ accuracy. These statements are valid if
a ``well-behaved'' renormalization scheme (e.g. \MS) is adopted.
They seem to hold also if improved lattice perturbation theory
is used.
Naive lattice perturbation theory gives much worser results.

\item[5)] The use of different resummation methods for the Wilson
coefficients seems to give a realistic idea of the systematic error
involved in their perturbative calculation.
\end{itemize}

These conclusions should be perhaps modified if 
the leading Wilson coefficient has a
power-like diverging behavior ($W(r)\sim 1/r$).
We studied a single case of this type, see
Sec. \ref{sec4.5}. In this case we were able to compute
the relevant Wilson coefficient up to two-loop order.
Nevertheless, the agreement between the OPE prediction and the
numerical results was not good as in the other examples. 
%
%**************************************************************************
%
%
%*****************************************************************************
%
%
%************************************************************************
%
%
%************************************************************************
%
%
%********************************************************************
%
%
%********************************************************************
%
%
%********************************************************************
%
%
%**********************************************************************
%
%
%********************************************************************
%
%
%**********************************************************************
%***********************************************************************
%
\clearpage
\chapter{Operator Product Expansion for Elementary Fields}
In this Chapter we shall consider the short-distance expansion of the 
product of two elementary fields for the $O(N)$ non-linear $\sigma$-model. 
The motivation for such a study is twofold.

From a numerical point of view, the task is simpler than in the 
previous Chapter. Computing the product of two fields implies fewer 
operations than computing the product of two Noether currents. This 
implies a significative speed up in the algorithms and, as a consequence, 
much better numerical data.
We will be able to discuss finer issues than in the previous Chapter, such as
lattice artifacts and next-to-leading terms in the OPE.

Unlike Noether currents, elementary fields require a non-trivial
renormalization.
This makes the exercise slightly more complicated from a conceptual point 
of view. We shall be forced to consider the intricacies of renormalization and
to compare various (perturbative and non-perturbative) renormalization methods.

As in the previous Chapter, we shall focus on matrix elements between 
one-particle states. Generally speaking, we shall consider the
following OPE:
\begin{eqnarray}
\<\pb|\sigma(x)\sigma(-x)|\pb\> \sim W_{\cal O}(x)\<\pb|{\cal
O}|\pb\>+ \sum_{\cal Q}W_{\cal Q}(x)\<\pb|{\cal Q}|\pb\>\, ,
\label{GenericOPEFields}
\end{eqnarray}
where, once more, we neglected $O(N)$ indices. They will be specified
in Sec. \ref{FieldsNumericalResults}. In the above equation ${\cal O}$ 
is the leading term of the OPE. The corresponding Wilson coefficient
$W_{\cal O}(x)$ is of order $r^0$ or $r$ (we recall that 
$r=|x|$), depending upon the specific example.
In the cases we considered, ${\cal O}$ does not mix with any other 
operator (it renormalizes multiplicatively). The other terms,
denoted generically as ${\cal Q}$, are power corrections of
relative order $r^2$.
The operators ${\cal Q}$ have a non-trivial mixing structure.

We evaluated the left-hand side of Eq. (\ref{GenericOPEFields})
in lattice simulations for $r\ltapprox \xi$. Notice that the
separation between the two field operators on the left-hand side of 
Eq. (\ref{GenericOPEFields}) is $2x$. This means that, in this Chapter,
we are probing larger distances than in the previous one.
We expect that power corrections, i.e. the ${\cal Q}$ terms in Eq. 
(\ref{GenericOPEFields}), will not be negligible on such distances.

In this Chapter we mimic what would be done in a physical (QCD)
application of the OPE method. We fit the numerical results for the 
l.h.s. of Eq. (\ref{GenericOPEFields}), using the r.h.s. as fitting
form, and keep the matrix elements $\<\pb|{\cal O}|\pb\>$
and $\<\pb|{\cal Q}|\pb\>$ as fitting parameters. 
Finally we compare the results for $\<\pb|{\cal O}|\pb\>$
with some independent prediction, either numerical or analytical, 
for the same quantity.

We restrict the fit
to the window $\rho\le |x|\le R$, and study the dependence of the
results upon the window. In particular we shall focus on the outer
limit $R$. We learned in the previous Chapter that the principal
source of error, in comparing the OPE with lattice simulations, 
is the perturbative truncation of the Wilson coefficients.
This systematic error depends upon $R$, and in particular vanishes as 
$R/\xi\to 0$ because of asymptotic freedom.
The crucial point is that we cannot easily take the limit
$R/\xi\to 0$ because of lattice artifacts (which force us to consider 
$R\gg 1$) and of finite-size effects (for avoiding them, we must take
$\xi\ll L$, $L$ being the linear size of the lattice). 

We shall also consider the evaluation of the matrix
elements $\<\pb|{\cal Q}|\pb\>$ of the non-leading operators in the
OPE.
It turns out that a distinction must be made among these operators.
Those which do not mix with the leading one (${\cal O}$) can be fixed  
through Eq. (\ref{GenericOPEFields}), adopting a perturbative determination
of the Wilson coefficients. 
Nevertheless the computation is, in practice, quite difficult.
The determination of the other operators, those that mix with
${\cal O}$, is instead impossible, even from a theoretical point of
view. Such a calculation would require a non-perturbative knowledge of
the leading Wilson coefficient $W_{\cal O}(x)$.

This Chapter is organized as follows. In
Sec. \ref{PerturbativeGeneralSection} we recall some well-known
properties of perturbative expansions in quantum field theory. 
We discuss the consequences of these properties on our
non-perturbative renormalization method. 
In Sec. \ref{WilsonFieldsSection} we write the structure of the OPE
for two elementary fields,
including $O(r^2)$ terms, and we list the perturbative results for the
Wilson coefficients. In Sec. \ref{ResonanceSection} we show how to
solve the RG equation
if a resonance occurs in the anomalous dimensions matrix, cf
Sec. \ref{RenormalizationGroupSection}.
In Sec. \ref{FieldsNumericalResults} we explain the details of the 
fitting procedure, and we present our numerical results.
Finally we summarize our conclusions in Sec. 
\ref{SummaryFieldsSection}.
%
%****************************************************************************
%
\section{Perturbative Expansions and OPE}
\label{PerturbativeGeneralSection}
As we explained above, in this Chapter we shall keep track of the 
next-to-leading terms in the OPE. 
Our interest is twofold. First of all this may improve the
determination of the leading operator. Moreover we want to understand 
if it is possible to estimate next-to-leading operators.

It turns out that there is some theoretical difficulty in estimating
next-to-leading operators in the OPE when they mix with the leading
one. This theoretical difficulties are related to the diseases
of perturbative expansions, namely to {\it renormalons}.
In this Section we try to describe these difficulties.

The Section is divided in two parts. In the first one we review 
some well-known facts concerning the perturbative series
of renormalizable asymptotically-free theories.
The intent is mainly pedagogical.
Rather than making general statements, we look at simple examples
taken from the $O(N)$ non-linear $\sigma$-model at large $N$.
This has been an important toy model for the study
of such problems. For a complete review on renormalons we refer to 
\cite{Beneke:1998ui}. In the second part of this Section we
discuss the possibility of evaluating the matrix elements of
next-to-leading operators which mix with the leading one.
%
%****************************************************************
%
\subsection{The Limits of Perturbative Expansions}
\label{PerturbativeLimitsSection}

For our discussion it is  convenient to introduce the widespread language 
of Borel transforms. Perturbation theory yields physical quantities in the
form of asymptotic series:
\begin{eqnarray}
F(g)\sim\sum_{n=0}^{\infty}F_n g^n\, .
\label{AsymptoticSeries}
\end{eqnarray}
In order to recover $F(g)$ from the r.h.s. of Eq. (\ref{AsymptoticSeries}) one
needs, in general, additional informations beyond (all) the coefficients
$F_n$. These informations usually concern the analyticity properties
of $F(g)$.
A simple example is the case in which $F(g)$ is analytic in a 
neighborhood of $g=0$. Then, $F(g)$ is obtained by summing the 
r.h.s. of Eq. (\ref{AsymptoticSeries}) for $|g|$ smaller than the 
convergence radius and by analytic continuation outside.

Unfortunately this case is not realized in any non-trivial 
field-theoretical example. 

The next simpler situation is realized in several 
interesting cases (e.g. $\phi^4_d$ theory with $d<4$) 
and is described by the hypothesis of the Nevalinna-Sokal theorem
\cite{Sokal:1980ey}:
\begin{itemize}
\item[I.]\label{NevalinnaAnaliticity} 
$F(g)$ is analytic inside a disc ${\cal K}(\rho)$ of radius $\rho>0$ and
center $g_0$ with $\Real (g_0)=\rho$ and $\Im (g_0)=0$.
\item[II.]\label{NevalinnaBound}
The remainder $R_M(g)\equiv F(g)-\sum_{n=0}^{M-1}F_n g^n$ satisfies the bound
\begin{eqnarray}
|R_M(g)|< A\mu^M M!|g|^M\quad\quad\forall\quad g\in {\cal K}(\rho)\, .
\label{NevalinnaBoundEq}
\end{eqnarray}
\end{itemize}
When the previous hypothesis are realized, the function $F(g)$ is uniquely 
determined by the coefficients $F_n$ through the following construction.
One defines the Borel transform of the series (\ref{AsymptoticSeries})
as the formal series given below
\begin{eqnarray}
F_{\cal B}(z)\sim\sum_{n=0}^{\infty}\frac{F_n}{n!} z^n\, .
\label{BorelTransform}
\end{eqnarray}
Under the hypotheses I and II
it can be proved that the sum on the r.h.s. of Eq. (\ref{BorelTransform})
converges for $|z|<\delta$, $\delta>0$.  
We can promote $F_{\cal B}(z)$ to the status of an analytic function.
Moreover it can be proved that $F_{\cal B}(z)$ is analytic inside the strip
$\{ \Im (z)<\delta', \Real (z)>0\}$ and that the integral
\begin{eqnarray}
{\widetilde F}(g)\equiv\frac{1}{g}\int_0^{\infty}\!dz\,F_{\cal B}(z)
\, e^{-z/g}
\label{BorelIntegral}
\end{eqnarray}
is finite and equal to $F(g)$ for $g\in{\cal K}(\rho)$.

The physically interesting case of four-dimensional
non-abelian gauge theories does not fit in the above picture.
In this case there is no exact result about the properties
of perturbative expansions. Here we recall the standard 
picture, which is based on heuristic calculations. 
The large-order behavior of the perturbative coefficients is 
$F_n\sim \mbox{const.}(\beta_0/2)^n \Gamma(n+1+2\beta_1/\beta_0^2)$,
where $-\beta_0$ and $-\beta_1$ are the first two coefficients
of the beta-function, see Eq. (\ref{BetaExpansion}).
This behaviour is 
compatible with the hypothesis II. However, 
the analiticity region of $F(g)$ is a wedge of zero opening angle with
the tip at the origin.
The essential ingredients for the above picture are renormalizability
and asymptotic freedom. 
As a consequence similar statements hold for the two-dimensional $O(N)$ 
non-linear $\sigma$-model.

In order to study the concepts outlined above in a simple model we shall 
consider the $O(N)$ nonlinear $\sigma$-model in the limit
$N\to\infty$. 
In this Section we
fix the field and coupling-constant renormalizations by requiring:
\begin{eqnarray}
\Gamma^{(2)}(p)_{ab}= Z^{-1}\left[m^2+p^2+O(p^4)\right]\,\delta^{ab}
\equiv Z^{-1} \Gamma_R^{(2)}(p)_{ab}\, ,
\end{eqnarray}
where $\Gamma^{(2)}(p)_{ab}$ is the two-point vertex function.
It is moreover convenient to define
the running coupling at the scale $\mu^2$ by using the renormalization-group 
equation $\mu\frac{\partial g(\mu)}{\partial \mu}=\beta(g(\mu))$. 
At leading order in $1/N$, $\beta(g)=-g^2/2\pi$, whence
\begin{eqnarray}
g(\mu) = \frac{4\pi}{\log \mu^2/m^2}+O(1/N)\, .
\label{RunningLargeN}
\end{eqnarray}

We start with a simple example of physical observable, and consider
its perturbative expansion.
We define the effective coupling $\hat{g}(p)$ 
as follows in terms of the four-point vertex function
$\Gamma^{(4)}(p_1,p_2,p_3,p_4)_{abcd}$\, :
\begin{eqnarray}
\Gamma^{(4)}(p_1,p_2,p_3,p_4)_{abcd} & = &
\delta^{ab}\delta^{cd}\Gamma(p_1,p_2|p_3,p_4)+
\delta^{ac}\delta^{bd}\Gamma(p_1,p_3|p_2,p_4)+\nonumber\\
&&+\delta^{ad}\delta^{bc}\Gamma(p_1,p_4|p_2,p_3)\, ,
\label{DefinitionGeff1}\nonumber\\
\Gamma(p/2,p/2|-p/2,-p/2) & \equiv &
-Z^{-2}\frac{\hat{g}(p)}{N}(p^2+m^2)
\, .
\label{DefinitionGeff2}
\end{eqnarray}
Being a renormalization group invariant quantity, $\hat{g}(p)$ will be a 
function of $p^2/m^2$. It can be rewritten as a function of the
the running coupling $g(p)$ at the scale $p^2$, see Eq. (\ref{RunningLargeN}).
The leading term in the $1/N$ expansion of $\hat{g}(p)$ is given by
\begin{eqnarray}
\hat{g}(p) = 4\pi\frac{\sqrt{1+4 e^{-4\pi/g(p)}}}{1+e^{-4\pi/g(p)}}
\left\{\log \left[
\frac{
\sqrt{1+4 e^{-4\pi/g(p)}}+1}
{
\sqrt{1+4 e^{-4\pi/g(p)}}-1
}
\right]\right\}^{-1}
+O(1/N)\, .
\label{LeadingGeff}
\end{eqnarray}
Let us make a few observation concerning this very simple result:
\renewcommand{\theenumi}{\arabic{enumi}}
\begin{enumerate}
\item For small positive $g(p)$ we have 
$\hat{g} = g+(g-g^2/2\pi)e^{-4\pi/g}+O(e^{-8\pi/g})$. The ``perturbative''
part of this expansion is trivial; it is obviously analytic in the whole 
complex plane. Nevertheless it does not determine $\hat{g}(p)$ uniquely.
The next terms are of order $e^{-4\pi/g(p)}=m^2/p^2$. These are the 
so-called ``power corrections''.
\item The singularity structure of $\hat{g}(p)$ is nontrivial, including:
\begin{enumerate}
\item Simple poles at $\frac{4\pi}{g} = (2n+1)i\pi$ with $n\in\mathbb{Z}$. 
These poles are of
``kinematical'' origin: they appear because we factored out the term 
$(p^2+m^2)$ in the definition (\ref{DefinitionGeff1}) of $\hat{g}(p)$.
\item Branching points at 
$\frac{4\pi}{g}=\log 4 +(2n+1)i\pi$, $n\in\mathbb{Z}$.
These singularities were predicted on general grounds by 't Hooft in
Ref. \cite{'tHooft:1977am}. 
They are the traces of the two-particle threshold at 
$p^2 = -4m^2$.
\end{enumerate}
\item The function on the r.h.s. of Eq. (\ref{LeadingGeff}) does not satisfies
the hypotheses of the Nevalinna-Sokal theorem. If they were satisfied, 
we could sum the 
perturbative series using the Borel procedure obtaining the wrong result 
$\hat{g}(p) = g(p)$. Indeed, although the function is analytic in any 
disc ${\cal K}(\rho)$ with $\rho<4\pi/\log 4$, it does not satisfy the bound
in Eq. (\ref{NevalinnaBoundEq}) for any $M\ge 1$. 
In fact, for $M >1$, we get $R_M(g) = \hat{g}(g)-g$ and
we remark that $\hat{g}(g)$ is 
periodic\footnote{More precisely, $\hat{g}(g_1)=\hat{g}(g_2)$
if $1/g_1 = 1/g_2+i/2$.} along the circles 
$1/g = 1/g_0 +i\theta$  with $-\infty<\theta<+\infty$
These circles pass through the origin $g=0$ and belong to the disc
${\cal K}(\rho)$ for $g_0< \rho$. Therefore, we can approach the origin
through one of these circles. The bound (\ref{NevalinnaBoundEq}) is
violated because of the periodicity of $\hat{g}(g)$.
\end{enumerate}

Let us now consider a less straightforward computation. The self-energy 
is defined as follows:
\begin{eqnarray}
\Gamma^{(2)}_R(p;m^2)\equiv p^2+m^2+\frac{1}{N}\Sigma(p;m^2)\, .
\end{eqnarray}
The leading term in the $1/N$ expansion of $\Sigma(p;m^2)$ is given by
\begin{eqnarray}
\Sigma(p;m^2) = \int\diq \left[\frac{q^2+m^2}{(p+q)^2+m^2}\right]_{(2)}
\hat{g}(q)+O(1/N)\, ,
\label{SelfEnergyIntegral}
\end{eqnarray}
where $[F(p,q)]_{(n)}$ denotes zero momentum subtraction up to the $n^{\rm th}$
order in the external momentum $p$. The integral in 
Eq. (\ref{SelfEnergyIntegral}) has been considered in 
Ref. \cite{Beneke:1998eq}.
One obtains, as a byproduct of its computation, the whole perturbative 
series for $\Sigma(p;m^2)$ which reads:
\begin{eqnarray}
\Sigma(p;m^2) &\sim & p^2\left\{\log\frac{g(p)}{4\pi}+{\rm const.}-
\frac{g(p)}{2\pi}+\right.\label{SelfEnergyPerturbative}\\
&&\phantom{p^2\left\{\right.}
\left.+\sum_{n=1}^{\infty}n!\left[(1+(-1)^n)\,\zeta(n+1)-2\right]
\left(\frac{g(p)}{4\pi}\right)^{n+1}\right\}\, ,
\nonumber
\end{eqnarray}
where $\zeta(z)\equiv\sum_{k=1}^{\infty}k^{-z}$ is the Riemann zeta function.
Let us quote some simple remarks:
\begin{enumerate}
\item The series on the r.h.s. of Eq. (\ref{SelfEnergyPerturbative}) has zero
radius of convergence. The coefficients have the general large order
behavior  $\Sigma_n\sim {\rm const.}\, [1-(-1)^n](\beta_0/2)^n\Gamma(n+1)$.
\item The function $\Sigma(p;m^2)$ does not satisfy the hypothesis of the
Nevalinna-Sokal theorem. 
If they were satisfied the Borel transform of the series in 
Eq. (\ref{SelfEnergyPerturbative}) would be analytic on the real axis. Indeed
we obtain (neglecting the $\log g(p)/4\pi$ and the constant):
\begin{eqnarray}
\Sigma^{\rm pert}_{\cal B}(z) = -\frac{p^2}{4\pi}
\left[\frac{2}{1-z/4\pi}+\psi(1+z/4\pi)+\psi(1-z/4\pi)+2\gamma 
\right\}\, ,
\end{eqnarray}
which has simple poles at $z = 4n\pi$ with $n\in\mathbb{Z}$, $n\neq 0$. 
Singularities occurring within this pattern are 
usually denoted as ``renormalons''.
\item One can ``resum'' the series (\ref{SelfEnergyPerturbative}) through
the integral (\ref{BorelIntegral}) by assigning a prescription on 
each singularities of $\Sigma^{\rm pert}_{\cal B}(z)$.
Examples of such prescriptions are: take the Cauchy principal value at each 
pole; move slightly upward (downward) in the complex planes all the poles;
move the pole at $z=4n\pi$ to $z_{\epsilon}=4n\pi+i(-1)^n\epsilon$,
and so on. 
Notice that, since the first pole is at $z=4\pi$ these prescriptions yield
resummations which differ by terms of relative order 
$\exp(-4\pi/g(p))=m^2/p^2$.
\end{enumerate}

The last of these observations is often rephrased by saying that the
perturbative expansion fixes physical quantities up to an ambiguity of 
order $m^2/p^2$.
This is the ``standard wisdom'' on the problem and is by no means 
self-evident. Indeed we could add to a given resummation 
a term of the type $\exp(-4\pi t/g(p))=(m^2/p^2)^t$ without modifying 
its asymptotic expansion. 

However it is commonly believed that the correct physical quantity
can be recovered by assigning a prescription
at the renormalon singularities.
We could associate to any perturbative expansion a family
of ``minimally ambiguous'' resummations, each one corresponding to a
well-defined prescription at renormalon singularities.
Any two of these resummed expansions differ by terms of order
$m^2/p^2$. The correct resummation lies among them but, in order to
recover it, some non-perturbative input is required.
%
%***********************************************************************
%
\subsection{The Definition of Composite Operators}
\label{OperatorDefSection}
Let us now consider a simple example of OPE:
\begin{eqnarray}
{\cal A}(x){\cal B}(-x) \sim W_{\cal O}(r){\cal O}+W_{\cal Q}(r)r^2
{\cal Q}+O(r^4)\, .
\label{GeneralPowerCorrection}
\end{eqnarray}
For sake of simplicity we considered the Wilson coefficients to be
rotationally invariant, i.e. to depend upon $x$
uniquely through its modulus $r$. Such a behavior can be enforced by averaging
over rotations. Moreover we made explicit the power-like $r$ dependence
of the Wilson coefficients. Both $W_{\cal O}(r)$ and $W_{\cal Q}(r)$ 
are of order $r^0$.  Finally let ${\cal O}$ and ${\cal Q}$ have the 
same (internal and Lorentz) symmetries. As a consequence 
they will mix under renormalization.

We suppose the {\it renormalized} operators ${\cal A}$ and ${\cal B}$
on the l.h.s. of Eq. (\ref{GeneralPowerCorrection}) to be 
non-perturbatively known. 
Hereafter we shall focus on the RGI 
operators ${\cal A}_{RGI}$, ${\cal B}_{RGI}$,
${\cal O}_{RGI}$, ${\cal Q}_{RGI}$, and the corresponding
Wilson coefficients $W_{{\cal O},RGI}$, $W_{{\cal Q},RGI}$. 
With a slight abuse of notation we shall drop the subscripts $RGI$ in this 
Subsection.

As usual, everything we know about the Wilson coefficients $W_{\cal O}(r)$ 
($W_{\cal Q}(r)$) is their $l$-loop (respectively, $m$-loop)
perturbative expansions.
If we resum the perturbative series using the renormalization
group, see Sec. \ref{RenormalizationGroupSection}, we obtain:
\begin{eqnarray}
W^{(l)}_{\cal O}(r) = \overline{g}(r)^{\Gamma_{\cal O}}
\sum_{k=0}^l W_{{\cal O},k}\,  \overline{g}(r)^k\, ,
\label{PerturbativeWilson}
\end{eqnarray}
and an analogous formula for $W^{(m)}_{\cal Q}(x)$.

Notice that
Eq. (\ref{GeneralPowerCorrection}) allows to define the composite
operator ${\cal O}$ regardless of the precise value of $l\ge 0$:
\begin{eqnarray}
{\cal O}(x)\equiv \lim_{\eta\to 0}\frac{{\cal A}(x+\eta) {\cal
B}(x-\eta)}{W^{(l)}_{\cal O}(|\eta|)}\, .
\label{CompositeOPEDefinition}
\end{eqnarray}
Both the left-hand and right-hand sides of the above equation must be
interpreted as inserted in a correlation function. 
This correlation function must be taken with elementary fields 
$\sigma^{a_1}(y_1),\dots ,\sigma^{a_n}(y_n)$
at space-time points $y_i$ distinct from $x$:
$\<(\, \cdot\, )\, \sigma^{a_1}(y_1)\dots\sigma^{a_n}(y_n)\>$ ($y_i\ne x$). 
Apart from this specification Eq. 
(\ref{CompositeOPEDefinition}) is an exact definition
because of asymptotic freedom.
The $\eta\to 0$ limit is approached with corrections
of relative order $|\log\Lambda \eta|^{-l-1}$. Equation
(\ref{CompositeOPEDefinition}) is the theoretical basis of the 
non-perturbative renormalization method studied in this thesis.

Let us now take a step further and see whether Eq.  
(\ref{GeneralPowerCorrection}) can be used to define the
next-lo-leading operator ${\cal Q}$. The naive approach would
be to fix ${\cal O}$ from Eq. (\ref{CompositeOPEDefinition}),
and then subtract its contribution from the OPE
(\ref{GeneralPowerCorrection}). In other words
one would define ${\cal Q}$ through the following short distance limit: 
\begin{eqnarray}
\lim_{\eta\to 0}\frac{
{\cal A}(x+\eta) {\cal B}(x-\eta) - W^{(l)}_{\cal O}(|\eta|){\cal O}(x)
}{\eta^2 W^{(m)}_{\cal Q}(|\eta|)}\, .
\label{NLCompositeDefinition}
\end{eqnarray}
This procedure is equivalent to using Eq. 
(\ref{GeneralPowerCorrection}) as a fitting form, restricting the fit
to the window $\rho\le r\le R$ and considering the $1\ll \rho,R\ll\xi$ regime.
This is what we do in Sec. \ref{FieldsNumericalResults}.

The problem with Eq. (\ref{NLCompositeDefinition}) is evident:
the $\eta\to 0$ limit diverges. The $\eta\to 0$ behavior of the ratio
in Eq. (\ref{NLCompositeDefinition}) is easily obtained using
Eq. (\ref{GeneralPowerCorrection}):
\begin{eqnarray}
\frac{W_{\cal O}(|\eta|)-W^{(l)}_{\cal O}(|\eta|)}
{\eta^2 W^{(m)}_{\cal Q}(|\eta|)}{\cal O}(x)\sim 
|\log \Lambda\eta|^{\widehat{\Gamma}}\eta^{-2}
{\cal O}(x)\, .
\end{eqnarray}
where $\widehat{\Gamma} = \Gamma_{\cal O}-\Gamma_{\cal Q}-l-1$.

The problem we encountered do not disappear if we push the perturbative
calculation of Wilson coefficients to high orders. 
Let us suppose, for instance,
that we know the coefficients $W_{{\cal O},k}$ for any $k$,
see Eq. (\ref{PerturbativeWilson}). The series 
(\ref{PerturbativeWilson}) with $l=\infty$ will diverge, as explained
in Sec. \ref{PerturbativeLimitsSection}. 
Nevertheless we can try to sum it, i.e. to find
a function $W^{(\infty)}_{\cal O}(\overline{g}(r))$ whose asymptotic expansion
for $\overline{g}\to 0$ coincides with the perturbative one.
We can moreover require $W^{(\infty)}_{\cal O}(\overline{g}(r))$
to have the minimum possible ambiguity. This
prescription should  be understood in the sense 
explained in the previous Subsection.

Even if we have such a minimally ambiguous Wilson coefficient
$W^{(\infty)}_{\cal O}(\overline{g}(r))$, we are left with a
great freedom. 
This freedom correspond to the choice of the prescription at the 
renormalon singularities. It produces an ambiguity of order $\Lambda^2r^2$.

Let us now repeat the construction
outlined in Eq. (\ref{NLCompositeDefinition}) using the new 
Wilson coefficient $W^{(\infty)}_{\cal O}(\overline{g}(r))$. We get:
\begin{eqnarray}
\frac{
{\cal A}(x+\eta) {\cal B}(x-\eta) - W^{(\infty)}_{\cal O}(|\eta|){\cal O}(x)
}{\eta^2 W^{(m)}_{\cal Q}(|\eta|)} \approx
\frac{W_{\cal O}(|\eta|)-W^{(\infty)}_{\cal O}(|\eta|)}
{\eta^2 W^{(m)}_{\cal Q}(|\eta|)}{\cal O}(x)+
\frac{W_{\cal Q}(|\eta|)}{W^{(m)}_{\cal Q}(|\eta|)}{\cal Q}(x)\, .\nonumber\\
\label{NewWilson}
\end{eqnarray}
According to a conjecture due to Parisi 
\cite{Parisi:1978bj,Parisi:1979az,Parisi:1979iq}, the ambiguities in the 
perturbative series are strictly related to the power corrections in
the OPE. In our case we get:
\begin{eqnarray}
W^{(\infty)}_{\cal O}(r) - W_{\cal O}(r) = {\rm const.}\, (\Lambda r)^2
|\log\Lambda r|^{-\Gamma_{\cal Q}}[1+O(|\log\Lambda r|^{-1})]\, .
\end{eqnarray}
Using this formula we can further elaborate Eq. (\ref{NewWilson}),
obtaining
\begin{eqnarray}
\lim_{\eta\to 0} \frac{
{\cal A}(x+\eta) {\cal B}(x-\eta) - W^{(\infty)}_{\cal O}(|\eta|){\cal O}(x)
}{\eta^2 W^{(m)}_{\cal Q}(|\eta|)} = {\cal Q}(x)+
{\rm const.}\, {\cal O}(x)\, .
\end{eqnarray}
Even if we know the whole perturbative series for the leading Wilson 
coefficient, we cannot fix the next-to-leading operator from the OPE
(\ref{GeneralPowerCorrection}). The renormalon ambiguity in the
leading Wilson coefficient is accompanied by the ambiguity 
of the additive renormalization of the next-to-leading operator.
David \cite{David:1982qv,David:1984gz,David:1985xj} 
studied this phenomenon  in  the $O(N)$ nonlinear $\sigma$-model at
large $N$.

Our discussion does not exclude the possibility of estimating the 
next-to-leading  operator ${\cal Q}$ from the OPE 
(\ref{GeneralPowerCorrection}). Nevertheless such a calculation
cannot be accomplished by naively substituting the coefficients
$W_{\cal O}(x)$ and $W_{\cal Q}(x)$ by their perturbative truncation.
A clever and accurate definition of the Wilson coefficients is
required.
This definition should be matched with the appropriate definition for
the composite operator ${\cal Q}$.
%
%*********************************************************************
%
\section{Perturbative Calculation of the Wilson Coefficients}
\label{WilsonFieldsSection}
In order to apply the OPE renormalization method, we have 
to compute the OPE of two elementary fields in perturbation theory.

The product of two fields can be decomposed in terms of irreducible 
representations of $O(N)$. We get a scalar, an antisymmetric
rank-2 tensor, and a symmetric traceless rank-2 tensor:
\begin{eqnarray}
\sigma^a(x)\sigma^b(-x) &=& \frac{\delta^{ab}}{N}\sg(x)\cdot   \sg(-x)+
\frac{1}{2}\left[\sigma^a(x)\sigma^b(-x)-\sigma^b(x)\sigma^a(-x)\right]+
\nonumber\\
&&+\frac{1}{2}\left[\sigma^a(x)\sigma^b(-x)+\sigma^b(x)\sigma^a(-x)-
\frac{2\delta^{ab}}{N}\sg(x)\cdot   \sg(-x)\right]\, .
\end{eqnarray} 
It is convenient to introduce the following notation for the
symmetrized and antisymmetrized products:
$\sigma^{[a}(x)\sigma^{b]}(y)=
\sigma^a(x)\sigma^b(y)-\sigma^b(x)\sigma^a(y)$,
and $\sigma^{\{ a}(x)\sigma^{b\}}(y)=
\sigma^a(x)\sigma^b(y)+\sigma^b(x)\sigma^a(y)$.
As we explained in the previous Chapter,
the form of OPE is dictated by $O(N)$ symmetry and Lorentz invariance.
Let us write it explicitly up to $O(x^2)$ terms:
\begin{eqnarray}
\sg(x)\cdot   \sg(-x)&=&F^{(0)}_0(x)\mbox{\boldmath $1$}+
F^{(0)}_1(x)T_{\mu   \rho}+\nonumber\\
&&\hspace{-1cm}+F^{(0)}_2(x)\opl (\partial\sg)^2\opr
+F^{(0)}_3(x)\opl\alpha \opr+O(x^4)\, ,\label{OPEFieldsScalar}\\
\sigma^{[a}(x)\sigma^{b]}(-x) & = & 
2gF^{(1)}_0 (x)\, j^{ab}_\mu+O(x^3)\, ,
\label{OPEFieldsAntisymmetric}\\
\hspace{-2cm}\frac{1}{2}\sigma^{\{a}(x)\sigma^{b\}}(-x)
-\hspace{2cm}&&\label{OPEFieldsSymmetric}\\
-\frac{\delta^{ab}}{N}\sg(x)\cdot   \sg(-x)  & = &
 F^{(2)}_0(x)\opl S_0 \opr
+\sum_{k=1}^7 F^{(2)}_k(x)\opl S_k\opr+O(x^4)\nonumber\, ,
\end{eqnarray}
where we defined
\begin{eqnarray}
S^{ab}_0 \equiv \sigma^a\sigma^b-\frac{1}{Z}\frac{\delta^{ab}}{N}\, .
\label{SymmetricTraceless}
\end{eqnarray}
The symmmetric traceless dimension 2 operators
$S_1,\dots,S_7$ are defined in
Eqs. (\ref{Symmetric1})--(\ref{Symmetric7}).
All the operators on the right-hand sides of Eqs. 
(\ref{OPEFieldsScalar})--(\ref{OPEFieldsSymmetric}) are
understood at the space-time position $x=0$.
For sake of simplicity we dropped the Lorentz indices 
of the Wilson coefficients in Eqs. 
(\ref{OPEFieldsScalar})--(\ref{OPEFieldsSymmetric}).
In Eq. (\ref{OPEFieldsSymmetric}) we neglected the Lorentz indices
also on the operators $S_1$, $S_2$ and $S_5$. The indices can be
restored without ambiguity. Summation over repeated indices is
understood.

The Wilson coefficients $F^{(n)}_i(x)=F^{(n)}_i(x;\mu,g)$ 
can be computed in perturbation theory. 
We computed the leading coefficient $F^{(n)}_0$ at two-loop order,
and the next-to-leading coefficients $F_i^{(n)}$, $i\ge 1$ at one-loop
order. 
Using the results for the anomalous dimensions given
in Sec. \ref{AnomalousSection}, we are able to resum the $F^{(n)}_0$ at 
next-to-next-to-leading log, and the $F^{(n)}_i$, $i\ge 1$
at next-to-leading log order. The outcomes of the resummation
procedure are given in Secs. \ref{ResonanceSection} and
\ref{FieldsNumericalResults}.

Let us begin from the scalar sector, see Eq. (\ref{OPEFieldsScalar}).
In this case the leading 
Wilson coefficient is known at three-loop order 
\cite{McKane:1980cm,Amit:1980kz}.
Since the field anomalous dimensions are known at four-loop order
\cite{Bernreuther:1986js,Wegner:1989ss}, 
see Sec. \ref{ModelSection}, we can resum this coefficient at
(next-to-)${}^3$leading log order, cf. Eq. (\ref{TwoPointResummed}). 
For greater convenience of the reader, we give below a complete list
of the perturbative results: 
\begin{eqnarray}
F^{(0)}_0(x;\mu,g) & = & 1-\frac{N-1}{2\pi}g\left(\gamma+\log\mu r\right)+
\frac{N-1}{8\pi^2}g^2\left(\gamma+\log\mu r\right)^2+
\label{F00Pert}\\
&&+g^3\left\{
\frac{(N-1)(N-3)}{48\pi^3}(\gamma+\log\mu r)^3
-\frac{(N-1)(N-2)}{16\pi^3}(\gamma+\log\mu r)^2-\right.\nonumber\\
&&\hspace{-0.5cm}\left.-\frac{3(N-1)(N-2)}{32\pi^3}(\gamma+\log\mu r)
+\frac{(N-1)(N-2)}{32\pi^3}\left(\zeta(3)-\frac{3}{2}\right)
\right\}+O(g^4)\, ,\nonumber\\
F^{(0)}_1(x;\mu,g) & = & -2gx_\mu   x_{\rho}
\left[1-\frac{1}{2\pi}g(\gamma+\log\mu r)+O(g^2)\right]\, ,\label{F01Pert}\\
F^{(0)}_2(x;\mu,g) & = & -x^2\left[1-\frac{1}{4\pi}g+O(g^2)\right]\, ,\\
F^{(0)}_3(x;\mu,g) & = & x^2\left[-\frac{N-1}{2\pi}g
\left(\gamma+\log\mu r-\frac{1}{2}\right)+O(g^2)\right]\, .
\label{F03Pert}
\end{eqnarray}

Next we consider the antisymmetric sector, see Eq. 
(\ref{OPEFieldsAntisymmetric}). Here we limit ourselves to the leading
term of the OPE:
\begin{eqnarray}
F^{(1)}_0 (x;\mu, g) & =& -x_\mu   \left[1-\frac{1}{2\pi}g(\gamma+
\log\mu r)-\frac{N-3}{8\pi^2}g^2(\gamma+\log\mu r)^2+\right.
\label{F10Pert}\\
&&\phantom{ 2x_\mu   \left[\right.}\left.
+\frac{N-2}{4\pi^2}g^2(\gamma+\log\mu r)-\frac{N-2}{16\pi^2}g^2
+O(g^3)\right] \nonumber\, . 
\end{eqnarray}
In Sec. \ref{AntisymmetricSection} we shall also consider the power
corrections (of relative order $r^2$) to this leading behavior. 
Since we did not compute them in perturbation theory, even in
leading-log approximation, we shall adopt a ``phenomenological''
point of view. We shall add all the terms with the correct dimension
and Lorentz symmetry, neglecting any logarithmic $x$
dependence\footnote{Something similar is done in Ref. \cite{Boucaud:2000nd}.}. 
This gives the feeling of how power corrections do affect the
estimates on the leading operator $j^{ab}_{\mu}$.

Finally we must consider rank-2 symmetric traceless $O(N)$-tensors,
see Eq. (\ref{OPEFieldsAntisymmetric}).
The list of Wilson coefficients is given below:
\begin{eqnarray}
F^{(2)}_0(x;\mu,g) & = & 1+\frac{1}{2\pi}g\left(\gamma+\log\mu r\right)+
\frac{N-1}{8\pi^2}g^2(\gamma+\log\mu r)^2+O(g^3)\, ,\label{F20Pert}\\
F^{(2)}_1(x;\mu,g) & = & x_\mu   x_{\rho}
\left[-1+\frac{1}{2\pi}g\left(\gamma+\log\mu r\right)+O(g^2)\right]\, ,\\
F^{(2)}_2(x;\mu,g) & = & x_\mu   x_{\rho}
\left[1+\frac{1}{2\pi}g\left(\gamma+\log\mu r\right)+O(g^2)\right]\, ,\\
F^{(2)}_3(x;\mu,g) & = & x^2\left[
-\frac{1}{2\pi}g\left(\gamma+\log\mu
r-\frac{1}{2}\right)+O(g^2)\right]\, ,\\
F^{(2)}_4(x;\mu,g) & = & x^2\left[O(g^2)\right]\, ,\\
F^{(2)}_5(x;\mu,g) & = & x_\mu   x_{\rho}
\left[\frac{1}{\pi}g\left(\gamma+\log\mu r\right)+O(g^2)\right]\, ,\\
F^{(2)}_6(x;\mu,g) & = & x^2\left[O(g^2)\right]\, ,\\
F^{(2)}_7(x;\mu,g) & = & x^2\left[\frac{1}{2\pi}g\left(\gamma+\log\mu r
-\frac{1}{2}\right)+O(g^2)\right]\, .
\label{F27Pert}
\end{eqnarray}

In Sec. \ref{FieldsNumericalResults} we shall compute, among the other
things, the renormalization constant for the dimension zero symmetric 
traceless operator $S_0^{ab}$, see Eq. (\ref{SymmetricTraceless}).
In order to obtain a non-perturbative estimate, we shall consider its 
two-point function, and apply an ``OPE-inspired'' procedure.
The first step in this procedure consists in computing perturbatively
the OPE for two operators $S_0^{ab}$. The structure of the OPE is the 
following
\begin{eqnarray}
\sum_{ab}\opl S_0^{ab}\opr \!(x)\opl S_0^{ab}\opr \!(-x)
= \frac{N-1}{N}\, E(x;\mu,g)+O(x^2) \, .
\end{eqnarray}
We computed this Wilson coefficient at three-loop order in
perturbation theory:
\begin{eqnarray}
E(x; \mu ,g) & = & 1 - \frac{N}{\pi}g\,(\gamma + \log \mu r) +
\frac{N(N+2)}{4\pi^2}g^2 (\gamma+\log\mu r)^2+
\label{PerturbativeSymmSymm}\\
&&+g^3\left\{ -\frac{N(N+2)}{6\pi^3}(\gamma+\log\mu r)^3
-\frac{N(N-2)}{8\pi^3}(\gamma+\log\mu r)^2 \right.\nonumber\\
&&\phantom{+g^3\left\{\right.}
\left.-\frac{3N(N-2)}{16\pi^3}(\gamma+\log\mu r)
-\frac{N(N-2)}{32\pi^3}[3-2\zeta(3)]
\right\} \, .\nonumber
\end{eqnarray}
%
%******************************************************************
%
\section{Solution of the RG Equations}
\label{ResonanceSection}

Solving the RG equations for the Wilson coefficients 
$F^{(2)}_1,\dots,F^{(2)}_7$
of the dimension 2 symmetric traceless operators, see Eq. 
(\ref{OPEFieldsSymmetric}), deserves some
unexpected technical difficulty. We anticipated this difficulty,
namely a resonance in the RG equations, already in
Sec. \ref{RenormalizationGroupSection}. Here we describe in detail how
to proceed if such a case occurs. In fact, we did not find any
reference to this problem in textbooks.
We study the concrete example we encountered, and refer to 
\cite{Wasow} for a general treatment of the subject.

The first step consists in choosing the most convenient basis of operators.
We shall adopt the basis of operators of definite spin
$\{ Q^{(1)R}_{\mu\nu},\dots,Q^{(7)R}\}$ defined in 
Section \ref{SymmetricAnomalousContinuum}, see Eqs.
(\ref{Q1Def})--(\ref{Q7Def}). The corresponding Wilson coefficients
${\cal F}^{(2)}_1 (x;\mu,g),\dots, 
{\cal F}^{(2)}_7 (x;\mu,g)$ are easily computed using
the results of Section \ref{WilsonFieldsSection}, see Eqs.
(\ref{F20Pert})--(\ref{F27Pert}).
In order to avoid the complications due to the tensor structure
of the Wilson coefficients, we shall adopt the following 
convention. Among the mentioned composite operators,
$Q^{(1)R}_{\mu\nu}$, $Q^{(3)R}_{\mu\nu}$ and $Q^{(4)R}_{\mu\nu}$ 
have spin 2, while $Q^{(2)R}$, $Q^{(5)R}$, $Q^{(6)R}$ and $Q^{(7)R}$
are Lorentz scalars. We give to the last ones 
two Lorentz indices in the obvious
way: $Q^{(i)R}_{\mu\nu}\equiv Q^{(i)R}\delta_{\mu\nu}$ for
$i=2,5,6,7$. 
We can now write all the Wilson coefficients in the form
${\cal F}^{(2)}_i(x;\mu,g) = x_\mu x_\nu  {\cal F}^{(2)}_i(\mu r,g)$.
Dropping the common factor $x_{\mu}x_{\nu}$, we get:
\begin{eqnarray}
{\cal F}^{(2)}_1 (\mu r;g) & = & -2 +O(g^2)\, ,\label{Pert22.1}\\
{\cal F}^{(2)}_2 (\mu r;g) & = & -1 
-\frac{1}{2\pi}g\left(\gamma+\log\mu r-2\right)+O(g^2)\, ,\\
{\cal F}^{(2)}_3 (\mu r;g) & = &  \frac{1}{\pi}g
\left(\gamma+\log\mu r\right)+O(g^2)\, ,\label{Pert22.3}\\
{\cal F}^{(2)}_4 (\mu r;g) & = &  1+
\frac{1}{2\pi}g\left(\gamma+\log\mu r\right)+O(g^2)\, , \\
{\cal F}^{(2)}_5 (\mu r;g) & = & \frac{1}{2}+
\frac{1}{4\pi}g\left(\gamma+\log\mu r-2\right) +O(g^2)\, ,\\
{\cal F}^{(2)}_6 (\mu r;g) & = &  \frac{1}{2\pi}g +O(g^2)\, ,\\
{\cal F}^{(2)}_7 (\mu r;g) & = & 
\frac{1}{2\pi}g\left(\gamma+\log\mu r-1\right)+O(g^2)\, .
\end{eqnarray}

The anomalous dimension matrix  for the Wilson 
coefficients ${\cal F}^{(2)}_1(\mu r;g),\dots,{\cal F}^{(2)}_7(\mu
r;g)$ is $\gamma_W^{(2,2)}(g) = [\gamma^{(2,2)}(g)]^T+\gamma(g)$,
where $\gamma^{(2,2)}(g)$ is given at two-loop order
by Eqs. (\ref{Gamma22Expansion})--(\ref{Gamma22g2}).
Because of the form (\ref{AnomalousForm22}) of $\gamma^{(2,2)}(g)$,
we obtain the following structure for  $\gamma_W^{(2,2)}(g)$:
\begin{eqnarray}
\gamma_W^{(2,2)}(g) = \left(
\begin{array}{ccc}
\gamma^W_{AA}(g) & 0 & 0\\
\gamma^W_{BA}(g) & \gamma^W_{BB}(g) & 0 \\
\gamma^W_{CA}(g) & 0 & \gamma^W_{CC}(g) 
\end{array}
\right)\, ,
\label{Splitting}
\end{eqnarray}
where we splitted the seven dimensional matrix as $3\oplus 2\oplus 2$
(in other words $\gamma^W_{AA}(g)$ is a $3\times 3$ matrix,
while both $\gamma^W_{BB}(g)$ and $\gamma^W_{CC}(g)$ are $2\times 2$).

From Eq. (\ref{Splitting}) it follows that the Wilson coefficients
${\cal F}^{(2)}_1(\mu r;g)$, ${\cal F}^{(2)}_2(\mu r;g)$, and
${\cal F}^{(2)}_3(\mu r;g)$ satisfy a ``reduced'' 
RG equation with anomalous dimensions matrix $\gamma^W_{AA}(g)$.
Recall that all the operators of our basis except 
$\{[Q^{(1)}_{\mu\nu}]_R,[Q^{(2)}]_R,[Q^{(3)}_{\mu\nu}]_R\}$ have vanishing
matrix element between on-shell states of equal momentum.
As we shall see in the next Section, we focused on such matrix
elements in our numerical simulations.
We can therefore limit ourselves to considering the ``reduced'' RG
equation for ${\cal F}^{(2)}_1(\mu r;g)$, ${\cal F}^{(2)}_2(\mu r;g)$, and
${\cal F}^{(2)}_3(\mu r;g)$.

We can further simplify the task, noticing that, while $[Q^{(1)}_{\mu\nu}]_R$
and $[Q^{(3)}_{\mu\nu}]_R$ are spin 2 operators, $[Q^{(2)}]_R$ is a scalar 
and thus renormalizes multiplicatively.
This observation was already made in
Sec. \ref{SymmetricAnomalousContinuum}, see
Eq. (\ref{AnomalousForm22}).
The computation of the RG improved Wilson coefficient for 
$[Q^{(2)}]_R$ is straightforward, and is accomplished along the lines
of Sec. \ref{RenormalizationGroupSection}. 
The final result for the renormalization group invariant 
Wilson coefficient reads
\begin{eqnarray}
{\cal F}_{RGI,2}^{(2)}(\overline{g}) = \overline{g}^{1/(N-2)}\left[
-1+\frac{N-5}{2\pi(N-2)}\overline{g}+O(\overline{g}^2)\right]\, .
\label{Wilson2}
\end{eqnarray}

Let us consider now the calculation of the Wilson coefficients of 
$[Q^{(1)}_{\mu\nu}]_R$ and $[Q^{(3)}_{\mu\nu}]_R$. It is convenient to write 
these two coefficients as a column vector:
$\widehat{\cal F}^{(2)}(\mu r;g) \equiv $ $[{\cal F}^{(2)}_1(\mu r;g),$ 
${\cal F}^{(2)}_3(\mu r;g)]^T$.
Since this vector satisfies a RG equation, see  Sec. 
\ref{RenormalizationGroupSection}, we can write it as follows:
\begin{eqnarray}
\widehat{\cal F}^{(2)}(\mu r;g) = U(g)\widehat{\cal F}_{RGI}^{(2)}(
\overline{g}(\Lambda_{\MMS} r))\, .
\label{RGI22}
\end{eqnarray}
The $2\times 2$ matrix $U(g)$ satisfies the equation:
\begin{eqnarray}
g\frac{\partial U(g)}{\partial g} = -\Gamma(g) U(g)\, ,
\label{DifferentialEquation}
\end{eqnarray}
with
\begin{eqnarray}
\Gamma(g) = \frac{g}{\beta(g)}\left[ 
\begin{array}{cc}
\gamma^{(2,2)}_{W,11}(g) & \gamma^{(2,2)}_{W,13}(g)\\
\gamma^{(2,2)}_{W,31}(g) & \gamma^{(2,2)}_{W,33}(g)
\end{array}\right] \, .
\end{eqnarray}
We are interested in calculating $\widehat{\cal F}_{RGI}^{(2)}(
\overline{g}(\Lambda_{\MMS} x))$. This can be done by solving 
Eq. (\ref{DifferentialEquation}) for $U(g)$, and by using 
Eq. (\ref{RGI22}), where $\widehat{\cal F}^{(2)}(\mu r;g)$
is substituted by its perturbative expansion, see Eqs.
(\ref{Pert22.1}) and (\ref{Pert22.3}).

Thanks to the perturbative results presented in 
Sec. \ref{SymmetricAnomalousContinuum}, we can write
the first two terms of the asymptotic expansion 
$\Gamma(g)\sim \sum_{k=0}^{\infty}\Gamma_kg^k$:
\begin{eqnarray}
\Gamma_0  = \frac{1}{N-2}\left[
\begin{array}{cc}
0 & 0 \\
-1 & N-1
\end{array}
\right]\, ,\quad
\Gamma_1  = \frac{1}{8\pi(N-2)}\left[
\begin{array}{cc}
-(N-3) & -4(N-4)\\
6N-7 & 4(3N-5)
\end{array} \right]\, .
\label{GammaTwoLoop}
\end{eqnarray}

In Section \ref{RenormalizationGroupSection} 
we wrote the solution of Eq. (\ref{DifferentialEquation})
as:
\begin{eqnarray}
U(g) \sim \left(\sum_{k=0}^{\infty}U_kg^k\right)\, g^{-\Gamma_0}\, ,
\quad U_0 =  1\, .
\label{UAsymptotic}
\end{eqnarray}
In computing $g^{-\Gamma_0}$ the eigenvalues of $\Gamma_0$ are needed.
In the case at hand a simple calculation yields:
\begin{eqnarray}
\Gamma_0^{({\rm I})} = \frac{N-1}{N-2}\, , \quad
\Gamma_0^{({\rm II})} = 0\, .
\end{eqnarray}
For $N>3$, the solution of Eq. (\ref{DifferentialEquation}) admits 
indeed the asymptotic expansion (\ref{UAsymptotic}).
The coefficients $U_k$ are obtained by plugging the expansion 
\reff{UAsymptotic} into Eq. \reff{DifferentialEquation} and matching 
the terms on the two sides order-by-order in $g$.
The relevant formulae at one-loop order have been given in Section 
\ref{RenormalizationGroupSection}.
In general we obtain the following recursive equation for the
coefficients $U_k$:
\begin{eqnarray}
k\, U_k-U_k\Gamma_0+\Gamma_0U_k = -\sum_{l=0}^{k-1}\Gamma_l U_{k-l}\, .
\end{eqnarray}
This equation can be easily solved with respect to $U_k$ if we adopt the basis
which diagonalizes $\Gamma_0$. Let us take $\Gamma_0 = V\Gamma_D
V^{-1}$, with $\Gamma_D = {\rm
diag}(\Gamma_{D,1},\Gamma_{D,2},\dots)$. In other words, $V$ is
the change of basis which diagonalizes $\Gamma_0$, and $\Gamma_{D,1}$,
$\Gamma_{D,2},\dots$ are the eigenvalues. In this basis we get
\begin{eqnarray}
(V^{-1}U_kV)_{ij} = -\frac{(V^{-1}\sum_{l=0}^{k-1}\Gamma_l U_{k-l}
V)_{ij}}{k-\Gamma_{D,i}+\Gamma_{D,j}}\, .
\label{USeriesCoefficients}
\end{eqnarray}

For $N=3$ the solution cannot be written in the form
(\ref{UAsymptotic}). The two eigenvalues 
$\Gamma_0^{({\rm I})} = 2$ and $\Gamma_0^{({\rm II})} = 0$ differ by a 
non-zero integer, and Eq. (\ref{USeriesCoefficients}) 
becomes meaningless for $k=2$ (the denominator vanishes).
The novel feature of the correct solution is that it contains terms 
of the type $g^{-2+k}\log g$, rather than simply $g^{-2+k}$
as prescribed by Eq. (\ref{UAsymptotic}).
Since we carried out our simulations for the $O(3)$ model,
hereafter we shall focus on the particular case $N=3$.

Since the vanishing denominator appears in
Eq. (\ref{USeriesCoefficients}) only for $k=2$, we expect that
one-loop calculations will not be affected by the resonance.
Nevertheless, it is instructive to proceed as in the general case.
The idea \cite{Wasow} is to transform Eq. (\ref{DifferentialEquation}) into an
equivalent one without a resonance. In the new equations
we will have two degenerate eigenvalues, rather than two eigenvalues
differing by a non-zero integer.
The transformation is accomplished through appropriate changes of
variables  (the so-called {\it shearing} transformations).

We start by writing the matrix $\Gamma(g)$ as
\begin{eqnarray}
\Gamma(g) = \left[
\begin{array}{cc}
g\phi_{11}(g)     &  g\phi_{12}(g)\\
-1+g\phi_{21}(g)  &  2+g\phi_{22}(g)
\end{array}
\right]\, ,
\end{eqnarray}
with $\phi_{ij}(g) = \phi_{ij}(0)+\phi'_{ij}(0)g+\dots$,
and apply the following transformation
\begin{eqnarray}
U(g) = \widehat{U}(g)X(g)\, , \quad
\widehat{U}(g) =\left[
\begin{array}{cc}
2 & \phi_{12}(0)/g \\
1 & 1/g^2 +\phi_{12}(0)/2 g
\end{array}
\right] \, .
\label{Trasformation}
\end{eqnarray}
The newly defined matrix $X(g)$ satisfies the equation:
\begin{eqnarray}
g\frac{\partial X(g)}{\partial g} = -\Omega(g)X(g)\, ,\quad
\Omega(g) \equiv \widehat{U}(g)^{-1}\Gamma(g)\widehat{U}(g)
+g\widehat{U}(g)^{-1}\frac{\partial\widehat{U}(g)}{\partial g}\, ,
\label{NoResonance}
\end{eqnarray}
which looks quite similar to our starting point 
Eq. (\ref{DifferentialEquation}), but now the resonance has
disappeared.
In fact
\begin{eqnarray}
\Omega(0) & = &\left[
\begin{array}{cc}
0 & \Omega_{12}(0)\\
0 & 0
\end{array}
\right]\, ,\\
\Omega_{12}(0) & = &\frac{1}{2}\left[
\phi_{11}(0)\phi_{12}(0)+\phi_{12}^2(0)-\phi_{12}(0)\phi_{22}(0)
+\phi'_{12}(0)\right]\, ,
\label{Omega12}
\end{eqnarray}
and, as promised, the two eigenvalues are now degenerate.
The ``miraculous'' matrix $\widehat{U}(g)$ given in
Eq. (\ref{Trasformation}) can be constructed through a
step-by-step procedure described in \cite{Wasow}. 
Alternatively it can be obtained by
imposing the degeneracy of eigenvalues in the new differential
equation (\ref{NoResonance}).

Notice that, since we know $\Gamma(g)$ only to $O(g)$, see Eq. 
(\ref{GammaTwoLoop}), we cannot compute $\Omega_{12}(0)$.
Using in Eq. \reff{Omega12} the known values of 
$\phi_{ij}(0)$, see Eq. (\ref{GammaTwoLoop}), 
we get $\Omega_{12}(0)=\phi'_{12}(0)-3/8\pi^2$.

The solution of Eq. \reff{NoResonance} has the form
\begin{eqnarray}
X(g) \sim \left(\sum_{k=0}^{\infty} X_k g^k\right) \, g^{-\Omega(0)}\, ,\quad
g^{-\Omega(0)} = \left[
\begin{array}{cc}
1 & -\Omega_{12}(0)\log g\\
0 & 1
\end{array}
\right]\, .
\end{eqnarray}
The coefficients $X_k$ can be computed by plugging this expression into Eq.
\reff{NoResonance}. Using the boundary condition $X_0=1$ we get
\begin{eqnarray}
X_1 = \left[
\begin{array}{cc}
-1/4\pi & * \\
0       & -7/4\pi
\end{array}
\right]\, ,\quad
X_2 = \left[
\begin{array}{cc}
* & * \\
0 & * 
\end{array}
\right]\, ,\quad
X_3 = \left[
\begin{array}{cc}
* & * \\
-3/2\pi & *
\end{array}
\right]\, ,
\end{eqnarray}
where we marked with a star ($*$) the entries which cannot be computed using
our two-loop perturbative results, see Eq. (\ref{GammaTwoLoop}).

The RG invariant Wilson coefficient $\widehat{\cal F}_{RGI}^{(2)}(g)$
is obtained by using Eq. (\ref{RGI22}) and the perturbative prediction
for $\widehat{\cal F}_{RGI}^{(2)}(g)$, see Eqs. (\ref{Pert22.1}) and
(\ref{Pert22.3}).
The final result is (for greater convenience of the reader 
we write here also the coefficient \reff{Wilson2} for $N=3$):
\begin{eqnarray}
{\cal F}_{RGI,1}^{(2)}(\overline{g}) & = & -1-\frac{1}{2\pi}\overline{g}+O(\overline{g}^2\log \overline{g})\, ,
\label{HTSpin21}\\
{\cal F}_{RGI,2}^{(2)}(\overline{g}) & = & -\overline{g}-\frac{1}{\pi}\overline{g}^2+O(\overline{g}^3)\, , 
\label{HTSpin22}\\
{\cal F}_{RGI,3}^{(2)}(\overline{g}) & = & \overline{g}^2+\frac{1}{4\pi}\overline{g}^3+O(\overline{g}^4)\, .
\label{HTSpin23}
\end{eqnarray}
Notice that, as expected, 
the non-analytic term $\overline{g}^2\log \overline{g}$ 
appears only in a two-loop 
calculation. Moreover, this term 
is present only in the coefficient
${\cal F}_{RGI,1}^{(2)}(\overline{g})$.
%
%*************************************************************************
%
\section{Numerical Results}
\label{FieldsNumericalResults}

In this Chapter we shall face a more involved task than in the 
previous one. We do not know the {\it renormalized} matrix elements
on the right-hand side of 
Eq. (\ref{GenericOPEFields}), and we would like to compute them using the 
OPE method.

We shall proceed as follows:
\renewcommand{\theenumi}{\arabic{enumi}}
\begin{enumerate}
\item We compute numerically a matrix element of the type
$\<1|\sigma^a_x\sigma^b_{-x}|2\>$ from properly
chosen lattice correlation functions.
This step yields a function $G_L(x)$.
\item \label{FieldRenormalization}
We  compute the field-renormalization constant $Z_L$.
\item \label{StepFit}
We fit the function $G_L(x)$ using the form 
$\sum_{\cal O} W^{(l_{\cal O})}_{\cal O}(x)\widehat{\cal O}$ 
and keeping the numbers 
$\widehat{\cal O}$ as parameters of the fit. 
The function $W^{(l_{\cal O})}_{\cal O}(x)$ is an 
$l_{\cal O}$-loop truncation of the Wilson coefficient
$W_{\cal O}(x)$ .
We restrict the fit to the region $\rho\le r\le R$, with $r = |x|$. 
The outcomes
of this step are the best fitting parameters $\widehat{\cal O}(\rho,R)$ 
together with an estimate of the statistical and systematic uncertainties.
\item We look for some range of $\rho$ and $R$ (in the regime 
$\rho, R\ltapprox \xi$) such that  $\widehat{\cal O}(\rho,R)$
remains constant within the above-mentioned uncertainties:
$\widehat{\cal O}(\rho,R)\approx \widehat{\cal O}^*$.
\item \label{FinalStep}
The $\<1|\,\cdot\,|2\>$ matrix element of the renormalized
operator ${\cal O}$ is obtained by taking into account the
renormalization of the bare lattice fields $\sigma_x$ and $\sigma_{-x}$:
$\<1|{\cal O}|2\>_{OPE} = Z_L^{-1}\widehat{\cal O}^*$.
\end{enumerate}
\begin{figure}
\centerline{
\hspace{0.0cm}\vspace{-1.5cm}\\
\hspace{-2.5cm}
\epsfig{figure=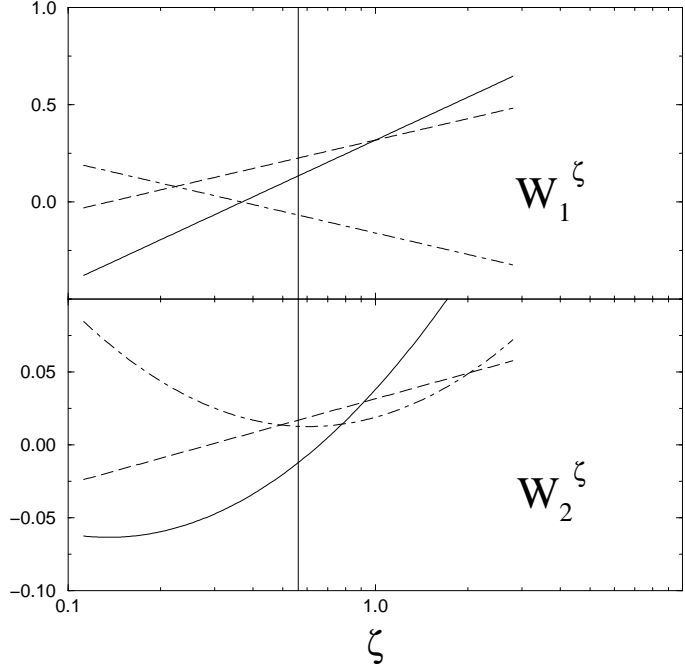,angle=-90,
width=0.6\linewidth}
}
\vspace{1.5cm}
\caption{The first two perturbative coefficients $W_1^{\zeta}$ and 
$W_2^{\zeta}$, see Eq. (\ref{GeneralWilson}), for various
Wilson coefficients. Here $N=3$. Continuous (dotted, dash-dotted) lines 
refer to the leading term of the OPE in the scalar (antisymmetric, symmetric)
sector. Vertical lines correspond to $\zeta e^{\gamma} =1$.}
\label{CoefficientFig}
\end{figure}

The above procedure is quite general. Let us now specify some of the
details.
In step \ref{StepFit} we use a minimum-squares fit. 
As it stands, step \ref{FieldRenormalization} is rather undefined
and could be accomplished using different methods. 
The calculation of $Z_L$ will be the object of Sec. \ref{FieldSection}.
We shall use once more the OPE. This will provide us with a further
check of our approach.

There is some ambiguity in choosing the perturbative truncation of the Wilson 
coefficients in step \ref{StepFit}.
We shall proceed as follows. We can define a one-parameter
family of running couplings $\gxz$, through the following 
equation\footnote{The factor $e^{\gamma}\approx 1.78107$ 
is introduced for future 
convenience. We shall in fact write down the RGI Wilson coefficients
for $\zeta = 1$. The factor $e^{\gamma}$ avoids the proliferation of
$(\log \gamma)$'s in these expressions.}
\begin{eqnarray}
\Lambda_{\MMS}x e^{\gamma}\zeta = \lambda_{\MMS}(\overline{g}_{\zeta}(x))\, ,
\label{Gzeta}
\end{eqnarray} 
where $\lambda_{\MMS}(g)$ is defined as in Eq. (\ref{lambdadig}).
$\zeta$ is a  positive real number which parametrizes the family of
couplings.
The solution of Eq. (\ref{Gzeta}) can be written as a series in inverse 
powers of $z = -\log(\Lambda_{\MMS}x e^{\gamma}\zeta)$. The 
knowledge of the beta function at four-loop order 
\cite{Hikami:1981hi,Bernreuther:1986js,Wegner:1989ss} allows us to
write this series up to the order $z^{-4}$
(cf. Eq. (\ref{GxExpansion}) for the first three terms of this expansion). 
This truncated 
series defines the coupling $\overline{g}^{(4)}_{\zeta}(x)$.
In the step \ref{StepFit} of our procedure
we shall use the RGI coefficients
(see Sec. \ref{RenormalizationGroupSection}) expanded in terms of
$\gxz$. They have, in general, the form
\begin{eqnarray}
W^{(l),\zeta}_{RGI}(\gxz) = 
\gxz^{\Gamma}\sum_{k=0}^{l}W_k^{\zeta} \, \gxz ^k \, .
\label{GeneralWilson}
\end{eqnarray}
Moreover, we shall substitute the coupling $\gxz$ with its 
four-loop approximation $\overline{g}^{(4)}_{\zeta}(x)$ defined above.
This completely specifies our truncation of the Wilson coefficients
for a given $\zeta$.

The perturbative coefficients $W_k^{\zeta}$ depend upon $\zeta$.
They can be expressed, for a generic $\zeta$, in terms
their values at $\zeta = 1$. The connection is obtained, up to
three-loop order, using the following formulae
\begin{eqnarray}
W^{\zeta}_0 & = & W^1_0\, ,\\
W^{\zeta}_1 & = & W^1_1+\Gamma c_1(\zeta) W^1_0\, ,\\
W^{\zeta}_2 & = & W^1_2+ (\Gamma+1) c_1(\zeta) W^1_1+
\left[\Gamma c_2(\zeta)+\frac{1}{2}\Gamma(\Gamma-1)c_1^2(\zeta)\right]W^1_0
\, ,\\
W^{\zeta}_2 & = & W^1_3+(\Gamma+2)c_1(\zeta)W^1_2+
\left[(\Gamma+1)c_2(\zeta)+\frac{1}{2}\Gamma(\Gamma+1)c_1^2(\zeta)\right]W^1_1+
\nonumber\\
&&+\left[\Gamma c_3(\zeta)+\Gamma(\Gamma-1)c_1(\zeta)c_2(\zeta)
+\frac{1}{6}\Gamma(\Gamma-1)(\Gamma-2)c_1^3(\zeta)\right] W^1_0\, ,
\end{eqnarray}
where
\begin{eqnarray}
c_1(\zeta) & = & -\beta_0\log \zeta\, , \\
c_2(\zeta) & = & \beta_0^2\log^2\zeta - \beta_1\log \zeta\, ,\\
c_3(\zeta) & = & -\beta_0^3\log^3\zeta+\frac{5}{2}\beta_0\beta_1\log^2\zeta-
\beta^{\MMS}_2\log\zeta\, .
\end{eqnarray}
The $\beta_i$ are the coefficients of the perturbative expansion of
the beta-function, see Eq. (\ref{BetaExpansion}). 

Notice that we use RGI Wilson 
coefficients. Therefore the outcomes of the fit at step \ref{StepFit} 
will be the matrix element of RGI operators.

\begin{table}
\begin{center}
\begin{tabular}{|c|c|c|c|}
\hline
 & $W_1^{\zeta}$ & $W_2^{\zeta}$ & $W_3^{\zeta}$\\
\hline
scalar        & $0.1345764$ & $-0.01204924$ & $-0.007041306$ \\
antisymmetric & $0.2264432$ & $0.01704183$ & $\ast$ \\
symmetric     & $-0.06728822$ & $0.01281617$ & $\ast$ \\
 \hline
\end{tabular}
\end{center}
\caption{Perturbative coefficients, see Eq. (\ref{GeneralWilson}),
of the leading contribution to the OPE in the scalar, antisymmetric 
and symmetric sectors. Here $N=3$ and $\zeta e^{\gamma}=1$.}
\label{CoefficientsTable}
\end{table}
We defined a whole family of running couplings parametrized by $\zeta$. 
Which value of $\zeta$ do we choose? If we knew exactly the Wilson
coefficient $W^{\zeta}_{RGI}(\overline{g})$ and the beta function,
then this choice would not matter.
By definition $W^{\zeta}_{RGI}(\gxz)$ is independent of $\zeta$. However
we must truncate the beta function to four-loop order and,
in most cases, 
$W^{\zeta}_{RGI}(\overline{g})$ to two-loop order
(for leading terms of the OPE), 
or one-loop order (for next-to-leading terms). This introduces a 
dependence upon $\zeta$ (mainly because of the truncation of
$W^{\zeta}_{RGI}(\overline{g})$).
Clearly, we must take $\zeta$ of order one for avoiding ``large logarithms''.
In order to have a more precise idea, we plot in Fig.  \ref{CoefficientFig}
$W^{\zeta}_1$ and $W^{\zeta}_2$ for the Wilson coefficients $F^{(0)}_0$,
$F^{(1)}_0$, and $F^{(2)}_0$.
The perturbative expansions for the Wilson coefficients are given
in Eqs. (\ref{F00Pert}), (\ref{F10Pert}) and (\ref{F20Pert}).
Their RGI counterparts are given by the
general formulae of Sec. \ref{RenormalizationGroupSection}. 
In all the cases considered in Fig.  \ref{CoefficientFig},
the coefficients $W^{\zeta}_1$ and $W^{\zeta}_2$ attain their minimum
absolute value around
$\zeta e^{\gamma}\approx 1$. This is a rather natural choice.
We shall compute our perturbative Wilson coefficients (and, consequently,
our estimates $\widehat{\cal O}(\rho,R)$ for the matrix elements)
using the running coupling defined by
$\zeta = e^{-\gamma} \approx 0.561459$. In
Tab. \ref{CoefficientsTable} we give the numerical values 
(for $\zeta = e^{-\gamma}$ and $N=3$) of
the first few perturbative coefficients $W^{\zeta}_k$ corresponding
to $F^{(0)}_0$, $F^{(1)}_0$, and $F^{(2)}_0$.
This gives a feeling of the range of validity of perturbation theory.

The above remarks suggest the following approach to the estimation 
of the systematic uncertainty on $\widehat{\cal O}(\rho,R)$. 
We shall repeat the calculation of $\widehat{\cal O}(\rho,R)$ 
for $\zeta$ varying in the range $1/\kappa \le \zeta e^{\gamma}\le
\kappa$. This yields a $\zeta$-dependent result
 $\widehat{\cal O}_{\zeta}(\rho,R)$. We estimate the systematic error
with the maximum deviation of $\widehat{\cal O}_{\zeta}(\rho,R)$
from the value taken at $\zeta e^{\gamma}=1$.

Again, the choice of $\kappa$ is rather 
arbitrary. In the following we take $\kappa = 2$. The consistency of
this choice can be checked by monitoring our estimates when 
the number of loops in the computation of the 
Wilson coefficients is varied, see in particular Sec. \ref{FieldSection}. 
A further check is obtained by studying 
cases  where an alternative evaluation of the matrix element 
$\<1|{\cal O}|2\>$ is available.

We conclude these introductory remarks by explaining some notations
which will be used in this Section.
We shall attribute to any best fitting parameter $\widehat{\cal O}(\rho,R)$
two types of errors: a systematic error (as defined above) 
and a statistical one (one standard deviation). 
In order to present both of them we shall
use the following convention. When writing a result we shall indicate
in parentheses the statistical error, and in brackets the
systematic one. For instance $1.00(2)[11]$ means $1.00$ with
a statistical error of $\pm 0.02$, and a systematic error of $\pm
0.11$.
In the graphs we shall often indicate by a vertical bar the systematic
error, and by horizontal ticks on the bar the statistical one. 
We finally notice that statistical errors will be typically smaller
than systematic ones.
%
%*******************************************************************
%
\subsection{The Observables}
\label{ObservablesSection}

Most of the simulations where done on the same lattices 
\ref{Lattice64x128}, \ref{Lattice128x256} and \ref{Lattice256x512}
employed in the previous Chapter. 
We shall be mainly concerned with the short-distance limit of
(normalized) four-point functions.
In order to study their scaling behavior, we added two more lattices
to the ones already considered (also in these cases we use periodic
boundary conditions):
\begin{enumerate}
\item[(A').] Lattice of size $128\times 64$ with $g_L^{-1} = 1.39838694$. 
 \label{Lattice64x128bis}
\item[(C').] \label{Lattice256x512bis} 
Lattice of size $512\times 256$ with  $g_L^{-1} = 1.66135987$.
\end{enumerate}
\def\64x128bis{(A')}
\def\256x512bis{(C')}
We used the same cluster algorithm as in the previous Chapter. In
order to obtain well decorrelated spin configurations, we evaluated
the observables every 15 or (for the majority of the observables
considered) 30 updatings.

In all our simulations we evaluated the standard energy per link
\begin{eqnarray}
E = \frac{1}{2LT}\sum_{x,\mu}\<\sg_x\cdot\sg_{x+\mu}\>
\end{eqnarray}
for each generated configuration. This quantity can be used for reweighting the
Monte Carlo results at a different bare coupling. Moreover in 
applying improved (``boosted'') perturbation theory, we shall need the 
value of the improved coupling $g_E\equiv 4(1-E)/(N-1)$.
We obtain the results $g_E = 0.875644(17)$, $0.768119(18)$, $0.693083(17)$,
respectively on lattices \ref{Lattice64x128},\ref{Lattice128x256},
\ref{Lattice256x512}, averaging over $N_{\rm conf} = 4\cdot 10^5$, 
$10^5$, $1.6\cdot 10^4$ independent configurations.
On lattices \64x128bis and \256x512bis we obtained, respectively, 
$g_E = 0.877004(16)$ and $g_E = 0.692316(21)$ using 
$N_{\rm conf} = 4.5\cdot 10^5$ and $N_{\rm conf} = 10^4$ configurations.
The precision of these computations can be easily increased but this
is useless for our purposes.

On lattices \64x128bis and \256x512bis
we computed the ``wall-to-wall'' correlation function
$C(\pb;t)$, see Eq. (\ref{TwoPointCorrelationFunction})
for momenta $\pb=2\pi n/L$, $n=0,\dots,3$, and time separations
$0\le t\le 40$ on lattice \64x128bis and
$0\le t\le 100$ on lattice \256x512bis. 
We generated $N_{\rm conf} \simeq 6\cdot 10^6 $ independent configurations 
on lattice  \64x128bis, 
and $N_{\rm conf} = 2.6\cdot 10^5$ independent 
configurations on lattice \256x512bis.

Among the other things, we shall employ the OPE method
for evaluating the renormalized matrix element of
the symmetric traceless operator $\opl S^{ab}_0\opr$, see Eq.
(\ref{SymmetricTraceless}).
In order to verify the result,
we repeated this calculation using a different method.
We computed the lattice matrix element and renormalized 
it in a successive step.
In particular we evaluated the three point function (obviously we
averaged over translations)
\begin{eqnarray}
C^{(2)}(\pb,\qb;2t)\equiv
\sum_{x_1,x_2=1}^Le^{i\pb x_1-i\qb x_2}\sum_{a,b}
\<\sigma^a_{-t,x_1}\left(
\sigma^a_{0}\sigma^b_{0}-\frac{\delta^{ab}}{N}
\right)\sigma^b_{t,x_2}\>\, ,
\label{SymmetricOperatorThreePoint}
\end{eqnarray}
on lattices \ref{Lattice64x128}, \ref{Lattice128x256} and 
\ref{Lattice256x512} for $\pb=\qb=2\pi n/L$, $n= 0,\dots,3$.
We used $N_{\rm conf} = 1.22\cdot 10^6$ configurations and $t=1,\dots,15$ 
on lattice \ref{Lattice64x128}, $N_{\rm conf} = 3.2\cdot 10^5$ 
configurations and $t=5,\dots,20$ on lattice \ref{Lattice128x256},
$N_{\rm conf} = 10^5$ 
configurations and $t=5,\dots,40$ on lattice \ref{Lattice256x512}.
The corresponding normalized function $\widehat{C}^{(2)}(\pb,\qb;2t)$
is defined analogously to Eq. (\ref{NormalizedCorrelation})

One possible approach to the computation of the renormalization
constant $Z^{\cal O}_L$ for a lattice operator ${\cal O}$,
consists in applying the OPE method to the corresponding two point
function. We applied this strategy to the elementary field
(${\cal O} = \sigma^a$), and to the symmetric traceless operator 
of dimension zero (${\cal O} = \sigma^a\sigma^b-\delta^{ab}/N$).
We evaluated numerically the corresponding two-point functions:
\begin{eqnarray}
G_V(t,x)\equiv \<\sg_{0,0}\cdot\sg_{t,x}\>\, ,\quad \quad
G_T(t,x)\equiv  \<(\sg_{0,0}\cdot\sg_{t,x})^2\>-1/N
\, .
\label{TwoPointFunctions}
\end{eqnarray}
We evaluated $G_V(t,x)$ ($G_T(t,x)$) from 
$N_{\rm conf} = 3.05\cdot 10^5$ ($N_{\rm conf} = 1.41\cdot 10^6$ resp.)
independent configurations on lattice \ref{Lattice64x128},
$N_{\rm conf} = 0.98\cdot 10^5$ ($N_{\rm conf} = 3.59\cdot 10^5$ resp.)
configurations on lattice \ref{Lattice128x256},
and $N_{\rm conf} = 1.6\cdot 10^4$ ($N_{\rm conf} = 1.9\cdot 10^4$ resp.)
configurations on lattice \ref{Lattice256x512}.
For $G_V(t,x)$ ($G_T(t,x)$) we considered $0\le t,x\le 9$, $17$, $34$ 
($0\le t,x\le 7$, $14$, $28$ resp.) on lattices
\ref{Lattice64x128}, \ref{Lattice128x256} and \ref{Lattice256x512}.

In order to study the OPE of the product of two fields, we considered
the following four-point correlation functions:
\begin{eqnarray}
G^{(0)}(t,x;\pb,\qb;2t_s) & = & \sum_{x_1,x_2}
\<(\sg_{t,x}\cdot\sg_{-t,-x})( \sg_{-t_s,x_1}\cdot\sg_{t_s,x_2})\>
\, e^{i\pb x_1-i\qb x_2}\, , \label{GScalar}\\ 
G^{(1)}(t,x;\pb,\qb;2t_s) & = & \sum_{x_1,x_2}\sum_{ab}
\<\sigma^{[a}_{t,x}\sigma^{b]}_{-t,-x}\,
\sigma^a_{-t_s,x_1}\sigma^b_{t_s,x_2}\>
\, e^{i\pb x_1-i\qb x_2}\, ,\\
G^{(2)}(t,x;\pb,\qb;2t_s) & = & \frac{1}{2}\sum_{x_1,x_2}\sum_{ab}
\< (\sigma^{\{a}_{t,x}\sigma^{b\}}_{-t,-x}-
\frac{2\delta^{ab}}{N}\sg_{t,x}\cdot\sg_{-t,-x} )\,
\sigma^a_{-t_s,x_1}\sigma^b_{t_s,x_2}\>
\, e^{i\pb x_1-i\qb x_2}\, .\nonumber\\
\end{eqnarray}
We computed the above functions in Monte Carlo simulations as follows:
\begin{itemize}
\item ${\rm Re}\,G^{(0)}(t,x;\pb,\pb;2t_s)$ using  
$N_{\rm conf} = 2.57\cdot 10^5$ configurations on lattice \ref{Lattice64x128},
$N_{\rm conf} = 4.64\cdot 10^4$ on lattice \ref{Lattice128x256}
and $N_{\rm conf} = 9.6\cdot 10^3$ on lattice \ref{Lattice256x512}.
\item ${\rm Re}\,G^{(1)}(t,x;\pb,\pb;2t_s)$ using 
$N_{\rm conf} = 4.35\cdot 10^4$ configurations on lattice \ref{Lattice64x128},
$N_{\rm conf} = 7.8\cdot 10^4$ configurations on lattice \ref{Lattice128x256},
and $N_{\rm conf} = 7.7\cdot 10^3$ configurations on lattice 
\ref{Lattice256x512}
and .
\item ${\rm Re}\,G^{(2)}(t,x;\pb,\pb;2t_s)$ using 
$N_{\rm conf} = 2.3\cdot 10^5$  independent configurations on lattice 
\ref{Lattice64x128}, $N_{\rm conf} = 7.1\cdot 10^4$  configurations on 
lattice \ref{Lattice128x256}, $N_{\rm conf} = 5\cdot 10^3$  configurations on 
lattice \ref{Lattice256x512}. In order to verify the relevance of 
scaling corrections we computed the same function also on lattices
\64x128bis ($N_{\rm conf}= 3.6\cdot 10^5$) and 
\256x512bis ($N_{\rm conf} = 8.6\cdot 10^3$).
\end{itemize}
In all the cases we consider $\pb = 2\pi n/L$ with $n=0,1,2$. Moreover 
$t_s = 8,9,10$ and $|t|\le 5,|x|\le 5$ on lattice \ref{Lattice64x128}, 
$t_s = 16,17,18$ and $|t|\le 8,|x|\le 8$ on lattice \ref{Lattice128x256}
$t_s = 30,34$ and $|t|\le 16,|x|\le 16$ on lattice \ref{Lattice256x512}.
Analogously to what is done in the previous Chapter, see 
Eq. (\ref{NormalizedFourPoint}), we
define the normalized functions $\widehat{G}^{(l)}(t,x;\pb,\qb;2t_s)$
with $l=0,1,2$, which have a finite limit for $t_s\to\infty$.
%
%*****************************************************************
%
\subsection{One-Particle States}
\label{OneParticleSectionBIS}
\begin{table}
\begin{center}
\begin{tabular}{|c|c|c|c|c|}
\hline
 &\multicolumn{2}{|c|}{lattice \64x128bis}&
\multicolumn{2}{|c|}{lattice \256x512bis}\\
\hline
$\pb$ & $\omega(\pb)$ & $Z(\pb)$ & $\omega(\pb)$ & $Z(\pb)$\\
\hline
$0$      & $0.146401(39)$ & $1.66247(82)$
         & $0.036597(32)$ & $1.1242(12)$\\
$2\pi/L$ & $0.176101(38)$ & $1.65954(73)$
         & $0.044128(31)$ & $1.1271(12)$\\
$4\pi/L$ & $0.244063(75)$ & $1.6504(13)$
         & $0.061256(56)$ & $1.1260(23)$\\
$6\pi/L$ & $0.32684(20)$ & $1.6430(32)$
         & $0.08213(11)$ & $1.1227(30)$ \\
 \hline
\end{tabular}
\end{center}
\caption{The one-particle spectrum and the field normalization
for lattices \64x128bis and \256x512bis.}
\label{LatticeSpectrumBis}
\end{table}
\begin{figure}
\begin{tabular}{cc}
\hspace{0.0cm}\vspace{-1.5cm}\\
\hspace{-2.5cm}
\epsfig{figure=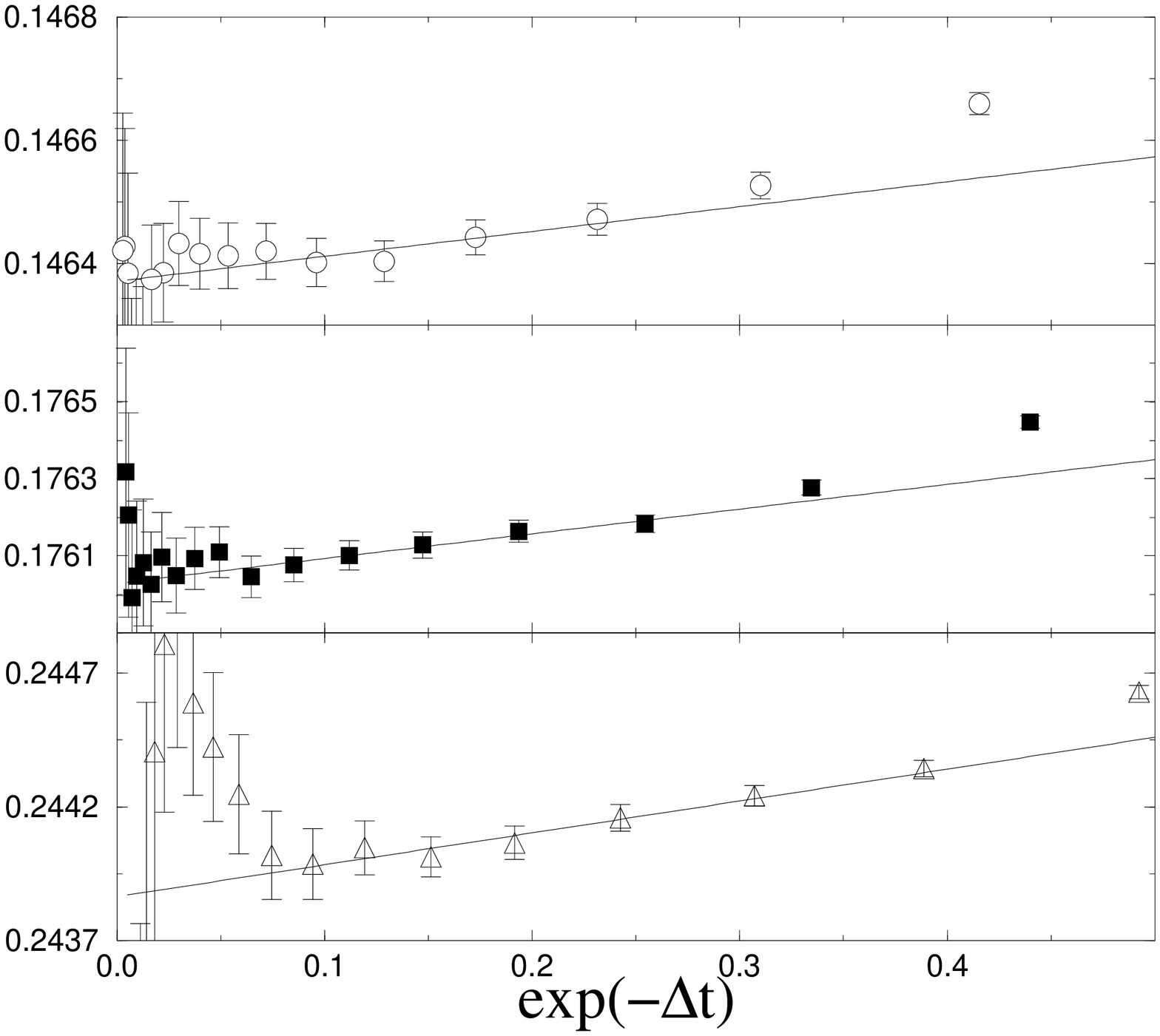,angle=-90,
width=0.6\linewidth}&
\hspace{-0.5cm}
\epsfig{figure=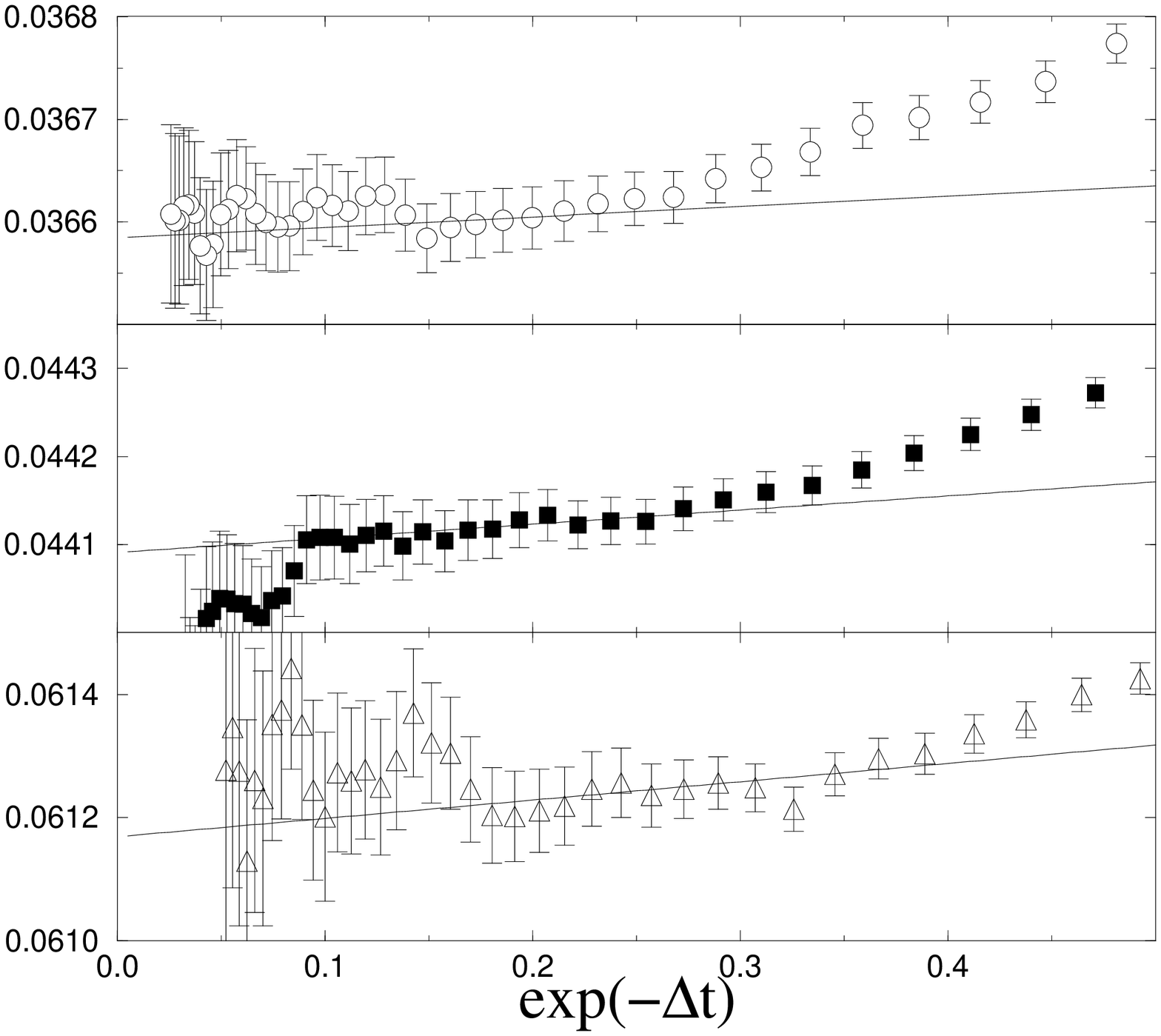,angle=-90,
width=0.6\linewidth}
\end{tabular}
\caption{The asymptotic behavior of $\omega_{\rm eff}(\pb,t)$ as 
$t\to\infty$ on lattices \64x128bis (left)
and \256x512bis (right).
Empty circles refer to $\pb = 0$, filled squares to $\pb = 2\pi/L$ 
and triangles to $\pb = 6\pi/L$. The continuous lines are the best fitting
curves of the form (\ref{SpectrumFit}).}
\label{MassAsymptotiaBIS}
\end{figure}
The one-particle spectrum $\omega(\pb)$ and the field normalization 
$Z(\pb)$ have been extracted from $C(\pb;t)$
on lattices \64x128bis and \256x512bis as explained in 
Sec. \ref{sec4.2}. For the relevant definitions see Eq.  
(\ref{AsymptoticCorrelation}).
The results of this computation are shown in Tab. \ref{LatticeSpectrumBis}.
We verified that the effective quantities $\omega_{\rm eff}(\pb,t)$
and $Z_{\rm eff}(\pb,t)$ do not depend upon $t$ for
$t\gtapprox\xi_{\rm exp}$.
The values reported in Tab. \ref{LatticeSpectrumBis}
correspond to $t= 8,\, 24$ respectively on lattices
\64x128bis and \256x512bis. 

In order to estimate the systematic error due to the fact that we use
a {\it finite} value of $t$ we fitted 
$\omega_{\rm eff}(\pb,t)$ and $Z_{\rm eff}(\pb,t)$, taking into 
account the first correction to the $t\to\infty$ behavior.
The procedure has been explained in the previous Chapter,
see Sec. \ref{sec4.2}.
The fitting form was of the type (\ref{SpectrumFit}).
The results of this fit are shown in Fig. \ref{MassAsymptotiaBIS}.
The corresponding estimates for the systematic errors on the 
values of $\omega(\pb=0)$  quoted in Tab. \ref{LatticeSpectrumBis} are 
about $0.3\div 0.4\cdot 10^{-4}$, on both lattices 
\64x128bis and \256x512bis. In general we verified the systematic
error to be of  the same order as (or smaller than) the statistical one.  

The exponential correlation length $\xi_{\rm exp} = m^{-1}$
obtained from the data of Tab. \ref{LatticeSpectrumBis} is
\begin{eqnarray}
\xi_{\rm exp} = 6.831(2),\; 27.325(24)\, ,
\end{eqnarray}
respectively for lattices \64x128bis and \256x512bis.
Notice that $\xi_{\rm exp}({\rm C'}) \approx 2 \xi_{\rm exp}({\rm B})$
and $\xi_{\rm exp}({\rm B}) \approx 2 \xi_{\rm exp}({\rm A'})$ with
deviations of (relative) order $10^{-3}$.
Comparing the numerical results obtained on these three lattices, we can
carefully verify the scaling of correlation functions.
%
%******************************************************************
%
\subsection{Corrections to Scaling}
\label{CorrectionsSection}

Let us consider the normalized functions 
$\widehat{G}^{(l)}(t,x;\pb,\qb;\infty)$, where the on-shell limit
$t_s\to\infty$ has been taken. We expect the 
quantity 
\begin{eqnarray}
\sqrt{4\omega(\pb) \omega(\qb)}
Z_L^{-1}\widehat{G}^{(l)}(t,x;\pb,\qb;\infty)
\end{eqnarray}
to have a finite continuum limit, i.e. $g_L\to 0$ with
$mx$, $mt$, $\pb/m$ and $\qb/m$ fixed.
This limit is approached with $O(a^2\log^p a)$ corrections 
\cite{Symanzik:1979ph,Symanzik:1983dc,Symanzik:1983gh}.
Renormalization group implies the following scaling form in the
continuum limit:
\begin{eqnarray}
\sqrt{4\omega(\pb) \omega(\qb)}\widehat{G}^{(l)}(t,x;\pb,\qb;\infty)
= V(g_L)\,{\cal G}^{(l)}(mt,mx;\pb/m,\qb/m) + O(m^2,t^{-2},x^{-2},\pb^2,\qb^2)
\, ,\nonumber\\
\label{ScalingForm}
\end{eqnarray}
where
\begin{eqnarray}
V(g_L) = g_L^{-\gamma^L_0/\beta_0}
\exp \left\{ -\int_0^{g_L}\left[
\frac{\gamma^L(z)}{\beta^L(z)}+ \frac{\gamma^L_0}{\beta_0 z}
\right]\right\}\, .
\end{eqnarray}

Let us now fix  $g_L$ and $g_L'$ in such a way that $m(g_L)/m(g_L') =
\theta$  is kept fixed. From
Eq. (\ref{ScalingForm}), and using the fact that 
$\omega(\pb)\to\sqrt{m^2+\pb^2}$ in the continuum limit, we get:
\begin{eqnarray}
\frac{\widehat{G}^{(l)}(\theta t,\theta x;\pb,\qb;\infty|g_L')}
{\widehat{G}^{(l)}( t, x;\theta \pb,\theta \qb;\infty|g_L)} & = & 
\theta\, U(g_L,g'_L)+\mbox{corrections}\; ,\label{Ratio}\\
U(g_L,g'_L) & \equiv &
\exp \left\{ -\int_{g_L}^{g_L'}\! dz\left[\frac{\gamma^L(z)}{\beta^L(z)}\right]
\right\}\, .
\end{eqnarray}
The r.h.s. of Eq. (\ref{Ratio}) is independent of $t$, $x$, $\pb$, $\qb$,
and even of the particular ($l=0,1,2$) correlation function, up to 
$O(a^2\log^p a)$ lattice  artifacts.
Since we adopted lattice regularization, Eq. (\ref{Ratio}) is meaningful
for some $t,x\in \mathbb{Z}$ only if $\theta$ is a rational 
number\footnote{For a generic value of $\theta$ we can 
give a meaning to Eq. (\ref{Ratio}) as follows. 
Let us, for sake of simplicity, drop all the indices and arguments but
the space-time ones, and consider the lattice function
$G_x$, with $x\in \mathbb{Z}^2$. Let us consider a ``smooth''
test function $\varphi(x)$ on $\mathbb{R}^2$, and form the 
``scalar product'' $\<G,\varphi\>_a\equiv\sum_x G_x\varphi(ax)$,
$a$ being the lattice spacing. The ratio on the l.h.s. of Eq. (\ref{Ratio})
can be substituted, for a generic $\theta = a/a'$, by
$\<G(g_L'),\varphi\>_{a'}/\<G(g_L),\varphi\>_{a}$.
However, in the following we shall not pursue this strategy.}.  

%
%*********************************************
%
\begin{figure}
\begin{tabular}{cc}
\hspace{0.0cm}\vspace{-1.5cm}\\
\hspace{-2.5cm}
\epsfig{figure=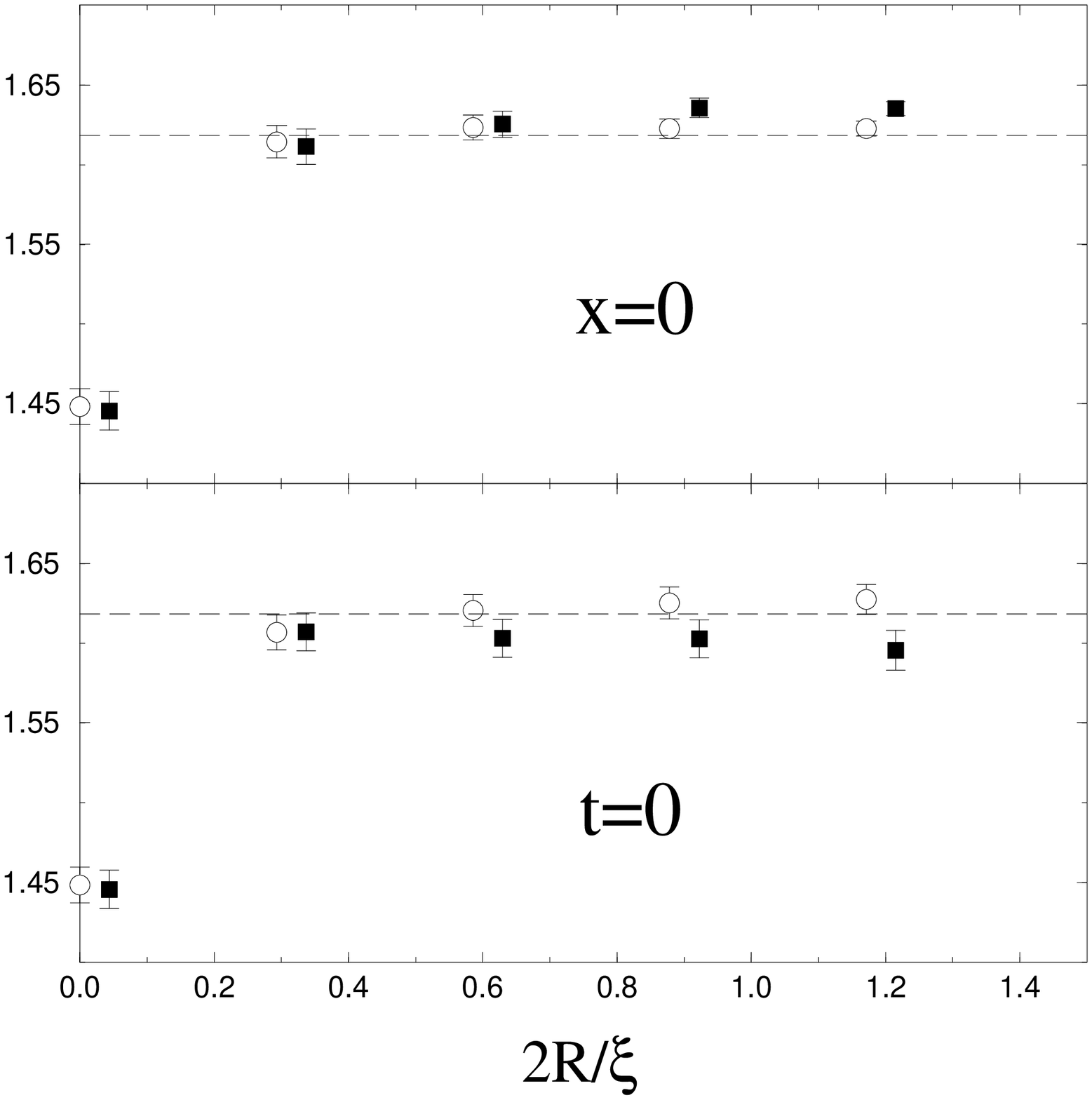,angle=-90,
width=0.6\linewidth}&
\hspace{-0.5cm}
\epsfig{figure=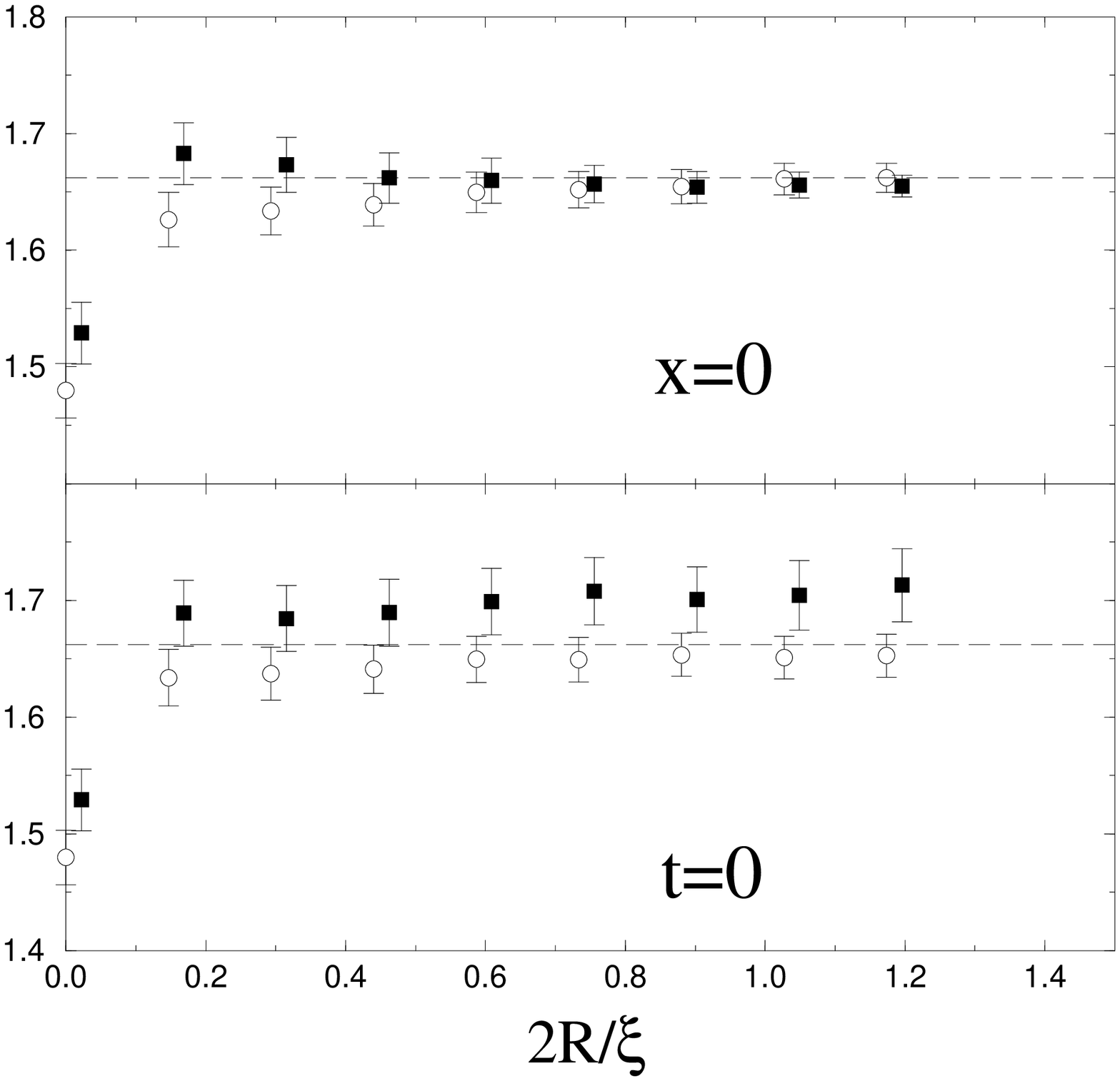,angle=-90,
width=0.625\linewidth}
\end{tabular}
\caption{The scaling ratio (\ref{Ratio}) for 
$l=2$ and $\theta=2$, along the directions $x=0$ and $t=0$.
We use lattices \64x128bis and \ref{Lattice128x256} on the left,
and lattices \ref{Lattice128x256} and \256x512bis on the right.
Empty circles ($\circ$) refer to $\pb = 0$, filled squares ($\blacksquare$)
to $\pb = 2\pi/L$.}
\label{Scaling.3112}
\end{figure}
%
%***************************************************
%
Let us now turn to the numerical results, which are shown in Fig. 
\ref{Scaling.3112}. 
We considered the ratio on the l.h.s. of Eq. (\ref{Ratio})
for $l=2$ and $\theta= 2$.
In the left column we use lattices \64x128bis 
and \ref{Lattice128x256} (i.e. $1/g_L=1.39838694$ and $1/g_L' = 1.54$);  
in the right column lattices \ref{Lattice128x256} and \256x512bis
(i.e. $1/g_L = 1.54$ and $1/g_L' = 1.66135987$).
We used $\pb=\qb=2\pi n/L$, $n= 0,1,2$.
In Eq. (\ref{Ratio}) we have to extrapolate 
the normalized correlation function $\widehat{G}^{(l)}(\dots;2t_s)$
for $t_s\to\infty$.
We verified that the ratio on the left-hand side of Eq. (\ref{Ratio}) 
does not depend upon the chosen
value of $t_s$ among the ones for which $\widehat{G}^{(2)}$ was computed,
see Sec. \ref{ObservablesSection}. The data presented refer to 
$t_s= 10,16,30$ respectively for lattices 
\64x128bis, \ref{Lattice128x256}, \256x512bis.
In Fig. \ref{Scaling.3112} we plot our numerical results
for the  ratio (\ref{Ratio}) along the directions $x=0$ and $t=0$. 

The plots obtained with lattices \ref{Lattice128x256} and 
\256x512bis (right column) show clear evidence of scaling at 
error bars level (about $1\div 2\%$ depending upon the chosen
$t,x$ and $\pb$) as soon as $(t,x)\ne (0,0)$. 
From Eq. (\ref{Ratio}), neglecting  corrections to scaling 
we obtain the estimate $U(g_L,g_L') = 0.83(2)$. Four-loop 
(three-loop, two-loop) lattice perturbation theory yields 
$U(g_L,g_L') = 0.84251$ ($0.84807$, $0.85310$).
Improved perturbation theory yields 
$U(g_L,g_L') = 0.81666(6)$ ($0.81751(6)$, $0.81614(6)$).

Lattices \64x128bis \ref{Lattice128x256} (Fig. \ref{Scaling.3112}
left column) show approximate scaling for $(t,x)\ne (0,0)$. 
The horizontal lines would imply $U(g_L,g_L') \sim 0.810(5)$ 
(we report the statistical error which is roughly independent of
the particular $(t,s)$ point, rather than the systematic error due to
scaling corrections).
Four-loop (three-loop, two-loop) lattice perturbation theory yields 
$U(g_L,g_L') = 0.79837$ ($ 0.80716$, $0.81570$).
With improved perturbation theory we get
$U(g_L,g_L') = 0.77277(4)$ ($0.77420(4)$, $0.77217(4)$).
Small scaling corrections could be suggested by the points
around $t=4$ or $x=4$. This discrepancies are not 
completely significant from a statistical point of view
(about $1\%$, while statistical errors are approximatively $0.5\%$).

Moreover, we remark that lattice perturbation theory gives unexpectedly
good estimates for the constant $U(g_L,g_L')$ on lattices 
\ref{Lattice128x256}-\256x512bis. This is probably due to the fact
that $U(g_L,g_L')$ is finite (indeed $U(g_L,g_L')\to 0$)
in the continuum limit ($g_L,g_L'\to 0$ at $m(g_L)/m(g_L')=\theta$ fixed).

Notice that both using lattices  \64x128bis and \ref{Lattice128x256},
and using lattices  \ref{Lattice128x256} and \256x512bis, the
relation $m(g_L)/m(g_L') = 2$ is not exactly satisfied,
the discrepancy being of order $10^{-3}$.
In order to check whether our conclusion could be changed by a better 
tuning of the bare lattice couplings, we reweighted our numerical data 
on lattice \64x128bis at $g_L^{-1}=1.39791766$.
At this coupling  we get $\xi_{\rm exp} = 6.816(2)$. The 
results for the scaling ratio (\ref{Ratio}) cannot be distinguished from 
the ones shown in Fig. \ref{Scaling.3112}.

In the following Subsections we shall use the OPE method for computing
renormalization constants and renormalized matrix elements from
lattice data. The results of the present Subsection give us a rough idea
of the relevance of lattice artifacts in these computations. 
As soon as we avoid products of lattice fields at coincident points,
we expect these errors to be about $1\%$ on lattice \ref{Lattice64x128},
and to be compatible with statistical uncertainties on lattices
\ref{Lattice128x256} and \ref{Lattice256x512}.
We shall see that these effects are negligible with respect to 
other sources of error (in particular the systematic error due to
the perturbative truncation of the Wilson coefficients).
As explained at the beginning of this Section, we shall put a short
distance cutoff $\rho$ on our fitting region (i.e. we take
$\rho<r<R$). In order to avoid fields products at coincident points, 
we shall always consider $\rho>0$. 
Varying $\rho$ (in particular we considered $\rho=0.5$ and $\rho=1.5$,
keeping $R>\rho+2$ fixed) does not change much the results of the fits.
These small changes (always much smaller than {\it estimated} 
systematic errors)
can be ascribed to asymptotic scaling corrections, rather
than to scaling corrections. In particular, increasing $\rho$
produce an effect of the same sign as increasing $R$.

Motivated by this discussion, we shall present, in the following
Subsections, the results obtained fixing $\rho = 0.5$.
%
%******************************************************************
%
\subsection{Field-Renormalization Constant}
\label{FieldSection}

We can apply the OPE approach to the computation of the 
field-renormalization constant. 
The method can be easily extended to any other composite operator
${\cal A}$. In Sec. \ref{RenormalizationSymmetricSection}
we shall apply it to the symmetric traceless operator 
$S_0^{ab}$, see Eq. (\ref{SymmetricTraceless}).
The idea is to compute numerically a short 
distance product of the type 
\begin{eqnarray}
\<1|{\cal A}(x){\cal A}(-x)|2\> \sim W_{\cal O}(x)\<1|{\cal O}|2\>+
\dots \, ,
\label{OPEAA}
\end{eqnarray}
such that the matrix element on the right-hand side
is known. Enforcing the validity of the OPE allows to compute
the renormalization constant of $Z_{L,{\cal A}}$.

In this Subsection we want to compute the field-renormalization
constant $Z_L$. We will apply the method described above with ${\cal
A} =\sigma^a$.
The simplest choice is to consider the two-point function. 
This is equivalent to choosing the states $\<1|$ and $|2\>$ in 
Eq. (\ref{OPEAA}) to be the vacuum state.
From Eq. (\ref{OPEFieldsScalar}) we obtain
\begin{eqnarray}
\<\sg_{RGI}(x)\cdot\sg_{RGI}(-x)\> = {\cal F}^{(0)}_{RGI, 0}(\grz;\zeta)+
\widehat{\cal F}^{(0)}_{RGI,2}(\grz;\zeta)r^2\<[(\partial\sg)^2]_{RGI}\>+
O(x^4)\, ,\nonumber\\
\label{TwoPointOPE}
\end{eqnarray} 
where we used Eq. (\ref{EnergyMomentumTrace})
to express the vacuum expectation value of the
energy-momentum tensor in terms of $\<(\partial\sg)^2\>$, and
Eq. (\ref{AlphaOnShell}) to eliminate the non-invariant operator $\alpha$.
The RGI Wilson coefficients are obtained
from Eqs. (\ref{F00Pert})--(\ref{F03Pert}) using the 
formulae of Sec. \ref{RenormalizationGroupSection}:
\begin{eqnarray}
{\cal F}^{(0)}_{RGI,0}(\overline{g};1) &=&  \overline{g}^{-(N-1)/(N-2)}
\left\{
1+\frac{N-1}{2\pi(N-2)}\overline{g}
-\frac{(N-1)(N^2-6N+6)}{16\pi^2(N-2)^2}\overline{g}^2+\right.\nonumber\\
&&
+\frac{(N-1)}{192\pi^3(N-2)^3}[N^4(8\zeta(3)-7)+N^3(-54\zeta(3)+56)+
\nonumber\\
&&\quad \quad +6N^2(20\zeta(3)-29)+N(-88\zeta(3)+256)-148]\overline{g}^3
\Bigg{ \} }\, ,\label{TwoPointResummed}\\
\widehat{\cal F}^{(0)}_{RGI,2}(\overline{g};1) &=&
-\overline{g}^{-(N-1)/(N-2)}\left\{
1+\frac{N^2-4N+5}{2\pi(N-2)}\overline{g}
\right\}\, .\label{TwoPointResummedBIS}
\end{eqnarray}
%
%********************************************
%
\begin{figure}
\centerline{
\epsfig{figure=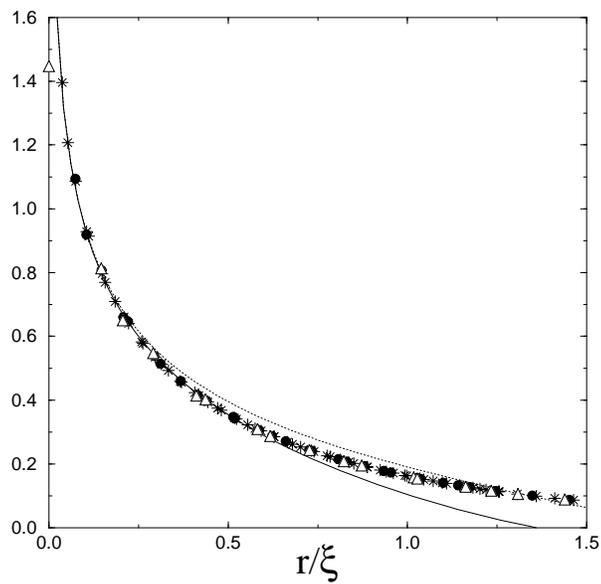,angle=-90,
width=0.6\linewidth}
}
\caption{Monte Carlo data for the isovector correlation function,
and OPE predictions. Different symbols refer to different lattices: 
empty triangles ($\triangle$) to lattice \ref{Lattice64x128},
filled circles ($\bullet$) to lattice \ref{Lattice128x256} and 
stars ($\ast$) to lattice \ref{Lattice256x512}. 
The continuous (dotted) line is the best fitting curve
including (not including) power corrections.}
\label{ZFieldScaling}
\end{figure}
%
%********************************************
%
\begin{figure}
\centerline{
\begin{tabular}{cc}
\hspace{0.0cm}\vspace{-3.5cm}\\
\epsfig{figure=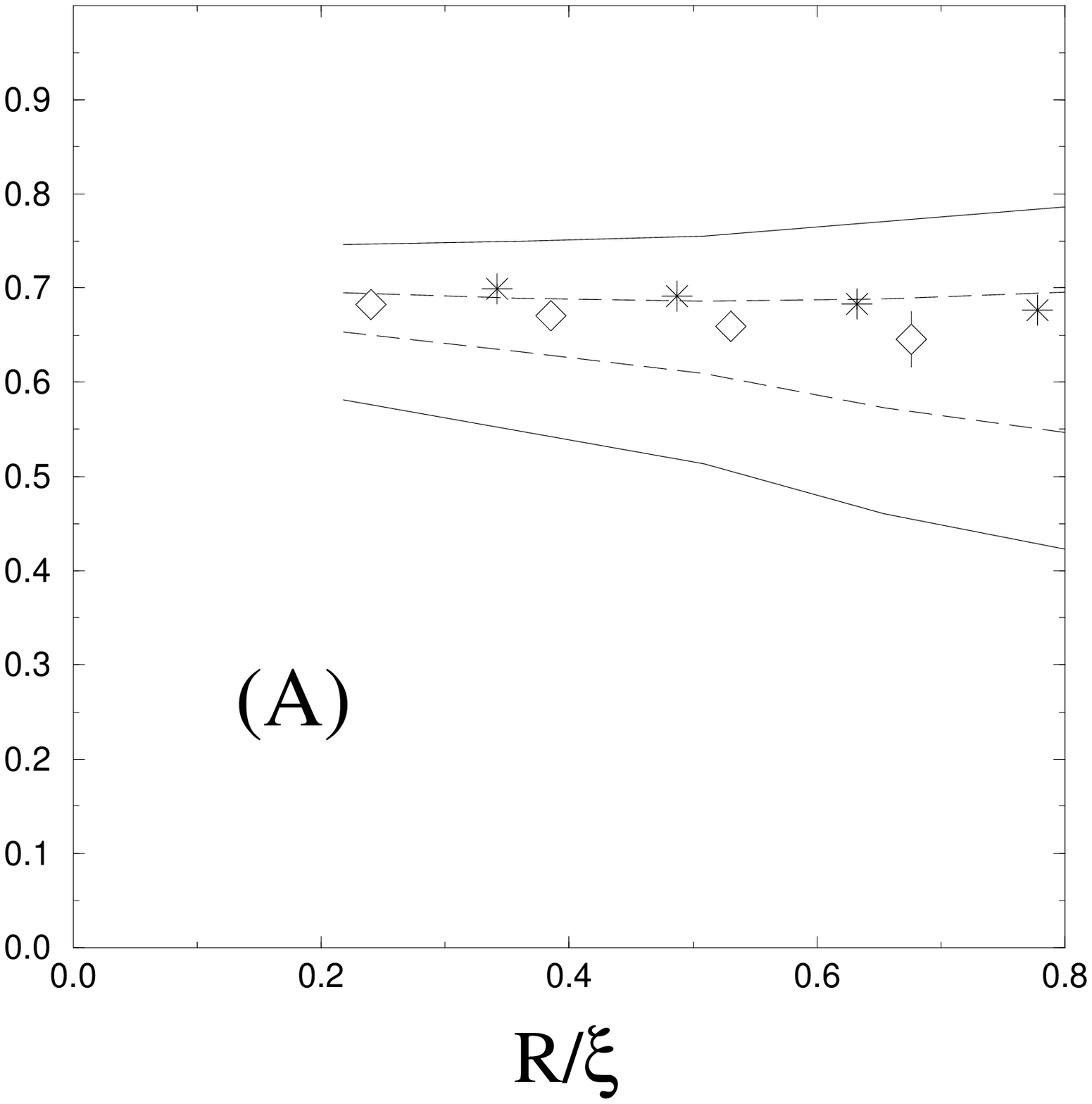,angle=-90,
width=0.45\linewidth}&
\hspace{-1.cm}
\epsfig{figure=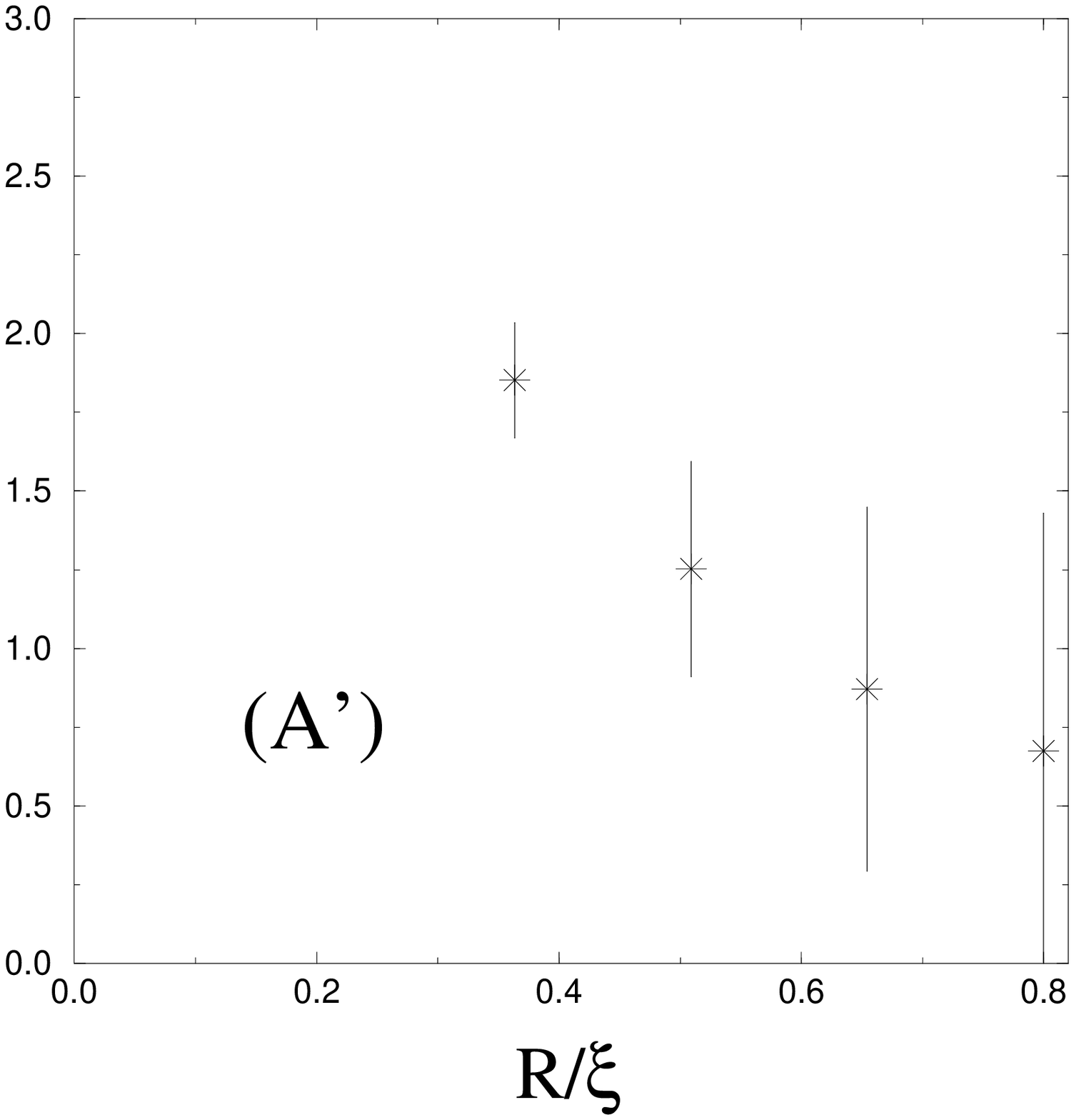,angle=-90,
width=0.45\linewidth}\\
\hspace{0.0cm}\vspace{-1.5cm}\\
\epsfig{figure=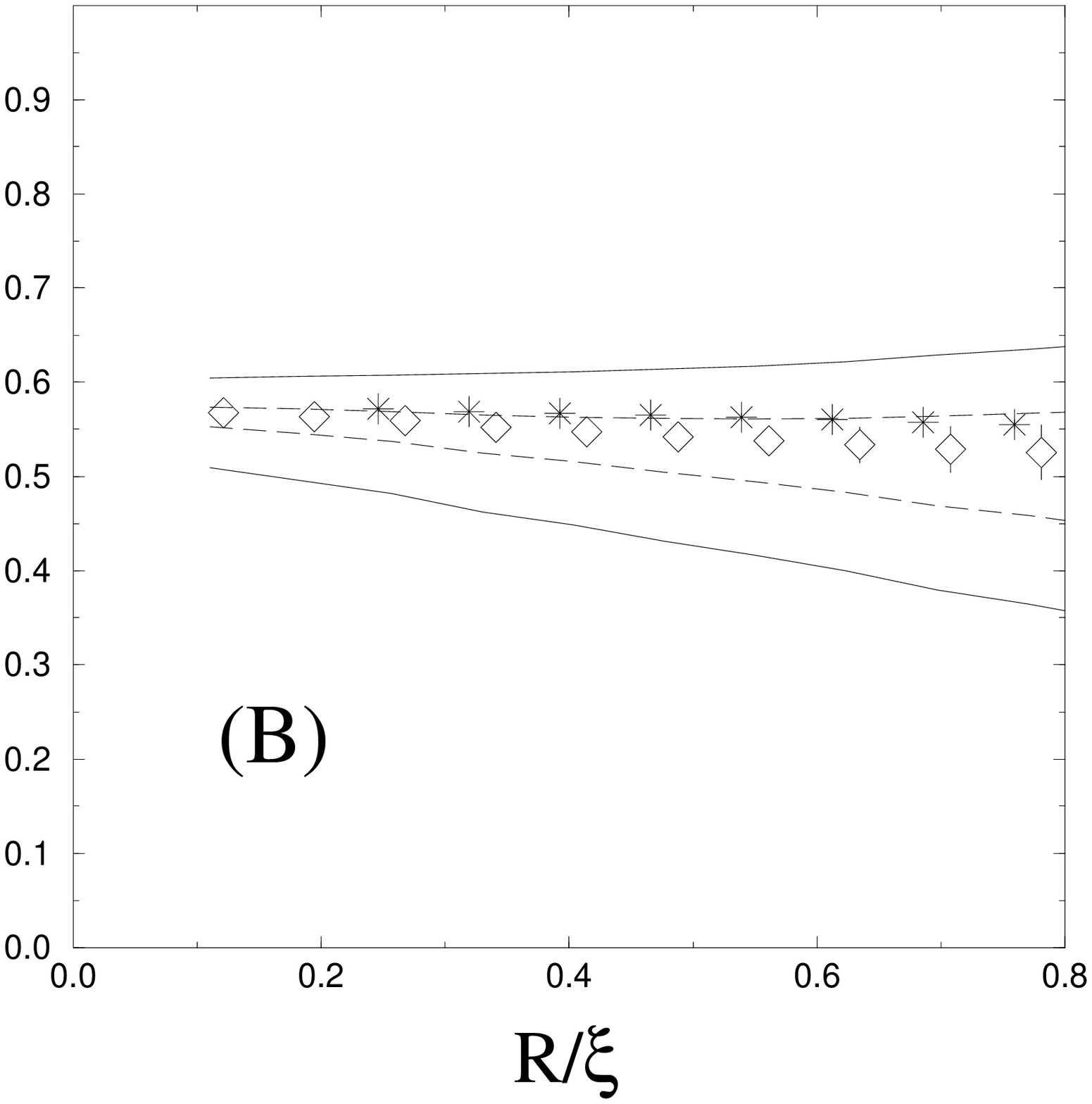,angle=-90,
width=0.45\linewidth}&
\hspace{-1.cm}
\epsfig{figure=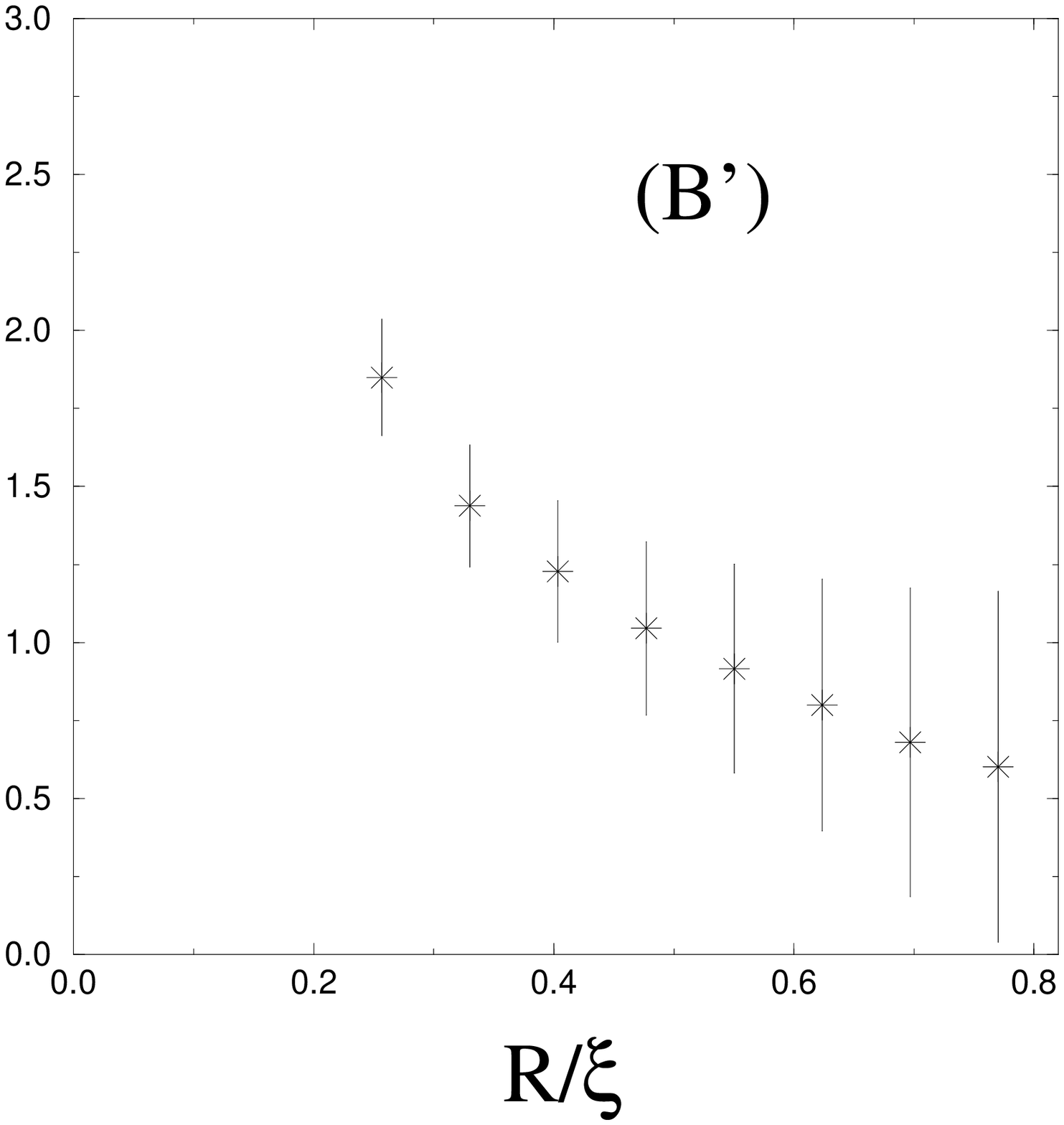,angle=-90,
width=0.45\linewidth}\\
\hspace{0.0cm}\vspace{-1.5cm}\\
\epsfig{figure=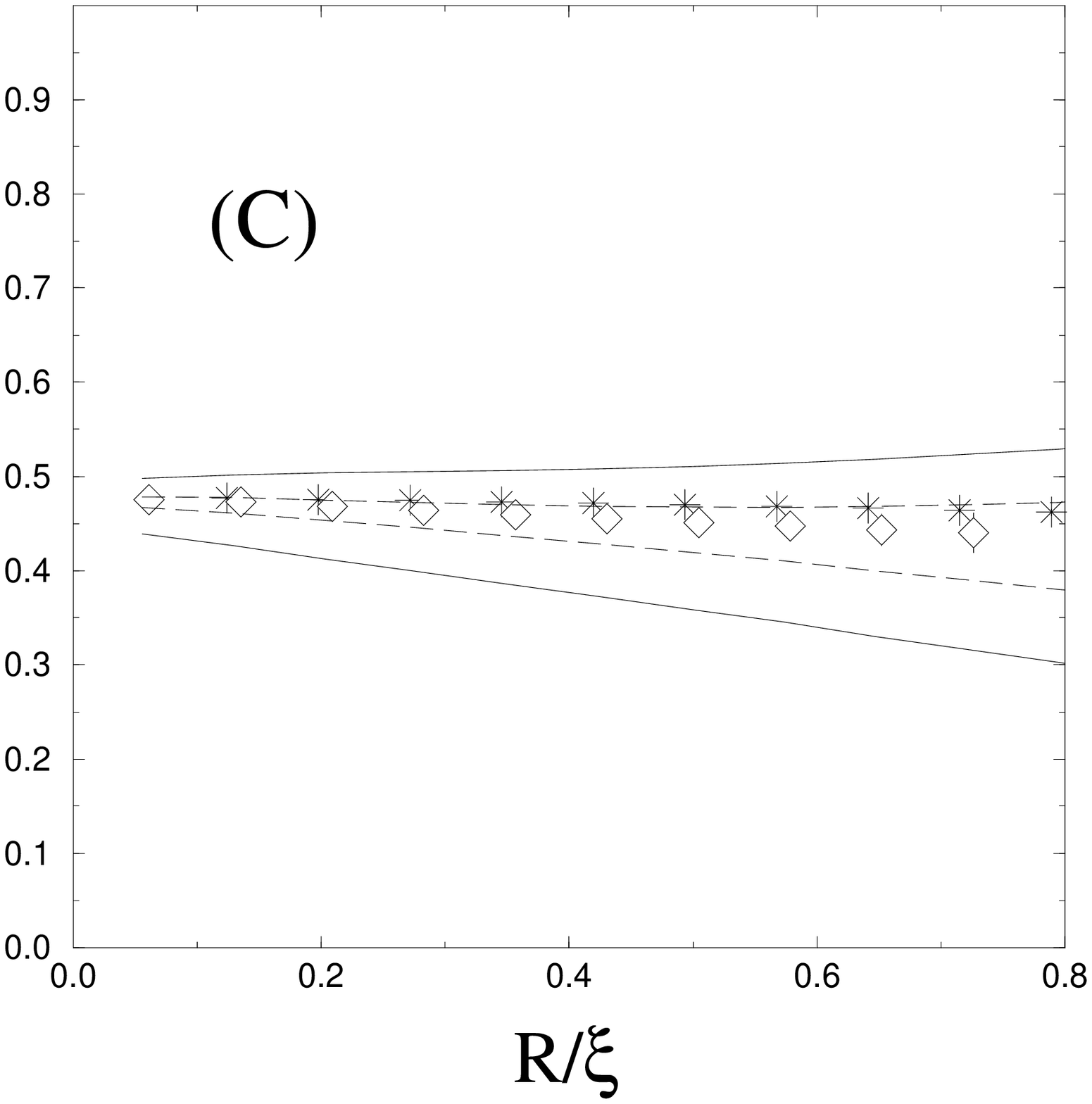,angle=-90,
width=0.45\linewidth}&
\hspace{-1.cm}
\epsfig{figure=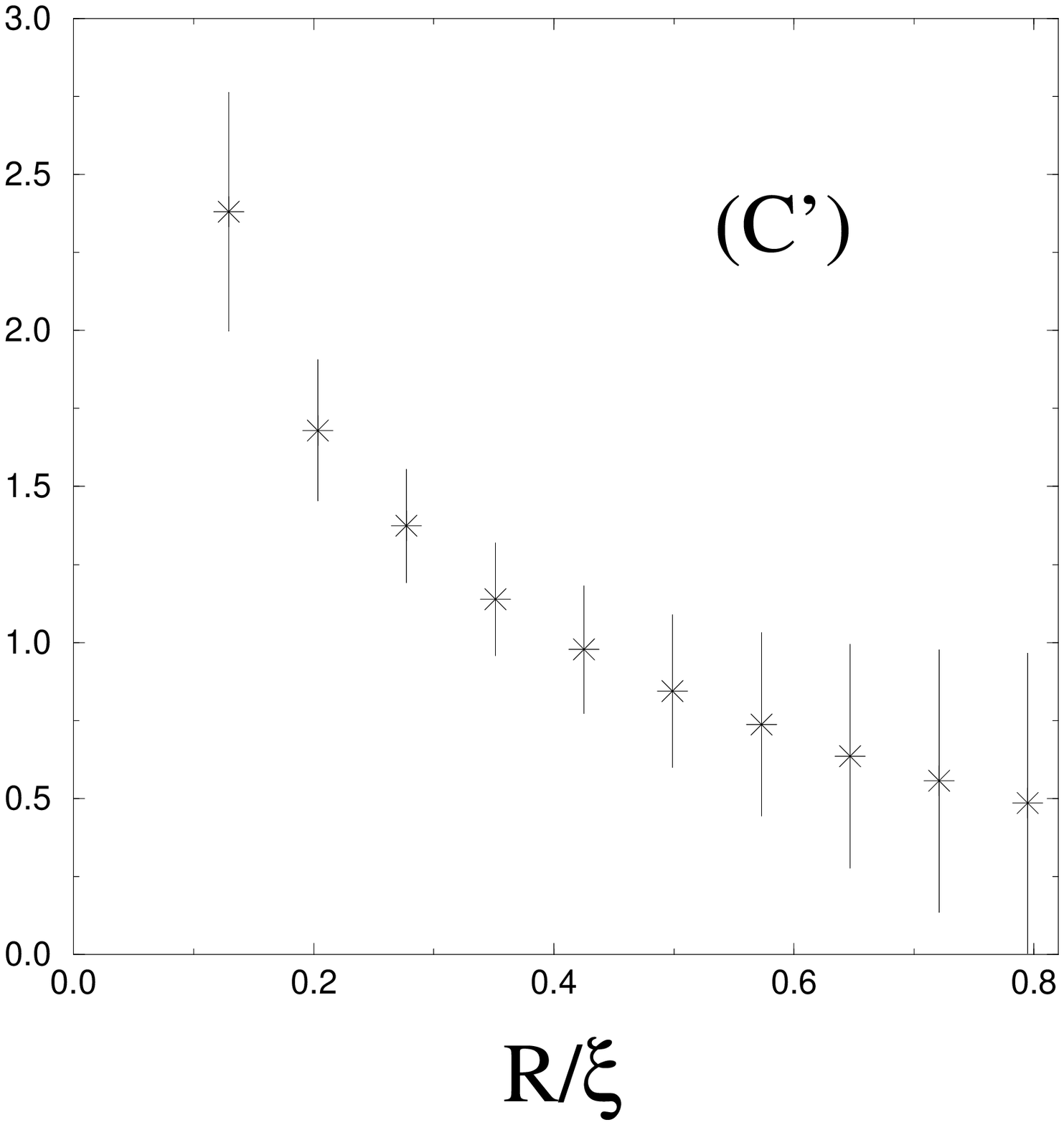,angle=-90,
width=0.45\linewidth}
\end{tabular}
}
\caption{The results of the fit of the two-point function.
In the left column (frames (A), (B) and (C)) we show 
${\cal Z}(\rho,R)$, in the right column (frames (A'), (B') and (C'))
${\cal W}(\rho,R)$. Graphs (A) and (A') refer to lattice \ref{Lattice64x128},
(B) and (B') to lattice \ref{Lattice128x256}, (C) and (C')
to lattice \ref{Lattice256x512}. Different symbols correspond to different
fitting form: stars ($\ast$) to Eq. (\ref{FittingForm0}) with power 
corrections; diamonds ($\Diamond$) to Eq. (\ref{FittingForm0}) 
without power corrections; 
dashed lines to Eq. (\ref{FittingForm0}) without power corrections and
two-loop Wilson coefficient; continuous lines to Eq. (\ref{FittingForm0}) 
without power corrections and one-loop Wilson coefficient.
In the first two cases we plot the value obtained at $\zeta e^{\gamma}=1$
and the systematic error bars, obtained by varying $\zeta$. 
In the other cases we show the
maximum and the minimum values obtained in the chosen range of
$\zeta$.
Statistical errors are negligible in these plots and we do not report
them.}
\label{ZField}
\end{figure}

The lattice fields renormalize as follows, see Sec. 
\ref{LatticeOperatorsSection},
\begin{eqnarray}
\sg(x) = Z_L^{-1/2}(g,g_L)\sg_x \, ,
\end{eqnarray}
where we emphasized the dependence of $Z_L$ upon $g_L$ and $g$,
i.e., in more physical terms, upon $\Lambda a$ and $\Lambda/\mu$.
This dependence can be predicted using RG.
In fact the following equations hold:
\begin{eqnarray}
\left[\beta(g)\frac{\partial}{\partial g}+ \gamma(g)\right]Z_L & = & 0\, ,\\
\left[\beta_L(g)\frac{\partial}{\partial g_L}-\gamma_L(g_L)\right]Z_L & = & 0
\, .
\end{eqnarray}
These are simply the definitions of the continuum and lattice 
anomalous dimensions $\gamma(g)$ and $\gamma_L(g_L)$.
The general solution of the above equations reads 
\begin{eqnarray}
\hspace{-2cm}
Z_L(g,g_L) = \widehat{Z} \cdot g^{-\gamma_0/\beta_0}
\exp\left[\int_0^g\! dx\left(\frac{\gamma(x)}{\beta(x)}
+\frac{\gamma_0}{\beta_0 x}\right)\right]\cdot
g_L^{\gamma_0/\beta_0}
\exp\left[-\int_0^{g_L}\! dx\left(\frac{\gamma_L(x)}{\beta_L(x)}
+\frac{\gamma_0}{\beta_0 x}\right)\right]\, .\hspace{-3cm}\nonumber\\
\label{RGZfield}
\end{eqnarray}
The constant $\widehat{Z}$ is easily fixed. Noticing that 
$Z_L(g,g_L)=1+O(g)$ and $g_L=g+O(g^2)$, we obtain
\begin{eqnarray}
\widehat{Z}=1\, .
\end{eqnarray}

As in the rest of this Chapter we have different attitudes towards 
different terms in Eq. (\ref{RGZfield}).
The lattice factor $g_L^{\gamma_0/\beta_0}\dots$ can be computed in 
perturbation theory. However, since lattice perturbation theory 
is not well behaved, we are interested in computing it non-perturbatively.
The continuum factor $g^{-\gamma_0/\beta_0}\dots$ can be computed in
perturbation theory too. In this case, we assume that 
we are interested in an energy scale $\mu$ 
which is high enough for making the perturbative calculation reliable
(recall that $g(\mu)\to 0$ as $\mu\to\infty$).

Notice that adsorbing the continuum factor  $g^{\gamma_0/\beta_0}\dots$
in the definition of $\sg$ yields the (finite) RGI field operator
$\sg_{RGI}(x)$. Motivated by the above discussion,
we shall compute the renormalization constant
$Z_{L,RGI}(g_L)$, defined by:
$\sg_{RGI}(x) = Z_{L,RGI}^{-1/2}(g_L)\sg_x$. Its explicit RG
expression is easily obtained from Eq. (\ref{RGZfield}):
\begin{eqnarray}
Z_{L,RGI}(g_L) = g_L^{\gamma_0/\beta_0}
\exp\left[-\int_0^{g_L}\! dx\left(\frac{\gamma_L(x)}{\beta_L(x)}
+\frac{\gamma_0}{\beta_0 x}\right)\right]\, .
\end{eqnarray}

Being written in terms of RGI fields, Eq. (\ref{TwoPointOPE}) 
gives access to $Z_{L,RGI}(g_L)$.
We use the following fitting form for the two point function:
\begin{eqnarray}
G(t,x) = {\cal F}^{(0)}_{RGI, 0}(\gr2z;\zeta) {\cal Z}+
\widehat{\cal F}^{(0)}_{RGI,2}(\gr2z;\zeta)(mr/2)^2{\cal W}\, .
\label{FittingForm0}
\end{eqnarray}
As always we restrict the fit to the region $\rho\le r\le R$.
The parameter ${\cal Z}$ gives an estimate of $Z_{L,RGI}(g_L)$.
The parameter ${\cal W}$ could be called a ``spin-wave'' condensate.
However we do not expect to be able to determine the
value of $\<[(\partial\sg)^2]_{RGI}\>$ from it.
In fact $(\partial \sg)^2$ mixes with the identity operator
and the remarks of Sec. \ref{OperatorDefSection} apply to this case.
We should determine the Wilson coefficient
${\cal F}^{(0)}_{RGI, 0}(\gr2z;\zeta)$ up to terms of order 
$m^2r^2$ for Eq. (\ref{TwoPointOPE}) to define $\<[(\partial\sg)^2]_{RGI}\>$
unambiguously. 

For each case we repeated the fit with and without the 
power-correction term ${\cal W}$. This gives a feeling of how good is 
the truncation of the OPE in Eq. (\ref{TwoPointOPE}).

In Fig. \ref{ZFieldScaling} we present a scaling plot of Monte Carlo
data for $G(t,x)$ together with the best fitting curves 
with and without power corrections. For sake of clarity we plotted 
only the values of $G(t,x)$ obtained along the time direction
$(t,x) = (t,0)$, and along the diagonal $(t,x)=(t,t)$.
We rescaled the data using the estimated values of $Z_{L,RGI}(g_L)$,
i.e. $\widehat{\cal Z}^*$, see Tab. \ref{ZFieldTable} and discussion below.
The data collapse on a single curve showing a clear evidence 
of scaling. The dotted curve (no power corrections) simply reports
${\cal F}^{(0)}_{RGI, 0}(\gr2z;\zeta)$. In the continuous
curve we add the power correction 
$\widehat{\cal F}^{(0)}_{RGI,2}(\gr2z;\zeta)(mr/2)^2
\widehat{\cal W}^*/\widehat{\cal Z}^*$, with $\widehat{\cal W}^*$
and $\widehat{\cal Z}^*$ obtained on lattice \ref{Lattice256x512}.

In Fig. \ref{ZField} we show the best fitting parameters
${\cal Z}(\rho,R)$ and ${\cal W}(\rho,R)$ on lattices \ref{Lattice64x128},
\ref{Lattice128x256} and \ref{Lattice256x512}.
In all the cases we kept $\rho = 0.5$ fixed (this excludes
only the point $(t,x)=(0,0)$ from the fit), 
and we varied $R$. Statistical errors
are negligible in these plots and we do not report them.

The estimates ${\cal W}(\rho,R)$ are reported in frames (A'), (B'), (C').
These graphs do not allow any reliable evaluation of 
the expectation value $\<[(\partial\sg)^2]_{RGI}\>$. For 
$R\gtapprox 0.7\, \xi_{\rm exp}$ systematic errors are of the same 
order as the estimate itself. For $R\ltapprox 0.7\, \xi_{\rm exp}$
systematic errors begin to shrink but ${\cal W}(\rho,R)$ shows a strong $R$
dependence. Indeed ${\cal W}(\rho,R)$ seems to diverge as $R\to 0$.
This can be easily understood if we assume that we are effectively fitting
higher loops (which go as $r^0$ as $r\to 0$) with a term of the type
${\cal W} r^2$.

\begin{table}
\centerline{
\begin{tabular}{|c|c|c|c|}
\hline
 & $\widehat{\cal Z}^*$ &
$\widehat{Z}_{4\, loop}\, \{
\widehat{Z}_{3\, loop},\, \widehat{Z}_{2\, loop}\}$ &
$\widehat{Z}^{bpt}_{4\, loop}\, \{
\widehat{Z}^{bpt}_{3\, loop},\, \widehat{Z}^{bpt}_{2\, loop}\}$\\
\hline
\hline
lattice \ref{Lattice64x128} & $0.67[1]$ & $0.957\, \{0.916,\, 0.861\}$ &
$1.075\, \{1.068,\, 1.082\}$\\
\hline
lattice \ref{Lattice128x256}& $0.559[5]$& $0.918\, \{0.889,\, 0.843\}$ &
$0.9990\, \{0.9944,\, 1.0048\}$\\
\hline
lattice \ref{Lattice256x512}& $0.468[3]$& $0.924\, \{0.900,\, 0.861\}$ &
$\,0.976 \{0.973,\, 0.981\}$\\
\hline
\end{tabular}
}
\caption{The OPE result $\widehat{\cal Z}^*$ for the field-renormalization 
constant $Z_{L,RGI}(g_L)$ and the corresponding perturbative estimates for the 
constant $\widehat{Z}$ defined in Eq. (\ref{ZhatComputation}). 
We used lattice perturbation theory in the third column 
and improved (boosted) perturbation theory in the fourth column.
These values should be compared with the exact result
$\widehat{Z}=1$.}
\label{ZFieldTable}
\end{table}  
In graphs (A), (B), (C) we report the results for ${\cal Z}(\rho,R)$.
The estimates obtained including the power-correction term
in Eq. (\ref{FittingForm0}) show a quite mild $R$ dependence and 
very small systematic errors which are roughly $R$-independent.
These values of ${\cal Z}(\rho,R)$ are not constant within systematic 
error bars. The reason is probably that the fitting parameter ${\cal W}$
mimics the effects of higher loops in ${\cal F}^{(0)}_{RGI, 0}(\gr2z;\zeta)$
and reduces the scheme dependence of the result.
The estimates obtained without power corrections show larger systematic 
errors and are flat within the systematic error bars.
Systematic errors\footnote{Notice that {\it relative} systematic errors are 
approximatively independent of $g_L$ at $R/\xi$ fixed.}
decrease as $R\to 0$.

In the same graphs we reported 
the analogous estimates (obtained without power corrections) with 
${\cal F}^{(0)}_{RGI, 0}(\gr2z;\zeta)$ computed in one-loop and two-loop 
perturbation theory. Our method to assess systematic errors seems to be 
consistent. Finally the difference between the results for 
${\cal Z}(\rho,R)$ obtained with or without power corrections
are quite small.

\begin{table}
\centerline{
\begin{tabular}{|c|c|c|c|c|}
\hline
$R$ & $R/\xi_{\rm exp}$ &
\multicolumn{3}{|c|}
{${\cal Z}(\rho,R)[\mbox{syst. }\kappa=1.5]\, [\kappa=2]\, [\kappa=2.5]$}\\
\hline
    &   & 3-loop & 2-loop & 1-loop \\
\hline
\hline
 $3.5$ & $0.13$ & $0.4726[14]\, [25]\, [33]$ & $0.469[4]\, [8]\, [12]$ & 
$0.46[2]\, [4]\, [5]$\\
\hline
 $5.5$ & $0.20$ & $0.468[2]\, [3]\, [5]$    & $0.464[5]\, [11]\, [16]$ & 
$0.46[2]\, [4]\, [7]$\\
\hline
 $7.5$ & $0.28$ & $0.464[3]\, [4]\, [7]$    & $0.459[7]\, [13]\, [20]$ &
$0.45[3]\, [5]\, [8]$\\
\hline
 $9.5$ & $0.35$ & $0.459[3]\, [6]\, [9]$    & $0.453[8]\, [17]\, [26]$ & 
$0.45[3]\, [6]\, [9]$\\
\hline
$15.5$ & $0.57$ & $0.447[6]\, [13]\, [21]$   & $0.439[13]\, [28]\, [46]$ & 
$0.43[4]\, [8]\, [13]$\\
\hline
\end{tabular}
}
\caption{We compare different estimates of the systematic error on the 
field renormalization constant. Here we consider the data obtained on lattice
\ref{Lattice256x512} and use the fitting form (\ref{FittingForm0}) without 
power corrections.}
\label{ZFieldTableSys}
\end{table} 
We summarize our results for $Z_{L,RGI}(g_L)$ in Tab. \ref{ZFieldTable}.
The values of $\widehat{\cal Z}^*$ correspond to ${\cal Z}(\rho,R)$
at $R=2.5$, $3.5$, $5.5$ on lattices \ref{Lattice64x128}, \ref{Lattice128x256}
and \ref{Lattice256x512}. In the third column we compute
the constant $\widehat{Z}$, see Eq. (\ref{RGZfield}), using
the relation
\begin{eqnarray}
\widehat{Z} =  Z_{L,RGI}(g_L)\cdot
g_L^{-\gamma_0/\beta_0}
\exp\left[\int_0^{g_L}\! dx\left(\frac{\gamma_L(x)}{\beta_L(x)}
+\frac{\gamma_0}{\beta_0 x}\right)\right]\,
\label{ZhatComputation}
\end{eqnarray}
and four-loop (three-loop, two-loop) perturbation theory.
In the last column we repeat the same calculation using improved 
perturbation theory. 
The results for $\widehat{Z}$ seem to scale well 
(i.e. they are approximatively $g_L$-independent)
for the lattices \ref{Lattice128x256} and \ref{Lattice256x512}.
A little discrepancy remains for lattice \ref{Lattice64x128}.
Nevertheless, even at four loops, the outcome of bare perturbation
theory  is about $7-8\%$ away from the correct
value $\widehat{Z} = 1$.
Improved perturbation theory yields a better agreement. For 
the largest lattice the discrepancy is about the $2\%$, 
which is not too far from the {\it estimated} systematic error
(approximatively the $1\%$).

Let us now try to judge our determination of systematic errors and
in particular our choice of $\kappa$. We shall concentrate on lattice 
\ref{Lattice256x512}, since it allows to investigate a larger 
range of distances. In Tab. \ref{ZFieldTableSys} we report 
the results for ${\cal Z}(\rho,R)$ and the corresponding
systematic error for several values of $R$ and $\kappa$.
The fitting form (\ref{FittingForm0}) without power correction was adopted.
The Wilson coefficient was computed in three-loop, two-loop, and one-loop
perturbation theory. 
If we look at a fixed value of $R$ in this table it seems that 
perturbation theory converges very well and that, fixing $\kappa = 2$,
we are overestimating the systematic errors:
$\kappa = 1.5$ could appear a more realistic choice.
However if we vary $R$ and consider the systematic error
obtained with $\kappa = 1.5$ we realize that ${\cal Z}(\rho,R)$ is by no
means flat. We deduce that $\kappa = 2$ is not too cautious and gives
a good (very rough) idea of the systematic errors.

%
%********************************************************************
%
\subsection{Symmetric Operator}
\label{SymmetricOperatorSection}
As we explained in Sec. \ref{ObservablesSection}, we 
are interested in computing the matrix elements of 
the {\it bare} lattice operator $\sigma^a\sigma^b-\delta^{ab}/N$ 
between one-particle states. In order to accomplish this task,
we considered the three-point function $C^{(2)}(\pb,\qb;2t)$ defined in Eq. 
(\ref{SymmetricOperatorThreePoint}).
The matrix element can be extracted from the corresponding normalized function 
$\widehat{C}^{(2)}(\pb,\qb;2t)$ as follows:
\begin{eqnarray}
\<\pb,c|\sigma^a\sigma^b-\frac{\delta^{ab}}{N}|\qb,d\> = 
T_N^{ab,cd}
\sqrt{4\omega(\pb)\omega(\qb)}\lim_{t\to\infty}\widehat{C}^{(2)}(\pb,\qb;2t)
\, ,
\label{OnShellLimitSymmetric}
\end{eqnarray}
where
\begin{eqnarray}
T_N^{ab,cd} = \frac{N}{(N-1)(N+2)}
\left( \delta^{ac}\delta^{bd}+\delta^{ad}\delta^{bc}-2\delta^{ab}\delta^{cd}/N
\right)
\, .
\end{eqnarray}
The numerical results for $\widehat{C}^{(2)}(\pb,\pb;2t)$ are constant
(within statistical errors) for $t\gtapprox \xi_{\rm exp}$.
\begin{table}
\begin{center}
\begin{tabular}{|c|c|c|c|}
\hline
$\pb$ & lattice \ref{Lattice64x128}& lattice \ref{Lattice128x256}&   
lattice \ref{Lattice256x512}\\
\hline
\hline
$0$      & $0.7888(13)$ & $0.5862(20)$ & $0.4309(29)$\\
\hline
$2\pi/L$ & $0.7877(20)$ & $0.5855(26)$ & $0.4311(36)$\\
\hline
$4\pi/L$ & $0.7915(95)$ & $0.573(12)$  & $0.421(17)$\\
\hline
$6\pi/L$ & $0.757(13)$ &  $0.581(19)$  & $0.462(23)$\\
 \hline
\end{tabular}
\end{center}
\caption{The numerical estimates of 
$2\sqrt{\pb^2+m^2}\,\widehat{C}^{(2)}(\pb,\pb;2t)$.
For $\pb = 0, \dots,4\pi/L$, $t = 9, 18, 36$ respectively on lattice 
\ref{Lattice64x128}, \ref{Lattice128x256} and \ref{Lattice256x512}.
For $\pb = 6\pi/L$, $t = 7, 14, 28$ on the same lattices.}
\label{LatticeSymmetricOperator}
\end{table}
In Tab.  we report the numerical estimates for
$2\sqrt{\pb^2+m^2}\,\widehat{C}^{(2)}(\pb,\pb;2t)$ obtained  on lattices 
\ref{Lattice64x128}, \ref{Lattice128x256} and \ref{Lattice256x512},
respectively for $t = 9$ ($t=7$ for $\pb = 6\pi/L$), $18$
($t=14$ for $\pb = 6\pi/L$) and $36$ ($t=28$ for $\pb = 6\pi/L$). 
\begin{figure}
\begin{tabular}{cc}
\hspace{0.0cm}\vspace{-1.5cm}\\
\hspace{-2.5cm}
\epsfig{figure=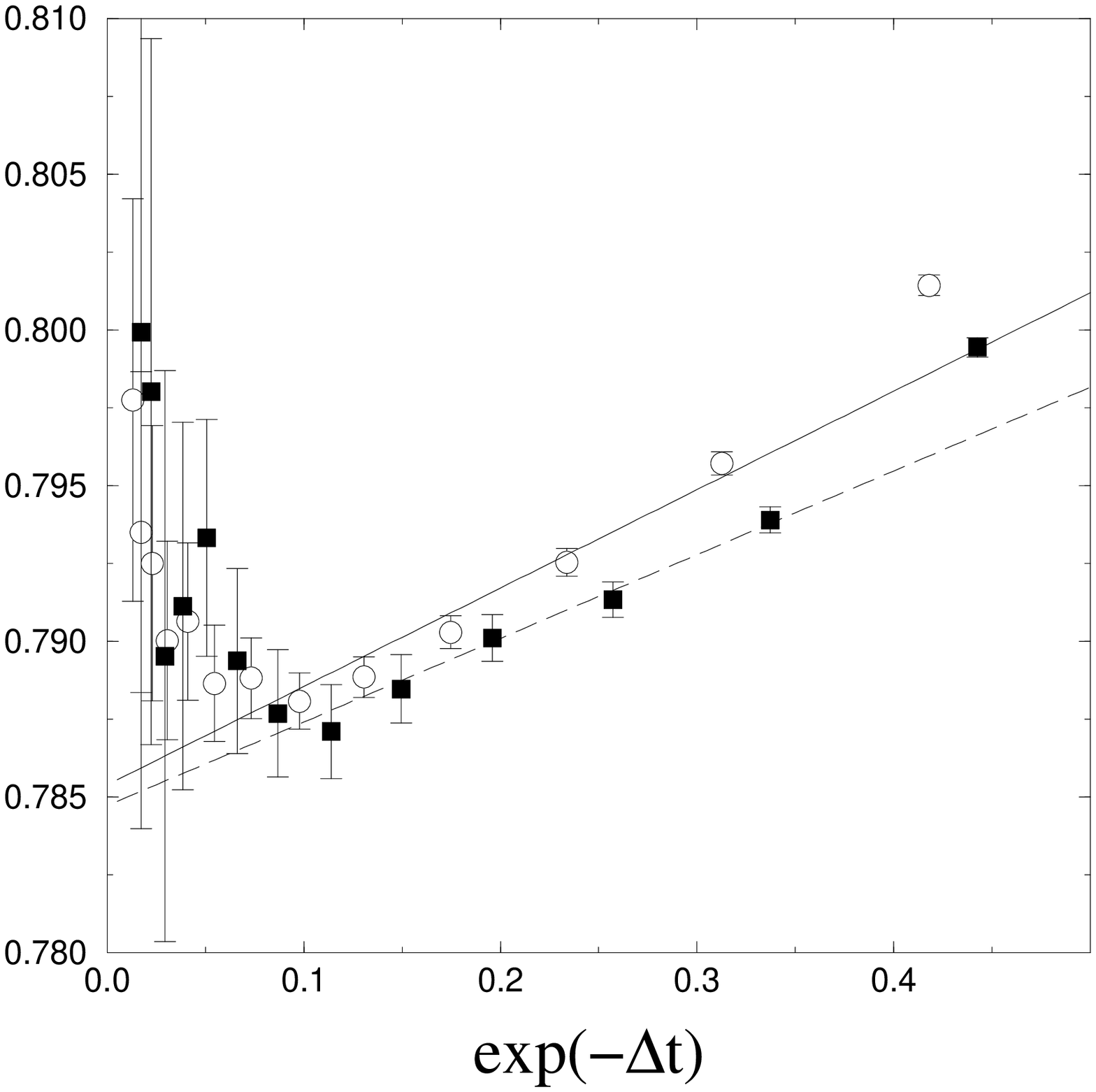,angle=-90,
width=0.6\linewidth}&
\hspace{-0.5cm}
\epsfig{figure=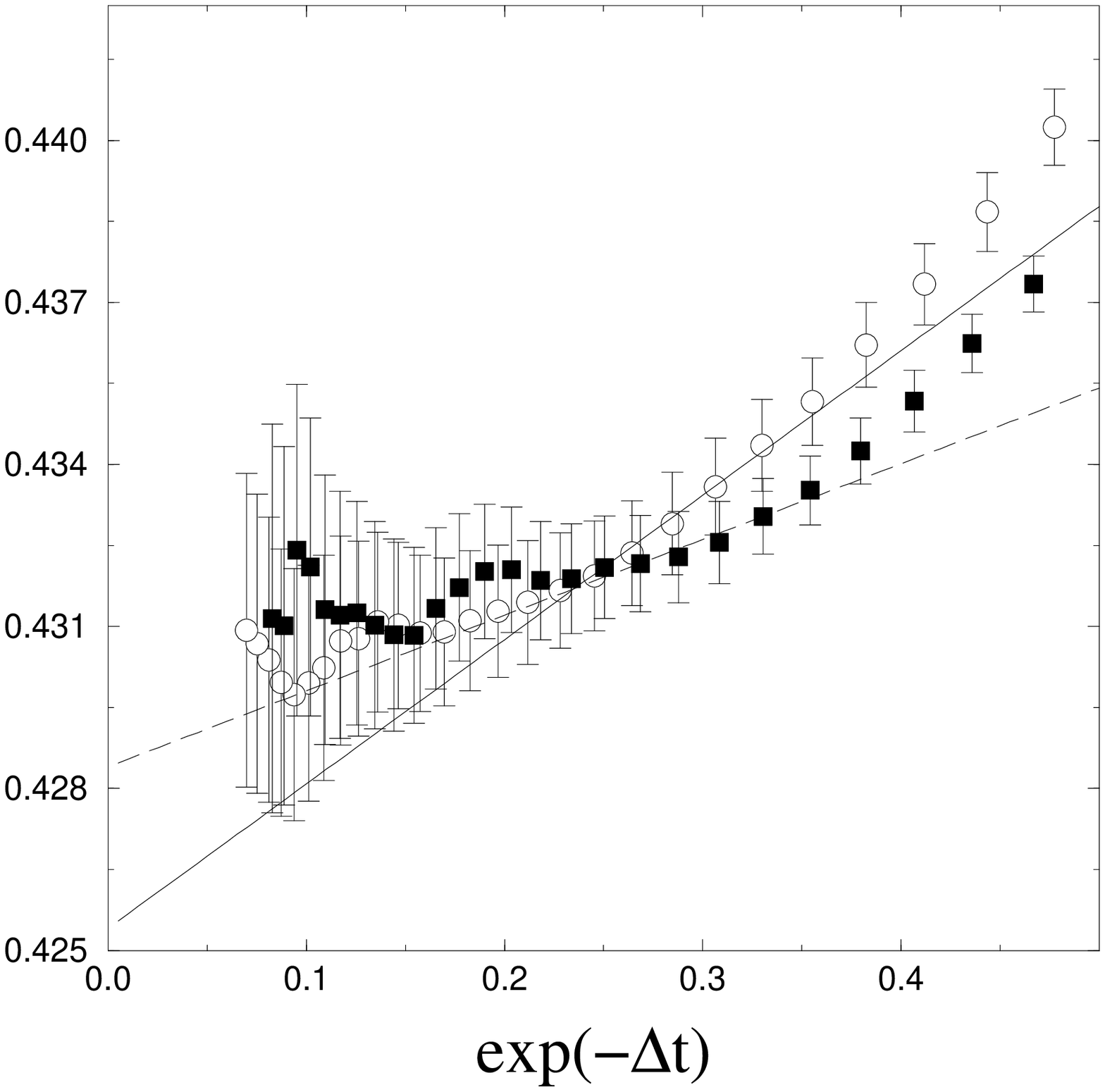,angle=-90,
width=0.6\linewidth}
\end{tabular}
\caption{The asymptotic behaviour of 
$2\sqrt{\pb^2+m^2}\,\widehat{C}^{(2)}(\pb,\pb;2t)$ on lattices 
\ref{Lattice64x128} (left) and \ref{Lattice256x512} (right). 
Empty circles refer to $\pb=0$, and filled squares to $\pb=2\pi/L$.
Continuous and dashed lines are the best fitting curves 
of the form (\ref{FittingFormSymmetric})
for (respectively) $\pb = 0$ and $\pb=2\pi/L$. For sake of clarity we show the 
data obtained on lattice \ref{Lattice256x512} only for $t\le 36$.}
\label{OperatorAsymptotia}
\end{figure}

In order to verify the ``contamination'' due to higher states,
we adopt the same method we used in the computation of the 
one-particle spectrum, see Sec. \ref{sec4.2}.
The expected behaviour of $\widehat{C}^{(2)}(\pb,\qb;2t)$ in the large
$t$ limit is
%mTOT.256x512.av180B1Jack.as 
\begin{eqnarray}
\widehat{C}^{(2)}(\pb,\qb;2t) = \widehat{C}^{(2)}(\pb,\qb;\infty)
+D(\pb,\qb;t) e^{-\Delta(\pb)t}+D(\qb,\pb;t)
e^{-\Delta(\qb)t}+\dots\, ,
\end{eqnarray}
where we  made the same approximations  as for the spectrum,
namely we neglected  terms of
order $e^{-\omega(\pb)T}$, and multi-particle states involving more than
three particles. 
The gap $\Delta(\pb)$ is given by Eq. (\ref{Gap}).
In the thermodynamic ($L\to\infty$) limit,
the coefficients $D(\qb,\pb;t)$ are slowly varying (power-like)
functions of $t$.
Analogously to what we did in Sec. \ref{sec4.2}, we shall neglect
the $t$-dependence of the coefficients $D(\qb,\pb;t)$. 
We fitted our data using the form
\begin{eqnarray}
2\sqrt{\pb^2+m^2}\,\widehat{C}^{(2)}(\pb,\pb;2t)
 = \widehat{C}^{(2)}_*(\pb,\qb;\infty)+
2D_*(\pb,\pb;t) e^{-\Delta(\pb)t}\, .
\label{FittingFormSymmetric}
\end{eqnarray}

In Fig. \ref{OperatorAsymptotia} we plot 
$2\sqrt{\pb^2+m^2}\,\widehat{C}^{(2)}(\pb,\pb;2t)$
versus $e^{-\Delta(\pb)t}$ on lattices \ref{Lattice64x128} and
\ref{Lattice256x512} for $\pb=2\pi n /L$, $n= 0,\, 1$. 
The estimated systematic error on the results of Tab. 
\ref{LatticeSymmetricOperator} is about
$2\cdot 10^{-3}$,$2\cdot 10^{-3}$,$3\cdot 10^{-3}$,$6\cdot 10^{-2}$,
respectively for $\pb = 0,\dots,6\pi/L$.

Notice that, for kinematical reasons, in the continuum limit
$\<\pb,c|\sigma^a\sigma^b-\delta^{ab}/N|\pb,d\>$ does not depend upon
$\pb$. 
The results of Tab. \ref{LatticeSymmetricOperator} verify this prediction
within the statistical errors. 
The only statistically significant discrepancy occurs at 
$\pb = 6\pi/L$ on lattice \ref{Lattice64x128}.
It is plausible to explain this discrepancy as a scaling correction.
%
%*****************************************************************
%
\subsection{Renormalization of the Symmetric Operator}
\label{RenormalizationSymmetricSection}

Let us now come to the problem of renormalizing the lattice
results obtained in the previous Section. 
This is a necessary step in order to check the results of
Sec. \ref{SymmetricSection}, where we shall adopt the OPE method
to compute the matrix elements of the renormalized operator
$\opl S_0\opr$, see Eq. (\ref{SymmetricTraceless}).

We have seen in Sec. \ref{FieldSection} that improved (boosted)
perturbation theory yields the field-renormalization constant
with $1-2\%$ of systematic error on lattice \ref{Lattice256x512}.
Now we have to compute the renormalization constant for
the symmetric operator $S_0$.
We shall use perturbation theory at first. Next, we 
shall switch to the OPE non-perturbative method, see
Sec. \ref{FieldSection}, in order to have more reliable results.
We shall see that, in this case, lattice perturbation theory
(even if improved) does not give an approximation as good as it does for the
field-renormalization constant.

We proceed as in Sec. \ref{FieldSection}.
Let us consider an operator ${\cal O}$ renormalizing multiplicatively:
$\opl {\cal O}\opr = Z_L^{\cal O}{\cal O}_L$. The corresponding
RGI operator ${\cal O}_{RGI}$ is easily given in terms of
its lattice counterpart: ${\cal O}_{RGI}=Z_{L,RGI}^{\cal O}{\cal O}_L$.
RG considerations yield:
\begin{eqnarray}
Z_{L,RGI}^{\cal O} = g_L^{\gamma_0/\beta_0}
\exp\left[-\int_0^{g_L}\! dx\left(\frac{\gamma^{\cal O}_L(x)}{\beta_L(x)}
+\frac{\gamma^{\cal O}_0}{\beta_0 x}\right)\right]\, ,
\label{RGZOperator}
\end{eqnarray}
where $\gamma^{\cal O}_L(g)$ are the {\it lattice} anomalous dimensions 
of the operator ${\cal O}$, see Sec. \ref{LatticeAnomalousSection}.

\begin{table}
\begin{center}
\begin{tabular}{|c|c|c|}
\hline
& ${\cal S}_{3\, loop}\, \{{\cal S}_{2\, loop} \}$
&$ {\cal S}^{bpt}_{3\, loop}\, \{{\cal S}^{bpt}_{2\, loop} \}$\\
\hline
\hline
lattice \ref{Lattice64x128} & $1.640(3)\,  \{1.800(3)\}$ & 
$1.302(2)\,  \{1.277(2)\}$ \\
\hline
lattice \ref{Lattice128x256}& $1.674(6)\,  \{1.809(6)\}$ &
$1.414(5)\, \{1.393(5)\}$ \\
\hline
lattice \ref{Lattice256x512}& $1.576(11)\, \{1.685(11)\}$ & 
$1.402(9)\,  \{1.384(9)\}$   \\
 \hline
\end{tabular}
\end{center}
\caption{The perturbatively-renormalized matrix element of 
the RGI symmetric operator, see Eq. (\ref{TableContent}).
While in the second column we use bare lattice perturbation theory,
in the third we use the improved expansion parameter $g_E$.}
\label{RGISymmetricOperator}
\end{table}
In Tab. \ref{RGISymmetricOperator} we report the results for 
\begin{eqnarray}
{\cal S}\equiv\frac{1}{N}\sum_{a,b} \<\pb,a|\left[S^{ab}_0 
\right]_{RGI}|\pb, b\> \, ,
\label{TableContent}
\end{eqnarray}
obtained from the data of Tab. \ref{LatticeSymmetricOperator} using
Eqs. (\ref{OnShellLimitSymmetric}) and (\ref{RGZOperator}). The 
lattice anomalous dimensions are known in three-loop 
perturbation theory \cite{Caracciolo:1994sk}. 
The bare matrix element has been obtained from the $\pb = 0$ data at 
of Tab. \ref{LatticeSymmetricOperator}.

%
%********************************************
%
\begin{figure}
\centerline{
\epsfig{figure=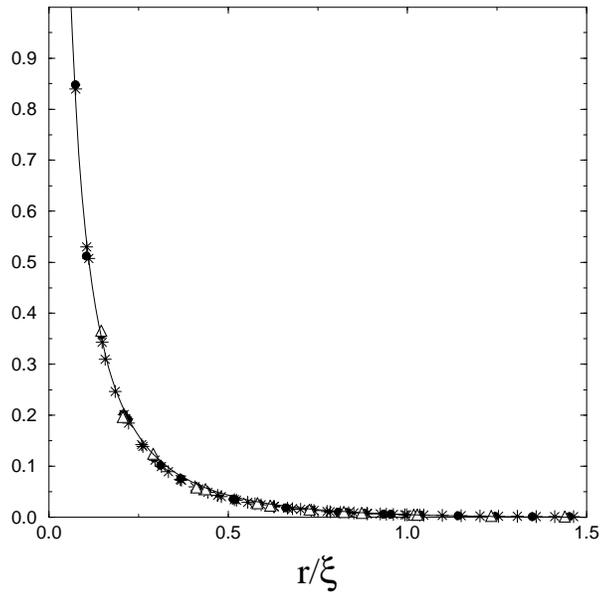,angle=-90,
width=0.6\linewidth}
}
\caption{Monte Carlo data for the isotensor correlation function,
and OPE predictions.
Empty triangles ($\triangle$) refer to lattice \ref{Lattice64x128},
filled circles ($\bullet$) to lattice \ref{Lattice128x256}, and 
stars ($\ast$) to lattice \ref{Lattice256x512}. The Monte Carlo data are
rescaled using the non-perturbative renormalization constant, see 
Tab. \ref{ZOPTab}, right column. The continuous line corresponds to
the leading term of the OPE.}
\label{ZOPScaling}
\end{figure}
%
%********************************************
%
\begin{figure}
\centerline{
\begin{tabular}{cc}
\hspace{0.0cm}\vspace{-1.5cm}\\
\hspace{-1.5cm}
\epsfig{figure=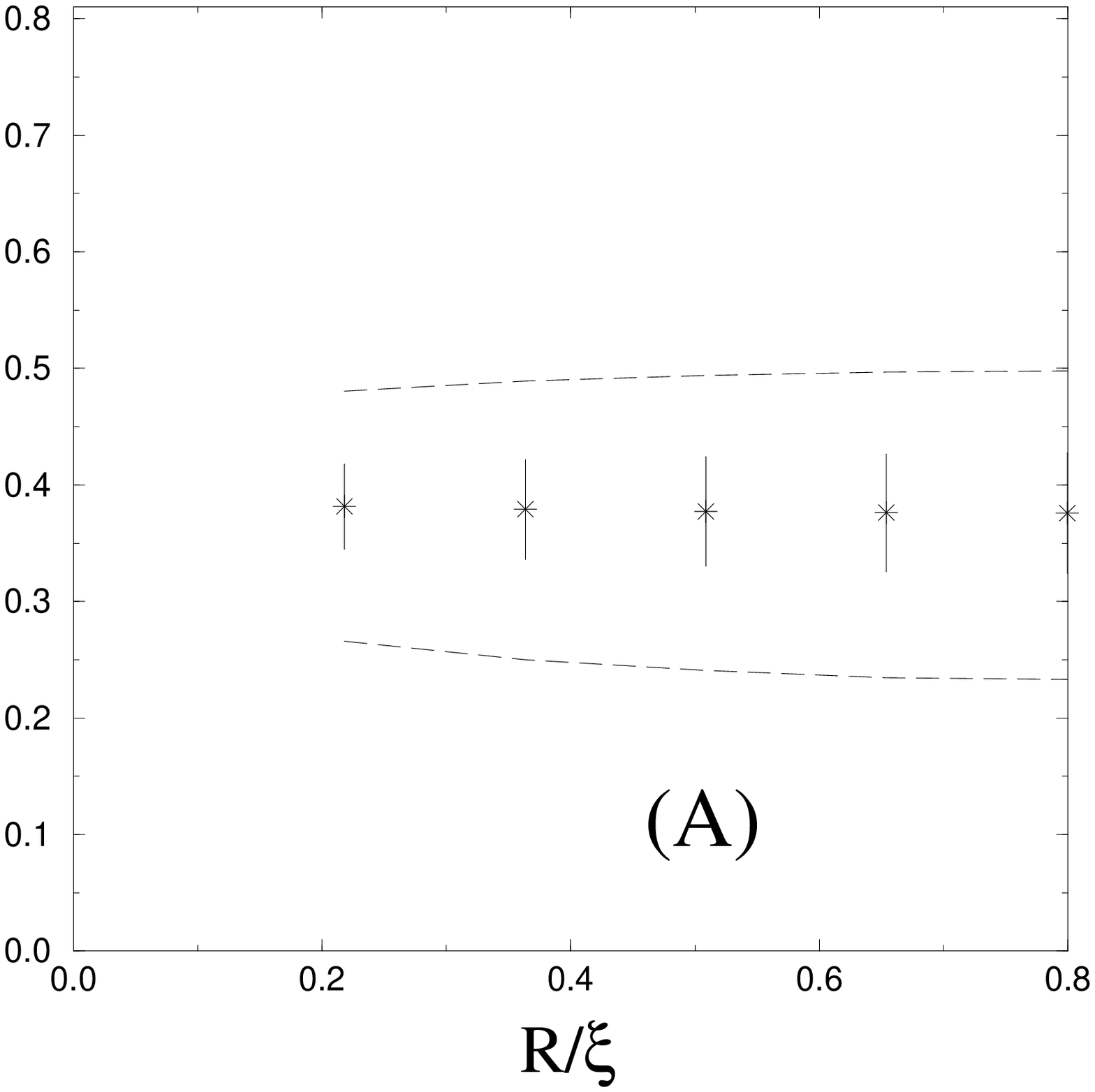,angle=-90,
width=0.5\linewidth}&\hspace{-0.5cm}
\epsfig{figure=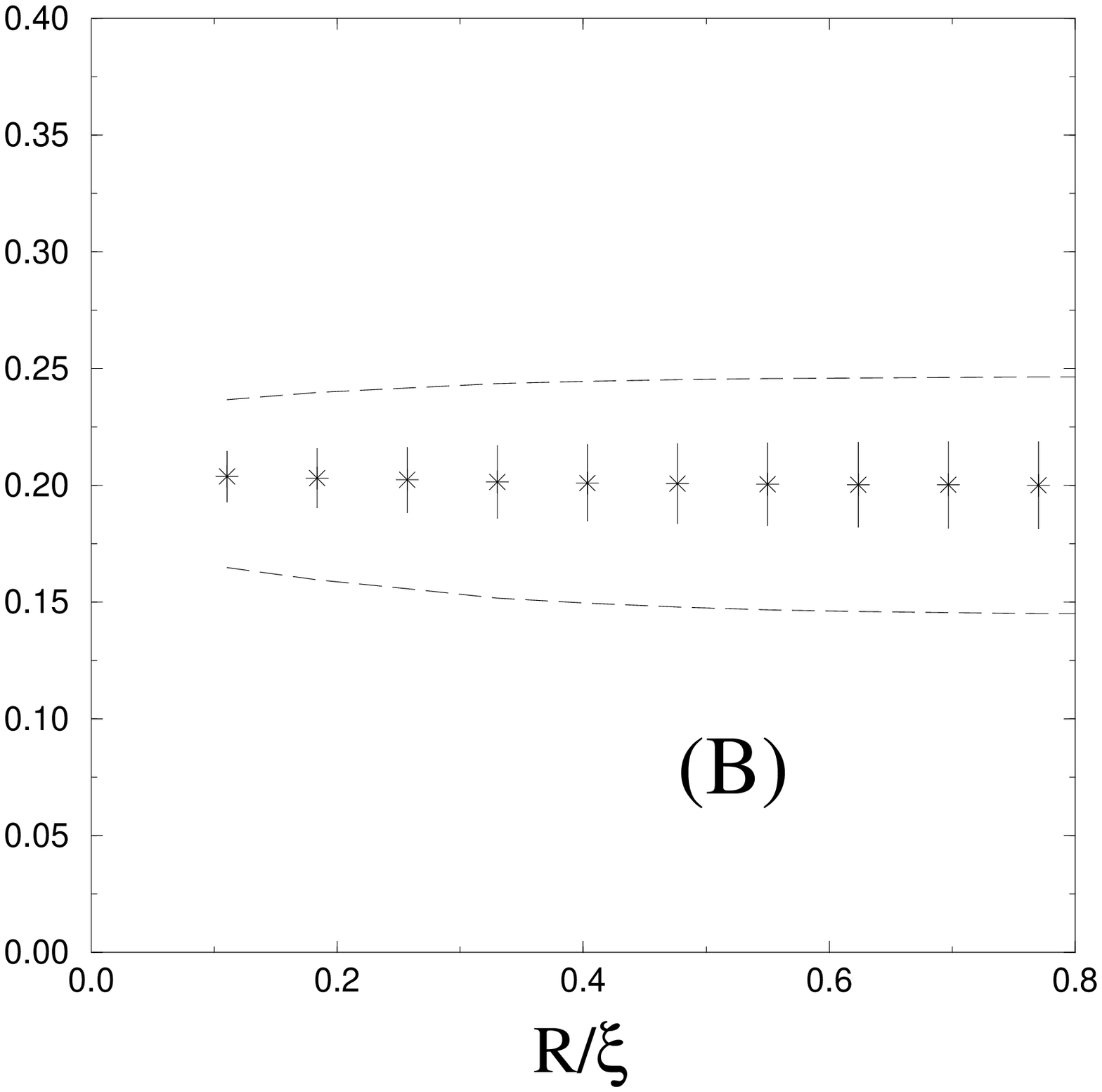,angle=-90,
width=0.5\linewidth}\\
\hspace{-1.5cm}
\epsfig{figure=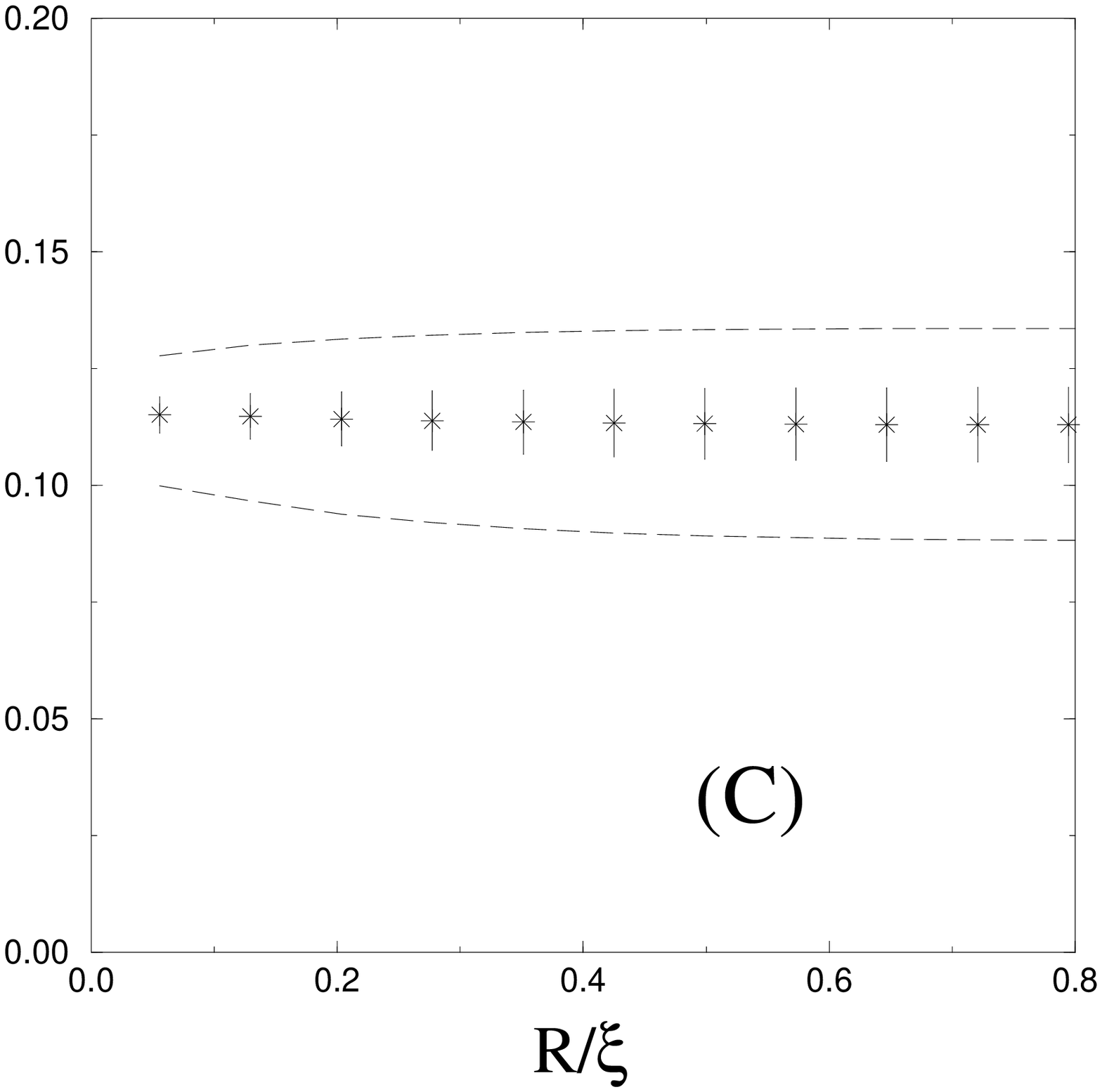,angle=-90,
width=0.5\linewidth}
\end{tabular}
}
\caption{The $R$ dependence of the fitting parameter 
$\widehat{\cal Z}^{(2)}(\rho,R)$, see Eq. (\ref{ZOPFittingForm}).
The three graphs (A), (B) and (C) have been obtained, respectively, on 
lattices \ref{Lattice64x128}, \ref{Lattice128x256} and \ref{Lattice256x512}.
We used three-loop (diamonds, $\Diamond$), and two-loop
(dashed lines) perturbation theory in the computation of the 
Wilson coefficient.
In the first case we plot the value obtained at $\zeta e^{\gamma}=1$
and the systematic error bars. In the second one we show the
maximum and the minimum values obtained in the chosen range of $\zeta$.}
\label{ZOPFig}
\end{figure}
%
%**********************************************
%
The renormalized matrix elements of Tab. \ref{RGISymmetricOperator} seem to
scale quite well. The results obtained with improved perturbation
theory change by $1\%$ when passing from lattice 
\ref{Lattice128x256} to lattice \ref{Lattice256x512}. 
They converge well when the order of the perturbative
calculation is increased from two to three loops.

However, we would like to have a nonperturbative control over
the renormalization constant of $S_0$. We shall 
consider the two-point function of this operator and proceed as in 
Sec. \ref{FieldSection}.
We shall limit ourselves to the first term of the OPE:
\begin{eqnarray}
\hspace{-0.5cm}\sum_{ab}
\left\<\left[S^{ab}_0\right]_{RGI}
\! (x) 
\left[S^{ab}_0\right]_{RGI}
\!(-x)
\right\> = \frac{N-1}{N} {\cal E}_{RGI}(\grz ;\zeta) + O(r^2) \, .
\label{TwoSpin2RGIOPE}
\end{eqnarray}
Using the perturbative result of Eq. (\ref{PerturbativeSymmSymm}),
and the general formulae of Sec. \ref{RenormalizationGroupSection},
we get 
\begin{eqnarray}
{\cal E}_{RGI}(\overline{g} ;1) & = & \overline{g}^{-2N/(N-2)}\left\{
1+\frac{N}{\pi(N-2)}\overline{g}-\frac{N(N^2-8N+4)}{8\pi^2(N-2)^2}
\overline{g}^2+\right.\\
&&+\frac{N}{96\pi^3(N-2)^3}[N^4(10\zeta(3)-7)+N^3(-64\zeta(3)+50)+
\nonumber\\
&&\hspace{0.75cm}+12N^2(11\zeta(3)-12)+N(-80\zeta(3)+264)-16(\zeta(3)+11)]
\overline{g}^3
\Bigg{ \} } \, .\nonumber
\end{eqnarray}
Let us recall our notation for the renormalization constants,
focusing on the case at hand:
\begin{eqnarray}
\left[\sigma^a\sigma^b-\frac{1}{Z}\frac{\delta^{ab}}{N}\right]_{RGI}=
Z_{L,RGI}^{(0,2)}(g_L)\left(\sigma^a_x\sigma^b_x-\frac{\delta^{ab}}{N}
\right)\, .
\label{ZOPRGIDefinition}
\end{eqnarray}
Equations (\ref{TwoSpin2RGIOPE}) and (\ref{ZOPRGIDefinition}) 
motivate the following fitting form for the isotensor correlation
function, see Eq. (\ref{TwoPointFunctions}):
\begin{eqnarray}
G_T(t,x) = \frac{N-1}{N}{\cal E}_{RGI}(\gr2z;\zeta){\cal Z}^{(2)}\, .
\label{ZOPFittingForm}
\end{eqnarray}
The fitting parameter ${\cal Z}^{(2)}$ gives an estimate of 
$Z_{L,RGI}^{(0,2)}(g_L)^{-2}$. As in the other cases we use a fitting 
window $\rho\le r\le R$ to extract the best fitting parameter 
$\widehat{\cal Z}^{(2)}(\rho,R)$.

%
%*******************************************
%
\begin{table}
\begin{center}
\begin{tabular}{|c|c|c|}
\hline
& $\widehat{\cal Z}^{(2),*}$
&$ {\cal S}^{np}$\\
\hline
\hline
lattice \ref{Lattice64x128} & $0.382[36]$&  $1.277(2)[61]$\\
\hline
lattice \ref{Lattice128x256}& $0.203[13]$&  $1.300(4)[40]$\\
\hline
lattice \ref{Lattice256x512}& $0.114[6]$ &  $1.275(8)[32]$\\
\hline
\end{tabular}
\end{center}
\caption{The OPE estimates $\widehat{\cal Z}^{(2),*}$ for the
constant $Z_{RGI,L}^{(0,2)}(g_L)^{-2}$, 
and the corresponding non-perturbatively renormalized
matrix element of the symmetric operator.}
\label{ZOPTab}
\end{table}
%
%*************************************************
%
In Fig. \ref{ZOPScaling} we show our Monte Carlo data for $G_T(t,x)$
in a scaling plot. The numerical results for $G_T(t,x)$ have been
rescaled using the estimated renormalization constant 
$\widehat{\cal Z}^{(2),*}$, see Tab. \ref{ZOPTab} and discussion
below. For sake of clarity we limit 
ourselves to showing the results obtained along the directions 
$(t,x)=(t,0)$ and $(t,x)=(t,t)$. Scaling is well verified on the three 
lattices. The continuous line corresponds to the leading OPE prediction 
$(N-1)/N\, {\cal E}_{RGI}(\gr2z;\zeta)$, with $\zeta e^{\gamma} = 1$.

In Fig. \ref{ZOPFig} we show the $R$ dependence of the best fitting parameter 
$\widehat{\cal Z}^{(2)}(\rho,R)$ on lattices \ref{Lattice64x128}, 
\ref{Lattice128x256} and \ref{Lattice256x512}. As in Sec. \ref{FieldSection},
we kept $\rho=0.5$ constant. In all the cases examined, the estimates
$\widehat{\cal Z}^{(2)}(\rho,R)$ are flat within the systematic error bars 
as soon as $R\ltapprox \xi$. Notice that the estimated systematic errors 
on $\widehat{\cal Z}^{(2)}(\rho,R)$ are larger than in Sec. 
\ref{FieldSection}. This is not unexpected. In fact, in the present case, the
Wilson coefficient is more strongly varying:
as $\overline{g}\to 0$,
${\cal E}_{RGI}(\overline{g};\zeta)\sim \overline{g}^{-6}$, 
while ${\cal F}^{(0)}_{RGI,0}(\overline{g};\zeta)
\sim \overline{g}^{-2}$ (these formulae hold for $N=3$). 
As a consequence, the parameter
$\widehat{\cal Z}^{(2)}(\rho,R)$ is more strongly dependent upon 
the perturbative truncation than its counterpart
$\widehat{\cal Z}(\rho,R)$.

In Tab. \ref{ZOPTab} we present our results for $Z_{L,RGI}^{(0,2)}(g_L)^{-2}$,
and the corresponding renormalized matrix elements for the symmetric 
operator, see Eq. (\ref{TableContent}). 
The values of $\widehat{\cal Z}^{(2),*}$ correspond
to $\widehat{\cal Z}^{(2)}(\rho,R)$ at $R = 1.5$, $2.5$, and $5.5$,
respectively
on lattices \ref{Lattice64x128}, \ref{Lattice128x256} and
\ref{Lattice256x512}. The renormalized matrix elements 
${\cal S}^{np}$ are obtained computing the renormalization constant
from $\widehat{\cal Z}^{(2),*}$, and using the
bare lattice matrix elements of  Tab. \ref{LatticeSymmetricOperator},
$\pb = 0$. These results scale, i.e. they are $g_L$ independent,
within systematic errors.
Systematic errors get reduced as the lattice becomes finer.
They are about the $3\%$ on lattice \ref{Lattice256x512}.

The results of Tab. \ref{ZOPTab}
should be compared with the ones of Tab. \ref{RGISymmetricOperator}
(we refer here to improved perturbation theory, i.e. to the 
rightmost column).
In both cases we present the matrix element 
${\cal S}$, defined in Eq. (\ref{TableContent}),
and we use the bare lattice data of Tab. \ref{LatticeSymmetricOperator}, 
obtained at $\pb=0$.
The only difference consists in the estimate of the renormalization constant.
In both cases, looking at the two larger lattices, 
\ref{Lattice128x256} and \ref{Lattice256x512}, the values
of ${\cal S}$ show scaling (i.e. they are independent of $g_L$)
at percent level.
There is, however, a discrepancy between the  determinations of 
Tab. \ref{RGISymmetricOperator} and of Tab. \ref{ZOPTab}.
This discrepancy is about the $10\%$ on lattice \ref{Lattice256x512}, 
and is not compatible with the  systematic error of the OPE method.

A $10\%$ disagreement between improved perturbation theory and 
non-perturbative results is not unfrequent at these
correlation lengths, see for instance Refs.
\cite{Caracciolo:1995ud,Caracciolo:1995ah,Alles:1997rr}.
Moreover the results of Tab. \ref{ZOPTab} agree with
the ones obtained by applying the OPE method to the computation of the matrix 
element, see Sec. \ref{SymmetricSection} and Tab.
\ref{SymmetricParametersRenorm}. This provides a strong check 
of the whole approach.
%
%******************************************************************
%
\subsection{OPE in the Scalar Sector}
\label{ScalarSection}
In this Subsection we study the short-distance product of two elementary
fields in the $O(N)$-scalar sector.

As a preliminary step, we rewrite the general form of the OPE,
see Eq. (\ref{OPEFieldsScalar}), with two changes:
we use RGI operators instead of \MS ones; we use operators with
definite spin.
Moreover,we focus on the $h\to 0$ limit of on-shell matrix elements.
Using Eq. (\ref{EnergyMomentumTrace}) for the trace of the 
energy-momentum tensor, we get (neglecting terms of order $r^4$):
\begin{eqnarray}
\sg_{RGI}(x)\cdot\sg_{RGI}(-x) & = &  
{\cal F}^{(0)}_{RGI,0}(\grz;\zeta)\mbox{\boldmath{1}}+\label{OPEScalar}\\
&&\widehat{\cal F}^{(0)}_{RGI,1}(\grz;\zeta) x_{\mu}x_\nu\widehat{T}_{\mu\nu}+
\widehat{\cal F}^{(0)}_{RGI,2}(\grz;\zeta) r^2 
\left[ (\partial\sg)^2\right]_{RGI}\, ,\nonumber
\end{eqnarray}
where $\widehat{T}_{\mu\nu}$ is the traceless energy-momentum tensor:
\begin{eqnarray}
\widehat{T}_{\mu\nu} = T_{\mu\nu}-\frac{1}{2}\delta_{\mu\nu}
\delta^{\alpha\beta} T_{\alpha\beta}\, .
\end{eqnarray}
The expressions
for ${\cal F}^{(0)}_{RGI,0}$ and  $\widehat{\cal F}^{(0)}_{RGI,2}$ 
have been already 
given in Sec. \ref{FieldSection}, see Eqs. (\ref{TwoPointResummed})
and (\ref{TwoPointResummedBIS}). The last Wilson coefficient 
is easily obtained from Eq. (\ref{F01Pert}):
\begin{eqnarray}
\widehat{\cal F}^{(0)}_{RGI,1}(\overline{g};1) = 
-2\overline{g}^{-1/(N-2)}\left[1+\frac{N-1}{2\pi(N-2)}\overline{g}
\right]\, .
\end{eqnarray}

We shall consider one-particle matrix elements of Eq. 
(\ref{OPEScalar}). Space-time symmetries impose several constraints
on the matrix elements of the operators on the right-hand side. 
This yields a further check of our calculation.
We adopt the following parametrization:
\begin{eqnarray}
\<\pb, a|\widehat{T}_{\mu\nu}|\pb,b\>&=&( 2p_{\mu}p_{\nu}-p^2\delta_{\mu\nu})
\delta^{ab}{\cal T}_R\, , \label{FitParameter01}\\
\<\pb, a|[(\partial \sg)^2]_{RGI}|\pb,b\> & = & -p^2{\cal E}_R\, ,
\label{FitParameter02}
\end{eqnarray}
where $p_{\mu}\equiv  (i\sqrt{\pb^2+m^2},\pb)$ and
$p^2 = p_0^2+p_1^2 = -m^2$.
Because of  kinematical considerations both ${\cal T}_R$ and ${\cal E}_R$
do not depend upon $\pb$.
Moreover,invariance under space and time inversions
implies that ${\cal T}_R$ and ${\cal E}_R$ are both real.
Finally, e know the exact value of the expectation
value of the energy-momentum tensor. Recalling the normalization
(\ref{OneParticleNormalization}) for one-particle states, 
we get ${\cal T}_R = 1$.

Let us make a few elementary remarks concerning  the status of 
the different terms appearing in Eq. (\ref{OPEScalar}).
The operator $(\partial\sg)^2$ has spin 0 and 
dimension 2. In the context of deep-inelastic scattering it would be called
a ``higher twist'' 
\footnote{Recall the definition  twist = dimension - spin.}.
It mixes under renormalization with the identity operator. 
The Lorentz structure of the corresponding 
Wilson coefficients is the same: they are rotationally invariant.
The remarks of Sec. \ref{OperatorDefSection} apply to this case.
The matrix elements of $(\partial\sg)^2$ cannot be determined
from the expansion (\ref{OPEScalar}) unless we fix the coefficient
${\cal F}^{(0)}_{RGI,0}$ up to $O(r^2)$. 
As a consequence, we do not expect to be able to compute the matrix
elements of $(\partial\sg)^2$, i.e. the parameter ${\cal E}$,
see Eq. (\ref{FitParameter02}), with our method\footnote{ 
Notice, however, that the  $\<\pb|\cdot|\pb\>$
matrix elements of the identity operator and 
of $(\partial\sg)^2$ have a different scaling with respect to 
the external momentum $\pb$. 
In the continuum limit, on a strip of spatial extent $L$,
we know that $\<\pb,a|{\bf 1}|\pb,b\>=2\delta^{ab}L\sqrt{\pb^2+m^2}$,
while $\<\pb,a|(\partial\sg)^2|\pb,b\>$ is $\pb$-independent.
This gives a clue to distinguish the two contributions.
We shall not pursue this strategy in this Section, since it
would require a statistical accuracy beyond the one of our 
Monte Carlo data.\label{FootSWave}}.

The traceless energy-momentum tensor $\widehat{T}_{\mu\nu}$ is instead a
leading twist (spin 2, dimension 2) and can be determined 
from the expansion (\ref{OPEScalar}), although it is only a power 
correction. We could, for instance, consider the quantity:
\begin{eqnarray}
S_{\mu\nu}(u) = \int\! \frac{d^2x}{2\pi x^2}
\,\delta(x^2-u^2) \left(2 x_{\mu}x_{\nu}
-x^2\delta_{\mu\nu}\right)\sg_{RGI}(x)\cdot\sg_{RGI}(-x)\, .
\label{Convolution}
\end{eqnarray}
From Eq. (\ref{OPEScalar}), it is easy to derive the following OPE:
\begin{eqnarray}
S_{\mu\nu}(u) = \frac{1}{4}u^2{\cal F}^{(0)}_{RGI,1}(
\overline{g}_{\zeta}(u);\zeta)
\widehat{T}_{\mu\nu}(0)\, ,
\end{eqnarray}
where $\widehat{T}_{\mu\nu}$ appears as the leading contribution.
In this particular case, however, we can use a more direct approach.
Since the leading term of the expansion (\ref{OPEScalar}) is proportional
to the identity operator, it cancels when considering 
connected correlation functions. In particular we could 
consider, see Eq. (\ref{GScalar}),
\begin{eqnarray}
\widehat{G}^{(0)}_c(t,x;\pb,\qb;2t_s) \equiv
\widehat{G}^{(0)}(t,x;\pb,\qb;2t_s)-G(2t,2x)\, .
\end{eqnarray}
From Eq. (\ref{OPEScalar}), it follows that 
$\widehat{G}^{(0)}_c(t,x;\pb,\qb;2t_s)$ is of order $r^2$ as $r\to 0$ 
(here $r=\sqrt{x^2+t^2}$). Nevertheless, in the following we shall study 
the whole OPE (\ref{OPEScalar}), without eliminating
the leading contribution.

\begin{figure}
\begin{tabular}{cc}
\hspace{0.0cm}\vspace{-1.5cm}\\
\hspace{-2.5cm}
\epsfig{figure=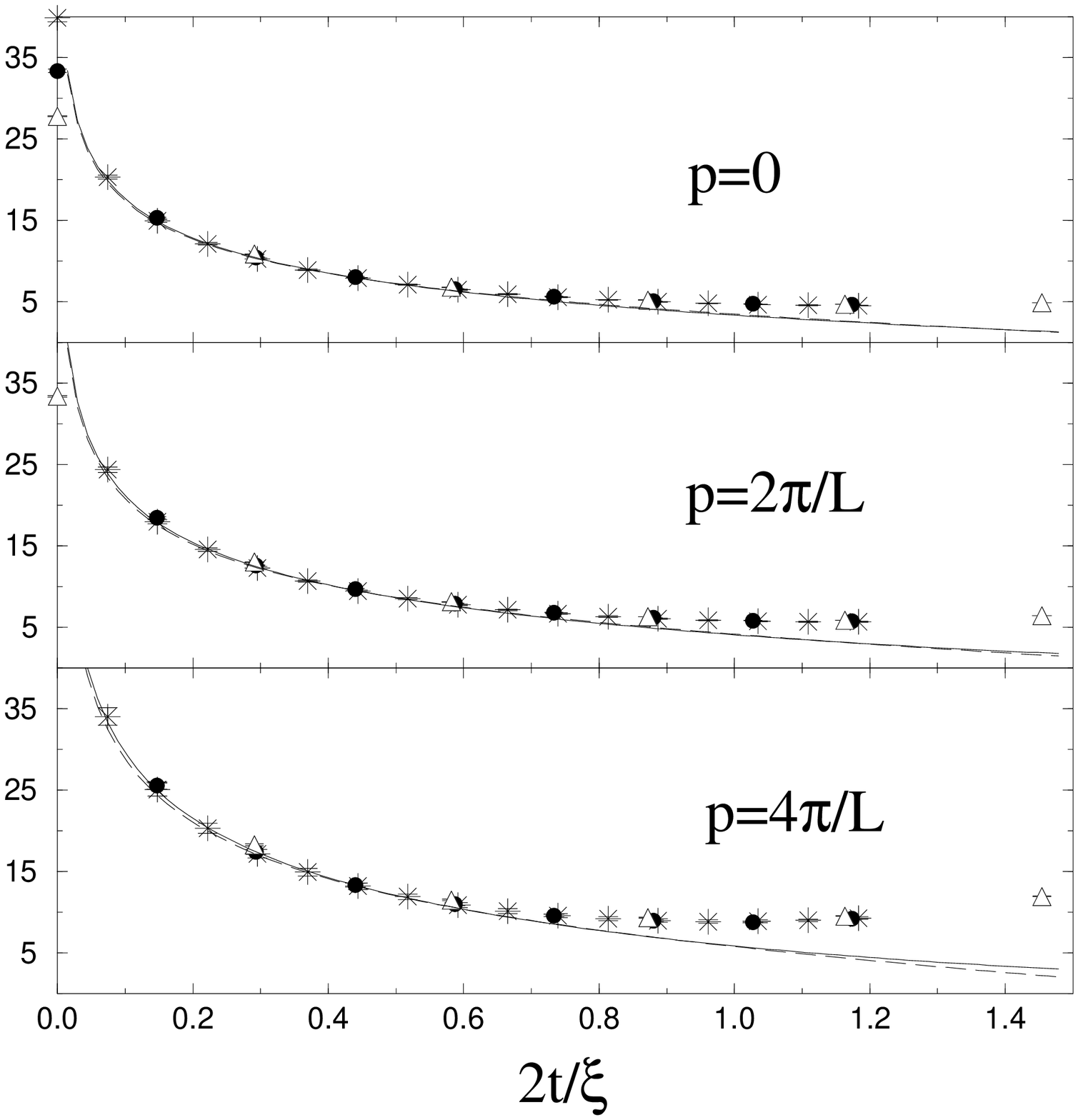,angle=-90,
width=0.6\linewidth}&
\hspace{-0.5cm}
\epsfig{figure=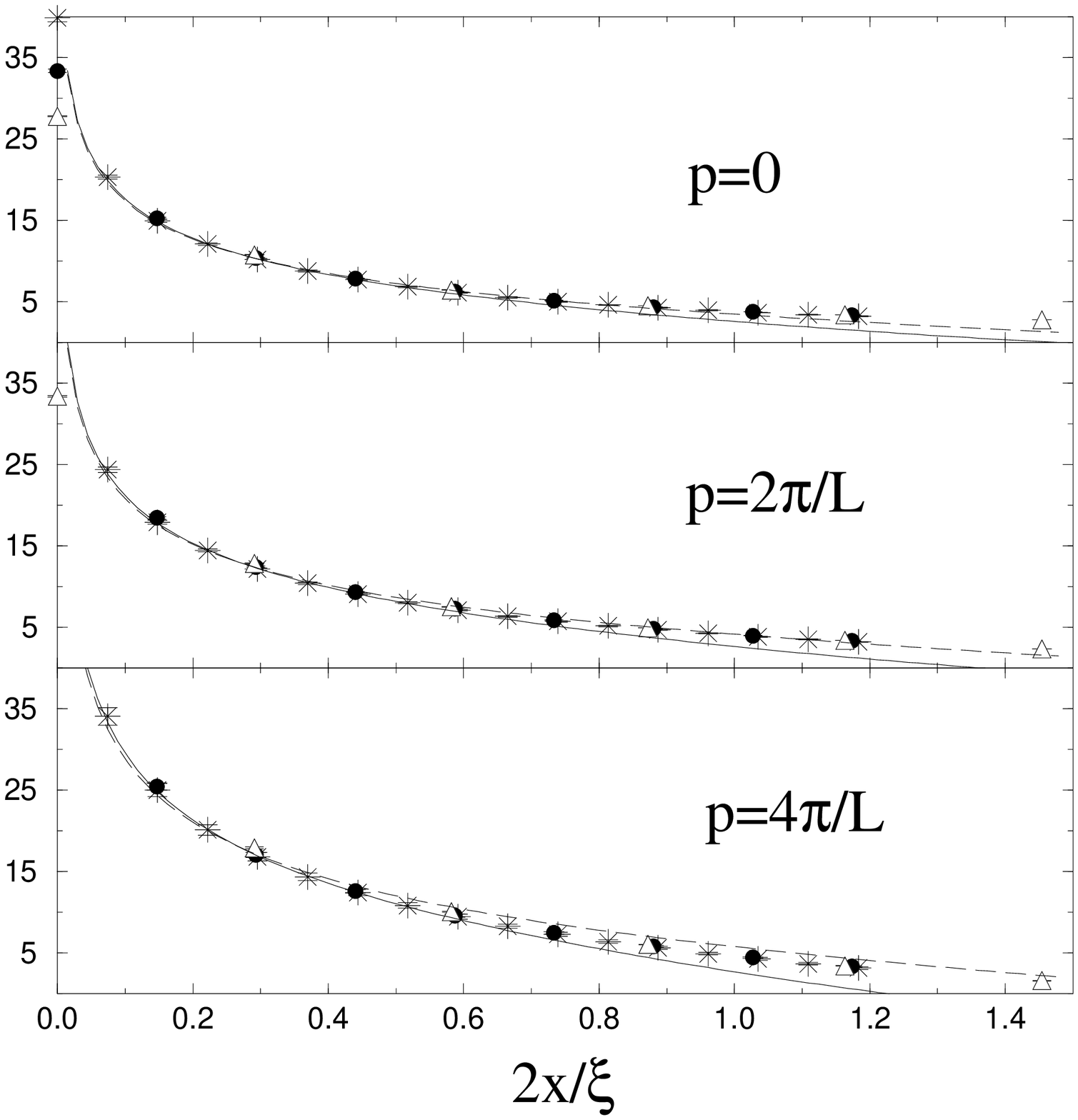,angle=-90,
width=0.6\linewidth}
\end{tabular}
\caption{The product of fields in the scalar sector (Monte Carlo results)
and the OPE prediction (best fitting curves).
We report $2\sqrt{\pb^2+m^2}Z_L^{-1}\widehat{G}^{(0)}(t,x; \pb;2t_s)$
on lattices \ref{Lattice64x128} (empty triangles, $\triangle$), 
\ref{Lattice128x256} (filled circles, $\bullet$) and 
\ref{Lattice256x512} (stars, $\ast$). We consider
$x=0$, $t\ne 0$ on the left, and $x\ne0$, $t= 0$ on the right.
The dashed curves are obtained with the leading term of the OPE, 
the continuous curves include power corrections.}
\label{ScalarFit}
\end{figure}
The one-particle matrix elements of the product on the 
l.h.s. of Eq. (\ref{OPEScalar}) can be obtained from the function 
$G^{(0)}(t,x;\pb,\qb;2t_s)$, see Eq. (\ref{GScalar}). Indeed we know that 
\begin{eqnarray}
\<\pb,a|\sg_{t,x}\cdot\sg_{-t,-x}|\qb,a\> =
\sqrt{4\omega(\pb)\omega(\qb)}\lim_{t_s\to\infty}
\widehat{G}^{(0)}(t,x;\pb,\qb;2t_s)\, .
\label{AsymptoticStatesScalar}
\end{eqnarray}
We computed $\widehat{G}^{(0)}(t,x;\pb,\pb;2t_s)$ for different values of $t_s$
(with $t_s\gtapprox \xi_{\rm exp}$, see Sec. \ref{ObservablesSection}) 
and verified it to be independent of $t_s$ in that range. 
This is compatible with the findings of Secs. \ref{OneParticleSectionBIS} and 
\ref{SymmetricOperatorSection}: the on-shell limit
for one-particle states is reached, with a good approximation,
at time separations $t_s\gtapprox \xi_{\rm exp}$.
We evaluated the limit on the r.h.s. of Eq. 
(\ref{AsymptoticStatesScalar}) using the lowest value of $t_s$
in the range considered in our Monte Carlo calculations. In particular 
we use $t_s = 8$ on lattice \ref{Lattice64x128},
$t_s = 16$ on lattice \ref{Lattice128x256}
and  $t_s = 30$ on lattice \ref{Lattice256x512}. The same procedure 
will be applied in the next Sections.
In Fig. \ref{ScalarFit} we compare the numerical results
obtained in this manner with the OPE fit.

The parametrization in Eqs. (\ref{FitParameter01}),
(\ref{FitParameter02}), and the OPE (\ref{OPEScalar}) imply the
following fitting form:
\begin{eqnarray}
2\sqrt{\pb^2+m^2}{\rm Re}\,\widehat{G}^{(0)}(t,x;\pb,\pb;2t_s) & = &
 2\sqrt{\pb^2+m^2}L{\cal F}_{RGI,0}^{(1)}(\grz;\zeta){\cal Z}'+
\label{FittingForm0p}\\
&&\hspace{-3cm}
+\widehat{\cal F}^{(1)}_{RGI,1}(\grz;\zeta)(m^2+2\pb^2)(x^2-t^2){\cal T}+
\widehat{\cal F}^{(1)}_{RGI,2}(\grz;\zeta)m^2r^2{\cal E}\, .
\nonumber
\end{eqnarray}
The renormalized parameters ${\cal T}_R$ and ${\cal E}_R$
are related to their unrenormalized counterparts ${\cal T}$ and ${\cal E}$
through the renormalization of the fields $\sigma$ on te l.h.s.
of Eq. (\ref{OPEScalar}). In particular we can estimate
${\cal T}_R$ and ${\cal E}_R$ using, respectively, $Z_L^{-1}{\cal T}$ and
$Z_L^{-1}{\cal E}$. The parameter ${\cal Z}'$ 
give access to the field-renormalization constant $Z_L$, analogously 
to the parameter ${\cal Z}$ in Sec. \ref{FieldSection}.

In Fig. \ref{ScalarFit} we 
plot $2\sqrt{\pb^2+m^2}Z_L^{-1}\widehat{G}^{(0)}(t,x; \pb;2t_s)$
on lattice \ref{Lattice64x128} ($t_s=8$), 
\ref{Lattice128x256} ($t_s=16$) and 
\ref{Lattice256x512} ($t_s=30$),
versus the separation between $\sigma(x)$ and $\sigma(-x)$
in physical units: $2r/\xi_{\rm exp}$. 
This function should have a finite $a\to 0$ limit
(at $2r/\xi_{\rm exp}$ fixed).
We used the values of $Z_L$ estimated with the OPE method
in Sec. \ref{FieldSection}, see 
Tab. \ref{ZFieldTable}, second column, and the results for 
$\xi_{\rm exp}$ of the previous Chapter, see Eq. (\ref{XiValues}).
The results obtained on lattices \ref{Lattice64x128}, 
\ref{Lattice128x256} and \ref{Lattice256x512} collapse
except for $r=0$, as expected. 
This fact indicates that we are in the scaling regime.

The best fitting curves shown in Fig. \ref{ScalarFit} have been obtained 
on lattice \ref{Lattice256x512}. We used a fitting window 
$\rho\le r\le R$, with $\rho = 0.5$ and $R=7.5$. The fit is quite good
for $2r/\xi\ltapprox 0.6$. 
We used the fitting form (\ref{FittingForm0p})
both with ${\cal Z}'$, ${\cal E}$, ${\cal T}$ free, and with
${\cal Z}'$ free and ${\cal E}={\cal T}=0$
(i.e. in this case we kept only the leading term of Eq. (\ref{FittingForm0p})).
The difference between the two
fitting procedure is hardly visible. The reason is that
the first term in Eq. (\ref{FittingForm0p}) is of order 
$L/\xi$ with respect to the other ones in the thermodynamic limit
($L\to\infty$ at fixed $g_L$).

\begin{figure}
\begin{tabular}{cc}
\hspace{0.0cm}\vspace{-1.5cm}\\
\hspace{-3.2cm}
\epsfig{figure=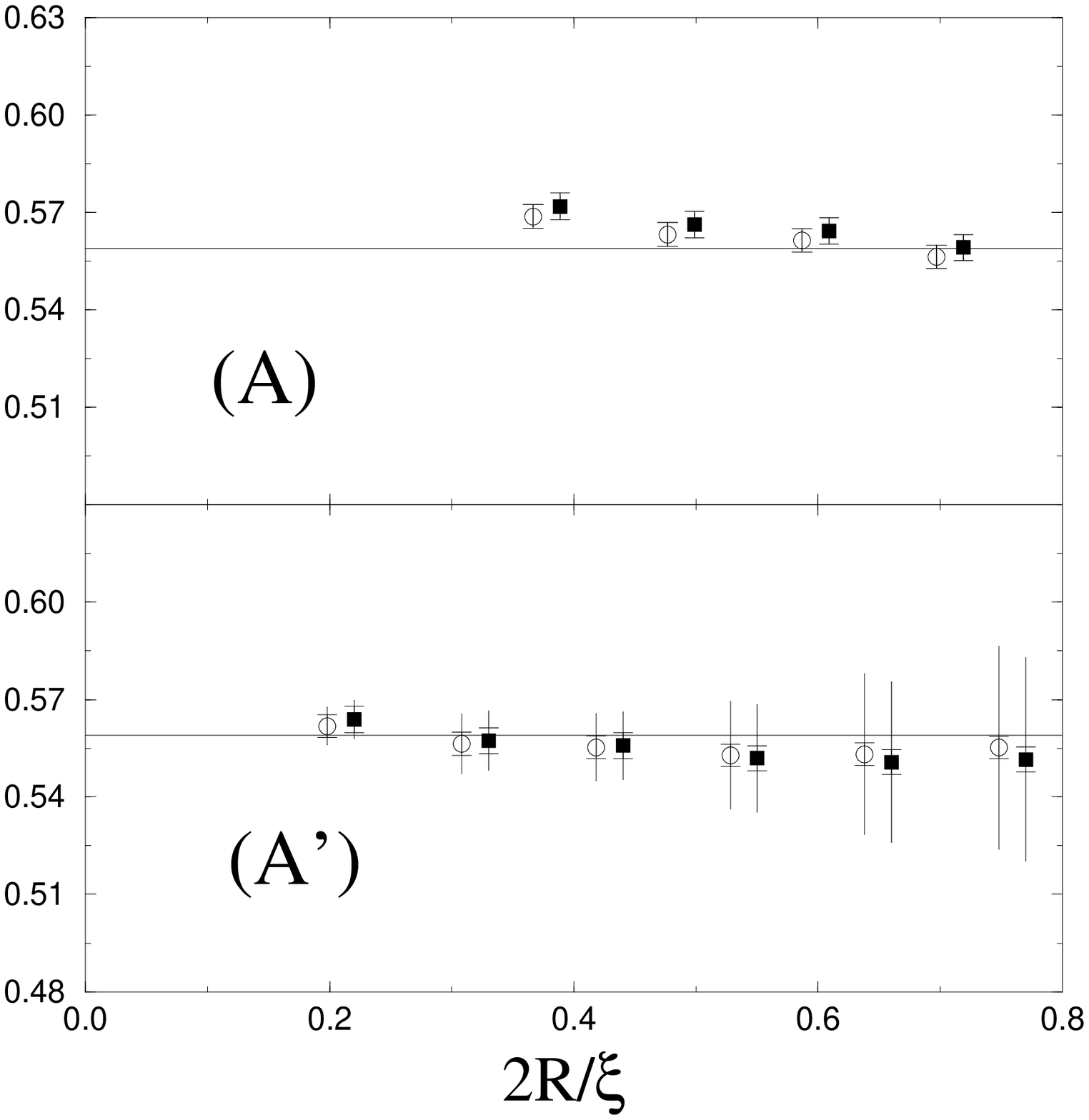,angle=-90,
width=0.7\linewidth}&
\hspace{-1.75cm}
\epsfig{figure=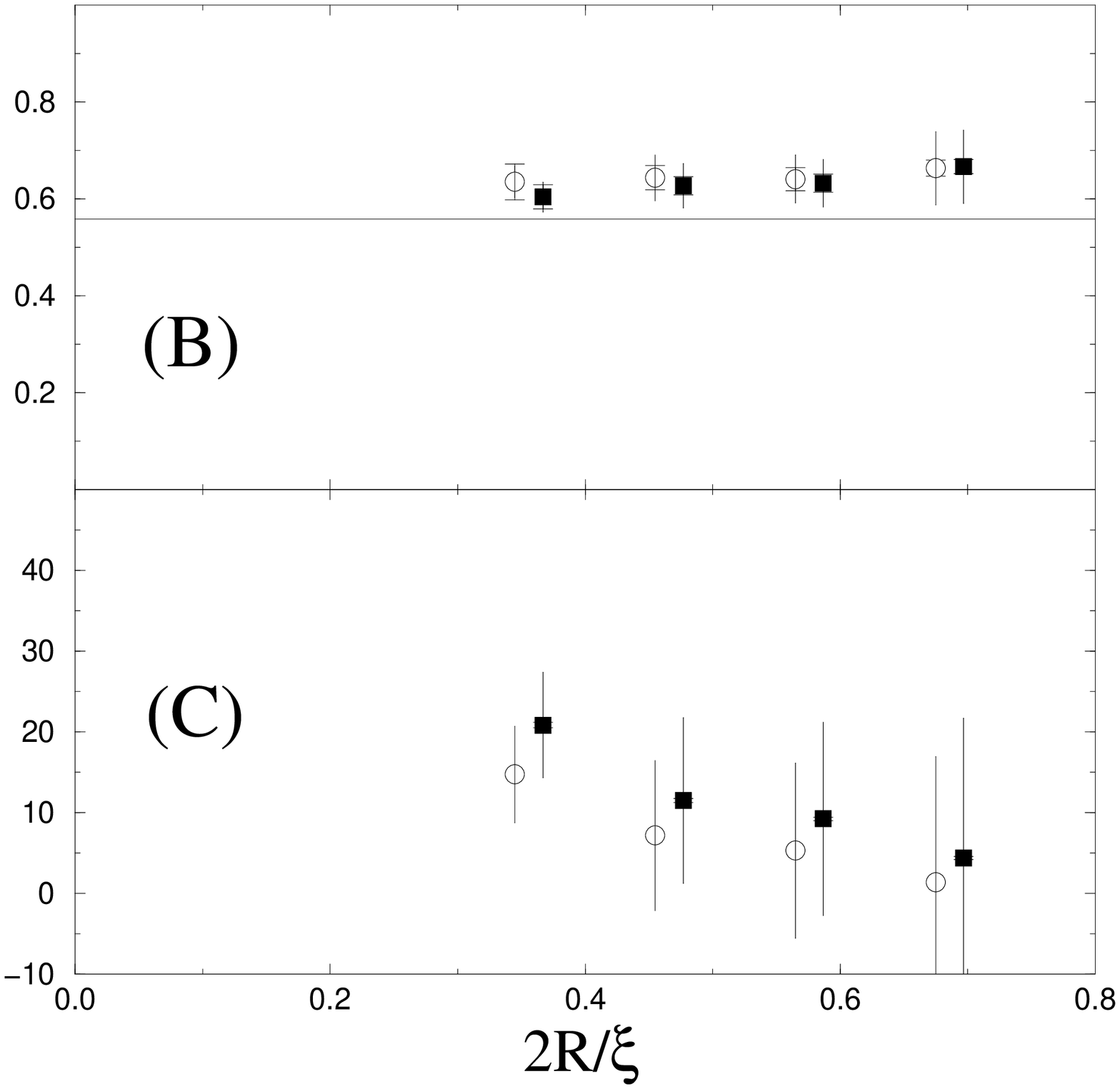,angle=-90,
width=0.6\linewidth}
\end{tabular}
\caption{The best fitting parameters 
$\widehat{\cal Z}'(\rho,R)$ (graphs (A) and (A')), 
$\widehat{\cal T}(\rho,R)$ (graph (B)) and  ${\cal E}(\rho, R)$ 
(graph (C)) on lattice \ref{Lattice128x256}. 
Different symbols refer to different external momenta:
$\pb = 0$ (empty circles) or $\pb = 2\pi/L$ (filled squares).
The continuous horizontal lines in graphs (A), (A') and (B) correspond
to the prediction ${\cal Z'}=Z_L$, ${\cal T}=Z_L$, 
with $Z_L\approx 0.559$ as estimated in Tab. \ref{ZFieldTable}.}
\label{ScalarParametersB}
\end{figure}
\begin{figure}
\begin{tabular}{cc}
\hspace{0.0cm}\vspace{-1.5cm}\\
\hspace{-2.5cm}
\epsfig{figure=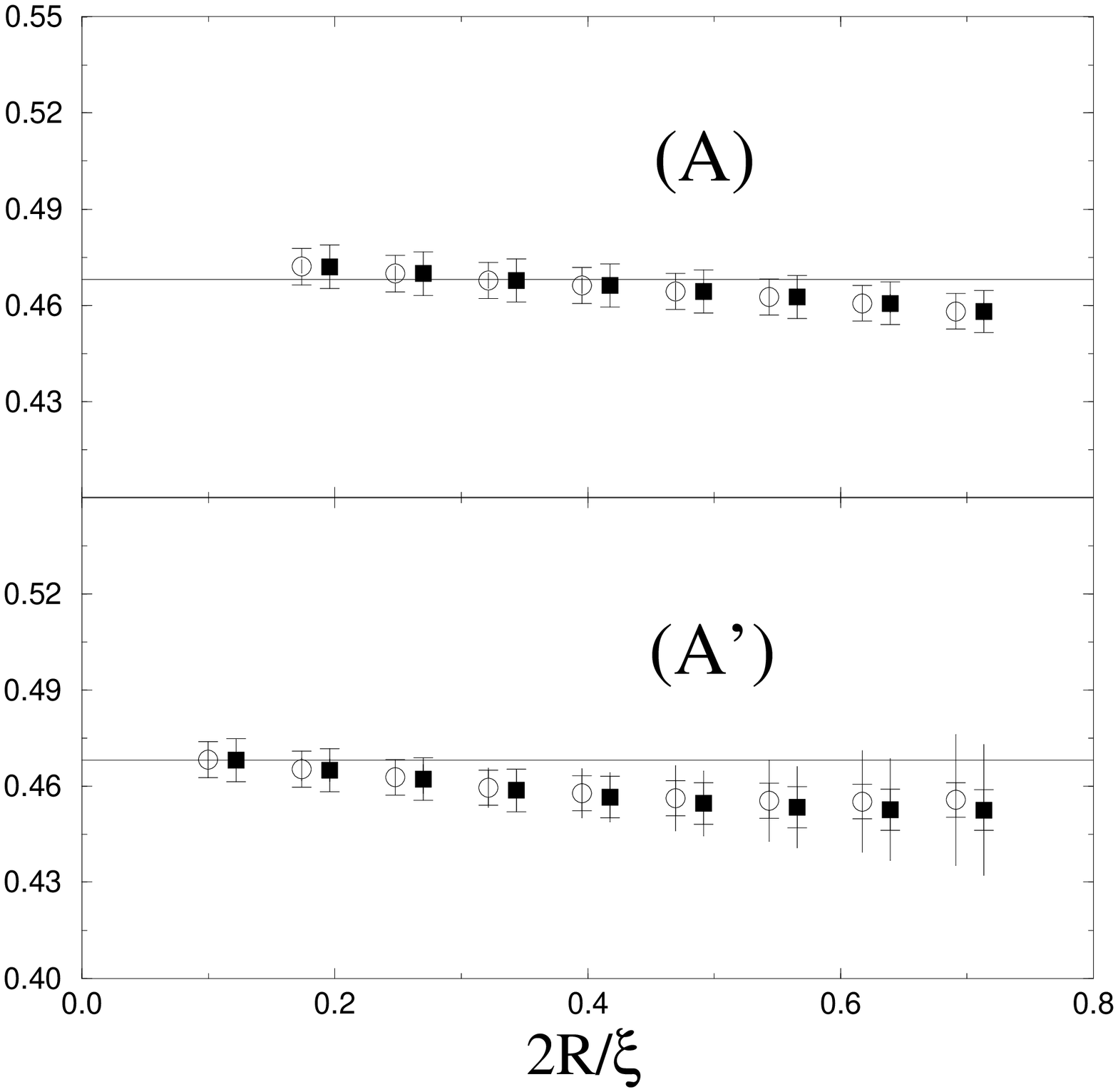,angle=-90,
width=0.6\linewidth}&
\hspace{-0.5cm}
\epsfig{figure=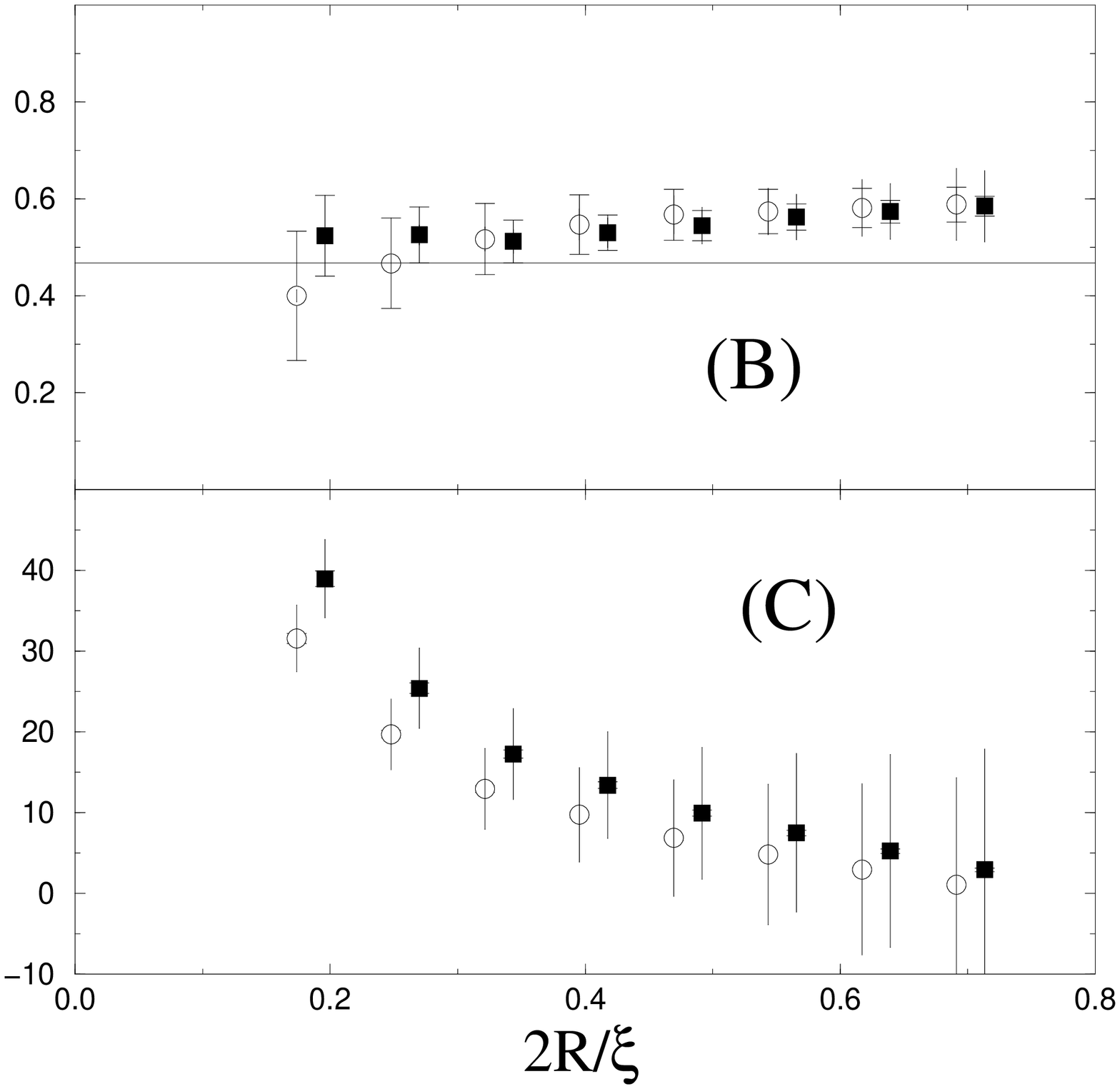,angle=-90,
width=0.6\linewidth}
\end{tabular}
\caption{As in Fig. \ref{ScalarParametersB} on lattice \ref{Lattice256x512}.}
\label{ScalarParametersC}
\end{figure}

In Figs. \ref{ScalarParametersB} and \ref{ScalarParametersC}
we study the dependence of the fit
parameters upon $R$. Our aim is to understand whether a 
window for asymptotic scaling exists. We plot the best fitting values 
obtained with $\zeta e^{\gamma} = 1$ together with the statistical and 
systematic uncertainties. For sake of clarity we limit ourselves to showing
the results obtained with $\pb=0,2\pi/L$ (the results for 
$\pb = 4\pi/L$ have larger statistical errors).
In graphs (A), (B), (C) we used the fitting form
(\ref{FittingForm0}) including power corrections.
In (A') we kept only the leading term of Eq. (\ref{FittingForm0}), i.e. 
${\cal Z}'$. 

The remarks formulated in Sec. \ref{FieldSection} can be repeated
here. Statistical errors on our numerical data are, also in this case, 
quite small. Systematic errors on the leading operator
are strongly reduced if power-correction terms are included 
in the fitting form. Nevertheless, as we already discussed in
Sec. \ref{FieldSection}, this is a somewhat ``spurious'' effect.
%
%******************************************************************
%
\subsection{OPE in the Antisymmetric Sector}
\label{AntisymmetricSection}

We consider now the antisymmetric product of two 
elementary fields at short distances.

We start by rewriting the form of the OPE in terms of 
RGI operators, cf. Eq. (\ref{OPEFieldsAntisymmetric}):
\begin{eqnarray}
\sigma^{[a}_{RGI}(x)\sigma^{b]}_{RGI}(-x) & = & 
-2x_{\mu} {\cal F}^{(1)}_{RGI}(\grz ;\zeta)\, j^{ab}_{\mu}(0)\, ,
\label{OPEAsymm}
\end{eqnarray}
where we neglected $O(r^3)$ terms.
Notice that the Noether current $j^{ab}_{\mu}$ is RGI 
(this happens for any regularization and
any renormalization scheme). As a consequence we did not add any 
subscript to it. The Wilson coefficient is easily obtained 
from Eq. (\ref{F10Pert}) using the formulae of Sec. 
\ref{RenormalizationGroupSection}:
\begin{eqnarray}
{\cal F}^{(1)}_{RGI}(\overline{g} ;1) & = & 
g^{-1/(N-2)}\left\{
1+\frac{N-1}{2\pi(N-2)}\overline{g}-
\frac{2N^3-13N^2+24N-14}{16\pi^2(N-2)^2}\overline{g}^2
\right\}\, .\nonumber \\
\label{WilsonAntiSymmetric}
\end{eqnarray}

Before continuing we remark that the $O(N)$ symmetry fixes the 
normalization of the Noether current. This can be seen 
by considering the $O(N)$ charges $Q^{ab}$, 
and requiring that $Q^{ab}$ generates the $O(N)$ transformations:
\begin{eqnarray}
Q^{ab}|\pb,c\> = \delta^{bc}|\pb,a\>-\delta^{ac}|\pb,b\>\, ,\quad
Q^{ab} \equiv \int_{-\infty}^{+\infty}\!\!\! dx\; j^{ab}_0(t,x)\, .
\end{eqnarray}
Using Lorentz invariance and the above condition we have:
\begin{eqnarray}
\frac{1}{N}\sum_{a,b}\<\pb,a|j^{ab}_{\mu}|\pb,b\> = -2ip_{\mu}(N-1)\, .
\label{CurrentNormalization}
\end{eqnarray}
This identity allows a tight check of the expansion (\ref{OPEAsymm}).

The OPE (\ref{OPEAsymm}) is quite different from the other examples 
studied in this Chapter. Since there exists no dimension-zero antisymmetric
operator, the leading term is of order $O(r|\log r|^p)$.
In the other cases we have a much weaker $r$ dependence:
$O(|\log r|^p)$.

%
%*******************************************************************
%
\begin{figure}
\begin{tabular}{cc}
\hspace{0.0cm}\vspace{-1.5cm}\\
\hspace{-2.5cm}
\epsfig{figure=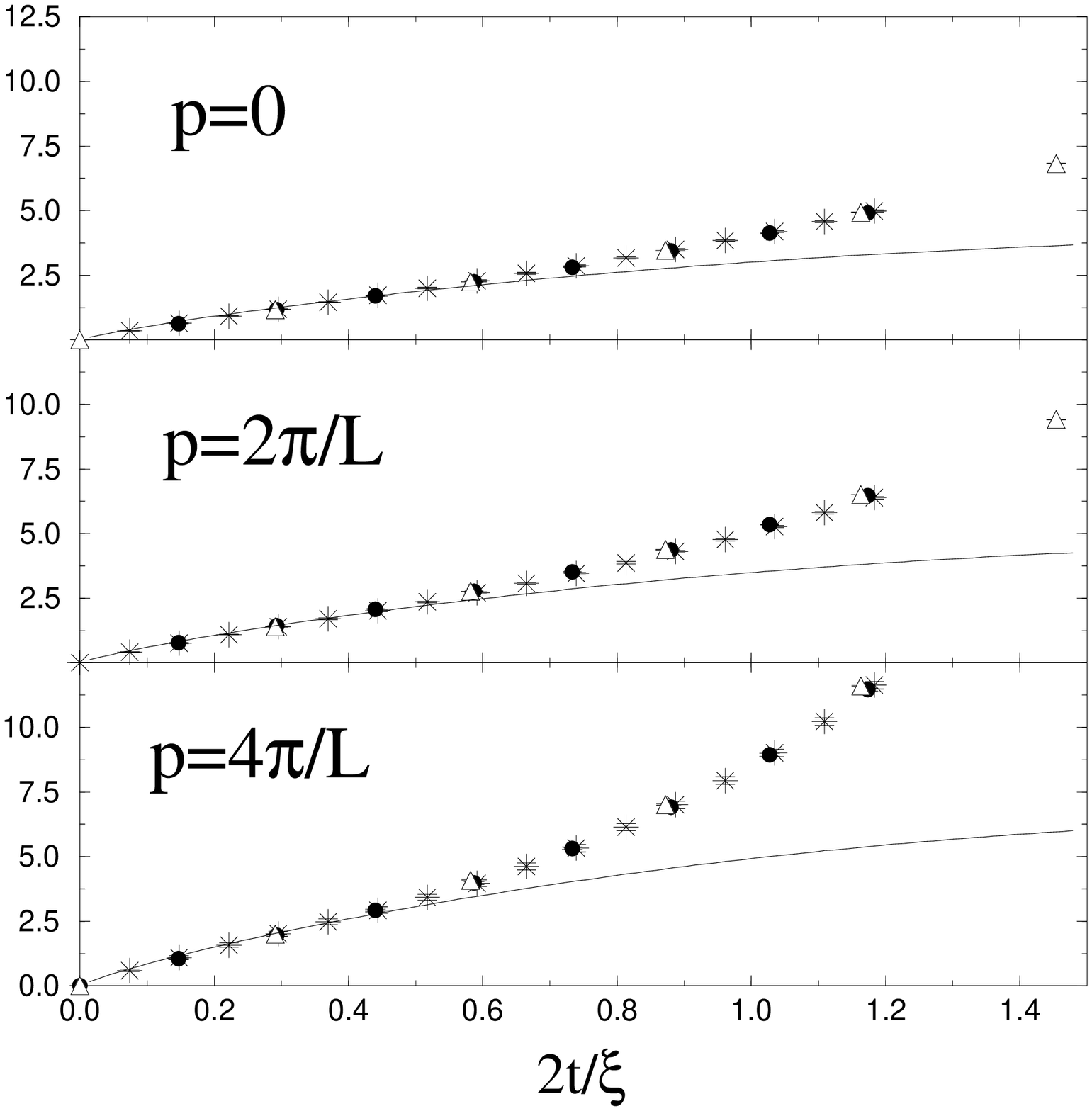,angle=-90,
width=0.6\linewidth}&
\hspace{-0.5cm}
\epsfig{figure=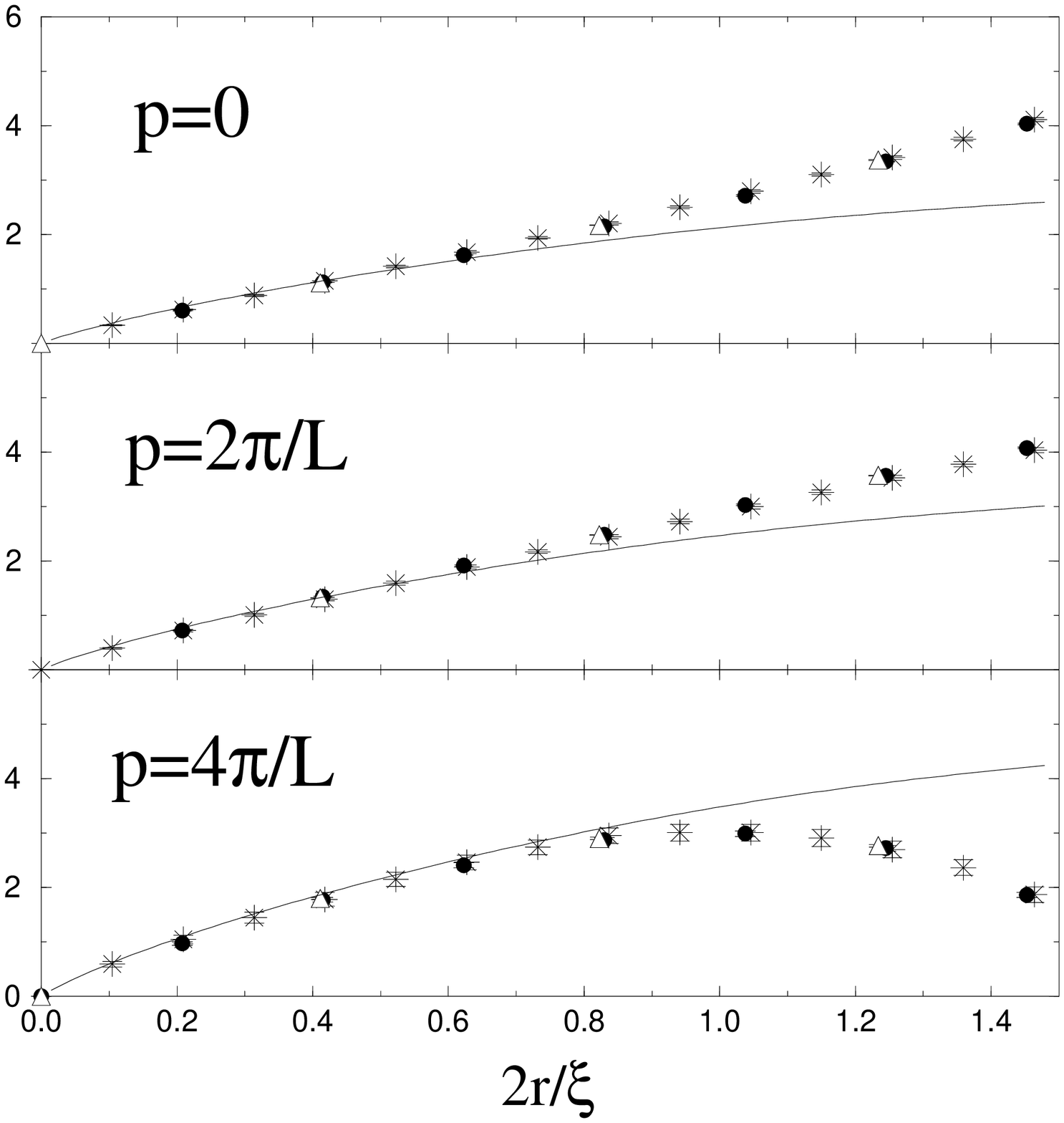,angle=-90,
width=0.6\linewidth}
\end{tabular}
\caption{Monte Carlo results and OPE fit for the product of fields 
in the antisymmetric sector.
We report $2\sqrt{\pb^2+m^2}Z_L^{-1}\widehat{G}^{(1)}(t,x; \pb;2t_s)$
on lattices \ref{Lattice64x128} (empty triangles, $\triangle$), 
\ref{Lattice128x256} (filled circles, $\bullet$) and 
\ref{Lattice256x512} (stars, $\ast$). We consider
$x=0$, $t\ne 0$ on the left, and $x =t =r/\sqrt{2}$ on the right.
The continuous curves are fits to the leading term of the OPE.}
\label{AntiSymmetricFit}
\end{figure}
%
%*********************************************************************
%
The one-particle matrix elements have been extracted as 
explained in the previous Subsection, see
Eq. (\ref{AsymptoticStatesScalar}).
We rewrite here the relevant equation in order to specify the correct 
normalization
\begin{eqnarray}
\sum_{a,b}\<\pb,a|\sigma^{[a}_{t,x}\sigma^{b]}_{-t,-x}|\qb,b\>=
N\sqrt{4\omega(\pb)\omega(\qb)}\lim_{t_s\to\infty}
\widehat{G}^{(1)}(t,x;\pb,\qb;2t_s)\, .
\end{eqnarray}

In Fig. \ref{AntiSymmetricFit} we show the numerical results for
the $\<\pb|\cdot|\pb\>$ matrix elements. We plot the function 
$2\sqrt{\pb^2+m^2}Z_L^{-1}{\rm Re}\,\widehat{G}^{(1)}(t,x;\pb,\pb;2t_s)$ 
along the  directions $(t,x)=(t,0)$ and $(t,x)=(t,t)$. 
In this case a plot along the direction $(t,x)=(0,x)$ would be trivial, since 
${\rm Re}\,\widehat{G}^{(1)}(0,x; \pb,\pb;2t_s)=0$.
The non-perturbative results of Sec. \ref{FieldSection},
see Tab. \ref{ZFieldTable}, have been used to estimate $Z_L$.

The OPE (\ref{OPEAsymm}) and the identity (\ref{CurrentNormalization}) 
imply the following fitting form
\begin{eqnarray}
2\sqrt{\pb^2+m^2}{\rm Re}\, \widehat{G}^{(1)}(t,x;\pb,\pb;2t_s) = 
2\sqrt{\pb^2+m^2}(N-1){\cal F}^{(1)}(\grz;\zeta)\, t\, {\cal Z}''\, ,
\label{AntiSymmetricFittingForm}
\end{eqnarray}
where ${\cal Z}''$ is an estimate of the field-renormalization constant
(analogously to ${\cal Z}$ in Sec. \ref{FieldSection}, and ${\cal Z}'$ in 
Sec. \ref{ScalarSection}). In Fig. \ref{AntiSymmetricFit} we show
the best fitting curves of the form (\ref{AntiSymmetricFittingForm}), 
as determined on lattice \ref{Lattice256x512}. 
We used a fitting window $\rho\le r\le R$ with $\rho = 0.5$ and 
$R=7.5$. The collapse of the data obtained on different lattices
indicates that scaling is well verified by our numerical results 
for $\widehat{G}^{(1)}$. 
Moreover, the fitting curves are in good agreement with numerical data up to
$2r/\xi\sim 0.5$.

%
%****************************************************************
%
\begin{figure}
\begin{tabular}{cc}
\hspace{0.0cm}\vspace{-1.5cm}\\
\hspace{-2.5cm}
\epsfig{figure=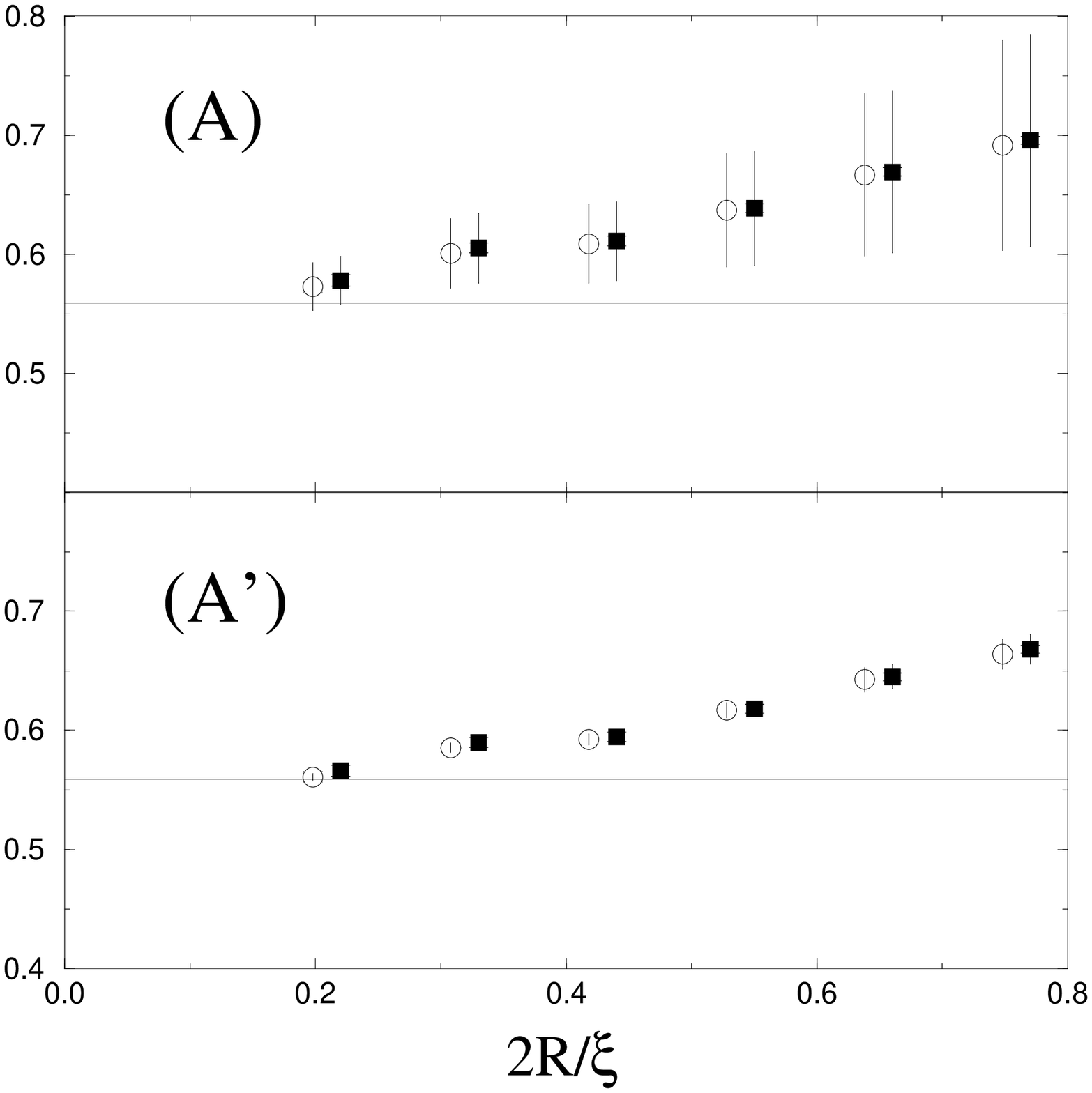,angle=-90,
width=0.6\linewidth}
&\hspace{-0.5cm}
\epsfig{figure=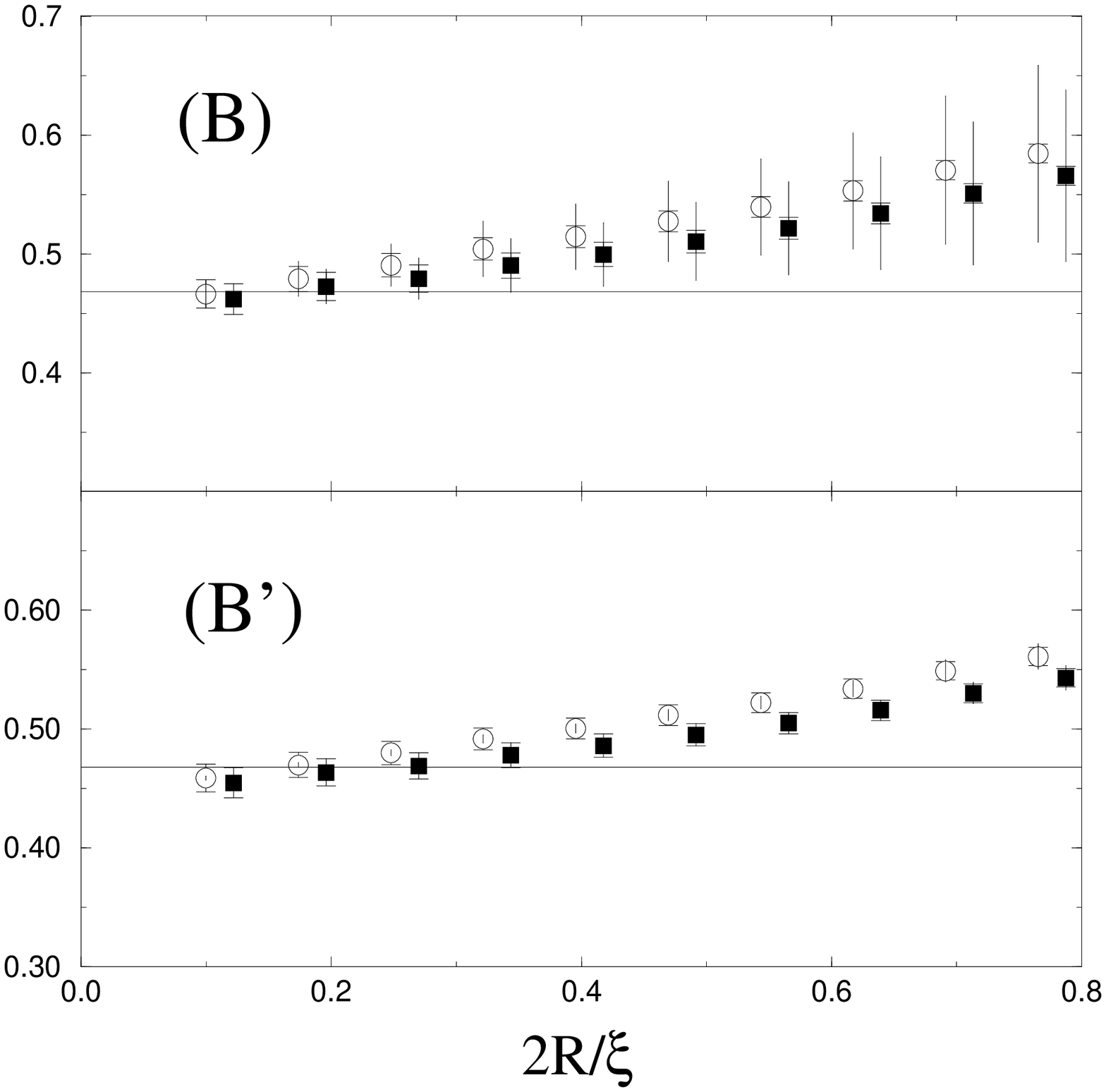,angle=-90,
width=0.6\linewidth}
\end{tabular}
\caption{
The best fitting parameter
$\widehat{\cal Z}''(\rho,R)$ on lattice \ref{Lattice128x256}
(graphs (A) and (A')), and lattice \ref{Lattice256x512}
(graphs (B) and (B')). 
We use one-loop perturbation theory in graphs (A) and (B),
and two-loop perturbation theory in (A') and (B').
Different symbols refer to different external momenta:
$\pb = 0$ (empty circles) or $\pb = 2\pi/L$ (filled squares).
The horizontal lines correspond to the theoretical prediction
${\cal Z}''(\rho,R) = \widehat{\cal Z}^*$, see Tab. \ref{ZFieldTable}.
}
\label{AntiSymmetricParameter}
\end{figure}
%
%*********************************************************************
%
In Fig. \ref{AntiSymmetricParameter} we study the $R$ dependence of 
the best fitting parameter $\widehat{\cal Z}''(\rho,R)$, keeping
$\rho = 0.5$ fixed. In graphs (A) and (B) we present the results obtained
with the one-loop Wilson coefficient, which is given by
dropping out the $O(\overline{g}^2)$ term in
Eq. (\ref{WilsonAntiSymmetric}). In graphs (A') and (B') we
use the two-loop Wilson coefficient (\ref{WilsonAntiSymmetric}).
The continuous lines refer to the field renormalization constant as 
computed in Sec. \ref{FieldSection}.

One-loop results, i.e. graphs (A) and (B), are almost flat within 
the systematic errors and agree with the prediction of
Sec. \ref{FieldSection} for the field-renormalization constant.
Our estimate of the systematic errors seems to be quite good in this case. 

\begin{table}
\centerline{
\begin{tabular}{|c|c|c|c|c|}
\hline
 &\multicolumn{3}{|c|}{$\widehat{\cal Z}''(\rho,R)$}
& $\widehat{\cal Z}^*$\\
\hline
 &$\pb=0$ &$\pb=2\pi/L$ &$\pb=4\pi/L$ & \\
\hline
\hline
lattice \ref{Lattice64x128} & $0.791(2)[14]$ & $0.787(1)[14]$ 
& $0.783(4)[14]$ & $0.67[1]$\\
\hline
lattice \ref{Lattice128x256}& $0.585(4)[4]$ & $0.590(4)[4]$ 
& $0.577(12)[4]$ &  $0.559[5]$\\
\hline
lattice \ref{Lattice256x512}& $0.480(10)[3]$ & $0.469(11)[2]$ 
& $0.484(26)[3]$ & $0.468[3]$\\
\hline
\end{tabular}
}
\caption{The field renormalization constant as computed from the 
numerical results in the antisymmetric sector with the
fitting form (\ref{AntiSymmetricFittingForm}). Notice that in this
case we found \underline{no} scaling window. 
We quote the best fitting parameters 
obtained with $R=2.5$ on lattices \ref{Lattice64x128} and \ref{Lattice128x256},
and $R=3.5$ on lattice \ref{Lattice256x512}.
For sake of comparison, we report in the fifth column the estimates of the 
field-renormalization constant obtained in Sec. \ref{FieldSection}, 
cf. Tab. \ref{ZFieldTable}.}
\label{AntiSymmetricParameterTable}
\end{table}  
Two-loop results, i.e. graphs (A') and (B') 
do not differ much from their one-loop counterparts, 
as far as the central value for $\widehat{\cal Z}''(\rho, R)$ 
(corresponding to $\zeta e^{\gamma}=1$) is concerned. 
However systematic errors are greatly reduced. 
They are much smaller than systematic errors obtained in the  
scalar or symmetric sectors with the same number of loops, 
at the same values of $R$. As a consequence the two-loop 
results for $\widehat{\cal Z}''(\rho, R)$ are no longer flat.
Looking at Fig. \ref{AntiSymmetricParameter}, graphs (A') and (B'),
we cannot find any ``scaling window''.
Nevertheless,  for small $R$, $\widehat{\cal Z}''(\rho, R)$
seems to converge to the field-renormalization constant computed in 
Sec. \ref{FieldSection}. 
In Tab. \ref{AntiSymmetricParameterTable}
we report the values of $\widehat{\cal Z}''(\rho, R)$ obtained with $R=2.5$ 
on lattices \ref{Lattice64x128} and \ref{Lattice128x256},
and $R=3.5$ on lattice \ref{Lattice256x512}.

%
%****************************************************************
%
\begin{figure}
\begin{tabular}{cc}
\hspace{0.0cm}\vspace{-1.5cm}\\
\hspace{-2.5cm}
\epsfig{figure=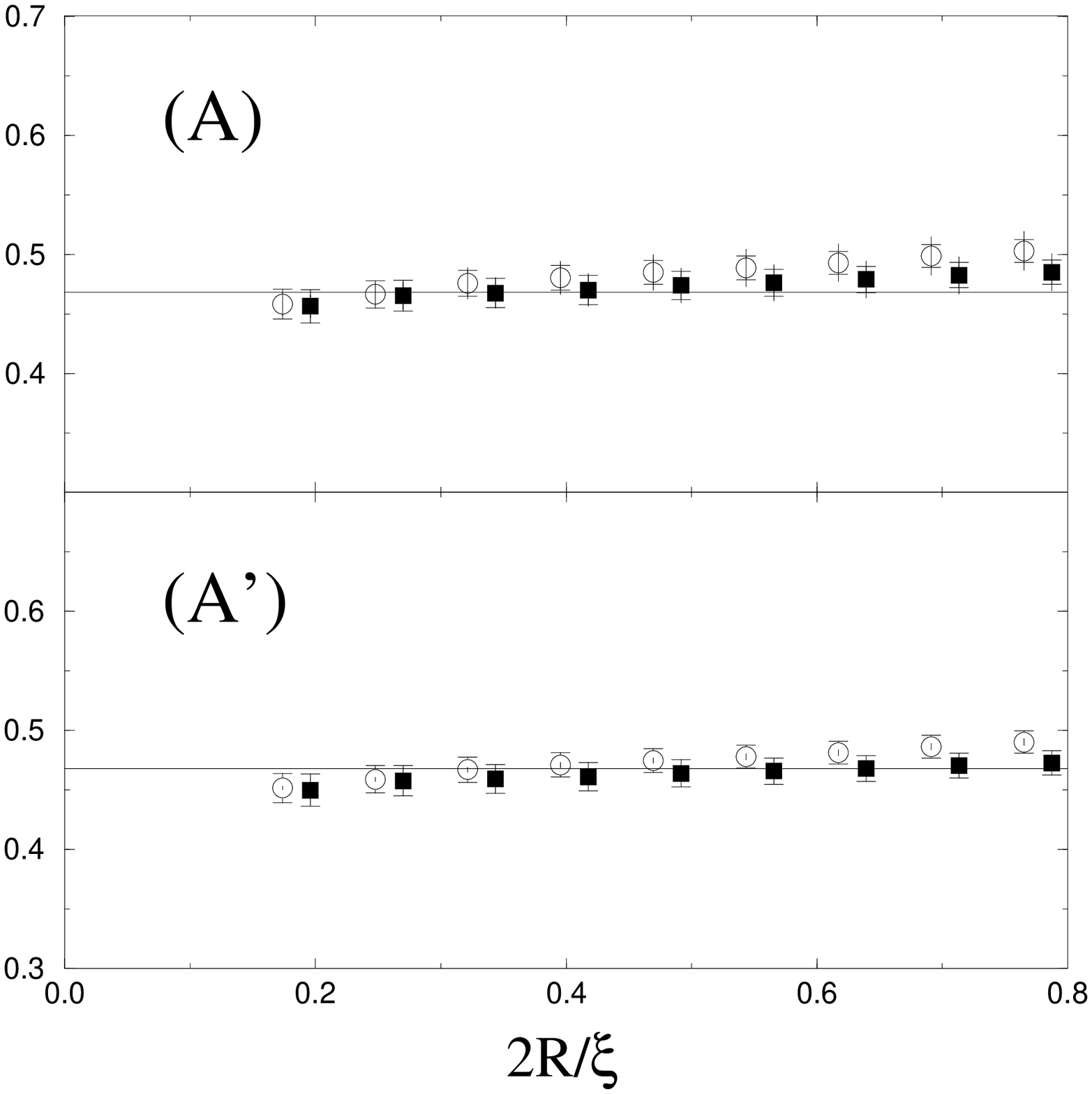,angle=-90,
width=0.6\linewidth}
&\hspace{-0.5cm}
\epsfig{figure=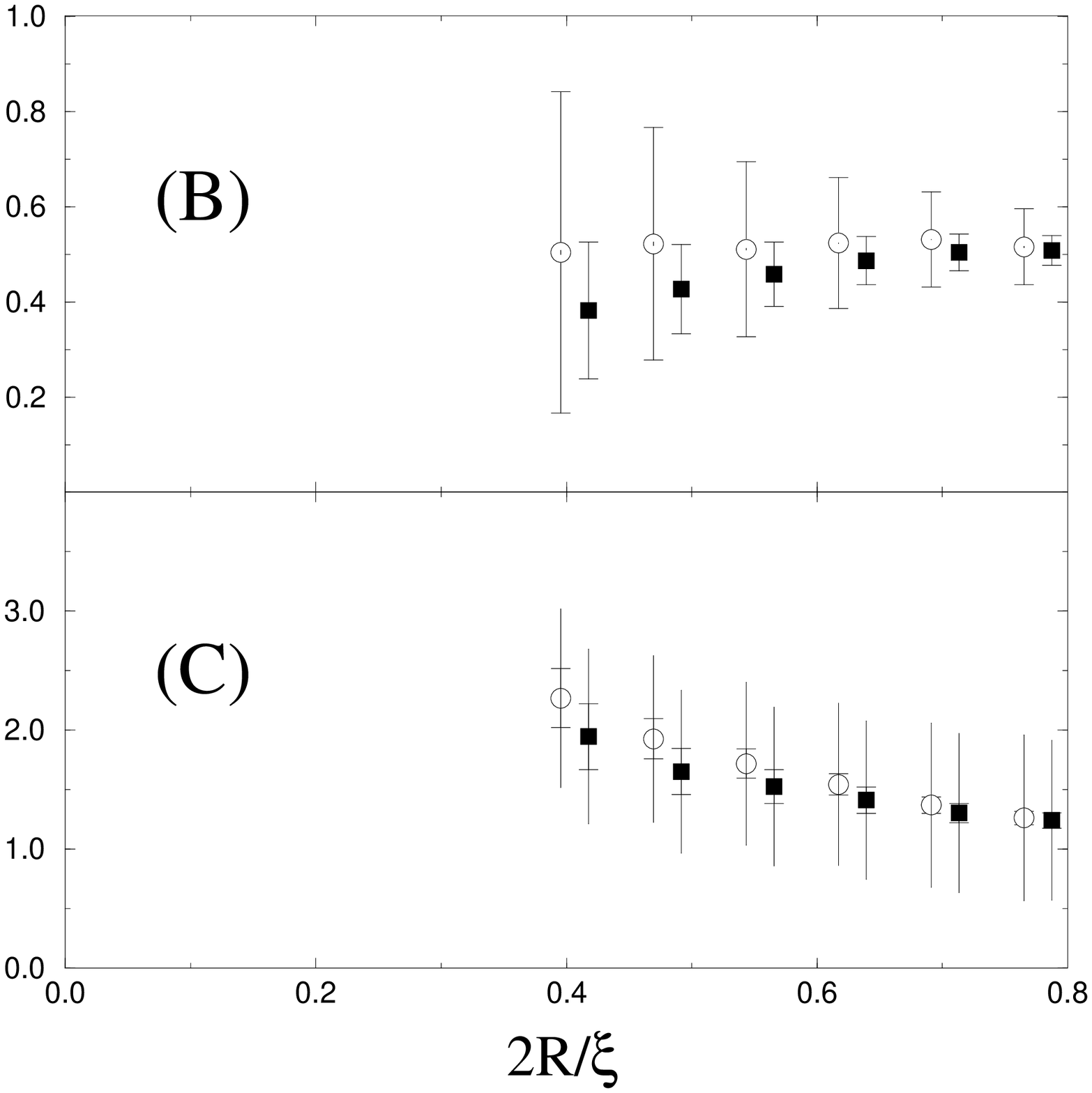,angle=-90,
width=0.6\linewidth}
\end{tabular}
\caption{
The best fitting parameters
$\widehat{\cal Z}''(\rho,R)$ (graphs (A) and (A')),
$\widehat{\cal M}(\rho,R)$ (graph (B)) and 
$\widehat{\cal N}(\rho,R)$ (graph (C)) corresponding to the fitting
form (\ref{Phenomenological}). These results have been obtained
on lattice \ref{Lattice256x512}. 
We use one-loop perturbation theory for the leading
Wilson coefficient in graphs (A), and two-loop perturbation theory in
(A'), (B) and (C).
Different symbols refer to different external momenta:
$\pb = 0$ (empty circles) or $\pb = 2\pi/L$ (filled squares).
The horizontal lines in (A) and (A') correspond to the theoretical prediction
${\cal Z}''(\rho,R) = \widehat{\cal Z}^*$, see Tab. \ref{ZFieldTable}.
}
\label{AntiSymmetricParameterBIS}
\end{figure}
%
%*********************************************************************
%
The discrepancy at larger values of $R$ can be attributed either to an
imperfect evaluation of systematic errors (an unlucky numerical coincidence
which makes them so small in this case), or to power-correction effects.
In order to better understand the problem, we tried to fit the numerical data
using the following ``phenomenological'' form: 
\begin{eqnarray}
2\sqrt{\pb^2+m^2}{\rm Re}\, \widehat{G}^{(1)}(t,x;\pb,\pb;2t_s) & = & 
2\sqrt{\pb^2+m^2}(N-1){\cal F}^{(1)}(\grz;\zeta)\, t\, {\cal Z}''+ 
\nonumber\\
&&+(\pb^2+m^2/4)\sqrt{\pb^2+m^2} \, t(t^2-3x^2)\, {\cal M}+\nonumber \\
&&+m^2 \sqrt{\pb^2+m^2} \, t(t^2+x^2)\,  {\cal N}\, .
\label{Phenomenological}
\end{eqnarray}
This fitting form is obtained as follows. We write down the
$O(r^3)$ terms of the OPE (\ref{OPEAsymm}), and single out the
Lorentz structure of the Wilson coefficients. The ``reduced''
Wilson coefficients depend logarithmically upon $r$ and are
rotationally invariant. We make the crude approximation 
of neglecting this logarithmic $r$ dependence.
Next, we single out the $\pb$ dependence of the $\<\pb|\cdot|\pb\>$
matrix elements of the composite operators of dimension 3.
The $\pb$ dependence can be easily deduced by using the space-time
symmetries. It depends uniquely upon the spin of the composite
operator. 
The result of this procedure has the form \ref{Phenomenological}.

The functions multiplying ${\cal M}$ and ${\cal N}$ have the same 
dimension and Lorentz structure as the Wilson coefficients
of the next-to-leading terms in the OPE (\ref{OPEAsymm}). The parameters 
${\cal M}$ and ${\cal N}$ correspond, respectively, to
dimension 3, spin 3 operators, and to dimension 3, spin 1 operators.

Equation (\ref{Phenomenological}) can be considered,
for what concerns power corrections, as a ``naive'' 
(i.e. not RG improved) tree-level approximation. 
Since power corrections in Eq. (\ref{Phenomenological})
do not have the correct $r\to 0$ limit, the parameters ${\cal M}$ and 
${\cal N}$ do not give access to well-defined matrix elements.
Anyway, by adopting the fitting form (\ref{Phenomenological}),
we gain some insight into the role of power corrections for the
determination of ${\cal Z}''$.

The results obtained on lattice \ref{Lattice256x512} are shown in
Fig. \ref{AntiSymmetricParameterBIS}. Notice that 
$\widehat{\cal Z}''(\rho, R)$ is much flatter than in 
Fig. \ref{AntiSymmetricParameterBIS}. The estimation at small $R$ does not
change much and is compatible with the one of Sec. \ref{FieldSection}.
For instance at $R=3.5$ on lattice \ref{Lattice256x512} we get 
$\widehat{\cal Z}''(\rho, R)= 0.459(11)[2]$ (here we quote the result at
$\pb=0$), cf. Tab. \ref{AntiSymmetricParameterTable}.
%
%******************************************************************
%
\subsection{OPE in the Symmetric Sector}
\label{SymmetricSection}

Our last example concerns the symmetric traceless product of two
elementary fields at short distances. 

We shall keep track of the $O(r^2)$ power corrections in the OPE.
As a consequence, we must take care of the non-trivial
mixing between dimension-2 symmetric traceless operators, see 
Sec. \ref{SymmetricZetaSection}.
For this task, it is convenient to adopt the basis of operators
with definite spin, see Eqs. (\ref{Q1Def})--(\ref{Q7Def}).

For sake of clarity  we shall restrict ourselves to the case of 
on-shell external states  of equal total momentum.
This allows two simplifications.
We can eliminate the operators $Q^{R(6)}$ and $Q^{R(7)}$ since they 
vanish on shell, see Eqs. (\ref{SymmetricEqMot1}) and 
(\ref{SymmetricEqMot1}). We can eliminate $Q^{R(4)}_{\mu\nu}$
and $Q^{R(5)}$, which are total space-time derivatives.
With these assumptions we get from Eq. (\ref{OPEFieldsSymmetric})
\begin{eqnarray}
\hspace{-1cm}\frac{1}{2}\sigma^{\{a}_{RGI}(x)\sigma^{b\}}_{RGI}(-x)-
\frac{\delta^{ab}}{N}\sg_{RGI}(x)\cdot\sg_{RGI}(-x) & = &
{\cal F}^{(2)}_{RGI,0}(\grz;\zeta)
\, \left[S_0^{ab}\right]_{RGI} +
\nonumber\\
&&\hspace{-2cm}+\sum_{i=1}^3{\cal F}^{(2)}_{RGI,i}(\grz;\zeta) x^{\mu}x^{\nu}
\left[Q^{(i)}_{\mu\nu}\right]_{RGI}\, ,
\label{OPESymm}
\end{eqnarray}
where we used, once more, renormalization-group invariant operators.
The coefficients ${\cal F}^{(2)}_{RGI,i}(\overline{g};1)$ for $i = 1,2,3$ 
are given in Eqs. (\ref{HTSpin21})-(\ref{HTSpin23})
for $N = 3$. The coefficient 
of the leading term can be calculated from Eq. (\ref{F20Pert}):
\begin{eqnarray}
{\cal F}^{(2)}_{RGI,0}(\overline{g};1) = 
\overline{g}^{1/(N-2)}\left[1-\frac{1}{2\pi(N-2)}\overline{g}+
\frac{N^2-4N+6}{16\pi^2(N-2)^2}\overline{g}^2\right]\, .
\end{eqnarray}

We shall be interested in the $\<\pb|\cdot|\pb\>$ 
matrix elements of Eq. (\ref{OPESymm}). 
Analogously to Sec. \ref{ScalarSection}, space-time symmetries 
constrain the matrix elements of the composite operators on the 
right-hand side of Eq. (\ref{OPESymm}).
Here we adopt the following parametrization:
\begin{eqnarray}
\frac{1}{N}\sum_{ab} \<\pb,a|\left[S_0^{ab}\right]_{RGI}
|\pb,b\> & = & {\cal A}_R\, ,
\label{FitParameter21}\\
\frac{1}{N}\sum_{ab} \<\pb,a|\left[Q^{(1)ab}_{\mu\nu}\right]_{RGI}|\pb,b\> 
& = & {\cal B}_R\, (p_{\mu}p_{\nu}-\frac{1}{2}\delta_{\mu\nu}p^2)\, ,\\
\frac{1}{N}\sum_{ab} \<\pb,a|\left[Q^{(3)ab}_{\mu\nu}\right]_{RGI}|\pb,b\> 
& = & {\cal C}_R\, (p_{\mu}p_{\nu}-\frac{1}{2}\delta_{\mu\nu}p^2)\, ,
\label{Q3CDefinition}\\
\frac{1}{N}\sum_{ab} \<\pb,a|\left[Q^{(2)ab}_{\mu\nu}\right]_{RGI}|\pb,b\> 
& = & {\cal D}_R\, p^2\delta_{\mu\nu}\, .
\label{FitParameter23}
\end{eqnarray}
The parameters ${\cal A}_R,\dots,{\cal D}_R$ are real numbers and
do not depend upon the external momentum $\pb$.

Let us make some remarks concerning the expansion 
(\ref{OPESymm}).
The operator $Q^{(2)}_{\mu\nu}=Q^{(2)}\delta_{\mu\nu}$
is a higher-twist operator and mixes under renormalization with the 
dimension-zero operator $S^{ab}_0$.
The warnings expressed in Sec. \ref{OperatorDefSection} 
apply also to this case\footnote{
In the analogous case of $(\partial\sg)^2$, see Sec.
\ref{ScalarSection}, it was possible 
to single out (at least in theory) the higher-twist contribution
by looking at the $\pb$ dependence, cf. footnote  \ref{FootSWave}
on page~\pageref{FootSWave}. In the present case no analogous trick is 
available.}.

Let us now consider the operators $Q^{(1)}_{\mu\nu}$ and $Q^{(3)}_{\mu\nu}$.
Since their canonical dimension is equal to their spin 
(dimension$=$spin$=2$), they are leading twists. 
As a consequence, they can be determined unambiguously from the expansion 
(\ref{OPESymm}).
This can be done analogously to what we explained in
Sec. \ref{ScalarSection} for $\widehat{T}_{\mu\nu}$,
cf. Eq. (\ref{Convolution}). 

There is, however, a practical difficulty in the determination
of $Q^{(1)}_{\mu\nu}$ and $Q^{(3)}_{\mu\nu}$. 
For sake of clarity we refer to the case $N=3$.
In this case ${\cal F}_{RGI,3}^{(2)}(\overline{g};1)$ (which is of order 
$\overline{g}^2$) is strongly suppressed with respect to 
${\cal F}_{RGI,1}^{(2)}(\overline{g};1)$ (of order 1). 
In order to disentangle the two contributions 
in Eq. (\ref{OPESymm}), we should compute 
${\cal F}_{RGI,1}^{(2)}(\overline{g};1)$ at least to order $\overline{g}^2$.
Since we have computed  ${\cal F}_{RGI,1}^{(2)}$ 
and ${\cal F}_{RGI,3}^{(2)}$ to one-loop  order,
we do not expect to obtain a good determination of
$Q^{(3)}_{\mu\nu}$. The best fitting ${\cal C}$ 
(see Eq. (\ref{Q3CDefinition})) will mimic
the higher-loop contributions in ${\cal F}_{RGI,1}^{(2)}(\overline{g};1)$. 

This difficulty is however quite different from the one described in
Sec. \ref{OperatorDefSection}. In the present case it would be 
``sufficient'' to push forward the perturbative calculation of 
${\cal F}_{RGI,1}^{(2)}(\overline{g};1)$, and
to perform numerical simulations at large enough correlation lengths,
in order to solve the problem.
 
%
%*********************************************
%
\begin{figure}
\begin{tabular}{cc}
\hspace{0.0cm}\vspace{-1.5cm}\\
\hspace{-2.5cm}
\epsfig{figure=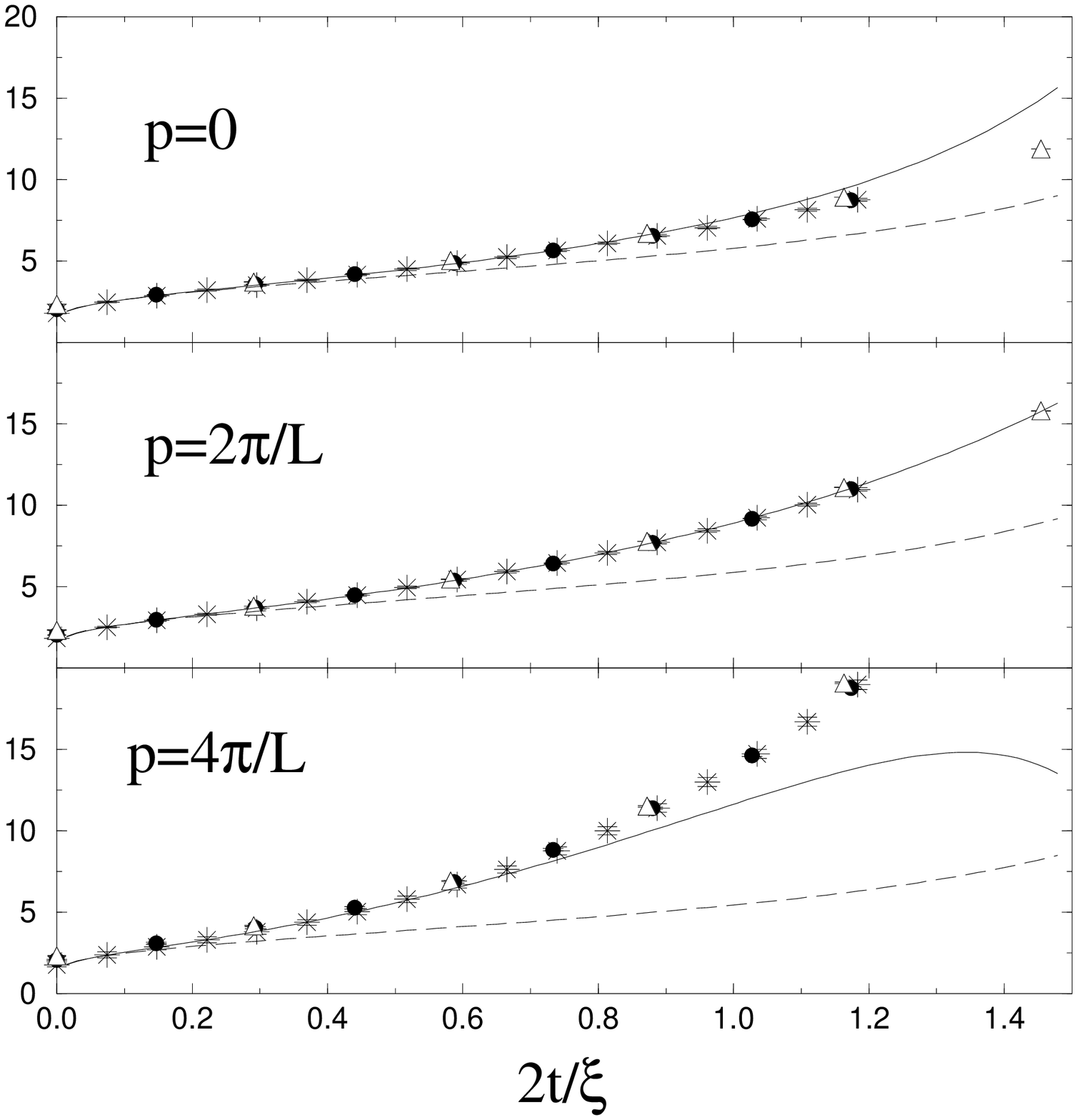,angle=-90,
width=0.6\linewidth}&
\hspace{-0.5cm}
\epsfig{figure=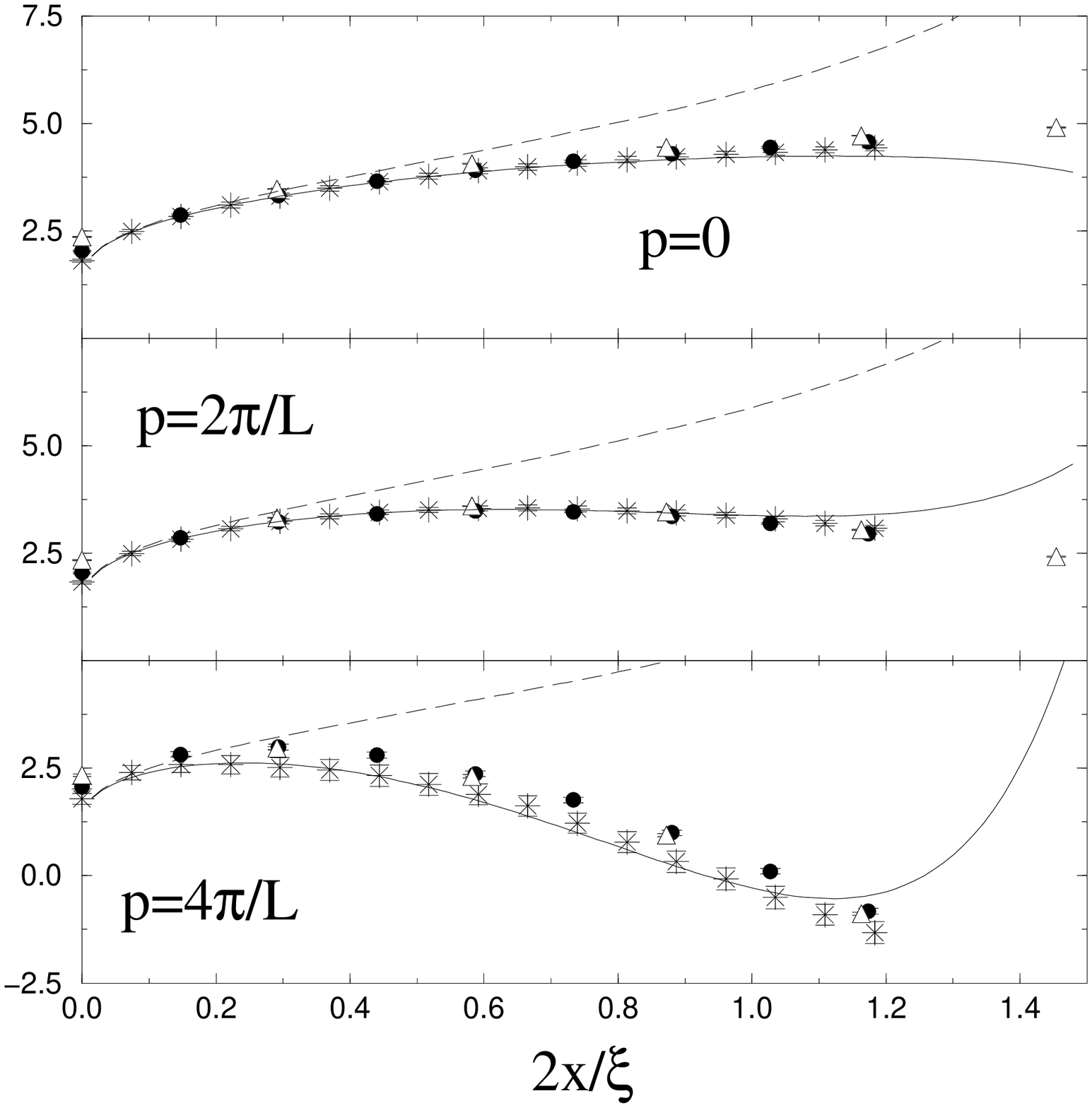,angle=-90,
width=0.6\linewidth}
\end{tabular}
\caption{Monte Carlo results and OPE fit for the product of fields 
in the symmetric sector.
We report $4\sqrt{\pb^2+m^2}Z_L^{-1}\widehat{G}^{(2)}(t,x; \pb;2t_s)$
on lattices \ref{Lattice64x128} (empty triangles, $\triangle$), 
\ref{Lattice128x256} (filled circles, $\bullet$) and 
\ref{Lattice256x512} (stars, $\ast$). We consider
$x=0$, $t\ne 0$ on the left, and $x\ne0$, $t= 0$ on the right.
The dashed curves are obtained with the leading term of the OPE, 
the continuous curves include power corrections.}
\label{SymmetricFit}
\end{figure}
%
%*********************************************
%

One-particle matrix elements have been extracted from the 
function  $\widehat{G}^{(2)}(t,x;\pb,\qb;2t_s)$ as explained
in Secs. \ref{ScalarSection} and \ref{AntisymmetricSection}.
The relation between one-particle matrix elements and
the $t_s\to \infty$ limit of $\widehat{G}^{(2)}(t,x;\pb,\qb;2t_s)$
is given by:
\begin{eqnarray}
\frac{1}{2}\sum_{ab}\<\pb,a|\sigma^{\{a}_{t,x}\sigma^{b\}}_{-t,-x}-
\frac{2\delta^{ab}}{N}\sg_{t,x}\cdot\sg_{-t,-x}|\qb,b\> =
N\sqrt{4\omega(\pb)\omega(\qb)}\lim_{t_s\to\infty}
\widehat{G}^{(2)}(t,x;\pb,\qb;2t_s)\, .\nonumber\\
\label{AsymptoticStates}
\end{eqnarray}

In Fig. \ref{SymmetricFit} we present the results of this computation.
We plot the function
$4\sqrt{\pb^2+m^2}Z_L^{-1}{\rm Re}\,\widehat{G}^{(2)}(t,x;\pb,\pb;2t_s)$ 
along the directions $(t,x)=(t,0)$ and $(t,x)=(0,x)$.
We used the non-perturbative estimates of $Z_L$ given in Sec. 
\ref{FieldSection}, see Tab. \ref{ZFieldTable}.
Together with the numerical data we show the best fitting curves. The form of
the fit is easily obtained from Eqs. (\ref{OPESymm}) and 
(\ref{FitParameter21})--(\ref{FitParameter23}):
\begin{eqnarray}
2\sqrt{\pb^2+m^2}\,{\rm Re}\,\widehat{G}^{(2)}(t,x;\pb,\pb;2t_s) &=&
{\cal F}^{(2)}_{RGI,0}(\grz;\zeta) {\cal A}-
{\cal F}^{(2)}_{RGI,2}(\grz;\zeta) m^2 r^2{\cal D}+\nonumber\\
&&\hspace{-1cm}
+{\cal F}^{(2)}_{RGI,1}(\grz;\zeta) (m^2/2+\pb^2)(x^2-t^2){\cal B}+
\label{FitForm2}\\
&&\hspace{-1cm}
+{\cal F}^{(2)}_{RGI,3}(\grz;\zeta) (m^2/2+\pb^2)(x^2-t^2){\cal C}\, .
\nonumber
\end{eqnarray}
The curves in Fig. \ref{SymmetricFit} are the best fitting curves
obtained on lattice \ref{Lattice256x512}.
For these curves we use a fitting window $\rho\le r\le R$,
with $\rho = 0.5$ and $R=8.5$.
As usual we try two types of fit: with (continuous line) 
and without (dashed line) power corrections. 

The fitting form which includes power corrections 
(i.e. the parameters ${\cal B}$, ${\cal C}$, ${\cal D}$,
see Eq. (\ref{FitForm2})) describes very 
well the numerical data for $2 r/\xi_{\rm exp}\ltapprox 1$. 
At distances $2 r/\xi_{\rm exp}\approx 1$ we expect both the OPE
and the perturbative expansion to break down.

The fit without power corrections (which amounts to dropping the
parameters ${\cal B}$, ${\cal C}$ and ${\cal D}$ in
Eq. (\ref{FitForm2})) correctly captures the small-$r$ 
behaviour of $\widehat{G}^{(2)}(t,x;$ $\pb,\pb;2t_s)$. 
The deviations from the numerical data become quite large as 
soon as $2 r/ \xi_{\rm exp}\gtapprox 0.3$ and depend upon the direction.
These deviations are mainly due to spin $2$ operators in 
the OPE (\ref{OPESymm}). Such terms are effectively averaged out when
considering a rotationally invariant fitting form
(like the one without power corrections).
As a consequence the best fitting value of ${\cal A}$
does not change very much whether or
not power correction terms are included in the fitting form.

%
%***************************************************
%
\begin{figure}
\begin{tabular}{cc}
\hspace{0.0cm}\vspace{-1.5cm}\\
\hspace{-2.5cm}
\epsfig{figure=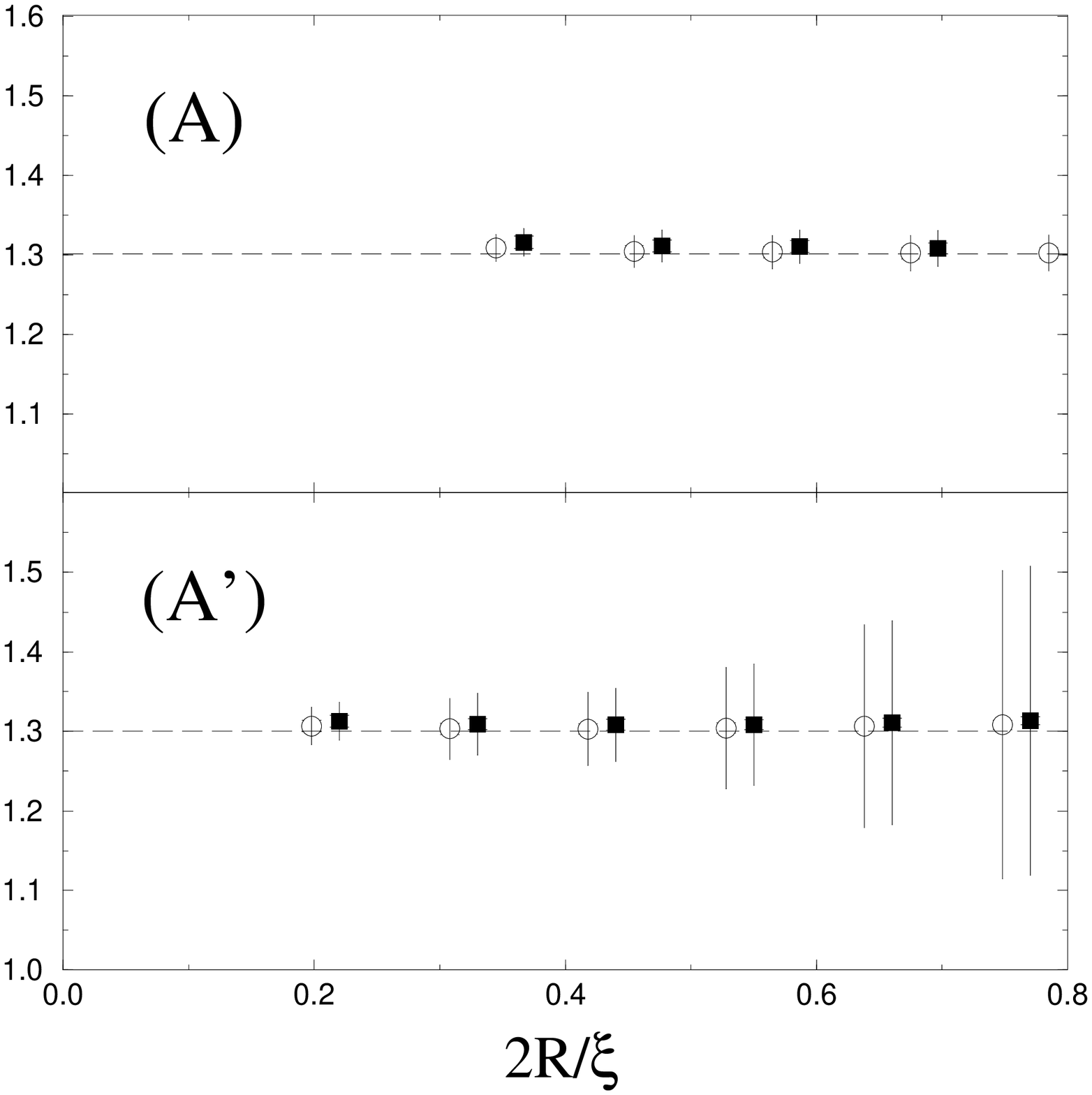,angle=-90,
width=0.6\linewidth}
&\hspace{-0.5cm}
\epsfig{figure=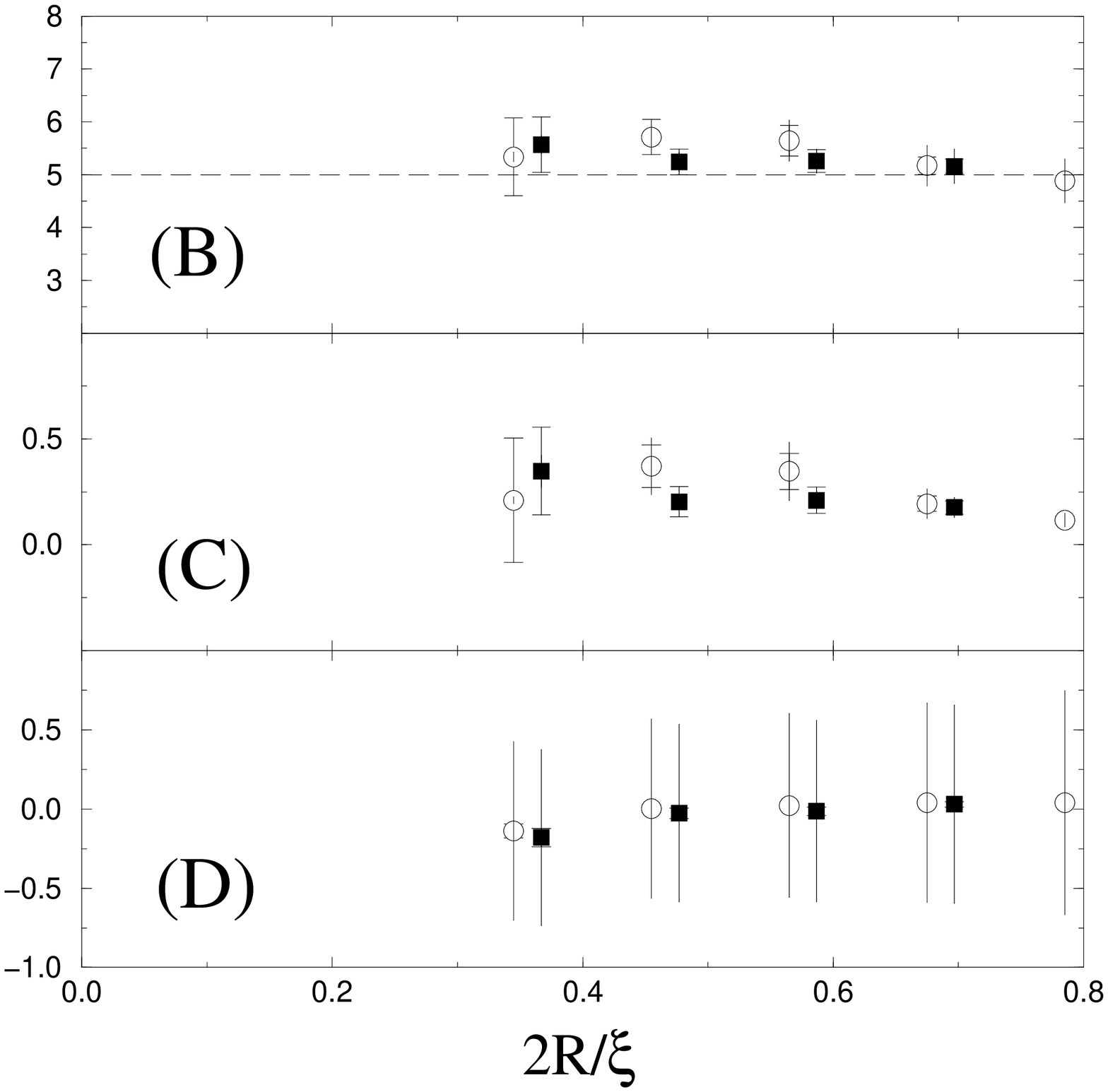,angle=-90,
width=0.6\linewidth}
\end{tabular}
\caption{The {\it renormalized} fitting parameters 
$\widehat{\cal A}_R(\rho,R)$ 
(graphs (A) and (A')), $\widehat{\cal B}_R(\rho,R)$ (graph (B)),
$\widehat{\cal C}_R(\rho,R)$ (graph (C)), and 
$\widehat{\cal D}_R(\rho,R)$ (graph (D)).
Empty circles and filled squares refer, respectively, to
$\pb = 0$ and $\pb = 2\pi/L$.
The dashed lines in graphs (A) and (A') (${\cal A}_R = 1.3$), and
in graph (B) (${\cal B}_R=5$) are a guide for the eye.
These data refer to lattice \ref{Lattice128x256}.}
\label{SymmetricParametersB}
\end{figure}
%
%***************************************************
%
\begin{figure}
\begin{tabular}{cc}
\hspace{0.0cm}\vspace{-1.5cm}\\
\hspace{-2.5cm}
\epsfig{figure=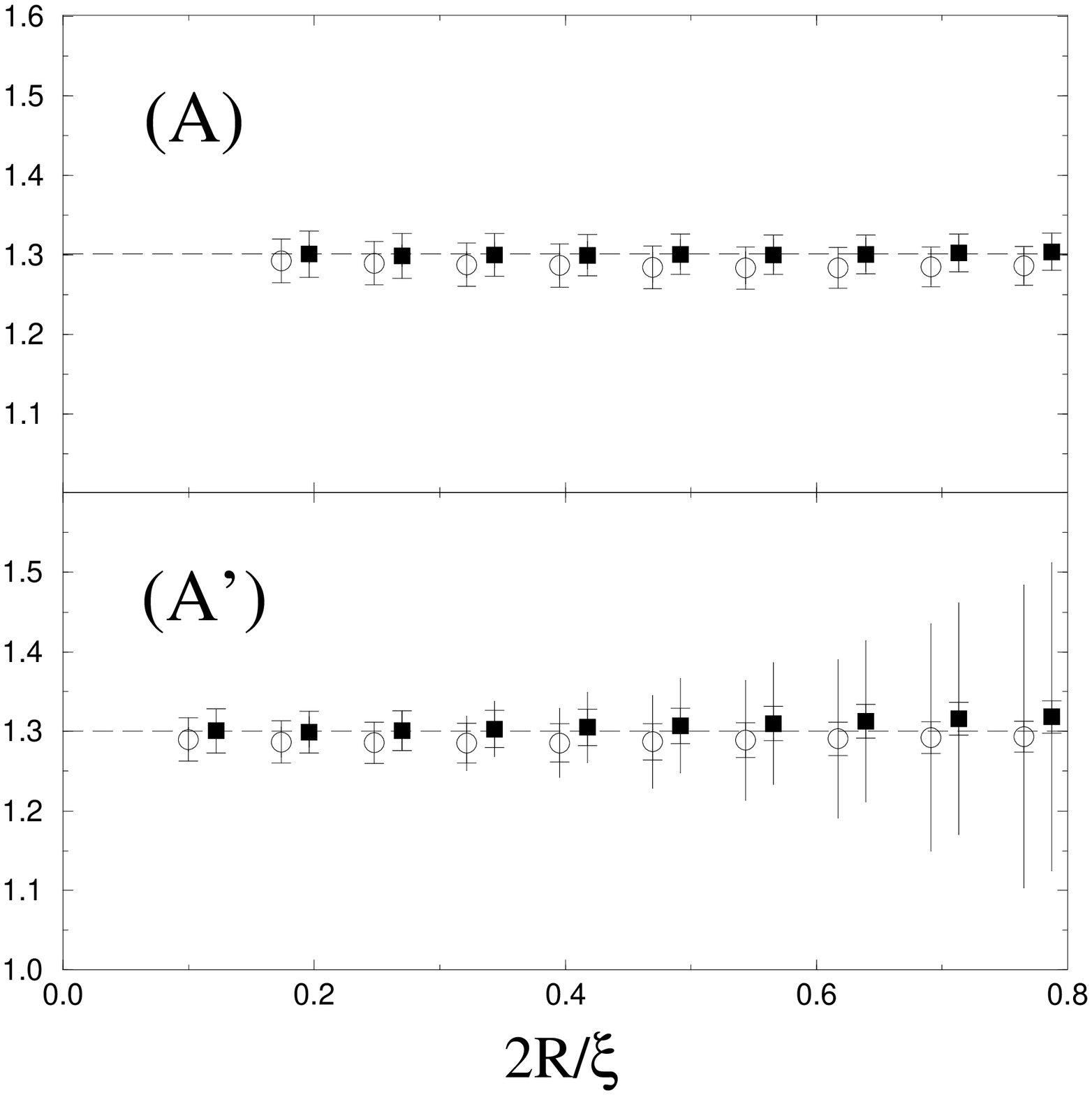,angle=-90,
width=0.6\linewidth}&
\hspace{-0.5cm}
\epsfig{figure=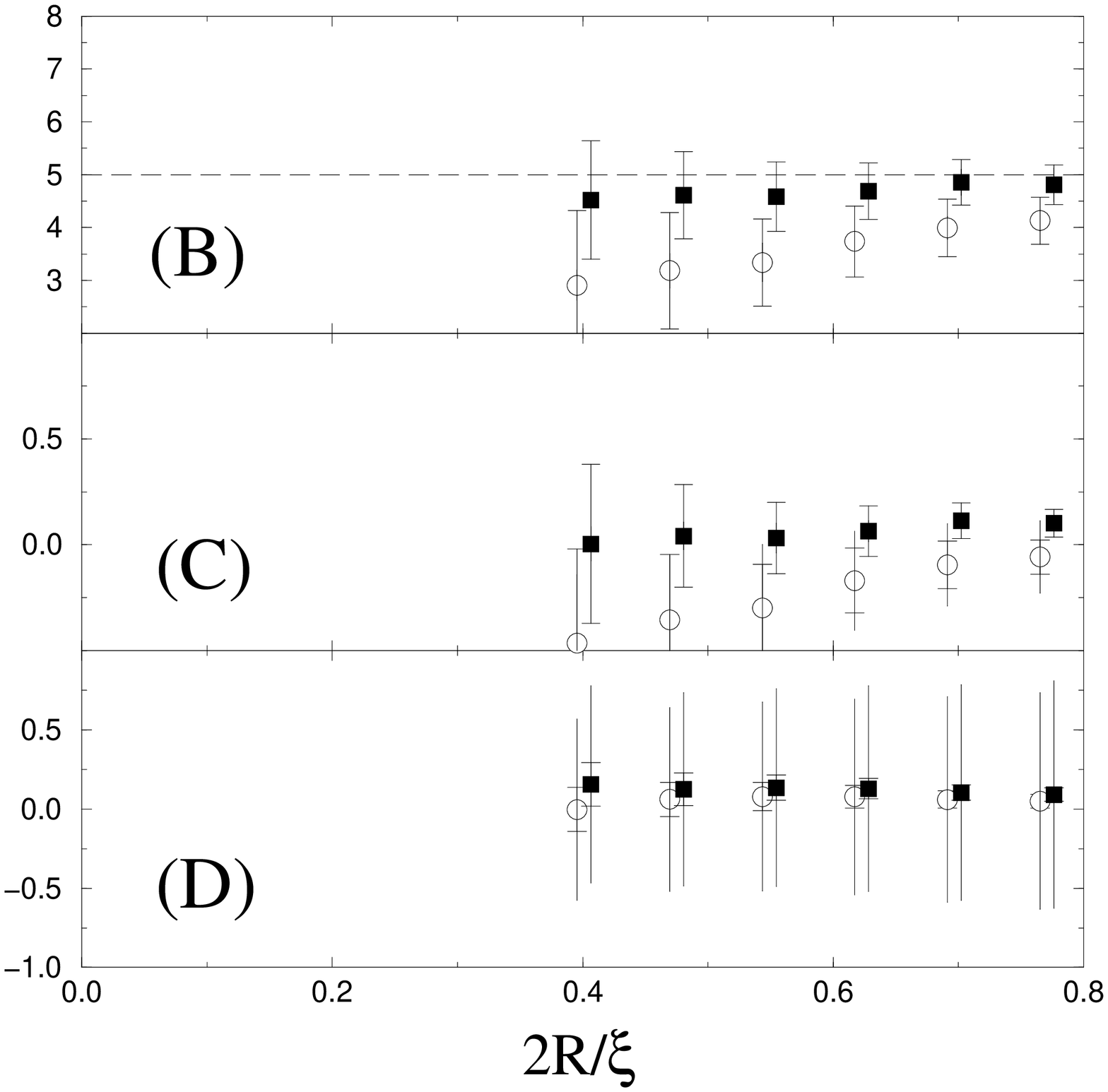,angle=-90,
width=0.6\linewidth}
\end{tabular}
\caption{The same as in Fig. \ref{SymmetricParametersB} on lattice
\ref{Lattice256x512}.}
\label{SymmetricParametersC}
\end{figure}
%
%***********************************************
%
In Figs. \ref{SymmetricParametersB} and \ref{SymmetricParametersC} 
we report the 
best fitting parameters as a function of the fitting window $\rho\le
r\le R$. We keep $\rho=0.5$ fixed and vary $R$. 
The kinematic scaling (independence of the parameters upon $\pb$) 
is verified within the statistical error.
The only exception is given by parameter ${\cal A}$ on 
lattice \ref{Lattice64x128}. However this effect is very small 
(about the $1.0\div 1.5\%$) and can be ascribed to scaling 
corrections\footnote{Notice that neither higher-order power
corrections, e.g., $O(r^4)$ terms, nor the approximate knowledge 
of the Wilson coefficients
can be the cause of this effect.}.

In the majority of the cases, statistical errors on the fitting
parameters are by far smaller than systematic ones. 
There are some exceptions to this rule.
Consider for instance parameters    
${\cal B}$ and ${\cal C}$ on lattices \ref{Lattice128x256} and 
\ref{Lattice256x512}, see Figs. \ref{SymmetricParametersB} 
and \ref{SymmetricParametersC}, graphs (B) and (C). 
As $R\to 0$ statistical errors become larger than systematic ones.
The reason of this fact is quite simple. 
At small $R$ the contribution of power corrections vanishes. The 
corresponding fitting parameters are fixed essentially by 
statistical fluctuations.

Let us now come to the cautious remarks concerning the evaluation
of $Q^{(2)}_{\mu\nu}$ and $Q^{(3)}_{\mu\nu}$ (i.e., respectively, the
parameters ${\cal D}$ and ${\cal C}$) formulated at the beginning of 
this Section. 

It is clear from Figs. 
\ref{SymmetricParametersB} and \ref{SymmetricParametersC},
graphs (D), that any sound estimate of ${\cal D}$ is hopeless. 
The central value $\widehat{\cal D}(r,R)$ (obtained with $\zeta
e^{\gamma}=1$) is approximatively 
zero (namely it is of order $10^{-1}\div 10^{-2}$), 
with systematic errors of order 1. This is true on all 
the lattices and for all the values of $R$ considered.
Let us suppose to repeat the calculation by fixing a particular
perturbative truncation of the Wilson coefficients.
We shall obtain, in general, values of $\widehat{\cal D}(\rho,R)$ of order one.
Nevertheless the results will depend strongly 
upon the chosen truncation. In our approach we can look at
$\widehat{\cal D}_{\zeta}(\rho,R)$, which depends 
depends upon $\zeta$. As $\zeta$ is varied the
neglected higher-loops contributions to 
${\cal F}^{(2)}_{RGI,0}(\grz;\zeta)$ vary. 
The value of $\widehat{\cal D}_{\zeta}(\rho,R)$ determined  by the fit 
mimics these higher-order terms. 

A subtle point in the computation of ${\cal D}$
is the following. Suppose to choose a particular value of $R/\xi_{\rm exp}$
and a particular perturbative truncation of the Wilson coefficients,
i.e., in our approach, a particular value of $\zeta$. Then compute 
$\widehat{\cal D}_{\zeta}(\rho,R)$ for several values of the bare coupling
$g_L$. This amounts to choosing several values of the lattice spacing.
It is not unlikely that the matrix elements of $Q^{(2)}_{\mu\nu}$ obtained
in this way will have (roughly) the correct scaling with the lattice spacing.
In other words one obtains $\widehat{\cal D}_{\zeta}(\rho,R)$ 
which is approximatively independent of $g_L$, instead of having
a power-like dependence upon the lattice spacing.
For example on lattices \ref{Lattice64x128}, \ref{Lattice128x256} and
\ref{Lattice256x512} at $\zeta = 1$ and $R/\xi_{\rm exp} = 0.367$
we get, respectively, $\widehat{\cal D} = -0.308(4)$, $-0.238(5)$ and 
$-0.189(17)$ (the quoted values refer to $\overline{p}=0$ and the corresponding
statistical error).
This fact could led to the conclusion that a genuine physical (continuum)
quantity has been estimated despite the theoretical warnings. The 
analysis in the previous paragraphs shows, however, that this estimate is 
unreliable.

Let us now consider the parameter ${\cal C}$, i.e. the operator 
$Q^{(3)}_{\mu\nu}$, whose $R$ dependence is shown in 
Figs. \ref{SymmetricParametersB}, \ref{SymmetricParametersC}, graphs (C).
In the general discussion above, we stressed that, for a sound 
estimate of ${\cal C}$, a two-loop calculation of the Wilson coefficient
${\cal F}^{(2)}_{RGI,1}$ is needed. 
Here, we shall ``guess'' the two-loop result. This will provide us
with an ``instructive'' estimate of ${\cal C}$.

From  Tab. \ref{CoefficientsTable} we learn that 
(if $\zeta e^{\gamma} = 1$ and 
$N=3$) one-loop coefficients are of order $10^{-1}$, while two-loop 
coefficients are of order $10^{-2}$. Let us suppose the same to be true for 
the unknown two-loop coefficient of  ${\cal F}^{(2)}_{RGI,1}$.
We can look at the values of $\widehat{\cal C}(\rho,R)$ presented
in Figs. \ref{SymmetricParametersB} and \ref{SymmetricParametersC}, 
graphs (C), as the sum of  two contributions: the genuine matrix element 
of $Q^{(3)}_{\mu\nu}$, and a spurious contribution coming
from the two-loop term of ${\cal F}^{(2)}_{RGI,1}$.
Using the value of ${\cal B}$ given below (${\cal B}\approx 3$),
we can estimate the spurious contribution to be about $10^{-2}\div 10^{-1}$.
We can now subtract this contribution from the best fitting values
$\widehat{\cal C}(\rho,R)$  reported in Figs.
\ref{SymmetricParametersB}, \ref{SymmetricParametersC}, 
graphs (C).  We obtain ${\cal C}$ of order $10^{-1}$: 
a conservative estimate is then ${\cal C}\ltapprox 1$.
It is interesting to notice that the systematic errors in
Figs. \ref{SymmetricParametersB}, \ref{SymmetricParametersC}, 
graphs (C) are about $0.2\div 0.5$ for intermediate values of
$2R/\xi$.
They correctly signal the effects due to the perturbative truncation of
the Wilson coefficients.

\begin{table}
\begin{tabular}{|c|c|c|c|c|c|c|}
\hline
 &\multicolumn{3}{|c|}{$\widehat{\cal A}^*$} 
&\multicolumn{3}{|c|}{$\widehat{\cal B}*$}\\
\hline
 &$\pb=0$ &$\pb=2\pi/L$ &$\pb=4\pi/L$ &$\pb=0$ &$\pb=2\pi/L$ &$\pb=4\pi/L$ \\
\hline
\hline
lattice \ref{Lattice64x128} & $0.91(0)[11]$ & $0.90(0)[11]$ & $0.90(1)[11]$ & 
$3.14(3)[29]$ & $3.16(2)[31]$ & $3.26(5)[36]$ \\
\hline
lattice \ref{Lattice128x256}& $0.73(0)[2]$ & $0.73(0)[2]$  & $0.74(1)[2]$ & 
$2.58(4)[35]$ & $2.53(4)[36]$ & $2.63(6)[37]$\\
\hline
lattice \ref{Lattice256x512}& $0.60(1)[1]$ & $0.61(1)[1]$ & $0.57(4)[1]$ & 
$2.1(1)[2]$& $2.2(1)[3]$& $2.4(2)[4]$
\\
\hline
\end{tabular}
\caption{The {\it unrenormalized} OPE estimates for the matrix
elements of  $\left[S_0\right]_{RGI}$ and $[Q^{(3)}_{\mu\nu}]_{RGI}$.}
\label{SymmetricParameters}
\end{table}  
We shall now consider the  leading operator  
$S_0$ (and the corresponding parameter ${\cal A}$).
In Figs. \ref{SymmetricParametersB} and \ref{SymmetricParametersC}, 
graphs (A), we report the values of  $\widehat{\cal A}(\rho,R)$
obtained with the fitting form (\ref{FitForm2}).
In the graphs (A') of the same figures, we 
repeat the fit using only the leading term
of the OPE, i.e. we drop out the terms ${\cal B}$, ${\cal C}$ and ${\cal D}$ 
in Eq. (\ref{FitForm2}).

The results for $\widehat{\cal A}(\rho,R)$ obtained including power
corrections, reported in graphs (A), are flat within statistical errors 
(approximatively $2\%$) as soon as $2R/ \xi_{\rm exp}\ltapprox 1$.
The estimates obtained without power corrections , see  graphs (A'), 
coincide with the previous ones within systematic errors. 
Systematic errors are strikingly different between graphs (A) and (A').
This phenomenon was already remarked in Sec. \ref{FieldSection}. 
Here we limit ourselves to underline a consequence of this fact.
The systematic error on the final evaluation of the parameter 
${\cal A}$ (i.e. on $\widehat{\cal A}^*$) strongly depends upon the 
lattice spacing, i.e. upon the bare coupling $g_L$.
For instance if we fix $R=3.5$ and $\pb=0$ we get
${\cal A} = 0.91[37]$, $0.73[4]$ and $0.60[1]$ respectively on lattices
\ref{Lattice64x128}, \ref{Lattice128x256} and \ref{Lattice256x512}.
Obviously systematic errors can be reduced on coarser lattices by taking 
smaller values of $R$, but one cannot reduce $R$ below the lattice 
spacing.

As we did in the previous Subsections, we estimate the
systematic error from the fit without power corrections.
Our final results for parameter ${\cal A}$ are given in Tab. 
\ref{SymmetricParameters}.
Here we used $R=2.5$ on lattices \ref{Lattice64x128} and \ref{Lattice128x256} 
while $R=3.5$ on lattice \ref{Lattice256x512}.

Finally we consider the evaluation of the parameter ${\cal B}$
(i.e. of the matrix element of $Q^{(3)}_{\mu\nu}$).
Here we cannot apply the same strategy as for ${\cal A}$,
that is taking smaller values of $R/\xi$ as $\xi$ gets larger 
(ideally $R/\xi \to 0$ and $R\to\infty$). 
As we explained above, if $R/\xi$ is small, the 
statistical error on ${\cal B}$ becomes large.
We shall keep $R/\xi_{\rm exp}$ roughly fixed.
Moreover  $R/\xi_{\rm exp}$ must be taken in the perturbative regime.
In Tab. \ref{SymmetricParameters} we present the corresponding estimates of
${\cal B}$ obtained with $R = 3.5$, $7$ and $13.5$, respectively on
lattices \ref{Lattice64x128},\ref{Lattice128x256} and \ref{Lattice256x512}.

\begin{table}
\centerline{
\begin{tabular}{|c|c|c|}
\hline
 &$\widehat{\cal A}_R^*$
&$\widehat{\cal B}_R^*$\\
\hline
\hline
lattice \ref{Lattice64x128} & $1.359(25)[157]$& $4.68(4)[44]$\\
\hline
lattice \ref{Lattice128x256}& $1.303(7)[39]$  & $4.62(7)[63]$\\
\hline
lattice \ref{Lattice256x512}& $1.286(26)[24]$ & $4.4(3)[4]$
\\
\hline
\end{tabular}
}
\caption{The {\it renormalized} fitting parameters
corresponding to the matrix elements of $\left[S_0\right]_{RGI}$
and $[Q^{(3)}_{\mu\nu}]_{RGI}$.}
\label{SymmetricParametersRenorm}
\end{table}  
We want to have an idea of the soundness of the systematic error bars
quoted in Tab.  \ref{SymmetricParameters}.
A possible check consists in looking at 
the results for the parameter ${\cal A}$ when the corresponding Wilson 
coefficient ${\cal F}^{(2)}_{RGI,0}(\grz;\zeta)$ 
is truncated to one-loop order.
Using the fitting form (\ref{FitForm2}) without power corrections,
$\pb = 0$, and the same values of $R$ as the ones used for Tab.
\ref{SymmetricParameters}, we obtain ${\cal A} = 0.95(0)[13]$, 
$0.75(0)[3]$ and $0.61(1)[2]$, respectively on lattices \ref{Lattice64x128}, 
\ref{Lattice128x256} and \ref{Lattice256x512}.

Finally, in Tab. \ref{SymmetricParametersRenorm}, we report the renormalized 
parameters corresponding to the matrix elements of 
$S_0$ and $Q^{(1)}_{\mu\nu}$ .
The bare values are taken from the $\pb = 0$ columns of
Tab. \ref{SymmetricParameters}. 
They have been renormalized using the field renormalization
constant computed in Sec. \ref{FieldSection}, cf. Tab. \ref{ZFieldTable}.
The results for $S_0$ can be compared with the ones
of Secs. \ref{SymmetricOperatorSection} and 
\ref{RenormalizationSymmetricSection}, 
reported in the rightmost column of Tab. \ref{ZOPTab}. 
We recall that in Secs.  \ref{SymmetricOperatorSection} and 
\ref{RenormalizationSymmetricSection}, we adopted a completely 
different method for the computation of the 
renormalized matrix elements of $S_0$.
The two calculations agree very well, yielding a strong consistency 
check of the OPE method.
%
%********************************************************************
%
\section{More Answers}
\label{SummaryFieldsSection}
The investigations described in this Chapter allow us to 
complete the partial conclusions of Sec. \ref{SummaryCurrentsSection}.

In this Chapter we considered short distance products of the form
$\sigma(x)\sigma(y)$. As in the previous Chapter, the leading behaviour
of this product was of the type $r^0$ (where $r=|x-y|$) in most 
of our examples (the only exception being the antisymmetric sector,
see Sec. \ref{AntisymmetricSection}). 
It seems that such a situation is more favorable and we shall focus
on it at first. Here we can make tighter statements than 
in Sec. \ref{SummaryCurrentsSection}, thanks to the smaller statistical
errors (a fraction of percent) reached in this Chapter. 
\begin{itemize}
\item[1)] Corrections to scaling determine the ultimate accuracy
achievable at a given value of the bare coupling. 
The product of two elementary fields scales quite well,
as long as the field operators are kept on different lattice sites. 
We were not able to reveal unambiguously corrections to scaling on
these observables. We obtained some indications of scaling
corrections (at the level of $1\%$) on lattice \64x128bis,  
at $\xi_{\rm exp} = 6.831(2)$.
\item[2)] Higher-twist operators, i.e. operators which
require power subtractions to be renormalized, cannot be determined
from the OPE (at least in our simple approach).
There are strong theoretical reasons pointing to this conclusion,
see Sec. \ref{OperatorDefSection}. The results obtained with the OPE 
method give a concrete support to these theoretical reasons.
\item[3)] Leading-twist operator can be determined (at least in
theory) from the OPE. They can be classified according to their
canonical dimension. Those with the lowest dimension yield the
leading contribution to the OPE. In all our examples there 
was a unique leading operator. We were able to determine 
its matrix elements with a few percent accuracy.
We used (next-to-)${}^2$leading-log Wilson coefficients and products 
$\sigma(x)\sigma(y)$ at distances $r/\xi\sim 0.2\div 0.3$ ($r=|x-y|$).
\item[4)] Higher-dimensional (leading-twist) operators 
appear as power corrections in the OPE.
They can be determined from the OPE, but this task presents some 
technical difficulties.
As the dimension of the operators increases, their mixings become
more and more complicated. Moreover, at small distances (where
perturbation theory is well behaved), their contribution 
to the OPE decreases quickly and statistical noise hides it.
\item[5)] Including or not power corrections in the OPE
fitting form seems not to be an important issue. 
\item[6)] As we already said in Sec. \ref{SummaryCurrentsSection},
the main problem is related to the need for asymptotic scaling.
Usually asymptotic scaling is judged with respect statistical errors.
If the perturbative prediction lies within statistical error bars, 
asymptotic scaling is considered to be reached. 
This point of view is not completely satisfactory. Among the other
things it heavily depends upon the statistical accuracy.
We proposed to estimate the convergence of perturbation theory by
assigning a systematic error to it.
There exists (to date) no proved good way to assign this systematic error.
However, our ``empirical'' definition was in rough agreement with all
our observations.
\end{itemize}
The example studied in Sec. \ref{AntisymmetricSection}, i.e. the 
antisymmetric product of two elementary fields, is somehow an
exception to these remarks. This is perhaps related to the fact that
the leading term of the OPE is, in this case, of order $r$, instead of
$r^0$. In this case we were not able to find a scaling window 
without adding power corrections to the OPE fitting form.
Taking care of power corrections at tree level allowed to improve 
the scaling behaviour. The final result was in good agreement
with an alternative calculation. Nevertheless the situation
is not fully satisfactory for what concerns this example.

\chapter*{Conclusions and Perspectives}
\addcontentsline{toc}{chapter}{\numberline{}Conclusions and perspectives.}

Renormalization of lattice composite operators is a central problem 
for applications of lattice QCD. The structure of hadrons can
be, for many purposes, encoded in matrix elements, which cannot be 
computed in perturbation theory. Lattice
computations are a non-perturbative, widely applicable tool for such problems.
Perturbation theory seems not to be the good method for translating the 
lattice results in the continuum language. This leads to the problem of 
designing well-chosen non-perturbative renormalization methods.

The authors of Ref. \cite{Dawson:1997ic} 
proposed to define renormalized composite operators by 
``splitting'' them into simpler operators (e.g., conserved currents).
Operator Product Expansion must then be used for recovering the operator
we were interested in.
This is essentially an ``infinite-volume scheme'' 
(see Chapter \ref{Introduction}) and has the disadvantage that many 
scales must be  separated on the same lattice. However it has some advantages.
\begin{itemize}
\item Renormalized operators are obtained in a massless scheme 
without any extrapolation to the chiral limit.
\item It is more direct: renormalization and computation of matrix elements 
are accomplished in the same step.
\item A simpler approach to operator mixing is possible.
In particular ``lattice-induced'' mixings can be disregarded since 
unrenormalized lattice operators are never used. 
\end{itemize}
Moreover the idea itself is quite appealing.
It is however far from obvious that it can applied in practice.

We completed a detailed feasibility study of this method on a simple 
two-dimensional model which can be simulated with very fast algorithms.
We gave here a thorough account of this work. The main result
is that the new method outlined above and in Section 
\ref{RenormalizationViaOPE} really works. We refer to
Secs. \ref{SummaryCurrentsSection} and \ref{SummaryFieldsSection}
for some technical highlights.

Here we recall the principal suggestions which come out from this work
and could be of help in a QCD application of the new method:
\begin{itemize}
\item The model studied in this thesis is affected by $O(a^2)$
lattice artifacts. The same type of corrections to scaling occurs
in $O(a)$-improved QCD.
We found the systematic errors due to these effects to be
under control even on quite coarse lattices. 
The largest lattice artifacts occur in fact at contact points,
and can be easily avoided in the OPE approach. 
On the coarsest lattice 
considered, the correlation length (inverse mass gap) was $\xi\approx
7a$.
\item The principal source of systematic error is the perturbative
truncation of the Wilson coefficients. We resummed the Wilson
coefficients using RG up to next-to-next-to-leading-log order.
We estimated the systematic error by varying the resummation
procedure and the perturbative order. Moreover, we recomputed the same 
renormalized matrix elements using different approaches.
The various estimates of the systematic error were roughly consistent.
The achievable precision depends upon the lattice spacing,
which control the shortest distance that can be reached on the
lattice considered.
\item Obviously, it is much simpler to estimate the leading operator
of the OPE. The best situation occur when
the corresponding Wilson coefficient behaves
logarithmically in the short distance limit.
\end{itemize}

Let us conclude by listing a few lines for further investigation:
\begin{itemize}
\item In QCD chiral symmetries are broken both spontaneously,
and ``softly'' by the quark-mass terms. These breakings
manifest themselves as power corrections in the OPE.
The role of these power corrections deserves some accurate investigation.
\item What is the effect of improvement? One of the advantages of the method 
exposed here is that improvement of composite operators is
straightforward. 
It would 
be interesting to study the efficiency of the method with an improved action 
and improved operators. 
\item The principal feature of the method studied in this thesis is
that it allows to employ continuum symmetries, rather than lattice
ones.
This is an advantage both on the ``standard'' infinite-volume methods, and
on the finite-volume ones.
Therefore it would be very interesting to find a finite-volume version of the
OPE method.
\end{itemize}

\bibliography{sigma,ope,qft}

\end{document}